%% file: EXO-23-016_temp.tex
\pdfoutput=1
\documentclass[11pt,twoside,a4paper,cmspaper,final,collab]{cms-tdr}
\def\svnVersion{1d8d4b3}\def\svnDate{2026/09/23}\def\cmsCernNoTag{CERN-EP-2025-282}\def\cmsCernDate{\today}\def\cmsMessage{Published in Physics Reports as \href{http://dx.doi.org/10.1016/j.physrep.2026.07.007}{\doi{10.1016/j.physrep.2026.07.007}.}}
\begin{document}\cmsNoteHeader{EXO-23-016}

\ifthenelse{\boolean{cms@external}}{\providecommand{\cmsLeft}{left\xspace}}{\providecommand{\cmsLeft}{left\xspace}}
\ifthenelse{\boolean{cms@external}}{\providecommand{\cmsRight}{right\xspace}}{\providecommand{\cmsRight}{right\xspace}}
\ifthenelse{\boolean{cms@external}}{\providecommand{\aref}[1]{\ref{#1}}}{\providecommand{\aref}[1]{Appendix~\ref{#1}}}

\newcommand{\PASGt}{{\HepAntiParticle{\PSGt}{}{}}\xspace}
\newcommand{\PS}{\ensuremath{\text{S}}\xspace}
\newcommand{\PY}{\ensuremath{\text{Y}}\xspace}
\newcommand{\PZD}{\ensuremath{\PZ_\text{D}}\xspace}
\newcommand{\Pchipm}{\PSGcpmDo}
\newcommand{\Pchiz}{\PSGczDo}
\newcommand{\Hdark}{\ensuremath{\PH_{\text{D}}}\xspace}
\newcommand{\mH}{\ensuremath{m_{\PH}}\xspace}
\newcommand{\mS}{\ensuremath{m_{\PS}}\xspace}
\newcommand{\mX}{\ensuremath{m_{\PX}}\xspace}
\newcommand{\WtoLNu}{\ensuremath{\PW \to \Pell\PGn}\xspace}
\newcommand{\PARchizero}{\ensuremath{\chi^{0}}\xspace}
\newcommand{\PARchipm}{\ensuremath{\chi^{\pm}}\xspace}
\newcommand{\abseta}{\ensuremath{\abs{\eta}}\xspace}
\newcommand{\dzero}{\ensuremath{d_{\text{0}}}\xspace}
\newcommand{\dz}{\ensuremath{d_{\text{z}}}\xspace}
\newcommand{\minpt}{\ensuremath{\text{min}(\pt)}\xspace}
\newcommand{\maxpt}{\ensuremath{\text{max}(\pt)}\xspace}
\newcommand{\mindzero}{\ensuremath{\text{min}(\dzero)}\xspace}
\newcommand{\mZD}{\ensuremath{m_{\PZD}}\xspace}
\newcommand{\cTau}{\ensuremath{c\tau}\xspace}
\newcommand{\cTauPS}{\ensuremath{\cTau_{\PS}}\xspace}
\newcommand{\mMuMu}{\ensuremath{m_{\PGm\PGm}}\xspace}
\newcommand{\Lxy}{\ensuremath{L_{\text{xy}}}\xspace}
\newcommand{\ireltrk}{\ensuremath{I}\xspace}
\newcommand{\Ecalo}{\ensuremath{E_{\text{calo}}}\xspace}
\newcommand{\Ndof}{\ensuremath{N_{\text{dof}}}\xspace}
\newcommand{\Nhits}{\ensuremath{N_{\text{hits}}}\xspace}
\newlength\cmsGapLength
\ifthenelse{\boolean{cms@external}}{\setlength\cmsGapLength{-0.30em}}{\setlength\cmsGapLength{-0.45em}}
\newcommand{\metNoMu}{\ensuremath{\pt^{\text{miss},\,\PGm\hspace{\cmsGapLength}/}}\xspace}
\newcommand{\instL}[1]{\ensuremath{#1\times10^{34}\,\text{cm}^{-2} \text{s}^{-1}}\xspace}
\newcommand{\PATATRACK}{\textsc{PataTrack}\xspace}
\newlength\cmsTabTinySkip\setlength{\cmsTabTinySkip}{0.5ex}
\newlength\cmsTabTinySkipInverse\setlength{\cmsTabTinySkipInverse}{\dimexpr\baselineskip - \cmsTabTinySkip\relax}
\newlength\cmsTabLittleSkip\setlength{\cmsTabLittleSkip}{1ex}
\newlength\cmsTabSkip\setlength{\cmsTabSkip}{3ex}
\newlength\cmsTabBigSkip\setlength{\cmsTabBigSkip}{5ex}
\newlength\cmsTabBiggerSkip\setlength{\cmsTabBiggerSkip}{6ex}
\newlength\cmsTabBiggestSkip\setlength{\cmsTabBiggestSkip}{8ex}
\providecommand{\cmsTable}[1]{\resizebox{\textwidth}{!}{#1}}
\newcommand{\cen}[1]{\multicolumn{1}{c}{#1}}

\cmsNoteHeader{EXO-23-016}
\title{Strategy and performance of the CMS long-lived particle trigger program in proton-proton collisions at \texorpdfstring{$\sqrt{s}=13.6\TeV$}{sqrt(s) = 13.6 TeV}}

\author*[cern]{The CMS Collaboration}

\date{\today}

\abstract{
In the physics program of the CMS experiment during the CERN LHC Run~3, which started in 2022, the triggers targeting beyond-the-standard-model long-lived particles have been improved and extended to expand the scope of the corresponding searches. These dedicated triggers and their performance are described in this paper, using several theoretical benchmark models that extend the standard model of particle physics. The results are based on proton-proton collision data collected with the CMS detector during 2022--2024 at a center-of-mass energy of 13.6\TeV, corresponding to integrated luminosities of up to 123\fbinv.
}

\hypersetup{
pdfauthor={CMS Collaboration},
pdftitle={Strategy and performance of CMS long-lived particle triggers in proton-proton collisions at sqrt(s) = 13.6 TeV},
pdfsubject={CMS},
pdfkeywords={CMS, Trigger, LLP, Long-lived particles, performance, Run 3}}

\maketitle 

\setcounter{tocdepth}{3}
\makeatletter
\renewcommand{\l@subsubsection}{\@dottedtocline{3}{7.3em}{3.5em}}
\makeatother
\tableofcontents

\newpage

\section{Introduction}
\label{sec:intro}

While the standard model (SM) of particle physics is a highly precise and predictive theory, it does not explain all observed phenomena, such as the nature of dark matter, the matter-antimatter asymmetry, and neutrino masses. The SM also suffers from several theoretical shortcomings, \eg, the hierarchy problem, the large number of arbitrary parameters, and the lack of an explicit reason leading to three generations of fermions. In addition, the SM does not include gravity. Thus, beyond-the-SM (BSM) phenomena are expected to exist and could manifest at energies accessible by the CERN LHC, although they have eluded detection so far.

Particles with macroscopically long lifetimes are an important possibility in the search for new phenomena and often appear in BSM theories, notably in models that describe the elementary particle nature of dark matter. Several physical mechanisms give rise to relatively long particle lifetimes in the SM, including approximate symmetries, small couplings, near-degenerate states, and heavy mediators~\cite{ParticleDataGroup:2024cfk}; such mechanisms are also present across a broad range of BSM theories~\cite{Alimena:2019zri,Lee:2018pag}. Long-lived particles (LLPs) are specifically motivated in many of these models~\cite{Curtin:2018mvb}. In the last several years, LLPs have become a major focus in the search for new physics at the LHC.

If LLPs are produced inside the CMS detector at the LHC, they may decay at a measurable distance from the primary proton-proton ($\Pp\Pp$) interaction point (IP) or may completely pass through the detector before decaying, in contrast to promptly decaying particles that result in products consistent with an origin at the primary $\Pp\Pp$ interaction. A number of signatures and decay topologies are possible, depending on the characteristics of the LLP. For example, neutral LLPs could travel a significant distance through the detector before decaying into displaced leptons, photons, or jets~\cite{ATLAS:2019fwx,ATLAS:2019kpx,LHCb:2019vmc,CMS:2025rtd,CMS:2021juv,CMS:2023arc,displacedJets2022Data,EXO-23-014,CMS:2019zxa}. Heavy, slow-moving LLPs could decay into particles with a measurable delay relative to the timing of the primary $\Pp\Pp$ interaction~\cite{CMS:2019qjk,CMS:2017kku,ATLAS:2021mdj}. A charged LLP could decay within the CMS tracker into a neutral particle and a charged particle whose momentum is too small for its corresponding track to be reconstructed, producing a ``disappearing'' track~\cite{CMS:2020atg,ATLAS:2017oal}, or into one detectable and one undetectable particle, producing a kinked track. Finally, magnetic monopoles could leave highly ionizing tracks in the detector because of their magnetic charge, and heavy stable charged particles could do so because of their electric charge, relatively low velocity, or both~\cite{ATLAS:2019wkg,CMS:2024nhn}.

\begin{sloppypar}Standard algorithms developed by the LHC experiments to trigger on and reconstruct ``physics objects'', such as leptons, photons, and jets, are usually inadequate for LLP searches because they are designed for promptly decaying particles, thus alternative techniques are often needed. The CMS trigger system is designed to collect data quickly, and therefore, the trigger algorithms generally assume promptly decaying particles because displaced object reconstruction is often more computationally intensive. For example, algorithms that reconstruct displaced tracks can have long processing times because of the high combinatorial complexity of finding tracks whose helical trajectories are not constrained to intersect the beam axis~\cite{CMS:2014pgm}.\end{sloppypar}

Many improvements and extensions for triggering on LLPs with the CMS detector were implemented before or during the LHC Run~3, which started in 2022 and is expected to end in 2026. In this paper, we present the CMS Run~3 LLP trigger program and the performance of these custom triggers, using $\Pp\Pp$ collision data corresponding to integrated luminosities of up to 123\fbinv, recorded at $\sqrt{s}=13.6\TeV$ and collected during 2022--2024. We focus on new LLP triggers introduced for Run~3 data taking, but also describe LLP triggers introduced earlier that have been maintained or improved since.

Dedicated trigger and reconstruction strategies for LLP signatures have also been developed by other LHC experiments. The ATLAS experiment has used specialized techniques for displaced tracking, disappearing tracks, hadronic decays in the calorimeters, and decays in the muon spectrometer, including large-radius tracking and ``CalRatio'' triggers~\cite{ATLAS:2024xna,ATLAS:2023nze,Mlinarevic:2024ntj,ATLAS:2019qrr}. The LHCb experiment has pursued a complementary approach based on a fully software trigger and real-time reconstruction, exploiting its precise vertexing capabilities for displaced signatures~\cite{LHCb:2018zdd,Kholoimov:2025cqe}. The CMS program described in this paper follows the same broad experimental goal of recovering signatures for which prompt-object triggers are inefficient, while using detector-specific information from the CMS tracker, calorimeters, muon system, and trigger architecture.

The paper is organized as follows: in Section~\ref{sec:detector}, we introduce the CMS detector, and in Section~\ref{sec:triggerSystem}, we briefly review the CMS trigger system in general. Section~\ref{sec:eventReconstruction} describes the standard ``online'' (trigger-level) and ``offline'' (after the data are recorded) event reconstruction. In Section~\ref{sec:dataMC}, we detail the data and simulations used. We describe the custom LLP trigger algorithms in Section~\ref{sec:LLPtriggers} and present the efficiency of each trigger in data and simulation. In Section~\ref{sec:trigPerf}, we compare the performance of the different LLP triggers, showing their complementary acceptance for benchmark LLP signals, in different fiducial regions of the detector. We provide an outlook of CMS LLP triggers at the future High-Luminosity LHC (HL-LHC) in Section~\ref{sec:HL_LHC}. The paper is summarized in Section~\ref{sec:summary}, and a glossary is provided in \aref{sec:glossary}. Tabulated results are provided in the HEPData record for this paper~\cite{hepdata}.

\section{The CMS detector}
\label{sec:detector}

\begin{figure}[!h]
  \centering
  \includegraphics[width=0.8\textwidth]{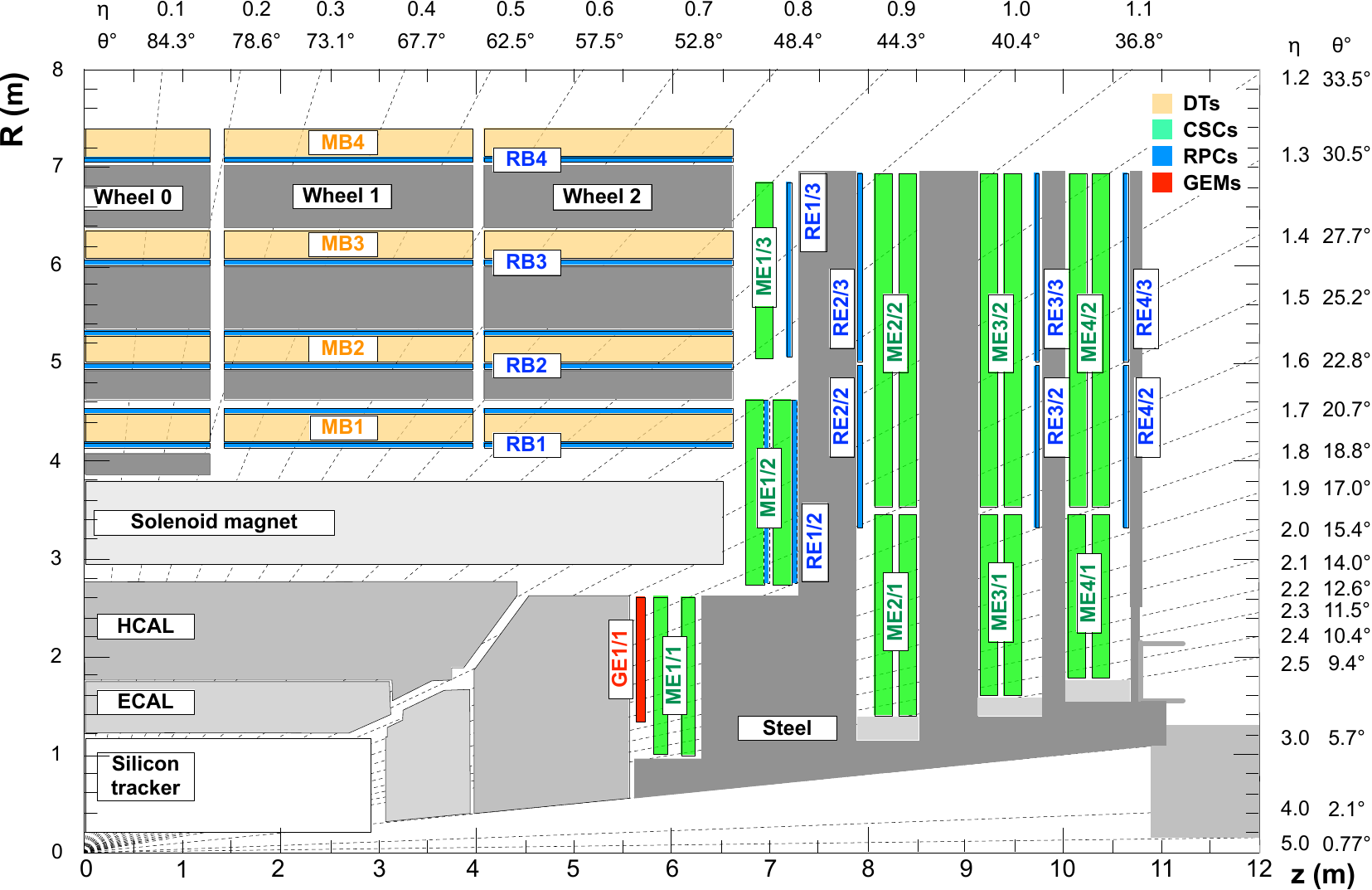}
  \caption{Schematic view in the $R$-$z$ plane of a CMS detector quadrant at the start of Run~3, with the axis parallel to the beam ($z$) running horizontally and the radius ($R$) increasing upward. The nominal interaction point (IP) is in the lower left corner. The locations of the various muon detectors and the steel flux-return yoke (dark areas) are shown, along with the silicon tracker, the electromagnetic calorimeter (ECAL), and the hadronic calorimeter (HCAL). The locations of the various muon detectors are shown in color: drift tubes (DTs) with labels MB, cathode strip chambers (CSCs) with labels ME, resistive plate chambers (RPCs) with labels RB and RE, and gas electron multipliers (GEMs) with labels GE. The M denotes muon, B stands for barrel, and E for endcap. Figure taken from Ref.~\cite{CMS:2023gfb}.}
  \label{fig:cms}
\end{figure}

The CMS apparatus~\cite{CMS:2008xjf,CMS:2023gfb} is a multipurpose, nearly hermetic detector, designed to trigger on~\cite{CMS:2020cmk,CMS:2016ngn,CMS:2024psu} and identify electrons, muons, photons, and (charged and neutral) hadrons~\cite{CMS:2020uim,CMS:2018rym,CMS:2014pgm}. Its central feature is a superconducting solenoid of 6\unit{m} internal diameter, providing a magnetic field of 3.8\unit{T}. Within the solenoid volume are a silicon pixel and strip tracker, a lead tungstate crystal electromagnetic calorimeter (ECAL), and a brass and scintillator hadron calorimeter (HCAL), each composed of a barrel and two endcap sections. Forward calorimeters extend the $\eta$ coverage provided by the barrel and endcap detectors. Muons are reconstructed using gas-ionization detectors embedded in the steel flux-return yoke outside the solenoid. More detailed descriptions of the CMS detector, together with a definition of the coordinate system used and the relevant kinematic variables, can be found in Refs.~\cite{CMS:2008xjf,CMS:2023gfb}. A quadrant of the CMS detector is shown in Fig.~\ref{fig:cms}.

Between the beginning of the LHC operation in 2009 and the start of Run~3 data taking in 2022, the CMS experiment underwent several changes and upgrades to enable the experiment to operate in conditions more challenging than assumed in its original design~\cite{CMS:2023gfb}. During this period, the LHC beam conditions evolved to provide increased luminosity, with several impactful effects. These include an increase in the average ``pileup'' (PU), the number of simultaneous $\Pp\Pp$ collisions per bunch crossing, from 10 in 2011, to 27 in 2016, and to 46 in 2022~\cite{CMS-LUM-17-003,CMS-PAS-LUM-17-004,CMS-PAS-LUM-18-002,CMS:LUM-22-001}, along with a corresponding increase in accumulated radiation dose.

The silicon tracker measures charged particles with $\abseta < 3.0$. During the LHC running period when the data used in this paper were recorded, the silicon tracker consisted of 1856 silicon pixel and 15148 silicon strip detector modules. The pixel detector was upgraded with the installation of a new detector in early 2017~\cite{Phase1Pixel}. In the new pixel detector, the number of barrel layers was increased from three to four, and the number of disks in each endcap from two to three. The innermost barrel layer was completely replaced to ensure optimal performance until the end of Run~3. At the same time, the detector material budget was reduced, creating better tracking performance for pixel detector tracks with $\abseta<3.0$. For nonisolated particles of $1 < \pt < 10\GeV$ and $\abseta < 3.0$, the track resolutions are typically 1.5\% in \pt and 20--75\mum in the transverse impact parameter \dzero~\cite{DP-2020-049}.

The ECAL measures the energy and timing of electrons and photons with $\abseta < 3.0$~\cite{CMS:2020uim}. It consists of a cylindrical barrel ($\abseta < 1.48$) and two endcaps. Scintillation light from the crystals is detected by photodetectors, digitized at 40\unit{MHz} by the readout electronics, and recorded as ten consecutive samples~\cite{CMS:2009onm}. The signal time is extracted from a fit to this distribution, with synchronization ensuring that particles from the interaction vertex yield a common reference time. Since Run~2, the signals in the ECAL crystals have been reconstructed by fitting the signal pulse with multiple template functions to mitigate the effect of out-of-time PU, arising from additional $\Pp\Pp$ interactions within previous or subsequent bunch crossings~\cite{ECAL_GPU,CMS:2020xlg}. Decays of \PZ bosons into electron-positron pairs allow the timing resolution to be accurately measured in data. For energies above 10\GeV, the ECAL barrel achieves a timing resolution better than $0.2\unit{ns}$~\cite{CMS:2024ppo}.

In the region $\abseta < 1.74$, the HCAL cells have widths of 0.087 in $\eta$ and 0.087 in $\phi$. In the $\eta$-$\phi$ plane, and for $\abseta < 1.48$, each HCAL cell maps onto $5{\times}5$ arrays of ECAL crystals to form calorimeter towers projecting radially outwards from close to the nominal IP. For $\abseta > 1.74$, the coverage of the towers increases progressively to a maximum of 0.174 in $\Delta \eta$ and $\Delta \phi$. Within each tower, the energy deposits in ECAL and HCAL cells are summed to define the calorimeter tower energies, which are subsequently used to provide the energies and directions of hadronic jets.

An upgrade of the HCAL was performed in stages that were installed between 2016 and 2019~\cite{CMS:TDR-010}. In the HCAL barrel and endcaps, the readout electronics were improved, providing a finer granularity readout. In addition, the segmentation was increased to allow for both layer-dependent corrections for the observed radiation damage to the scintillating tiles~\cite{CMS:2020mce} and better rejection of energy deposits from PU interactions. The previous generation of photosensors was replaced by silicon photomultipliers, which measure the scintillator light output with a better signal-to-noise ratio. The readout electronics were upgraded to support the increased channel count, improve the precision, and add signal timing information. When combining information from the entire detector, the jet energy resolution amounts typically to 15--20\% at 30\GeV, 10\% at 100\GeV, and 5\% at 1\TeV~\cite{CMS:2016lmd}.

The CMS muon system consists of four types of gas-ionization detectors: drift tube chambers (DTs), cathode strip chambers (CSCs), resistive-plate chambers (RPCs), and triple gas electron multiplier chambers (GEMs). The DT and CSC detectors are located in the regions $\abseta < 1.2$ and $0.9 < \abseta < 2.4$, respectively, and are complemented by the RPCs in the range $\abseta < 1.9$. The GEMs are located in the region $1.55 < \abseta < 2.18$. The chambers are arranged to maximize the coverage and to provide some overlap wherever possible. In both the barrel and endcap regions, the chambers are grouped into four ``muon stations'', where each station is progressively further from the center of CMS and separated by layers of the steel absorber of the flux-return yoke. A detailed description of these detectors, including the gas composition and operating voltage, is reported in Ref.~\cite{CMS:2018rym}. 

The GEM detector, consisting of four gas gaps separated by three GEM foils, was added in the endcaps before the start of Run~3~\cite{CMS:TDR-016,CMS:2023gfb}. The other subsystems, namely the DTs, CSCs, and RPCs, underwent several upgrades. The upper part of the CMS detector was covered with a neutron shield to reduce the background in the upper external DT chambers. An outer ring of CSCs (ME4/2) was added after data taking ended in 2012 (at the end of Run~1), and in preparation for the HL-LHC, the bulk of the CSC electronics upgrades that required chamber access were already performed during the long shutdown before Run~3 data taking. Outer rings of the RPC chambers in station four (RE4/2 and RE4/3) were also added. 

To cope with increasing instantaneous luminosities, the CMS data acquisition system underwent multiple upgrades~\cite{CMS:2023gfb}. A new optical readout link with a higher bandwidth of 10\unit{Gb/s} was developed. The bulk of the data acquisition system downstream from the detector-specific readout benefited from advances in technology to achieve a much more compact design, while doubling the event building bandwidth.

\section{The CMS trigger system}
\label{sec:triggerSystem}

The CMS detector selects events of interest using a two-tiered trigger system that reduces the input event rate of 40\unit{MHz} to several~kHz. The first level (L1), composed of custom hardware processors, uses information from the calorimeters and muon detectors to select events at a rate of around 100\unit{kHz} within a fixed latency of approximately 4\mus~\cite{CMS:2020cmk}. The second level, known as the high-level trigger (HLT), consists of a farm of processors running a version of the full event reconstruction software optimized for fast processing and reduces the event rate to 2--20\unit{kHz}, depending on the output data format, before data storage~\cite{CMS:2016ngn}.

After Run~1, the L1 trigger (L1T) hardware was entirely upgraded and has operated successfully since 2016~\cite{CMS:TDR-012}. For Run~3, although no major hardware changes have been made, the L1T performance has been improved using new algorithms. Thanks to the modular architecture of the L1T system and the flexibility of its design, new features are straightforward to deploy online. Some of the new algorithms in Run~3 are based on machine-learning techniques. Dedicated software, such as \textsc{hls4ml}~\cite{Duarte:2018ite}, facilitates the use of these machine-learning techniques in field-programmable gate arrays (FPGAs). Developments for Run~3 within the L1T primarily focus on expanding the physics reach of CMS through improved object measurement and calibration, as well as the introduction of dedicated triggers for LLPs and other exotic signatures. Some of these new triggers only became possible because of the enhanced capabilities of the global trigger logic and the increased trigger information provided by the calorimeters and muon systems. Furthermore, in recent years, the output rate of the L1T has been increased to accommodate more triggers tailored to BSM searches and \PB physics. These L1Ts include triggers for displaced and delayed jets using the HCAL (Section~\ref{sec:displacedjetshcal}), displaced muons (Section~\ref{sec:displaceddimuon}), high-multiplicity showers in the CSCs (Section~\ref{sec:mdsshowers}), vector boson fusion, \PB physics with inclusive dimuons and dielectrons~\cite{CMS:2024zhe}, as well as anomaly detection triggers~\cite{CMS-DP-2023-086,CMS-DP-2024-121,CMS-DP-2024-059,CMS-DP-2025-061}.

The data processing of the HLT is structured around the concept of an HLT ``path'', which is a set of algorithmic processing steps run in a predefined order to both reconstruct physics objects and apply selection criteria to these objects based on the physics requirements. At the start of the execution of an HLT path, each path requires events to pass specific L1Ts. These L1Ts are said to ``seed'' the rest of the HLT algorithm. The trigger objects generally used in the L1T are muons, $\Pe/\PGg$ objects (so-called since electrons and photons are indistinguishable at the L1T, as currently no tracking information is available there), jets, hadronically decaying tau leptons, and event-level information, such as \HT and \ptmiss, explained in Section~\ref{sec:eventReconstruction}. Then, each HLT path is implemented as a sequence of reconstruction and filtering steps, generally of increasing complexity, reconstruction refinement, and physics sophistication. For example, the full track reconstruction is usually performed only after some initial reconstruction and selection using data from the calorimeters and muon detectors. In case a path has requirements on two or more different kinds of physics objects, these different parts of the path are referred to as ``legs'' of the trigger. The reconstruction modules and selection filters of the HLT use the same software framework, called CMSSW~\cite{Jones:2015soc}, used for offline reconstruction and analyses.

The L1T and HLT include primary triggers for analyses, as well as triggers for calibration, efficiency measurements, control region measurements, \etc{} that typically have looser requirements than the primary triggers. These latter triggers are often ``prescaled'', meaning that they select only a fraction of the events that satisfy their conditions, to limit the storage rate. In contrast, an ``unprescaled'' trigger selects all events that satisfy the conditions of the trigger algorithm.

The HLT refines the purity and further reduces the rate of events previously selected by the L1T to an average rate of several kHz for offline storage and prompt reconstruction of standard $\Pp\Pp$ collision events. Additional rate beyond the nominal few kHz is allocated to triggers that collect ``parked'' data. The offline reconstruction of these parked data is postponed until computing resources are available to process the data. In addition, a higher rate of reduced-size events can be acquired using a technique known as ``data scouting'', where only high-level physics objects, such as jets or leptons, reconstructed at the HLT, are stored on disk. In data scouting, no raw data from the detector are stored for later offline analysis. In total, CMS employs three main approaches to data collection: the standard data stream, which uses the full event reconstruction performed offline immediately after data taking; the parked data stream, where events are saved for delayed offline reconstruction when computing resources permit; and the scouting data stream, in which no offline reconstruction is performed, and only the output of the HLT-level reconstruction is stored and used for analysis. The performance of parking and scouting is described further in Ref.~\cite{CMS:2024zhe}. Related real-time data-processing strategies are also used at other LHC experiments; in particular, LHCb employs a fully software trigger with real-time alignment, calibration, and reconstruction, enabling selections based on displaced tracks and vertices already in the online processing~\cite{LHCb:2018zdd,Aaij:2019zbu}.

Since 2016, the HLT has been operated using multithreaded event processing software, which minimizes memory requirements by increasing the number of processes running in parallel~\cite{CMS:2023gfb}. Also, since the start of Run~3, the HLT has made use of graphical processing units (GPUs) in the trigger farm. Substantial improvements were achieved in the physics performance and speed of the software, as well as in the computing infrastructure, including direct remote data access. For example, the L1T frontend boards have been newly designed to allow remote programming of their FPGAs~\cite{CMS:2023gfb}. Algorithms implemented to run on both central processing units (CPUs) and GPUs are automatically directed to run on a GPU if one is available; otherwise, the CPU-based version of the algorithm is executed. The HLT can offload the pixel detector track reconstruction and parts of the calorimeter reconstruction to GPUs. For example, the \PATATRACK project~\cite{Bocci:2020pmi} developed parallelized versions of pixel detector track and vertex reconstruction algorithms that can run on NVIDIA GPUs, while a collaboration between CMS and OpenLab ported the electromagnetic and hadronic calorimeter local reconstruction algorithms to GPUs~\cite{ECAL_GPU,HCAL_GPU}. These efforts have reduced the overall event processing time by about 40\%~\cite{2024HLTtimingDPNote}.

\begin{sloppypar}In preparation for Run~3, the ECAL trigger-primitive formation and calibration algorithms were optimized to cope with higher instantaneous luminosity and PU~\cite{CMS:2023gfb}. The trigger-primitive thresholds were reoptimized to suppress spurious signals produced by ionizing particles traversing the avalanche photodiode photodetectors, which have become relatively more frequent since the start of Run~2. Additionally, the calibration update frequency at the L1T and HLT was increased from twice per week to once per LHC fill to correct for dynamic variations in crystal and photodetector responses during collisions.\end{sloppypar}

Several changes were implemented for the Run~3 data-taking period to improve the reconstruction of physics objects at the HLT. For example, the tracking in the pixel and strip trackers, which is generally discussed in Section~\ref{sec:trackingVertexing}, was significantly revised. The pixel detector tracks are now reconstructed by the \PATATRACK algorithm mentioned earlier, which offers improved performance over the HLT pixel detector tracking used in 2018~\cite{Bocci:2020pmi}. This improvement allowed the HLT tracking to be typically performed using a single global iteration~\cite{Run3HLTTrackRecoDPNote}, as opposed to the three iterations that were used in Run~2 (2015--2018). Since 2025, a parallel and vectorized Kalman filter-based algorithm called ``mkFit''~\cite{Lantz:2020yqe} has been employed for track pattern recognition at the HLT, similar to the procedure used in the offline reconstruction since the start of Run~3~\cite{CMS-DP-2022-018}. The \PATATRACK and mkFit algorithms are further described in Section~\ref{sec:trackingVertexing}. Furthermore, the identification of \PQb jets at the HLT is essential to collect events containing such jets with lower \pt thresholds than those of the standard lepton, jet, or \ptmiss triggers. Two new neural network taggers, \textsc{DeepJet}~\cite{Bols:2020bkb} and \textsc{ParticleNet}~\cite{Qu:2019gqs}, were deployed in 2022, with improved performance over Run~2. In addition to tracks, the \textsc{DeepJet} algorithm also uses information from neutral and charged particle-flow (PF)~\cite{CMS:2017yfk} jet constituents, while the graph neural network-based \textsc{ParticleNet} algorithm provides concurrent jet-flavor classification for categories of \PQb, \PQc, and light quarks, gluons, and hadronically decaying tau leptons. The global PF algorithm~\cite{CMS:2017yfk}, which aims to reconstruct all individual particles in an event by combining information provided by each CMS subdetector, is explained in Section~\ref{sec:particleflow}.

In Run~3, data parking still targets \PB physics, as in Runs~1 and 2, but it also allows a wide range of other physics topics to be investigated~\cite{CMS:2024zhe}. At the end of 2022, the parked data recorded events with a single displaced muon, events with two low-\pt muons, and events with two low-\pt, central electrons. In 2023, the parking strategy was extended by improving the purity of the dielectron triggers and adding parked data triggers for events with two \Pb-tagged jets, Higgs boson events produced via vector boson fusion and selected with forward jet triggers, and many LLP signatures.

In Run~3, the HLT scouting data were recorded at a rate as high as $30\,(22)\unit{kHz}$ in 2022 (2023), greatly increased compared to Run~2, with an event size of about 7\unit{kB}, compared to the raw event size of about 1\unit{MB} for fully reconstructed events. In 2022--2023, a special version of the PF reconstruction algorithm was used for HLT scouting, using pixel detector tracks reconstructed with \PATATRACK~\cite{CMS:2024zhe}. In 2024, the algorithm was updated to use entire tracks, as is done for fully-reconstructed events at the HLT, to improve the momentum and spatial resolution. In addition to the PF objects that were already stored during Run~2, HLT scouting in Run~3 was expanded to include electrons and photons for the first time, as well as enhanced muon and track information with respect to Run~2.

In addition, a scouting system is being prepared for the L1T~\cite{Zabi:2020gjd,James:2852916,CMS:2023ukn,Ardino:2024gcz,Migliorini:2024zdp}. This system collects and stores the reconstructed particle primitives and intermediate information of the L1T processing chain at the full 40\unit{MHz} bunch crossing rate. The L1 scouting will provide vast amounts of data for detector diagnostics, luminosity measurements, and the study of otherwise inaccessible signatures, such as LLPs. A demonstrator L1 scouting system first took data in 2018 and has since been expanded with new FPGAs and inputs. The first analysis using the demonstrator L1 scouting system was a search for LLPs~\cite{CMS:2026vft}.

\section{Offline and online event reconstruction}
\label{sec:eventReconstruction}

In this section, we describe the standard event and object reconstruction algorithms, both for the offline reconstruction and the online reconstruction at the L1T and HLT. The details of the online reconstruction algorithms and their performance during Run~2 can be found in Refs.~\cite{CMS:2020cmk} and \cite{CMS:2024psu} for the L1T and the HLT, respectively. We explain here how the offline reconstruction algorithms are simplified and sped up for online use; in particular, online reconstruction algorithms usually assume that particles are promptly produced, allowing online objects to be reconstructed within the available CPU budget. We point out where the standard online reconstruction algorithms are inefficient for displaced objects, motivating the dedicated triggers described in Section~\ref{sec:LLPtriggers}.

\subsection{The particle-flow algorithm}
\label{sec:particleflow}

A PF algorithm~\cite{CMS:2017yfk} aims to reconstruct and identify each individual particle in an event, with an optimized combination of information from the various elements of the CMS detector. The energy of photons is obtained from the ECAL measurement. The energy of an electron is determined from a combination of the electron momentum at the primary interaction vertex, as determined by the tracker, the energy of the corresponding ECAL cluster, and the energy sum of all bremsstrahlung photons spatially compatible with originating from the electron track. The energy of a muon is obtained from the curvature of the corresponding track as measured by the muon system and the silicon tracker. The energy of a charged hadron is determined from a combination of the momentum measured in the tracker and the matching ECAL and HCAL energy deposits, corrected for the response function of the calorimeters to hadronic showers. Finally, the energy of neutral hadrons is obtained from the corresponding corrected ECAL and HCAL energies. The primary vertex (PV) is taken to be the vertex corresponding to the hardest scattering in the event, evaluated using tracking information alone, as described in Section 9.4.1 of Ref.~\cite{CMS-TDR-15-02}. The online PF reconstruction features a simplified version of the offline PF algorithm, particularly for the electron and photon reconstruction, to meet the timing constraints for online reconstruction.

\subsection{Tracking and vertexing}
\label{sec:trackingVertexing}

Charged particle tracks in the tracker are reconstructed from hits in the pixel and strip tracker using a Kalman filtering technique~\cite{Fruhwirth:1987fm}. The collection of reconstructed tracks is produced by multiple passes (iterations) of the track reconstruction sequence in a process called ``iterative tracking''~\cite{CMS:2014pgm}. The basic idea of iterative tracking is that the initial iterations search for tracks that are easiest to find (\eg, of relatively large \pt, and produced near the interaction region). After each iteration, the hits associated with the tracks are removed, thereby reducing the combinatorial complexity and simplifying subsequent iterations that search for more difficult classes of tracks (\eg, low-\pt or significantly displaced tracks).

\begin{figure}[htb!]
\centering
\includegraphics[width=0.49\textwidth]{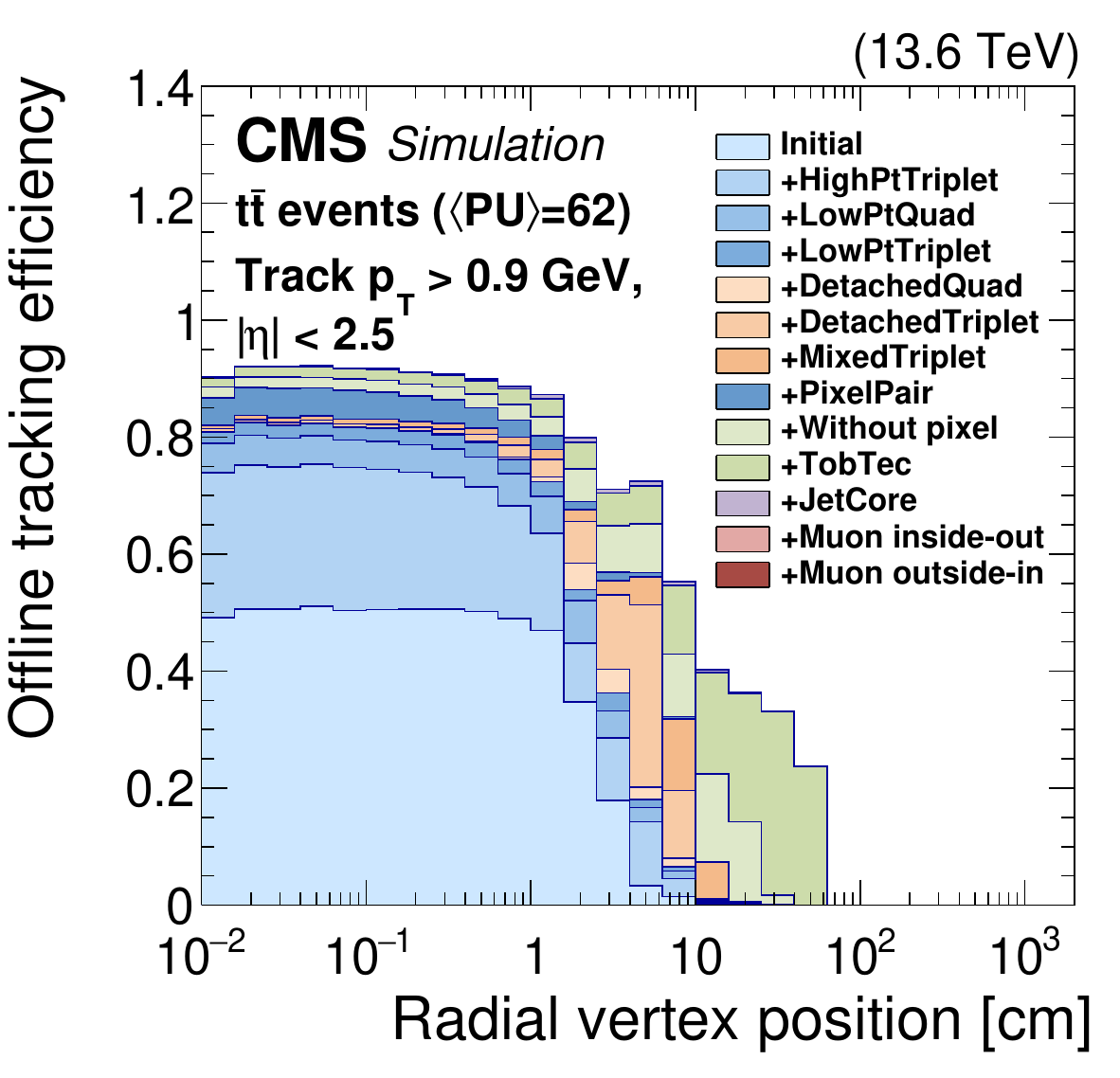} 
\includegraphics[width=0.49\textwidth]{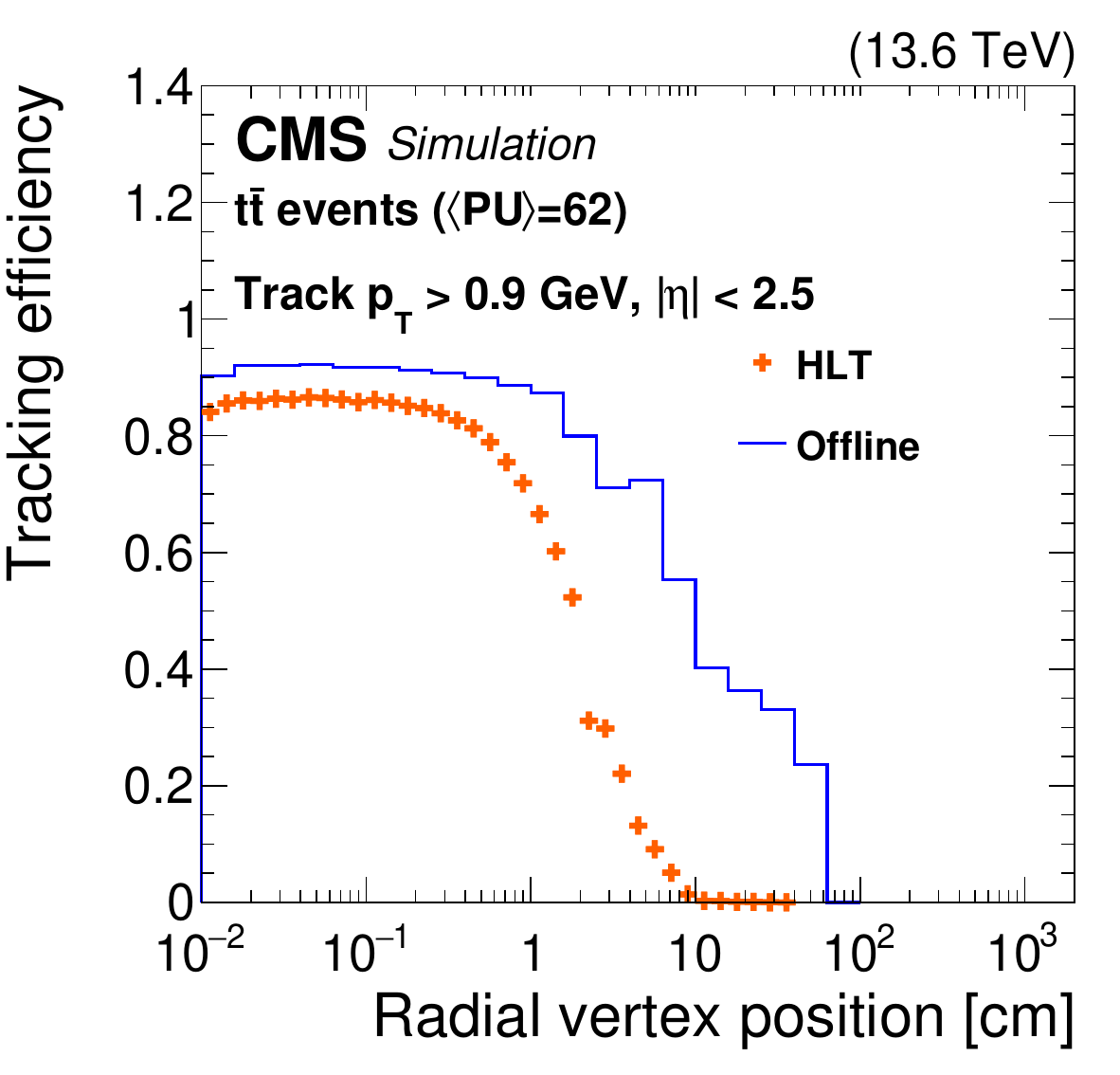}
\caption{The offline standard tracking efficiency during Run~3 (\cmsLeft) for different tracking iterations, as a function of simulated radial position of the track production vertex. The overall standard tracking efficiency at the HLT (orange crosses) and offline (blue line) during Run~3 (\cmsRight), as a function of the simulated radial position of the track production vertex. The standard tracking at the HLT includes the nominal \PATATRACK global iteration plus the additional iteration that mitigates the pixel detector issues. For both plots, a \ttbar simulation under 2025 conditions with an average PU of 62 is used, and the tracks are required to have $\pt>0.9\GeV$ and $\abseta<2.5$. The tracking efficiency is defined as the number of simulated tracks (with the aforementioned selection requirements) geometrically matched to a reconstructed track, divided by the total simulated tracks passing the selections.}
\label{fig:trackingEffVsR}
\end{figure}

The first iteration (the ``initial step''), which is the source of most reconstructed tracks, is designed to reconstruct prompt tracks originating near the $\Pp\Pp$ IP that have at least four pixel detector hits. The following iterations are used to recover tracks with fewer hits in the pixel tracker, lower \pt, and/or larger displacements. Figure~\ref{fig:trackingEffVsR} (\cmsLeft) shows the tracking efficiency for different iterations as a function of the radial position of the track production vertex, for the standard offline Run~3 tracking based on a simulation of top quark-antiquark pair production (\ttbar). The offline tracking misidentification rate, defined as the ratio between the number of reconstructed tracks not matched to simulation to the total number of reconstructed tracks, grows from 0.02 to 0.30 as the distance of closest approach of the track to the beam axis increases from 0 to 10\cm.

Each iteration starts with a ``seed'' that defines the initial estimate of the trajectory parameters and their uncertainties, using only a few hits. Then, the Kalman filter extrapolates the seed trajectory along the expected path of the charged particle, searching for additional hits that can be assigned to the track candidate. The next step in each iteration is performed by the track-fitting module, which provides the best possible estimate of the parameters of each trajectory using a Kalman filter and track refit to optimize the track parameters at every detector layer (not just the last layer). Finally, the track selection sets quality flags and discards tracks that fail criteria specific to each iteration. The main differences among the iterations are in the configuration of the seed generation and the final track selection. Furthermore, in Run~3, the highly parallel and vectorized mkFit algorithm~\cite{Lantz:2020yqe,CMS-DP-2022-018} was introduced, enabling the time for the overall track reconstruction to be reduced by about 25\% with respect to the previous algorithm, while retaining strong physics performance.

The standard HLT tracking~\cite{Run3HLTTrackRecoDPNote,CMS-DP-2023-028} is similar to that used in the offline reconstruction, except that in Run~3, HLT tracking is typically performed using a single global iteration targeting prompt tracks, as discussed in Section~\ref{sec:triggerSystem}. An additional tracking iteration, seeded by pixel detector hit pairs, has been used since 2024 to mitigate issues in specific regions of the pixel detector that had significantly reduced the efficiency of the standard iteration~\cite{CMS-DP-2024-013}. Figure~\ref{fig:trackingEffVsR} (\cmsRight) shows the HLT tracking efficiency, compared to the total offline tracking efficiency, as functions of the radial position of the track production vertex. Starting in 2025, the mkFit algorithm has also been employed at the HLT. The offline iterations that target displaced tracks are not used in HLT tracking except in dedicated triggers such as the displaced-tau paths discussed in Section~\ref{sec:displacedtau} and the displaced-jets paths discussed in Section~\ref{sec:displacedjetstracking}. As a result, the HLT tracking efficiency drops quickly for vertex positions greater than approximately 1\cm from the beam axis. The HLT tracking misidentification rate grows from 0.08 to 0.35 as the distance of closest approach of the track to the beam axis increases from 0 to 10\cm.

\begin{sloppypar}Vertices are produced from clusters of tracks, and the same deterministic annealing algorithm~\cite{726788} is used in the offline and HLT reconstruction~\cite{CMS:2014pgm}. The vertex position is fitted using an adaptive vertex fitter~\cite{Fruhwirth:1027031}. While the offline reconstruction uses standard tracks for vertexing, the HLT vertices, called ``pixel vertices'', are produced from a subset of the pixel detector tracks described in Section~\ref{sec:triggerSystem}. The HLT pixel vertices algorithm has been modified in Run~3 to run on GPUs, where available.\end{sloppypar}

\subsection{Muons}
\label{sec:muons}

Muon tracks in the offline analysis are first reconstructed independently in the tracker and in the muon system using Kalman filtering techniques~\cite{Fruhwirth:1987fm}. In the muon system, ``standalone muons'' are reconstructed by combining the information from the DT, CSC, RPC, and GEM detectors. The signals from these four types of muon detectors are first combined into track segments, which are thereafter used as the input to a Kalman filtering algorithm. As described in Section~\ref{sec:trackingVertexing}, in the tracker, tracks are reconstructed iteratively, with each iteration seeded by a different combination of pixel or strip detector hits, to ensure all relevant track topologies are reconstructed~\cite{CMS:2014pgm}. In the same way, dedicated iterations seeded by standalone muons maximize the efficiency of muon track reconstruction. Tracker tracks and standalone muon tracks with compatible momentum, direction, and position in the transverse plane are combined into ``global muons'', and a combined fit to all associated tracker and muon system signals is performed to determine the parameters of the global track. Within the tracker, the global muon trajectory is built ``outside-in'', in that it starts with the outer tracker layers and propagates inwards towards the IP. For muons with $10<\pt<100\GeV$, the \pt resolution of global muons is 1--3\%, while for standalone muons, it is about an order of magnitude worse~\cite{CMS:2018rym,CMS:2012nsv}. However, standalone muons, particularly those reconstructed without constraining the track to the PV (``displaced standalone muons''), are often used in offline analyses to allow the efficient reconstruction of muons with displacements up to a few meters. The trajectories of muons with very low \pt are deflected in the magnetic field to such a degree that they do not reach all stations of the muon system. To recover such muons, for which a standalone muon might not be reconstructed, tracker tracks are extrapolated to the muon segments, and a loose geometrical matching to DT and CSC segments is performed. If at least one compatible segment is found, the track is classified as a ``tracker muon''. The tracker muon trajectory is reconstructed ``inside out'' from the center of the detector towards the muon system. The offline muon reconstruction and performance are described in more detail in Refs.~\cite{CMS:2018rym,MUO-24-001}.

In the trigger system, three types of muons are defined, referred to as L1, L2, and L3 muons~\cite{MUO-19-001}. An L1 muon is identified by the L1T system using only information from the muon detectors. The L1T uses three regional track finders to reconstruct muons: the barrel muon track finder (BMTF) 
covering $0 < \abseta < 0.83$, the overlap muon track finder (OMTF) covering $0.83 < \abseta < 1.24$, and the endcap muon track finder (EMTF) covering $1.24 < \abseta < 2.40$. The track finder algorithms have implicit minimum \pt requirements ranging from about 2 to 4\GeV, since the muons need enough momentum to reach the muon system, and since the \pt is assigned via a different method in each algorithm. The global muon trigger collects reconstructed muons from the three track finders and removes geometrically overlapping tracks using information based on \pt and quality, sending the remainder to the L1 Global Trigger, which decides whether to keep the event for processing by the HLT. The L1 muon reconstruction is described in greater detail in Refs.~\cite{CMS:2016ngn,CMS:2020cmk,MUO-19-001}. 

In Run~3, there are beam axis-constrained and beam axis-unconstrained versions of each of these three L1 muon track finder algorithms~\cite{CMS:2023gfb}. The beam axis-unconstrained track finder algorithms are designed to measure displaced muons, while the beam axis-constrained algorithms are inefficient for displaced muons. The beam-axis-constrained BMTF uses a Kalman filter algorithm that is constrained to the beam-axis, while the beam-axis-constrained OMTF and EMTF algorithms include implicit beam-axis constraints because they are optimized for prompt muons with machine-learning algorithms trained on prompt samples. The beam-axis-constrained algorithms underestimate the \pt of the displaced muons, while the beam-axis-unconstrained algorithms measure the \pt of displaced muons closer to their true values. As a result of using the beam axis-unconstrained track finder algorithms, the trigger efficiency is increased significantly for muons with $\pt > 10\GeV$ when the transverse displacements are larger than 20\cm. For lower-\pt muons, the performance of the displaced-muon algorithms is similar to that of the prompt algorithms, since the large bending of these muons allows for a more accurate \pt assignment, regardless of their displacement. The developments for displaced L1 muons and their performance are further explored in Section~\ref{sec:displaceddimuon}.

The L2 and L3 muons are defined within the HLT: an L2 muon is identified using information from the muon detectors alone, analogous to the offline standalone muon reconstruction algorithm, while an L3 muon additionally matches the muon detector track to a track from the tracker, analogous to the offline tracker and global muons, with both inside-out and outside-in trajectories. Displaced muons are found with both the inside-out and outside-in L3 muons. As is the case with their offline counterpart reconstruction algorithms, L3 muons have better \pt resolution than L2 muons. Standard muon HLT paths usually use the full L3 muon reconstruction, while muon HLT paths targeting displaced signatures, including some paths described in Sections~\ref{sec:displaceddimuon}--\ref{sec:dimuonscouting} and \ref{sec:nobptxmuons}, often stop after the L2 muon reconstruction algorithm. The L3 muons can still have sensitivity to displaced muons, especially with the L3 muon reconstruction algorithms for displaced muons that are described in Sections~\ref{sec:displaceddimuon}--\ref{sec:displacedL3muonandphoton}.

\subsection{Electrons and photons}
\label{sec:electronsPhotons}

Electron and photon reconstruction is based on clusters of ECAL energy deposits~\cite{CMS:2020uim}. In the offline reconstruction, these clusters are formed by grouping nearby energy deposits typically exceeding 80\MeV in the ECAL barrel and 300\MeV in the ECAL endcaps. The times associated with the energy deposits must be within 5\unit{ns} of the collision, corrected for the time of flight of a particle. A useful concept in the reconstruction of calorimeter objects is the transverse energy \ET, which is the energy of the object in question multiplied by $\sin{\theta}$, where $\theta$ is the polar angle with respect to the center of the CMS detector. Placing requirements on this quantity is an efficient way to reduce the number of objects reconstructed to a manageable level, and this approach is used when only calorimeter energy is available for a reconstructed element, without any associated primary vertex. A seed cluster is then defined as the one containing most of the energy deposited in any specific region with $\ET>1\GeV$. Clusters of ECAL deposits within a certain geometric area around the seed cluster are combined into ``superclusters'' to include photon conversions and bremsstrahlung losses. A dedicated tracking algorithm, based on the Gaussian sum filter (GSF)~\cite{GSF}, is used for electrons to estimate the track parameters. This GSF tracking step is seeded by both pixel detector seeds that are compatible with the supercluster position and the trajectory of an electron, as well as all tracker tracks with $\pt>2\GeV$ that are compatible with an electron trajectory. A dedicated algorithm~\cite{CMS:2015myp} is used to find the tracker tracks that are likely to originate from photons converting into electron-positron pairs. Then, ECAL clusters, superclusters, GSF tracks, as well as conversion tracks and associated clusters, are all used in the PF algorithm, which resolves the inputs into electron and photon objects. There is no dedicated offline reconstruction algorithm for displaced electrons, but GSF electrons benefit from the displaced tracking iterations performed in the iterative tracking (Section~\ref{sec:trackingVertexing}).

In the online reconstruction, the electron and photon candidates at the L1T are based on ``trigger towers''. As mentioned in Section~\ref{sec:detector}, in the barrel, these trigger towers are defined by 5$\times$5 ECAL crystals that are grouped with the HCAL tower directly behind them, while in the endcaps, the trigger towers are formed from groups of 5--25 crystals depending on their $\eta$-$\phi$ position~\cite{CMS:2020cmk}. The trigger tower with the largest \ET is clustered together with adjacent towers, using a procedure that trims the energy deposits to include only contiguous towers, to match the electron or photon signature in the calorimeter. To form an L1 candidate, energy clusters must satisfy additional identification criteria and, optionally, isolation requirements. This L1T reconstruction of the electron and photon candidates is called ``$\Pe/\PGg$ dynamic clustering''.

The HLT electron and photon identification begins with a regional reconstruction of the energy deposited in the ECAL crystals around the L1 candidates~\cite{CMS:2024psu}. Superclusters are then built using the same reconstruction algorithm as used offline~\cite{CMS:2015xaf}. Not all the energy of the incident particle is captured by the superclustering algorithm because of energy threshold requirements on individual crystals or because the energy is deposited in an uninstrumented area of the detector. Therefore, in both the online and offline reconstruction, the supercluster energy is corrected for such effects. At the HLT, this correction is designed to be more tolerant of changing detector conditions and the presence of background, rather than aiming for the ultimate achievable precision.

The resulting superclusters may then undergo additional selections, such as those based on kinematic variables, the shower shape, or isolation, depending on the HLT path. Electron identification can be enhanced by looking for pixel detector hits that match the trajectory the electron took through the detector based on its supercluster position, in a process known as pixel matching. If further background rejection is required, a GSF track can be reconstructed using the trajectory of any matched pixel detector hits as a seed. In contrast to the offline reconstruction, GSF tracks are not seeded by generic tracker tracks at the HLT. Thus, any electron track at the HLT must have hits in the pixel detector and therefore can not be significantly displaced. As a result, analyses that target displaced electron signatures often do not require a matched electron track at the HLT (Sections~\ref{sec:delayeddiphoton}, \ref{sec:displacedphotonandht}, and \ref{sec:displacedL3muonandphoton}). Since 2017, the pixel matching algorithm has required three pixel detector hits rather than two to maximize early background rejection, while a configuration with two pixel detector hits in a row is accepted only if the trajectory passes through a maximum of three active modules.

The electron momentum is estimated by combining the energy measurement in the ECAL with the momentum measurement in the tracker. The momentum resolution for electrons with $\pt \approx 45\GeV$ from $\PZ \to \Pe \Pe$ decays ranges from 1.6 to 5\%. It is generally better in the barrel region than in the endcaps, and also depends on the bremsstrahlung energy emitted by the electron as it traverses the material in front of the ECAL~\cite{CMS:2020uim,CMS-DP-2020-021}.

In the barrel section of the ECAL, an energy resolution of about 1\% is achieved for unconverted or late-converting photons in the tens of GeV energy range. The energy resolution of the remaining barrel photons is about 1.3\% up to $\abseta = 1$, changing to about 2.5\% at $\abseta = 1.4$. In the endcaps, the energy resolution is about 2.5\% for unconverted or late-converting photons, and between 3 and 4\% for the other ones~\cite{CMS:2015myp}.

\subsection{Jets}
\label{sec:jets}

Offline, hadronic jets are clustered from the energy deposits in the calorimeter towers using the infrared- and collinear-safe anti-\kt algorithm~\cite{Cacciari:2008gp, Cacciari:2011ma} with a nominal distance parameter of 0.4. Jet momentum is determined as the vectorial sum of all particle momenta in the jet, and is found from simulation to be, on average, within 5 to 10\% of the true momentum over the whole \pt spectrum and detector acceptance. When combining information from the entire detector, the jet energy resolution amounts typically to 15--20\% at 30\GeV, 10\% at 100\GeV, and 5\% at 1\TeV~\cite{CMS:2016lmd}. The PU interactions can contribute additional tracks and calorimetric energy deposits to the jet momentum. The PU per particle identification algorithm (PUPPI)~\cite{Sirunyan:2020foa,Bertolini:2014bba} is used to mitigate the effect of pileup at the reconstructed particle level, making use of local shape information, event pileup properties, and tracking information. A local shape variable is defined, which distinguishes between collinear and low-\pt diffuse distributions of other particles surrounding the particle under consideration. Collinear distributions are attributed to particles originating from the hard scatter, while low-\pt diffuse distributions are attributed to particles originating from pileup interactions. Charged particles identified to be originating from pileup vertices are discarded. For each neutral particle, a local shape variable is computed using the surrounding charged particles compatible with the primary vertex within the tracker acceptance ($\abseta < 2.5$), and using both charged and neutral particles in the region outside of the tracker coverage. The momenta of the neutral particles are then rescaled according to their probability to originate from the primary interaction vertex deduced from the local shape variable, superseding the need for jet-based pileup corrections~\cite{Sirunyan:2020foa}.

Jets are reconstructed at the L1T by summing the energies of a 9$\times$9 window of trigger towers centered on a jet seed that must have an energy greater than 4\GeV, which corresponds to approximately the same jet size as for jets reconstructed offline within a cone of size $\DR \equiv \sqrt{\smash[b]{(\Delta\eta)\,^2 + (\Delta\phi)\,^2}} < 0.4$ within the barrel calorimeter. The energy contribution from pileup is corrected by considering the four 3$\times$9 regions on the boundaries of the jet and subtracting the sum of the three lowest energies in these regions. The resulting jet energy is then calibrated. For Run~3, an LLP jet identification algorithm is implemented that uses the HCAL timing and depth information (Section~\ref{sec:displacedjetshcal}).

Jets are reconstructed at the HLT~\cite{CMS:2024psu} using the anti-\kt clustering algorithm. The inputs for the HLT jet algorithm can be either calorimeter towers or reconstructed objects from the PF algorithm. Most HLT jet paths use the PF inputs (``PF jets''), whereas calorimeter jets (``calo jets'') are used as a first step to identify jet signatures and initiate the PF reconstruction. To account for detector effects and collision conditions, corrections are applied to the estimated PF hadron energies, average PU energy, and jet energy scale.

Several HLT \Pb-tagging algorithms designed for promptly produced \Pb jets, such as the \textsc{DeepJet} and \textsc{ParticleNet} neural networks discussed in Section~\ref{sec:triggerSystem}, can provide complementary contributions for displaced-jet signatures with proper decay lengths $\cTau\lesssim 1\mm$, where $\tau$ is the proper lifetime and $c$ is the speed of light. However, the HLT \PQb-tagging algorithms take standard HLT tracks and secondary vertices as inputs, which are tuned for \Pb tagging and not efficient for LLPs with significant displacements or large masses. Furthermore, the \textsc{DeepJet} and \textsc{ParticleNet} neural networks are trained to identify the prompt \Pb jets produced in SM processes. Therefore, their efficiency decreases rapidly for $\cTau\gtrsim 1\mm$, necessitating the development of dedicated triggers targeting displaced and delayed jets. Sections~\ref{sec:displacedjetstracking}, \ref{sec:displacedjetshcal}, \ref{sec:delayedjetsecal}, and \ref{sec:nobptxjets} describe triggers that target displaced- and delayed-jet signatures.

\subsection{Tau leptons}
\label{sec:taus}

Tau leptons decay predominantly to one or three charged hadrons, together with zero, one, or two neutral pions (\tauh)~\cite{ParticleDataGroup:2024cfk}. The final-state particles may produce clusters in the calorimeters that are separated in $\phi$ because of the magnetic field. At the L1T, the \tauh reconstruction is based on an adaptation of the $\Pe/\PGg$ dynamic clustering (Section~\ref{sec:electronsPhotons}), which is used to reconstruct single clusters around local maxima or seeds that can subsequently be merged into a single \tauh candidate. Both offline and at the HLT, hadronic \PGt lepton decays are reconstructed from jets using the hadrons-plus-strips (HPS) algorithm~\cite{CMS:2018jrd}, which combines one or three tracks from the strip tracker with energy deposits in the calorimeters to identify the tau lepton decay modes. Neutral pions are reconstructed from reconstructed electrons and photon candidates as strips with dynamic size in $\eta$-$\phi$, where the strip size varies as a function of the \pt of the electron or photon candidate. The HLT \tauh algorithm is composed of three steps. The first two steps, called L2 and L2.5, are the same as those used in the online \tauh reconstruction in Run~2, and they are described in detail in Ref.~\cite{CMS:2018jrd}. The L2 step uses the energy deposits in the calorimeter towers, while the L2.5 step complements these energy deposits with the information from the pixel detector. The third step, L3, applies the HPS algorithm, which receives the necessary inputs from the PF event reconstruction. The tracking reconstruction step within the PF algorithm is inefficient for displaced tau leptons because it assumes prompt tracks, and it has been modified in Run~3 for the displaced \tauh use case, as discussed in Section~\ref{sec:displacedtau}. In addition, the standard HPS algorithm requires that the \tauh constituents originate from the IP, making it inefficient for displaced \tauh objects. Overcoming this challenge at the HLT is also discussed in Section~\ref{sec:displacedtau}.

It is possible for light leptons and jets originating from the hadronization of quarks or gluons to be erroneously reconstructed as \tauh candidates. The deep neural network \textsc{DeepTau}~\cite{CMS:2022prd} was developed to distinguish genuine \tauh from misidentified \tauh candidates in the offline reconstruction. Information from all individual reconstructed particles near the \tauh axis is combined with properties of the \tauh candidate and the event. The rate for a jet to be misidentified as \tauh by \textsc{DeepTau} depends on the \pt and quark flavor of the jet. In simulated events from \PW boson production in association with jets, it has been estimated to be around 3\% for a genuine \tauh identification efficiency of 60\%, for jets with $20 <\pt <100\GeV$. The misidentification rate for electrons (muons) is $2.60\,(0.03)\%$ for a genuine \tauh identification efficiency of $80\,({>}99)\%$. Section~\ref{sec:displacedtau} describes triggers that target displaced \tauh signatures, and Section~\ref{sec:delayedjetsecal} describes triggers that make explicit use of L1 tau seeds.

\subsection{Hadrons, energy sums, and missing transverse momentum}
\label{sec:hadronsEnergySumsMET}

Charged hadrons are identified as energy clusters associated with charged particle tracks that are neither identified as electrons nor as muons. Neutral hadrons are identified as HCAL energy clusters not linked to any charged-hadron trajectory, or as a combined ECAL and HCAL energy excess with respect to the expected charged-hadron energy deposit.

The total hadronic transverse momentum \HT is defined as the scalar \pt sum of all jets that meet certain selection criteria. While the details of the selection may vary among different offline analyses and different HLT paths, a common definition includes all jets with $\PT > 30\GeV$ and $\abseta < 3.0$.

The missing transverse momentum vector \ptvecmiss is computed as the negative vector sum of the transverse momenta of all the PF candidates in an event, and its magnitude is denoted as \ptmiss~\cite{CMS:2019ctu}. The \ptvecmiss is modified to account for corrections to the energy scale of the reconstructed jets in the event. The PUPPI algorithm is applied to reduce the PU dependence of the \ptvecmiss observable. The \ptvecmiss is computed from the PF candidates weighted by their probability to originate from the PV~\cite{CMS:2019ctu}.

At the L1T, \ptmiss is calculated as the negative vector sum of the trigger tower energies across the full calorimeter. In contrast, the HLT calculates \ptmiss while accounting for the instrumental effects of detector noise, beam-induced backgrounds, and anomalous forward calorimeter signals to keep the rates of the \ptmiss triggers within operational limits. Thus, additional filtering algorithms are applied during the HLT \ptmiss reconstruction to achieve lower rates for \ptmiss triggers~\cite{CMS:2024psu}. Calorimeter deposits consistent with a noise signature or beam halo interaction are removed from the energy sum computation at the HLT. Online jet energy corrections can also be propagated to the calculation of the HLT \ptmiss, in a similar way to the offline procedure. An alternative \ptmiss trigger is based on a calculation that uses all the reconstructed PF objects except for muons, leading to the ``\metNoMu'' HLT paths. In this approach, events with high-\pt muons are also assigned large online \ptmiss, whereas for events with no reconstructed muons, the two calculations coincide. This path is mainly used for searches for new physics in final states with only jets and \ptmiss. A dedicated trigger to select disappearing tracks (Section~\ref{sec:disappearingtracktriggers}) requires an isolated track in addition to \metNoMu, thereby allowing a lower minimum threshold on \metNoMu.

\section{Data and simulation}
\label{sec:dataMC}

The trigger performance studies use data collected with the CMS detector during 2022--2024. In addition, simulations of both SM background and LLP BSM signal processes are used. The simulation of collision events is implemented in two steps. In the first, a fixed-order perturbative quantum chromodynamics (QCD) calculation of up to four noncollinear high-\pt partons is performed. This is followed by a simulation of the underlying event, parton showering, and hadronization. The first step is usually performed by a matrix element calculator and event generator, and the \MGvATNLO (v2.73 and v2.9.13, depending on the sample)~\cite{Alwall:2014hca} package is used for almost all the studies presented in this paper; the latter step is usually implemented by the \PYTHIA8.306~\cite{Bierlich:2022pfr} generator. Combining the two steps is a delicate procedure; a matching procedure must be implemented to avoid double-counting of processes in the combination, with the exact recipe depending on the order of the perturbative calculation. The MLM merging~\cite{Alwall:2007fs} is used for leading-order (LO) calculations, while the FxFx~\cite{Frederix:2012ps} method is used for next-to-LO (NLO). Parton distribution functions (PDFs) must be used to map the simulated colliding protons to the initial-state partons that are present in the matrix element calculations. The \PYTHIA parameters must be adjusted to a set of values, called a tune, that best describe the observed dynamics of the $\Pp\Pp$ collisions. The samples used here employ the NNPDF3.1 next-to-NLO PDFs~\cite{NNPDF:2017mvq} and the CP5 tune~\cite{CMS:2019csb}. The detector response to simulated particles is modeled using the \GEANTfour software~\cite{Agostinelli:2002hh}. Simulated minimum bias interactions are superimposed on each hard scattering to simulate the effect of additional PU interactions within the same or neighboring bunch crossings.

To demonstrate the performance of the triggers, a set of LLP signal processes described below is used. A wide variety of models with different mass and lifetime hypotheses are employed, as the regions of phase space targeted by the triggers and the corresponding searches vary significantly.

\begin{figure}[!ht]
\centering  \includegraphics[width=0.45\textwidth]{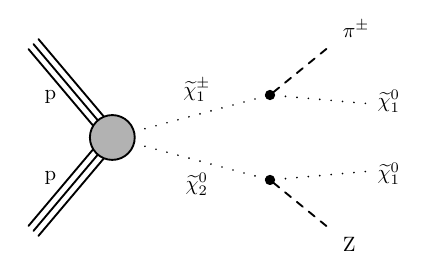}
\includegraphics[width=0.45\textwidth]{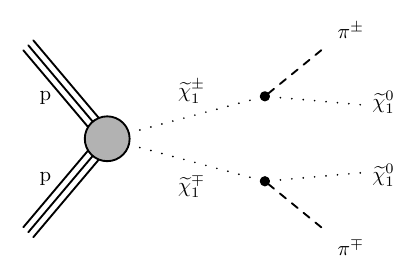}
\includegraphics[width=0.45\textwidth]{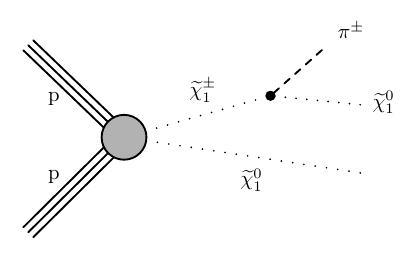}
\caption{Diagrams for the AMSB $\Pchipm \to \Pchiz+\PX$ process. The \PX particle is shown here as a charged pion \PGppm, but \PX can also be an electron or a muon, depending on the type of neutralino \Pchiz.}
\label{fig:AMSB_feynmanDiagrams}
\end{figure}

Supersymmetry (SUSY) is a class of models that extend the SM with a new symmetry that introduces a fermionic (bosonic) superpartner, called a ``sparticle'', for each boson (fermion) in the SM. In SUSY, the superpartners of the leptons are called ``sleptons'', and the superpartners of the gauge and Higgs bosons (called ``winos'', ``binos'', and ``higgsinos'') mix to form ``charginos'' and ``neutralinos''. Figure~\ref{fig:AMSB_feynmanDiagrams} shows diagrams for an anomaly-mediated SUSY breaking (AMSB) model~\cite{Giudice:1998xp, Randall:1998uk} in which a chargino \Pchipm is nearly mass-degenerate with a neutralino \Pchiz, and the decay of the chargino is of the form $\Pchipm \to \Pchiz+\PX$, where \PX is an SM particle that can be a charged pion \PGppm, an electron, or a muon, depending on the superpartner composition of the lightest neutralino state. The \Pchipm is generated for masses between 100 and 1200\GeV and \cTau between 0.1 and 10\unit{m}.

\begin{figure}[!ht]
\centering  \includegraphics[width=0.5\textwidth]{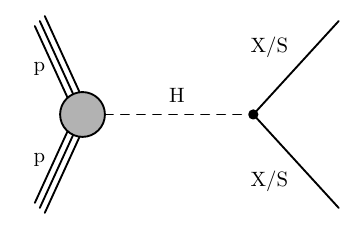}
\caption{Diagram for the $\PH \to \PX\PX$ or $\PH \to \PS\PS$ process.}
\label{fig:HtoXX_feynmanDiagrams}
\end{figure}

Figure~\ref{fig:HtoXX_feynmanDiagrams} shows the diagram for the $\PH \to \PX\PX$ or $\PH \to \PS\PS$ process. This scenario is a simplified model motivated by the twin Higgs scenario~\cite{Chacko:2005pe,Curtin:2015fna,Cheng:2015buv}. Generically, we define \PH to be an SM-like Higgs boson with a mass of 125\GeV (that undergoes an exotic decay) or a heavier BSM Higgs boson. The \PX particle can be a generic LLP. In most $\PH \to \PX\PX$ scenarios discussed in this paper, however, the \PX is specifically a long-lived scalar particle, denoted \PS. The masses of the \PH and \PX/\PS, as well as the lifetime and decay mode of the \PX/\PS, are specified, according to the context, in Sections~\ref{sec:LLPtriggers} and \ref{sec:trigPerf}.

\begin{figure}[!ht]
\centering  \includegraphics[width=0.5\textwidth]{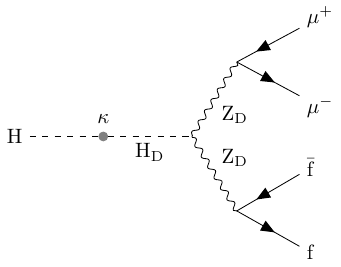}
\caption{Diagram for the $\PH \to \PZD\PZD$ process.}
\label{fig:H_darkPhotons_feynmanDiagrams}
\end{figure}

Figure~\ref{fig:H_darkPhotons_feynmanDiagrams} shows the diagram for the $\PH \to \PZD\PZD$ process, which arises in a hidden Abelian Higgs model (HAHM) \cite{Curtin:2014cca}. Here, \Hdark is the dark Higgs boson (generated with a mass of 400\GeV), which mixes with the 125\GeV SM Higgs boson via the mixing parameter $\kappa$, taken here to be 0.01, and gives mass to the long-lived dark photon \PZD. The \PZD is generated with masses between 10 and 60\GeV, and the $\cPZ$--$\PZD$ kinetic mixing parameter $\epsilon$ is varied between $10^{-7}$ and $2 \times 10^{-9}$. One \PZD is required to decay into two muons, but the decay of the other \PZD is not restricted. The branching fraction of $\PZD \to \Pgm\Pgm$ depends on the mass; for the masses considered in this paper, it is around 10\%. Thus, in about 10\% of the generated $\PH \to \PZD\PZD$ events, the second \PZD decays into muons as well, providing events with four muons.

\begin{figure}[!ht]
\centering  \includegraphics[width=0.5\textwidth]{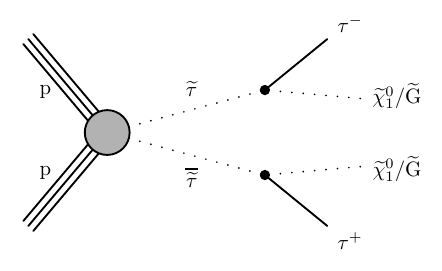}
\caption{Diagram for the direct \PSGt pair production process, with each \PSGt decaying to a \PGt lepton and a neutralino \PSGczDo (or a gravitino \PXXSG).}
\label{fig:staustau_feynmanDiagram}
\end{figure}

Figure~\ref{fig:staustau_feynmanDiagram} shows the diagram for a gauge-mediated SUSY breaking (GMSB) model where a pair of tau sleptons (\PSGt) is directly produced, and each \PSGt subsequently decays to a \PGt lepton and a neutralino \PSGczDo or gravitino \PXXSG. The \PSGt is generated for masses between 100 and 500\GeV and \cTau values of 0.01 to 1\unit{m}.

\begin{figure}[!ht]
\centering  \includegraphics[width=0.5\textwidth]{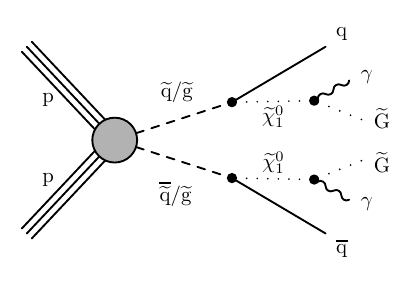}
\caption{Diagram for the GMSB SPS8 benchmark model, where pair-produced squarks \PSQ and gluinos \PSg undergo cascade decays and eventually produce a gravitino \PXXSG.}
\label{fig:GMSBSPS8_feynmanDiagram}
\end{figure}

Figure~\ref{fig:GMSBSPS8_feynmanDiagram} shows an example diagram for another GMSB SUSY model, commonly referred to as the ``Snowmass Points and Slopes 8'' (SPS8) benchmark model~\cite{Allanach:2002nj}, in which pair-produced squarks and gluinos undergo cascade decays and eventually produce a gravitino \PXXSG, which is the lightest SUSY particle, and as such is stable and weakly interacting. In this benchmark model, the lightest neutralino \PSGczDo is the next-to-lightest SUSY particle, and its nature largely determines the phenomenology of the cascade decay chains. The mass of the \PSGczDo is linearly related to the effective scale of SUSY breaking $\Lambda$. Depending on the value of $\Lambda$, the coupling of the \PSGczDo to the \PXXSG could be weak and lead to long \PSGczDo lifetimes. A benchmark $\Lambda$ value of 100\TeV is generated for \cTau values of 10 and 1000\cm. The \PSGczDo dominantly decays to \PXXSG and a photon, resulting in a final state with one or two photons and \ptmiss. The dominant squark-pair and gluino-pair production modes also result in additional energetic jets.

\begin{figure}[!ht]
\centering  \includegraphics[width=0.5\textwidth]{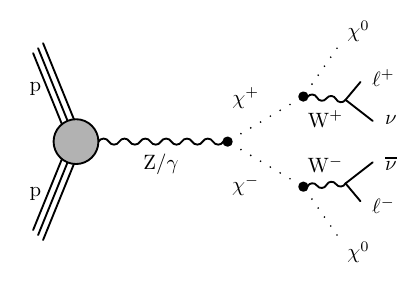}
\caption{Diagram for the singlet-triplet Higgs portal dark matter benchmark model, where a charged dark partner particle \PARchipm in a compressed dark sector decays via an off-shell \PW boson and produces a stable neutral dark matter candidate \PARchizero.}
\label{fig:singletTripletHiggsPortalDM_feynmanDiagram}
\end{figure}

Figure~\ref{fig:singletTripletHiggsPortalDM_feynmanDiagram} shows a diagram for a compressed dark-sector scenario containing a dark matter candidate \PARchizero and a charged dark partner \PARchipm with a small mass splitting~\cite{Blekman:2020hwr}. The small mass splitting leads to soft visible decay products in the decay $\PARchipm \to \PARchizero \Pell \PGn$ via an off-shell \PW boson, while the long lifetime is controlled by the small coupling or mixing that governs the transition between the charged partner and the dark matter state. At the LHC, this benchmark model would produce low-\pt particles (\eg, leptons, as used here) and \ptmiss. Simulated samples are generated for a benchmark \PARchipm mass of 220\GeV, a mass splitting of 20\GeV, and \PARchipm \cTau values of 3, 30, and 300\cm.

\section{Long-lived particle trigger algorithms and efficiencies}
\label{sec:LLPtriggers}

In this section, we describe the CMS LLP trigger program, namely, the trigger algorithms specifically designed to collect events that could contain LLPs. The rates provided throughout this publication correspond to an instantaneous luminosity of \instL{2.1}, unless otherwise specified. The ``pure'' rate of a trigger, corresponding to events that only pass this trigger and do not pass other triggers, is sometimes indicated, but usually the ``total'' rate of a trigger, corresponding to all events that pass this trigger, regardless of whether or not they are also selected by other triggers, is presented.

The total and pure rates of LLP triggers are shown in Table~\ref{tab:triggerRates}. The rates are calculated from a single run taken on October 4, 2024 (run 386593), which had a constant instantaneous luminosity of \instL{2.1} and a mean PU of 63.6. These instantaneous luminosity and mean PU values were typical of the maximum output of the LHC during a $\Pp\Pp$ collision fill in 2024, although the peak instantaneous luminosity achieved in this year was \instL{2.3}. This run lasted 4 hours and 17 minutes, during which time an integrated luminosity of 299\pbinv was recorded. In this table, the pure rates correspond to the unique events added from the logical OR of all the dedicated LLP HLT paths, on top of the rate from the logical OR of the rest of the CMS HLT paths. The pure rate, therefore, quantifies the rate of completely new events recorded by these triggers that would otherwise not be collected by CMS. The total rate of the LLP triggers accounts for about 10, 5, and 18\% of the total rate collected by CMS in the standard, parking, and scouting data-taking streams, respectively. Most of the rate saved by LLP triggers is pure, constituting around 79, 78, and 90\% of the total rate of LLP triggers in the standard, parking, and scouting data-taking streams, respectively, further demonstrating the extension of the CMS physics sensitivity achieved with these triggers. In addition to providing the total and pure rates in the standard, parking, and scouting data-taking streams, Table~\ref{tab:triggerRates} reports the rates for the logical OR of the standard and parking data-taking streams, which represents the rate of fully reconstructed events. The overlap between standard and parking LLP triggers, which is concentrated in the displaced- and delayed-jet paths, has been removed for recent data taking. In particular, all displaced- and delayed-jet paths that were previously in the standard data-taking stream were moved to the parking data-taking stream by the start of data taking in 2025.

\begin{table}[htbp]
\centering
\topcaption{The total and pure HLT rate of all dedicated LLP triggers, as calculated from 2024 data for an instantaneous luminosity of \instL{2.1}, with a mean PU of 63.6. Rates are reported for triggers in the standard, parking, and scouting data-taking streams separately, as well as for the combined standard and parked data to show the rate of all events that are fully reconstructed. The total rates include events that may or may not have been selected by other triggers, while the pure rates correspond to events that pass dedicated LLP HLT paths and do not pass non-LLP HLT paths. The pure rates are measured separately with standard and parked data. The combined rates are less than the sum of the separate standard and parking rates because some events overlap. Each rate shown has a statistical uncertainty of less than 1\%. \label{tab:triggerRates}}
\begin{tabular}{lrr}
Data stream & \cen{Total rate [{Hz}]} & \cen{Pure rate [{Hz}]}\\ [\cmsTabTinySkip] \hline &\\[-\cmsTabTinySkipInverse]
Standard & 393 & 311 \\
Parking & 234 & 182 \\
Scouting & 4200 & 3800 \\ [\cmsTabSkip]
Full reconstruction: standard or parking & 586 & 389 \\
\end{tabular}
\end{table} 

The various LLP trigger algorithms are outlined in Table~\ref{tab:triggerlist}. In this section, we describe the algorithms and any important updates or new triggers introduced for the Run~3 data-taking period, show their rates and efficiencies, and mention the analyses that make use of these triggers. The trigger efficiencies shown throughout this section are usually measured in simulation or control regions in data, and the efficiencies are typically shown with respect to offline object quantities, such as jet \pt or muon \dzero. Later, Section~\ref{sec:trigPerf} gives the acceptance of the different LLP triggers in signal simulation within a defined fiducial region as a function of LLP \cTau or decay position, to demonstrate the complementarity of the different triggers. We first summarize the new LLP triggers introduced for Run~3 in Section~\ref{sec:newTriggersForRun3}. We then provide the descriptions of the algorithms, organized into four subsections: tracker-based algorithms (Section~\ref{sec:trackerLLPtriggers}), calorimeter-based algorithms (Section~\ref{sec:caloLLPtriggers}), muon spectrometer-based algorithms (Section~\ref{sec:muonLLPtriggers}), and ``No-BPTX'' algorithms that are active in bunch crossings without proton bunches detected by the beam pickup timing device (BPTX), which measures the timing of proton bunches as they pass on either side of the CMS interaction point (Section~\ref{sec:NoBPTXtriggers}).

\begin{table}[htbp]
\centering
\topcaption{The LLP triggers and their total HLT rates in Run~3, calculated from 2024 data for an instantaneous luminosity of \instL{2.1}, with a mean PU of 63.6. Triggers implemented for the first time in Run~3 are indicated by a dagger ($^\dagger$). Rate values in parentheses correspond to the parked data rates; all others are standard data rates, except for dimuon scouting. Nearly all rates shown have a statistical uncertainty of less than 1\unit{Hz}. ``Disp.'' is an abbreviation for ``displaced''; ``req.'' for ``requirement''; and ``NHEF'' for ``neutral hadron energy fraction''. The terms used in this table are explained in the corresponding subsection within Section~\ref{sec:LLPtriggers}.}
\cmsTable
{\renewcommand{\arraystretch}{1.1}
\begin{tabular}{lll}
Triggered signature &  Trigger description & HLT rate [{Hz}]\\ [\cmsTabTinySkip] \hline &\\[-\cmsTabTinySkipInverse]
\hyperref[sec:disappearingtracktriggers]{Disappearing track} & \begin{tabular}[l]{@{}l@{}} $\ptmiss > 105\GeV$ + ${\geq}1$ isolated track ($\pt>50\GeV$) \end{tabular} & 4 \\ [\cmsTabSkip]
\hyperref[sec:displacedtau]{Disp. tau}  & \begin{tabular}[l]{@{}l @{}} ${\geq}2$ disp. \tauh ($\pt>32\GeV$, $\dzero > 0.005\cm$)$^\dagger$ \\ ${\geq}1$ disp. \tauh ($\pt>24\GeV$) + ${\geq}1$ \PGm ($\pt>24\GeV$)$^\dagger$ \\ ${\geq}1$ disp. \tauh ($\pt>34\GeV$) + ${\geq}1$ \Pe ($\pt>34\GeV$)$^\dagger$ \end{tabular} & 36 \\ [\cmsTabBigSkip]
\hyperref[sec:displacedjetstracking]{Disp. jet} &  \begin{tabular}[l]{@{}l@{}} ${\geq}2$ jet ($\pt>40\GeV$, inclusive tagging req.) + $\HT>430\GeV$ \\ ${\geq}2$ jet ($\pt>40\GeV$, disp. tagging req.) \\ + $\HT>240\GeV$ + ${\geq}1$ L1 \Pgm ($\pt>6\GeV$) \end{tabular}   & 53 (163) \\[\cmsTabBigSkip]
\hyperref[sec:displacedjetshcal]{\begin{tabular}[l]{@{}l@{}}HCAL-based disp. \\ and delayed jet\end{tabular}} & \begin{tabular}[l]{@{}l@{}} ${\geq}2$ jet ($\pt>40\GeV$, displ. tagging req.) + $\HT>170\GeV^\dagger$ \\ ${\geq}2$ jet ($\pt>40\GeV$, inclusive tagging req.) + $\HT>200\GeV^\dagger$ \\ ${\geq}1$ jet ($\pt > 60\GeV$, NHEF $> 0.7$) + $\HT>200\GeV^\dagger$\end{tabular} & 35 \\[\cmsTabBigSkip]
\hyperref[sec:delayedjetsecal]{ECAL-based delayed jet} & ${\geq}1$ inclusive and trackless jet$^\dagger$ & 37 (77) \\[\cmsTabLittleSkip]
\hyperref[sec:delayeddiphoton]{Delayed diphoton} & ${\geq}2$ ECAL superclusters (time ${>}1\unit{ns}$)$^\dagger$ & 15 \\[\cmsTabLittleSkip]
\hyperref[sec:displacedphotonandht]{Disp. photon + \HT} & ${\geq}1$ \PGg ($\pt>60\GeV$) + PF $\HT>350\GeV$ & 12 \\[\cmsTabLittleSkip]
\hyperref[sec:displaceddimuon]{\begin{tabular}[l]{@{}l@{}}Disp. single-muon \\ and dimuon\end{tabular}} &  \begin{tabular}[l]{@{}l@{}} ${\geq}2$ L2 \PGm ($\pt>10\GeV$, $\dzero>1\cm$)$^\dagger$ \\ ${\geq}2$ L3 \PGm ($\pt>16, 10\GeV$, $\dzero>0.01\cm$)$^\dagger$ \\ ${\geq}2$ L2 \PGm ($\pt>23\GeV$) \\ ${\geq}1$ L2 \PGm ($\pt>50\GeV$, $\dzero>1\cm$)$^\dagger$ \\ ${\geq}1$ L3 \PGm ($\pt>30\GeV$, $\dzero>0.01\cm$)$^\dagger$ \end{tabular}  & 165 \\[\cmsTabBiggestSkip]
\hyperref[sec:displacedL3muon]{Double disp. L3 muon} & ${\geq}2$ L3 \Pgm ($\pt>43\GeV$) & $2$ \\[\cmsTabLittleSkip]
\hyperref[sec:displacedL3muonandphoton]{Disp. L3 muon + photon} & \begin{tabular}[l]{@{}l@{}} ${\geq}1$ L3 \Pgm ($\pt>43\GeV$) + \PGg ($\pt>43\GeV$) \\ ${\geq}1$ L3 \Pgm ($\pt>38\GeV$, $\dzero>1\cm$) + \PGg ($\pt>38\GeV$) \end{tabular} & $5$ \\[\cmsTabSkip]
\hyperref[sec:dimuonscouting]{Dimuon scouting} &  ${\geq}2$ scouting \Pgm ($\pt>3\GeV$)  & 4200 \\[\cmsTabLittleSkip]
\hyperref[sec:mdsshowers]{MDS in CSCs} &\begin{tabular}[l]{@{}l@{}}
${\geq}1$ CSC cluster (${\geq}200/500$ hits in outer/inner rings)$^\dagger$\\
${\geq}2$ CSC clusters (${\geq}75$ hits)$^\dagger$\\
\end{tabular}  & 14 \\[\cmsTabSkip]
\hyperref[sec:mdsshowers]{MDS in CSCs + X} &\begin{tabular}[l]{@{}l@{}}
${\geq}1$ CSC cluster (${\geq}100$ hits) + ${\geq}1$ \Pe ($\pt > 5\GeV$)$^\dagger$ \\
${\geq}1$ CSC cluster (${\geq}100$ hits) + ${\geq}1$ L3 \Pgm ($\pt > 5\GeV$)$^\dagger$ \\
${\geq}1$ CSC cluster (${\geq}100$ hits) + ${\geq}1$ \tauh ($\pt > 10\GeV$)$^\dagger$ \\
${\geq}1$ CSC cluster (${\geq}50$ hits) + ${\geq}1$ \PGg ($\pt > 20\GeV$)$^\dagger$ \\
\end{tabular}  & 14 \\[\cmsTabBiggerSkip]

\hyperref[sec:dtcluster]{MDS in DTs} & 
\begin{tabular}[l]{@{}l@{}}
L1 $\ptmiss > 150 \GeV$ + ${\geq}1$ DT cluster (${\geq}50$ hits)$^\dagger$ \\
${\geq}1$ L1 CSC cluster  + ${\geq}1$ DT cluster (${\geq}50$ hits)$^\dagger$ \\

\end{tabular}
& 9 \\[\cmsTabSkip]
\hyperref[sec:nobptxjets]{Jet No-BPTX} & ${\geq}1$ out-of-time jet ($E > 60\GeV$) & 1 \\[\cmsTabLittleSkip]
\hyperref[sec:nobptxmuons]{Muon No-BPTX} & ${\geq}1$ out-of-time L2 \Pgm ($\pt > 40\GeV$) & 7 \\
\end{tabular}
}
\label{tab:triggerlist}
\end{table}

\subsection{Summary of new long-lived particle triggers for Run~3} \label{sec:newTriggersForRun3}

At the start of Run~3, a few hundred Hz of additional trigger rate at the HLT was allocated for new LLP triggers. The overall CMS physics program was significantly expanded to explore this type of new and unconventional physics signatures. We mention which triggers are new for Run~3 in the following, and then describe the entire LLP trigger program in detail, starting from Section~\ref{sec:trackerLLPtriggers}. Additional information can be found in Refs.~\cite{LLPRun3TriggerDPNote,LLPRun3HCALDPNote,HMTnote,HMTnote22To23,DisplacedPhotonsPlusHT2017DPNote}.

The HLT paths for hadronically decaying displaced tau leptons are new for Run~3. These paths, which feature a dedicated reconstruction algorithm, are described in Section~\ref{sec:displacedtau}.

In addition, there are now several flavors of displaced- and delayed-jet triggers available at the HLT in Run~3. As described in Section~\ref{sec:displacedjetstracking}, some of the displaced-jet triggers were already available to a certain extent in Run~2, but have benefited from major improvements over time, while the delayed-jet triggers, described in Sections~\ref{sec:displacedjetshcal} and \ref{sec:delayedjetsecal}, are completely new for Run~3. The signal efficiency for the displaced-jet triggers was improved (Section~\ref{sec:displacedjetstracking}), especially for low-mass LLPs by making particular use of the new L1Ts that use HCAL timing and depth information (Section~\ref{sec:displacedjetshcal}). There are also new HLT paths that exploit the timing of the electromagnetic calorimeter (Section~\ref{sec:delayedjetsecal}). Since these new features and variables are introduced for the first time at the trigger level in Run~3, they are largely not employed together in a single trigger. For example, the delayed-diphoton triggers do not use ECAL shower shape variables alongside the ECAL timing variable. This allows for the effects of these variables to be disentangled and studied separately. Optimizing the possible use of multiple new variables in a single trigger may be studied in the future.

Triggers for displaced muons at the HLT have also been improved in Run~3. Displaced dimuon high-level triggers take L1 muons with low \pt thresholds and new displaced double muon triggers as input. These L1 muons feed into several types of displaced dimuon triggers at the HLT, designed to cover a wide range of displacements and improve the signal efficiency over that of Run~2. These displaced dimuon triggers are further described in Section~\ref{sec:displaceddimuon}.

In addition, as described in Section~\ref{sec:dimuonscouting}, the dimuon scouting triggers were improved in Run~3 to increase their efficiency for displaced muons.

Neutral LLPs with particularly long lifetimes could decay beyond the calorimeters, creating a high-multiplicity shower in the muon system. Such a muon detector shower (MDS) is expected to consist of hundreds of hits with no reconstructed tracks or jets in the direction of the cluster of hits. New MDS triggers at L1 have been developed for Run~3 to collect these high-multiplicity events in the CSCs and provide input to triggers at the HLT. Several MDS HLT paths were developed, reconstructing clusters in both the CSCs (described in Section~\ref{sec:mdsshowers}) and the DTs (described in Section~\ref{sec:dtcluster}).

\subsection{Tracker-based algorithms}
\label{sec:trackerLLPtriggers}

\begin{sloppypar}Here, LLP triggers that primarily employ tracking algorithms are described. These include \ptmiss-based triggers for disappearing tracks (Section~\ref{sec:disappearingtracktriggers}), displaced-tau triggers (Section~\ref{sec:displacedtau}), and displaced-jet triggers (Section~\ref{sec:displacedjetstracking}). The ATLAS experiment has also developed Run~3 inner-detector trigger strategies for unconventional tracking signatures, including displaced tracks, displaced jets, and short disappearing tracks~\cite{ATLAS:2024xna,Mlinarevic:2024ntj}.\end{sloppypar}

\subsubsection{\texorpdfstring{The \ptmiss}{Missing transverse momentum}-based disappearing-track triggers}
\label{sec:disappearingtracktriggers}

\begin{figure}[htb!]
\centering
\includegraphics[width=0.4\textwidth]{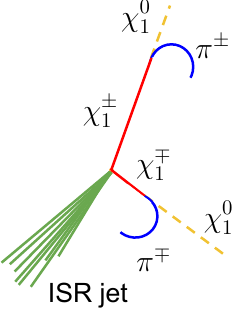} 
\caption{Diagram of a disappearing-track event. Each \Pchipm (solid red lines) is an LLP that decays in the tracker and recoils off an ISR jet (multiple solid green lines), while the \Pchiz (dashed yellow line) is undetected, and the \PGppm (blue curve) is not reconstructed because of its low \pt. The recoil of the disappearing track against the ISR jet produces measurable calorimeter and PF \ptmiss.}
\label{fig:disTrkEvDisplay}
\end{figure}

For the disappearing-track analysis~\cite{CMS:2020atg}, specialized trigger algorithms were developed in Run~2. The analysis searches for a disappearing-track signature, \ie, a track made by a charged particle that lives sufficiently long to decay into undetectable particles inside the volume of the tracker. This track can recoil off a jet produced from initial-state radiation (ISR) and is highly isolated, producing a small amount of deposited energy in the calorimeter (\Ecalo) associated with the track, calculated in a cone of $\DR <0.5$ around it, and at least three missing hits in the outermost region of the tracking detector. Tracks are described as having a missing hit if they are reconstructed as passing through a functional tracker layer, but no hit in that layer is associated with the track. A missing hit is described as ``inner'' if the missing layer is between the IP and the innermost hit of the track, ``middle'' if between the innermost and outermost hits of the track, and ``outer'' if it is beyond the outermost hit of the track. It is possible to have more than one hit per layer. Offline, the disappearing track must have a large number of missing outer hits. Given that there is no tracking in the L1T, a disappearing-track trigger must rely on only calorimeter \ptmiss (calo \ptmiss), and a combination of \ptmiss and an isolated track at the HLT. A diagram of a signal event is shown in Fig.~\ref{fig:disTrkEvDisplay}. Disappearing-track signatures are also targeted by ATLAS using triggers based on large \ptmiss, typically in association with high-\pt ISR jets~\cite{ATLAS:2022rme}. For Run~3, ATLAS has also developed inner-detector tracking triggers for unconventional signatures, including short tracks that disappear within the tracking volume~\cite{Kerridge:2023fns}.

The CMS disappearing-track triggers in Run~3 use the logical OR of several L1 seeds to select events with relatively high \ptmiss, that is, $\ptmiss>70\GeV$. The HLT algorithm starts with the calo \ptmiss reconstruction and requires $\ptmiss > 105\GeV$ to reject a large fraction of events and increase the speed of the algorithm. For events that pass this requirement, isolated tracks are reconstructed and required to have:

\begin{itemize}
    \item $\Ecalo < 100\GeV$;
    \item relative isolation (\ireltrk), quantified by requiring the scalar sum of the \pt of all other tracks within $\DR < 0.3$ of the candidate track must be ${<}5\%$ of the candidate track \pt ($\ireltrk<0.05$);
    \item number of missing inner hits $=0$;
    \item number of missing middle hits $=0$;
    \item total number of hits $>5$;
    \item $\pt > 50\GeV$.
\end{itemize}

Thus, the HLT path only requires an isolated, high-\pt track associated with \ptmiss; it does not require that the track ``disappears'' and has missing outer hits. The total rate of this HLT path is 4\unit{Hz}.

The efficiency of the HLT path is evaluated by selecting events that contain significant \ptmiss (coming from the neutrino) and an isolated track (coming from the muon), using data containing at least one muon, and a simulated \WtoLNu sample. The efficiency is also calculated with the AMSB signal benchmark model, whose associated diagram is shown in Fig.~\ref{fig:AMSB_feynmanDiagrams} and described in Section~\ref{sec:dataMC}, for three different values of LLP \cTau. Given the small difference between the chargino and the neutralino masses ($\Delta m(\Pchipm,\Pchiz) \sim m_{\PGppm}$), the outgoing SM particle does not have sufficient energy for it to be reconstructed as a track in the detector, and the chargino \cTau is large enough for it to travel through the inner layers of the tracker subdetector. In addition, the neutralino does not interact with the CMS detector material. This signature is the target of the disappearing-track analysis: a charged particle track that vanishes inside the volume of the tracker.

To calculate the efficiency, charginos with a mass of 900\GeV, the smallest chargino mass not yet excluded by the disappearing-track search, are used, together with three distinct \cTau values: 10, 100, and 1000\cm. The efficiency is calculated as the number of events passing a given HLT path, divided by the total number of events. Table~\ref{tab:muonTrigSel} shows the offline requirements applied when calculating the efficiency in data and simulation. To study the trigger efficiency, requirements are applied to data and simulated \WtoLNu events to select a control sample of muons. These \WtoLNu events are chosen since they contain an isolated, high-\pt track associated with \ptmiss, which reproduces the HLT path selection criteria. Different requirements are applied to the signal simulation to obtain a sample of candidate disappearing tracks. This second set of requirements emulates the offline analysis requirement for the events to have an isolated, high-\pt track. In the table, the tight muon identification~\cite{CMS:2018rym} refers to global muons that have a certain number of muon chamber hits included in the global-muon track fit, pixel tracker layers, tracker layers, and matched muon station segments, specifically ${>}1$, ${>}1$, ${>}10$, and ${>}2$, respectively. Tight muons are also required to be consistent with the PV by having $\dzero<0.2\cm$ and $\chi^{2}/\Ndof<10$ with respect to the global muon fit, where \Ndof is the number of degrees of freedom of the fit. The identification of tracks with high purity is performed with a deep neural network~\cite{Giannini:2023vjo}. This deep neural network algorithm makes use of several kinematic and geometric parameters of the track and its hits, and its identification value is assigned based on the score of the network.

\begin{table}[htbp]
\centering
\topcaption{Offline event requirements for the disappearing-track trigger efficiency calculation, for data and \WtoLNu simulation (\cmsLeft) and signal simulation (\cmsRight). The selection criteria are applied sequentially over the given objects, and an event is selected if at least one object per event passes all requirements.}
\begin{tabular}{llll}
\multicolumn{2}{l}{Data and \WtoLNu simulation} & \multicolumn{2}{l}{Signal simulation} \\
Object & Selection & Object & Selection \\[\cmsTabTinySkip]
\hline & \\[-\cmsTabTinySkipInverse]
${\geq}1$ HLT muon & $\pt>24\GeV$  & ${\geq}1$ track & $\abseta < 2.5$ \\
 & passing isolation &   & passing high purity ID \\
${\geq}1$ muon & $\pt > 55\GeV$ &   & $\dzero <0.02\cm$  \\
 & $\abseta < 2.1$ &  & $\dz <0.5\cm$ \\
 & passing tight muon ID  &   & hits ${\geq} 4$ \\
 & $\ireltrk < 0.15$ &  & missing inner hits $=0$ \\
 & missing inner hits $=0$ &  & missing middle hits $=0$ \\
 & missing middle hits $=0$ &  & $\ireltrk < 0.01$ \\
 & match HLT muon ($\DR < 0.1$) & & \\
\end{tabular}
\label{tab:muonTrigSel}
\end{table}

\begin{figure}[htb!]
\centering
\includegraphics[width=0.47\textwidth]{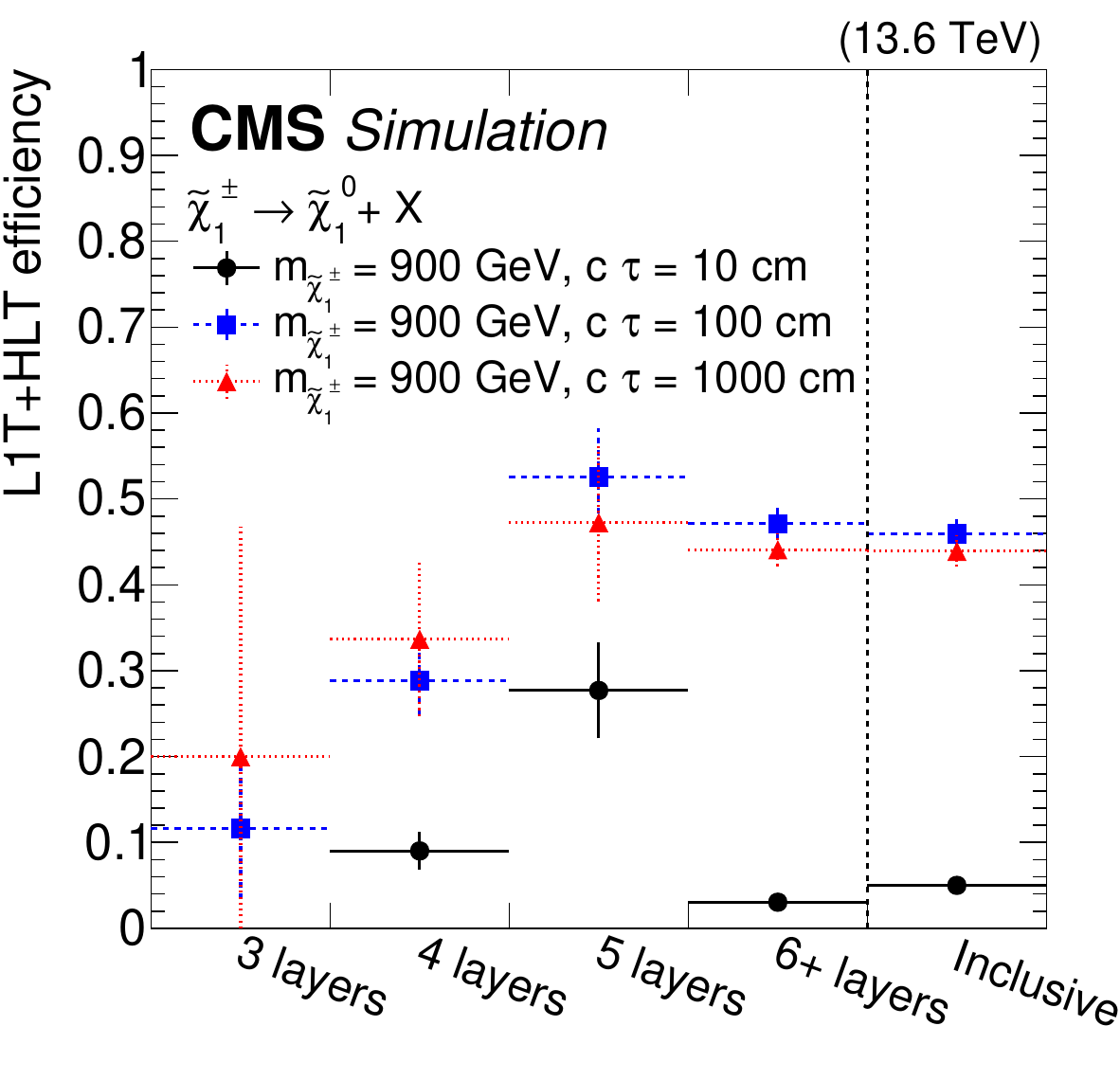} \hspace{1em}
\includegraphics[width=0.46\textwidth]{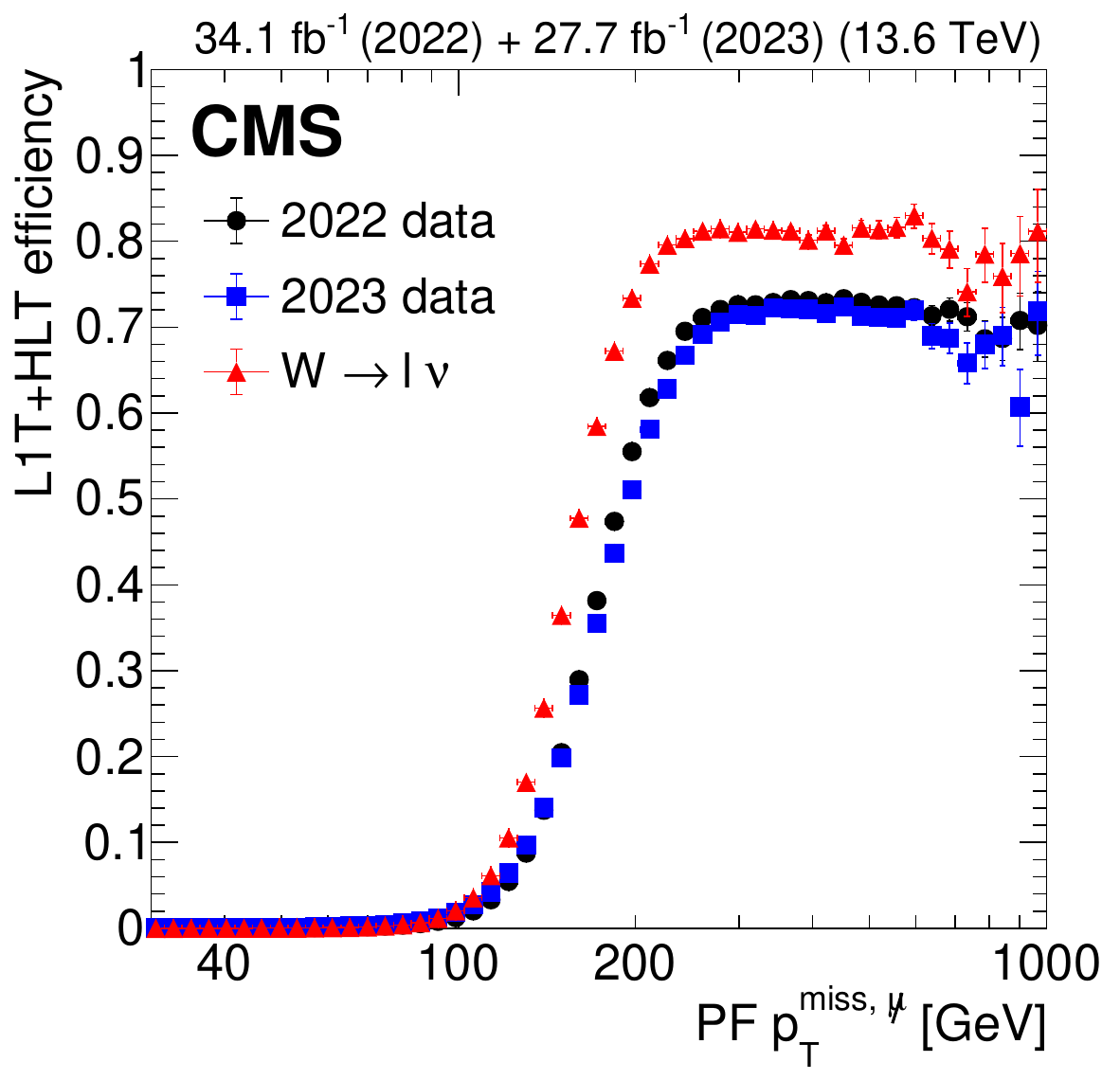} 
\caption{The L1T+HLT efficiency of the disappearing-track trigger: Efficiency as a function of the number of tracker layers with valid measurements for tracks that pass the offline requirements described in Table~\ref{tab:muonTrigSel}, in $\Pchipm \to \Pchiz+\PX$ simulated events for 2022 conditions, where $m_{\Pchipm}=900\GeV$ and \Pchiz is nearly mass-degenerate with \Pchipm (\cmsLeft). The efficiency is shown for LLPs with $\cTau=10$, 100, and 1000\cm in black circles, blue squares, and red triangles, respectively. Comparison of efficiencies calculated with 2022 data (black circles), 2023 data (blue squares), and \WtoLNu simulation (red triangles), as a function of offline reconstructed PF \metNoMu (\cmsRight). The efficiency rises to a plateau of less than 100\% because of the isolated track leg of the algorithm.}
\label{fig:disTrkEffs}
\end{figure}

Figure~\ref{fig:disTrkEffs} (\cmsLeft) shows the overall efficiency of the main disappearing-track HLT path in $\Pchipm \to \Pchiz+\PX$ simulated events, for $\cTau=10$, 100, and 1000\cm. The efficiency is shown without applying corrections that improve the agreement between the simulation and the data. The events in this figure are required to have $\metNoMu> 120\GeV$ (described in Section~\ref{sec:hadronsEnergySumsMET}) to replicate the disappearing-track search selection. The efficiency is plotted as a function of the number of layers with valid measurements of the track that pass the offline selection. This illustrates the different efficiencies for tracks of different sizes, reflecting the subdivisions of the signal region in the search. The figure shows that the chargino proper lifetime has a large impact on the efficiency. For the 10\cm \cTau line in particular, the efficiency is smaller in the ``4 layers'' than in the ``5 layers'' case, even though most of these tracks would disappear before the fifth layer of the tracker detector, in contrast to the behavior observed for larger \cTau values. This smaller efficiency comes from requiring a minimum of 5 hits in the online tracks. The efficiencies are estimated with offline tracks that are reconstructed using more sophisticated algorithms. This step can cause some of the offline tracks to have 4 layers with measurements, even when the HLT track has 5 hits.

Figure~\ref{fig:disTrkEffs} (\cmsRight) shows the overall efficiency of the main disappearing-track HLT path in 2022 and 2023 data and in the \WtoLNu background simulation. The overall efficiency is defined as the number of events that pass the offline selection, specified in Table~\ref{tab:muonTrigSel}, combined with events satisfying the disappearing-track HLT path (including the L1T, HLT \ptmiss, and HLT isolated-track requirements), divided by the number of events that pass the offline selection alone. Because the disappearing-track triggers use calo \ptmiss, the efficiency is measured with respect to the offline \metNoMu. 

\begin{figure}[htb!]
\centering
\includegraphics[width=0.48\textwidth]{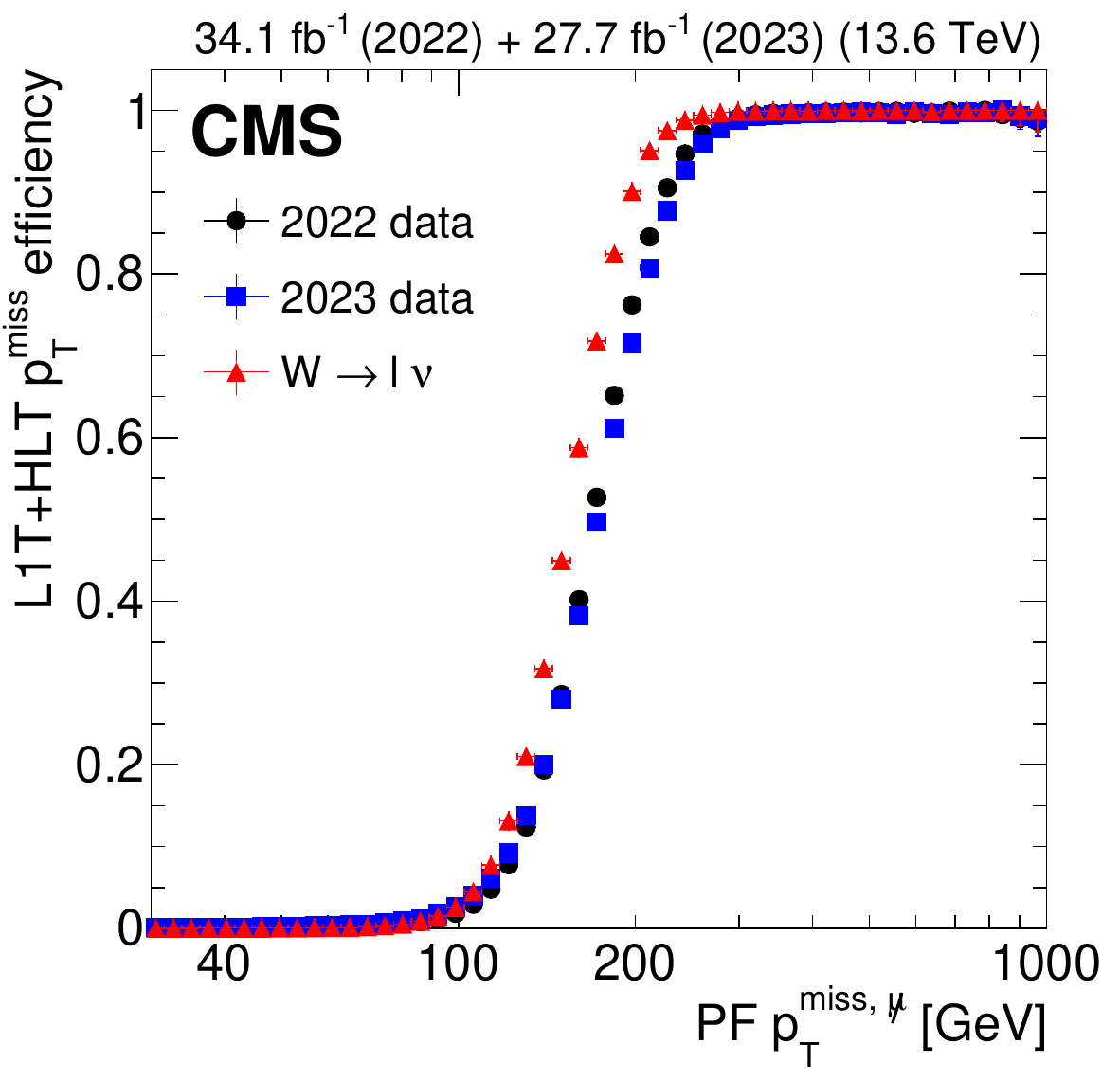} 
\includegraphics[width=0.475\textwidth]{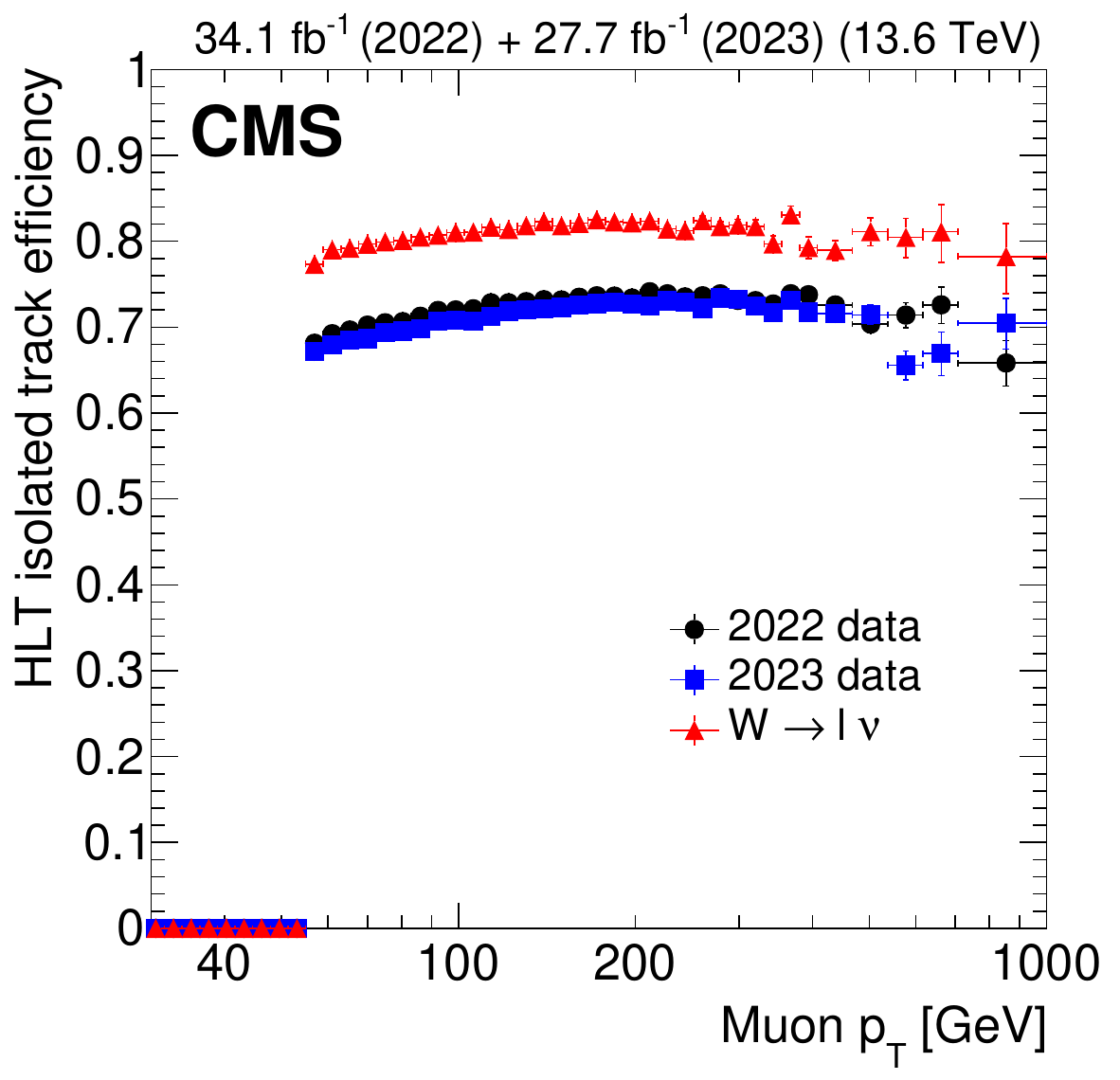}
\caption{The L1T+HLT efficiency of each leg of the disappearing-track trigger in 2022 data (black circles), 2023 data (blue squares), and \WtoLNu simulation (red triangles). Efficiency of the L1T+HLT \ptmiss leg as a function of offline reconstructed PF \metNoMu (\cmsLeft). Efficiency of the full HLT path, taking into account only events that already passed through the \ptmiss leg, as a function of the selected muon \pt (\cmsRight).}
\label{fig:disTrkIndivEffs}
\end{figure}

The overall efficiency is the convolution of the per-leg efficiencies, which are shown in Fig.~\ref{fig:disTrkIndivEffs}. The L1T+HLT \ptmiss-leg efficiency as shown in Fig.~\ref{fig:disTrkIndivEffs} (\cmsLeft) is defined as the number of events that pass the offline selection from Table~\ref{tab:muonTrigSel} and the L1T and HLT \ptmiss requirements, divided by the number of events that pass the offline selection from Table~\ref{tab:muonTrigSel}. The isolated track leg efficiency as shown in Fig.~\ref{fig:disTrkIndivEffs} (\cmsRight) is defined as the number of events that pass the offline selection from Table~\ref{tab:muonTrigSel} and the disappearing-track HLT path (including the L1T, HLT \ptmiss, and HLT isolated-track requirements), divided by the number of events that pass the offline selection from Table~\ref{tab:muonTrigSel} and the L1T and HLT \ptmiss requirements.

The overall efficiency curve in Fig.~\ref{fig:disTrkEffs} (\cmsRight) does not reach 100\% because of the isolated track leg of the HLT path, as shown in Fig.~\ref{fig:disTrkIndivEffs} (\cmsRight). This leg includes the online track reconstruction and the subsequent isolated track selections; the HLT tracking efficiency itself is also below 100\%~\cite{CMS-DP-2024-013}.

The efficiencies in 2022 and 2023 data are very similar, but there is a large discrepancy between data and simulation in Fig.~\ref{fig:disTrkEffs} (\cmsRight) and Fig.~\ref{fig:disTrkIndivEffs} (\cmsRight). This difference comes from higher efficiencies in simulation in the HLT selections on the track isolation and the number of missing inner and middle hits. When measuring these properties offline, 9.2\% more simulated events than data events pass these requirements. The differences between the trigger efficiencies in the simulation and the data are corrected in the offline analysis.

The disappearing-track signature may also be triggered by standard \metNoMu triggers, since those also target events containing a large amount of calo \ptmiss. However, the disappearing-track triggers increase the efficiency to trigger on the disappearing tracks, in part because they lower the threshold on \metNoMu from 110\GeV in the standard \metNoMu triggers to 105\GeV. Using standard \metNoMu triggers in logical OR with the disappearing-track triggers to select simulated signal events provides access to additional parameter space for a small increase in rate, with respect to using standard \metNoMu triggers alone.

\subsubsection{Displaced-tau triggers}
\label{sec:displacedtau}

\begin{figure}[htb!]
\centering
\includegraphics[width=0.7\textwidth]{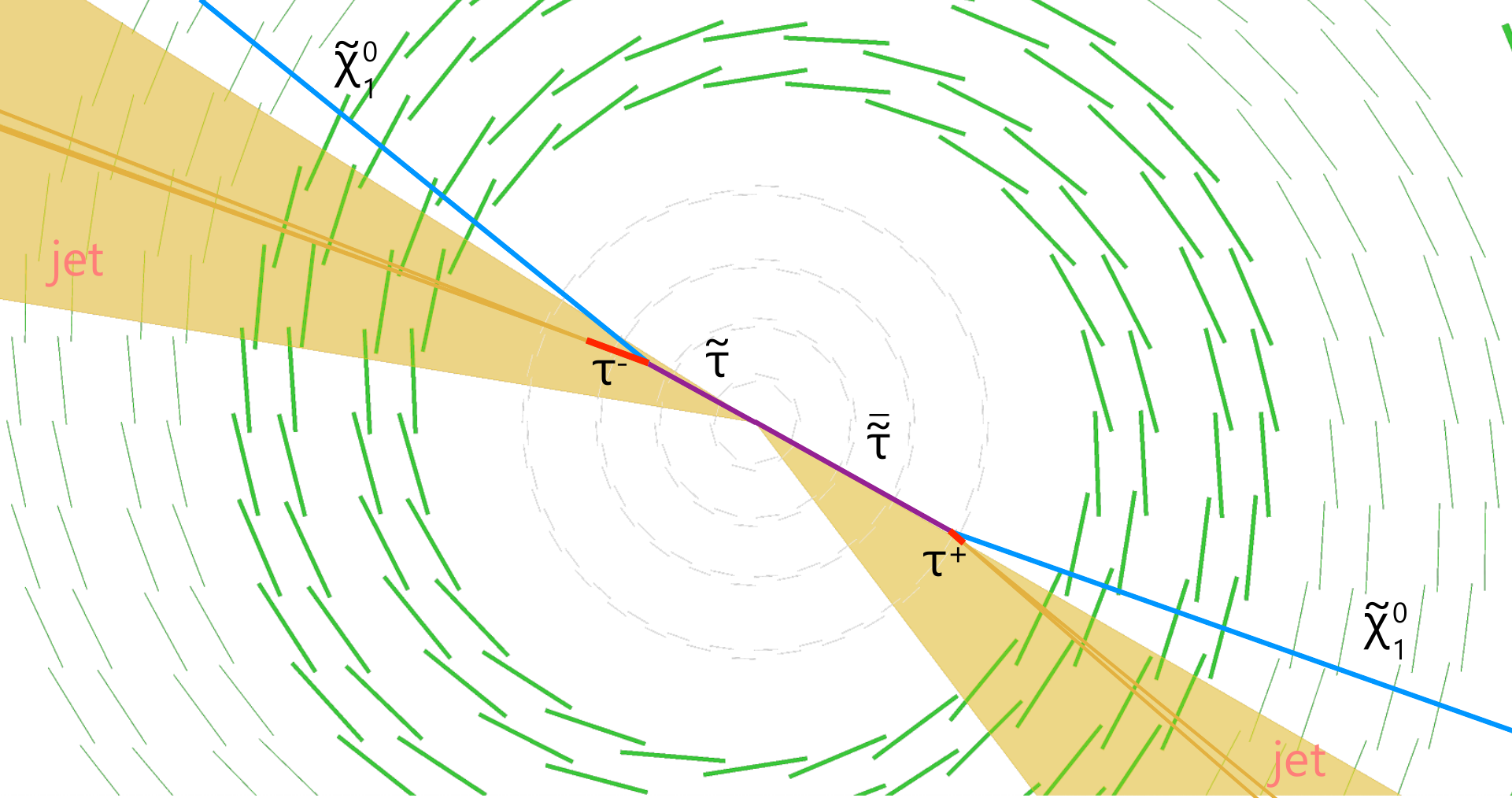} 
\caption{Event display of simulated \PSGt pair production in a GMSB benchmark model, followed by the decay of each \PSGt to a \PGt lepton and a neutralino \PSGczDo, with the \PGt leptons decaying hadronically. The magenta lines indicate the \PSGt particles, the light blue lines indicate the neutralinos, and the red lines indicate the \PGt leptons decaying hadronically. The shaded dark yellow cones show the two reconstructed jets, and the orange lines inside the jets are the hadrons from the \PGt decay. 
}
\label{fig:disTauEventDisplay}
\end{figure}

Displaced tau leptons represent a well-motivated signature in many BSM theories, particularly some GMSB models~\cite{Buican:2008ws,Meade:2008wd,GIUDICE1999419} that include the process described in Section~\ref{sec:dataMC} and shown in Fig.~\ref{fig:staustau_feynmanDiagram}. About one-third of the time, the tau lepton will decay into a displaced muon or displaced electron, and two neutrinos. Dedicated HLT reconstruction techniques are available for displaced muons, as described in Sections~\ref{sec:displaceddimuon} and \ref{sec:displacedL3muon}, while for displaced electrons, the options available at the HLT are to select superclusters as described in Section~\ref{sec:delayeddiphoton} or photons as proxies for displaced electrons, as described in Section~\ref{sec:displacedL3muonandphoton}. Almost all the remaining tau lepton decay possibilities result in final states containing hadrons and a neutrino. The reconstruction of displaced \tauh candidates is particularly challenging, both at the offline and trigger levels. There was no online \tauh reconstruction sequence in CMS addressing this topology in Run~2, and the previous search~\cite{CMS:2022syk} had to rely on a combination of \ptmiss-based triggers and prompt di-$\tauh$ triggers to select events with at least one of the final state tau leptons decaying hadronically. Figure~\ref{fig:disTauEventDisplay} shows an event display of a simulated GMSB benchmark signal event, with two $\tauh$ particles in the final state.

A dedicated set of HLT paths targeting the direct reconstruction of displaced \tauh has been designed and included for the Run~3 data taking. This displaced \tauh algorithm uses a combination of a tracking iteration that uses only hits in the silicon strip detector and not the pixel detector to seed the track reconstruction, the standard tracking, and a modified \tauh reconstruction algorithm.

\begin{figure}[htb!]
\centering
\includegraphics[width=0.49\textwidth]{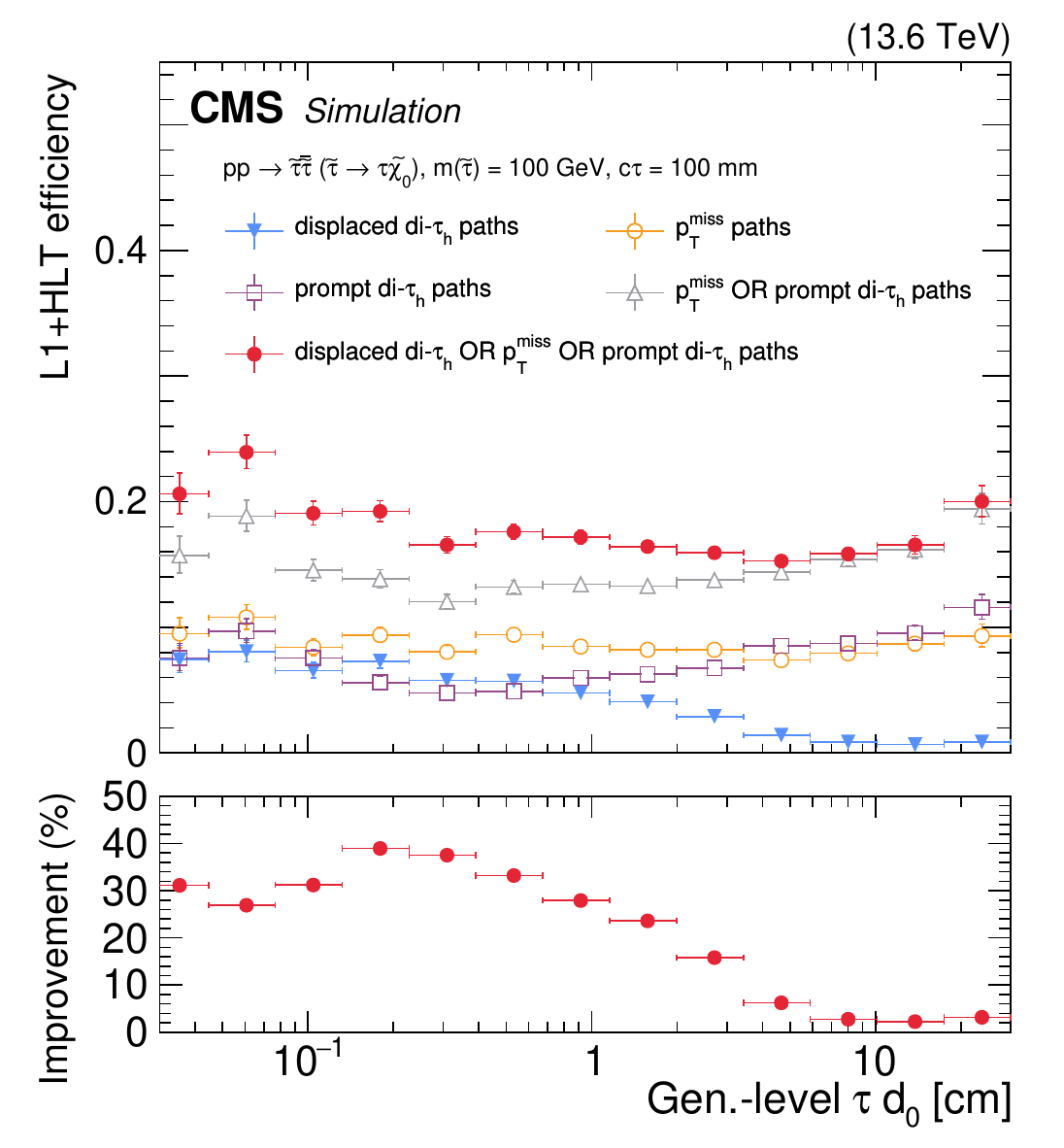}
\includegraphics[width=0.49\textwidth]{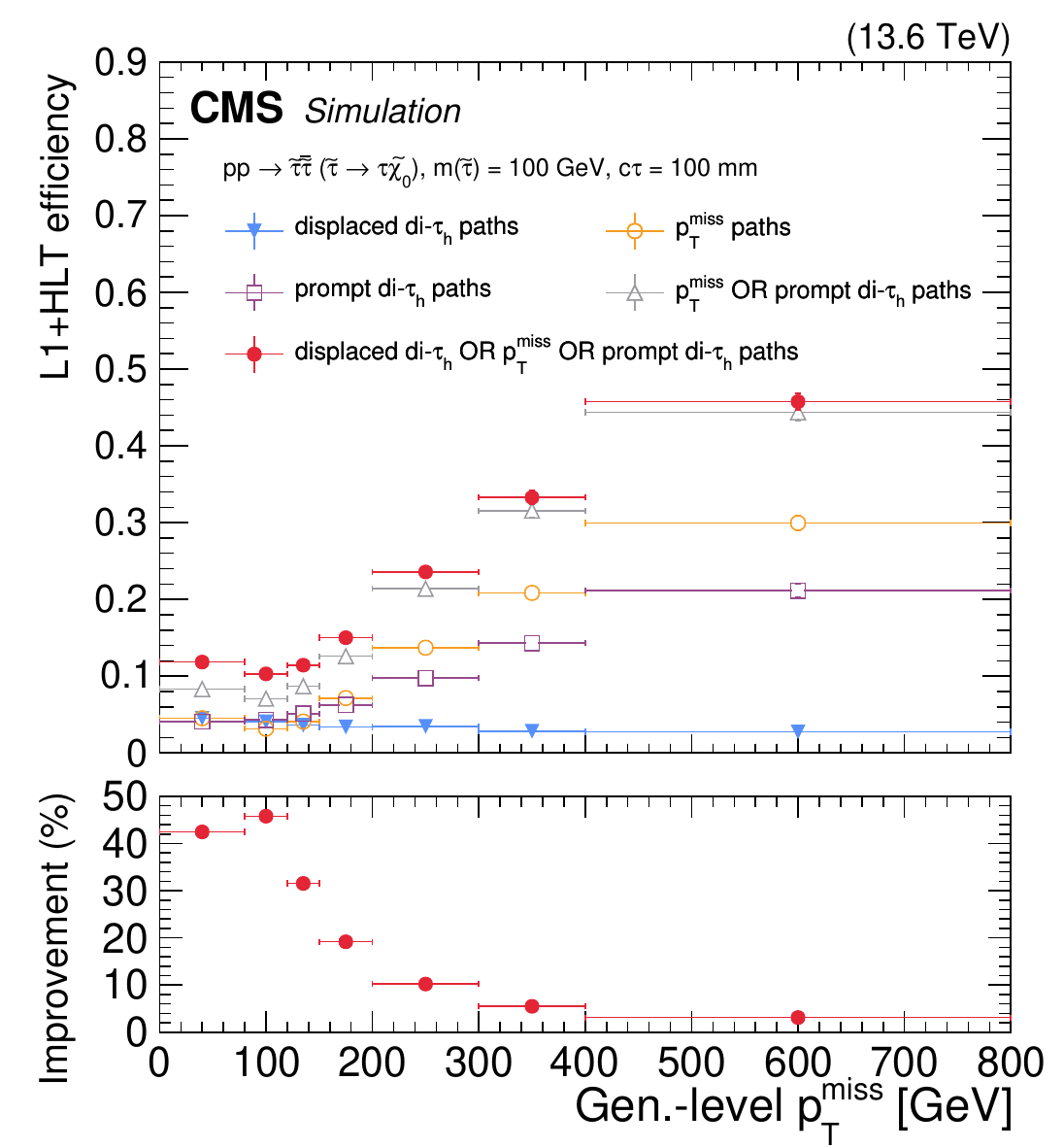}
\caption{The L1T+HLT efficiency of the displaced \tauh trigger, for simulated $\Pp\Pp \to \PSGt\PASGt\,(\PSGt \to \PGt\PSGczDo)$ events, where the \PSGt has $\cTau = 10\cm$ and each \PGt decays hadronically. The efficiency is shown as a function of the \dzero of the highest \pt \PGt lepton in the event (\cmsLeft) and as a function of \ptmiss (\cmsRight). The efficiency is shown for the displaced di-\tauh trigger path (blue filled triangles), the previously available \ptmiss-based paths (orange open circles), the previously available prompt di-\tauh paths (purple open squares), the combination of the \ptmiss-based and prompt di-\tauh paths (gray open triangles), and the combination of the \ptmiss-based, prompt di-\tauh, and displaced di-\tauh paths (red filled circles), using 2022 data-taking conditions. The efficiency is evaluated with respect to events that contain at least two generator-level \PGt leptons, where the visible component of the \PGt lepton \pt is greater than 30\GeV and $\abseta < 2.1$. The lower panels show the ratio (improvement in \%) of the trigger efficiency given by the combination of the displaced di-\tauh trigger path with the \ptmiss-based and prompt di-\tauh paths to that of the combination of the previously available \ptmiss-based and prompt di-\tauh paths.
\label{fig:displacedtau_mc_eff} 
}
\end{figure}

\begin{figure}[hbt!]
\centering
\includegraphics[width=0.49\textwidth]{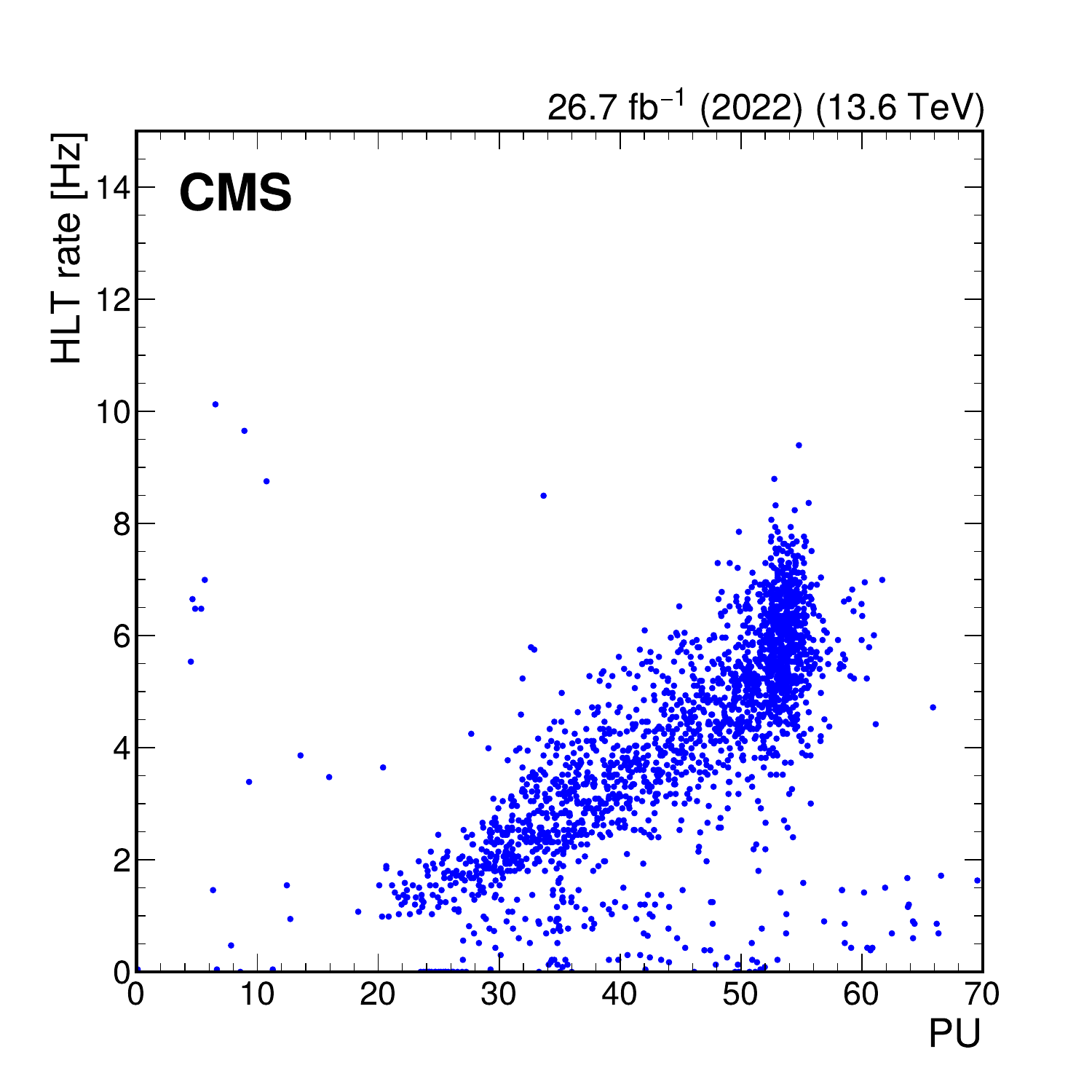}
\includegraphics[width=0.49\textwidth]{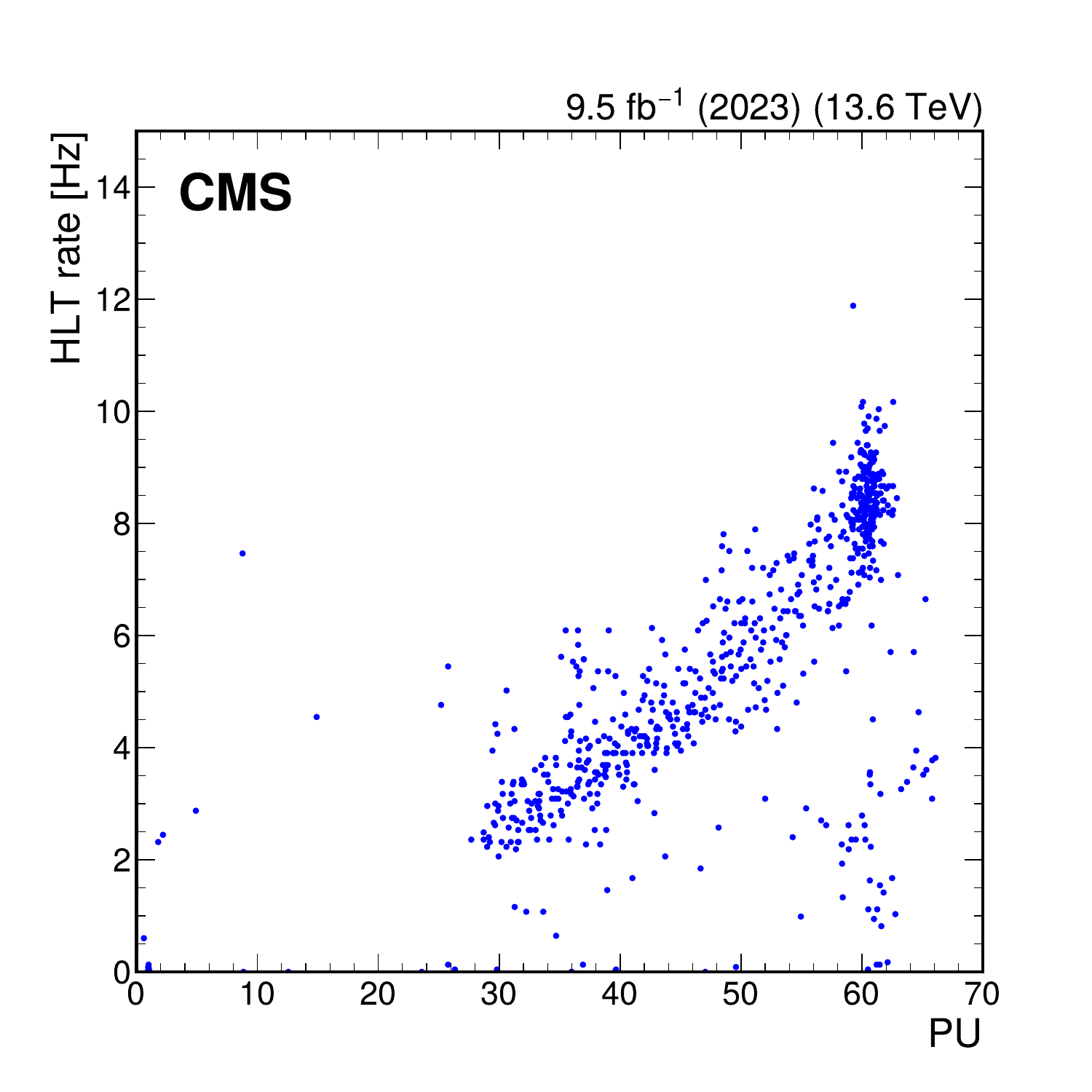}
\caption{Total rate of the displaced \tauh trigger for a few representative runs in 2022 (\cmsLeft) and 2023 (\cmsRight) data, as a function of PU.}
\label{fig:diTauRate}
\end{figure}

Studies performed on the standard L1 \tauh reconstruction algorithm have shown that it was not necessary to develop a dedicated displaced \tauh L1 seed. This is because the L1 \tauh reconstruction algorithm does not depend on tracking information. As long as the \tauh decays before the surface of the ECAL, it will leave energy deposits in the calorimeter, and the L1 \tauh reconstruction will be efficient. Therefore, the HLT displaced reconstruction is seeded by standard L1 \tauh objects.

The dedicated displaced reconstruction is a modification of the standard HPS \tauh reconstruction algorithm, which is described in Section~\ref{sec:taus} and is used in the third step of the online reconstruction of promptly decaying \tauh particles. To enable the HPS algorithm to reconstruct displaced \tauh particles, the standard tracking algorithm has been extended to include a dedicated displaced iteration that starts from a seed consisting of a combination of hits in the strip tracker alone and builds tracks without assuming their provenance from the IP. To meet the CPU constraints, this iteration is only performed in regions around isolated L2 tau leptons. Modifications to the HPS algorithm itself have also been introduced. The standard algorithm includes requirements that the \tauh constituents (charged and neutral hadrons) should originate from the IP, effectively preventing the reconstruction of displaced \tauh objects. These requirements have therefore been loosened.

A suite of trigger paths using the modified HPS algorithm has been developed, targeting final states with two \tauh or with one \tauh and one displaced lepton. In the di-\tauh path, each \tauh is required to have $\dzero > 0.005\cm$ and $\pt > 32\GeV$. The mixed-object paths, which require \tauh $\pt > 24\,(34)\GeV$ and muon (electron) $\pt > 24\,(34)\GeV$, have been available online since 2024. The total rate of these HLT paths is 36\unit{Hz}.

Figure~\ref{fig:displacedtau_mc_eff} shows the trigger efficiency in simulated $\Pp\Pp \to \PSGt\PASGt\,(\PSGt \to \PGt\PSGczDo)$ events, for 2022 data-taking conditions, with the \PSGt having $\cTau = 10\cm$ and each \PGt decaying hadronically. The efficiencies are given for the previously available Run~2 \ptmiss-based and prompt di-\tauh paths, the displaced di-\tauh, and the combination of these paths. A relative improvement of around 30\% in efficiency is found for \dzero values up to 1\cm, progressively decreasing to close to zero for a \dzero of 10\cm; at low values of \ptmiss, where the previously available triggers are not efficient, the improvement is as large as 45\%, gradually decreasing for $\ptmiss > 100\GeV$. The displaced di-\tauh path is optimized for \dzero values up to 1\cm.

Figure~\ref{fig:diTauRate} shows the rates of the trigger path in 2022 and 2023, as functions of PU. The plots indicate a linear rate response with respect to PU.

\subsubsection{Displaced-jet triggers using the tracker}
\label{sec:displacedjetstracking}

\begin{figure}[htbp!]
\centering
\includegraphics[width=0.7\textwidth]{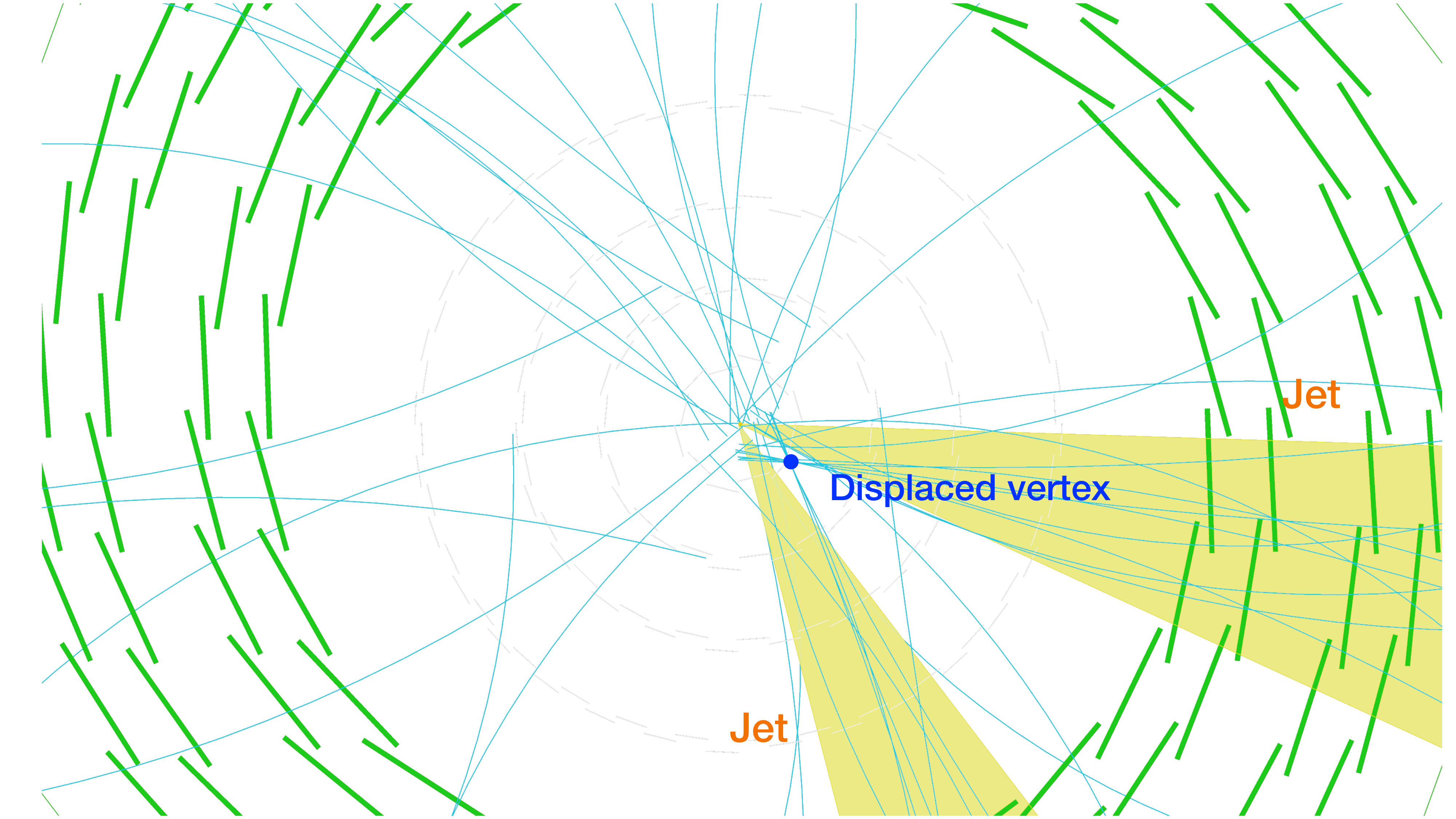}
\caption{Event display of a simulated pair of displaced jets arising from an LLP decay, producing displaced vertices and tracks. The simulated process is $\PH \to \PS\PS \to \qqbar\qqbar$. The blue curves indicate the reconstructed tracks. The yellow cones indicate two reconstructed jets, with a number of associated displaced tracks. The jets are reconstructed using the standard algorithms described in Section~\ref{sec:jets}, which assume they originate from the PV. The blue point indicates the displaced vertex induced by the LLP decay.}
\label{fig:dj_display}
\end{figure}

A suite of displaced-jet triggers was already available in Run~2~\cite{CMS:2020iwv}, utilizing the tracking information available at the HLT. Displaced jets are identified, or ``tagged'', in the HLT as jets with a small number of prompt tracks or with the presence of displaced tracks. With these triggers, the Run~2~\cite{CMS:2020iwv} and Run~3~\cite{displacedJets2022Data} displaced-jet searches set stringent limits on a large number of BSM models with hadronically decaying LLPs and with LLP masses as low as $\mathcal{O}(100\GeV)$, including split SUSY~\cite{ArkaniHamed:2004fb,Giudice:2004tc,Hewett:2004nw,ArkaniHamed:2004yi,Gambino:2005eh,Arvanitaki:2012ps,ArkaniHamed:2012gw},  GMSB~\cite{GIUDICE1999419,Meade:2008wd,Buican:2008ws}, and $R$-parity violating SUSY~\cite{Fayet:1974pd,Farrar:1978xj,Weinberg:1981wj,Hall:1983id,Barbier:2004ez,Csaki:2013jza,Csaki:2015fea}, covering a variety of final-state topologies. Such displaced-jet signatures are well-motivated in many BSM scenarios, especially those involving Higgs-portal hidden sectors. Figure~\ref{fig:dj_display} displays a simulated $\PH \to \PS\PS \to \qqbar\qqbar$ event with $\mS=55\GeV$, where one long-lived \PS decays inside the tracker, producing a pair of displaced jets accompanied by displaced tracks and a displaced vertex. The diagram for this process is shown in Fig.~\ref{fig:HtoXX_feynmanDiagrams}.

In Run~3, major improvements to the displaced-jet triggers were implemented to increase the efficiencies for low-mass LLPs with masses smaller than 100\GeV, such as those expected in the exotic Higgs boson decays. The main improvements are in the tracking-based displaced-jet tagging requirements. Displaced jets are tagged at the HLT using two types of requirements. The first requires calo jets with zero or one associated prompt tracks, referred to as the ``inclusive'' tagging requirement. Prompt tracks are tracks that have $\pt>1\GeV$ and \dzero measured with respect to the PV smaller than 0.5\mm. Tracks are associated with the calo jets by requiring the angular distance \DR between a given track and a given jet to be smaller than 0.4. The second requirement, referred to as the ``displaced'' tagging requirement, starts with the inclusive tagging requirement and additionally requires that if there is exactly one associated prompt track, there must also be at least one associated displaced track with $\pt>1\GeV$ and $\dzero>0.3\mm$. The two requirements are motivated by two complementary characteristics of exotic LLPs: a minimal number of prompt particles accompanying the LLP production and the presence of displaced tracks from the LLP decays. 

Two sets of displaced-jet triggers are implemented in standard data taking: a main set, which provides most of the data and signal efficiency, and an auxiliary set, which was added to increase the signal efficiency. The main displaced-jet trigger is seeded by an \HT L1T and then requires $\HT>430\GeV$ at the HLT, where \HT is the scalar sum of \pt for all calo jets satisfying $\pt>40\GeV$ and $\abseta<2.5$. The trigger also requires the presence of at least two calo jets with $\pt>40\GeV$ and $\abseta<2.0$ that pass the inclusive tagging requirement. Since 2023, the HLT \HT threshold of this trigger has been lowered to 390\GeV, with correspondingly lower L1 \HT thresholds, and this set of triggers has been moved to the parked data-taking stream. 

The auxiliary set of displaced-jet triggers in standard data taking is seeded by an L1T that requires $\HT>240\GeV$ and the presence of a muon with $\pt>6\GeV$, to improve the efficiency for LLPs with heavy flavor decays. This trigger further requires the presence of at least two calo jets with $\pt>40\GeV$ and $\abseta<2.0$ that pass the displaced tagging requirement. 

\begin{figure}[htb!]
\centering
\includegraphics[width=0.49\textwidth]{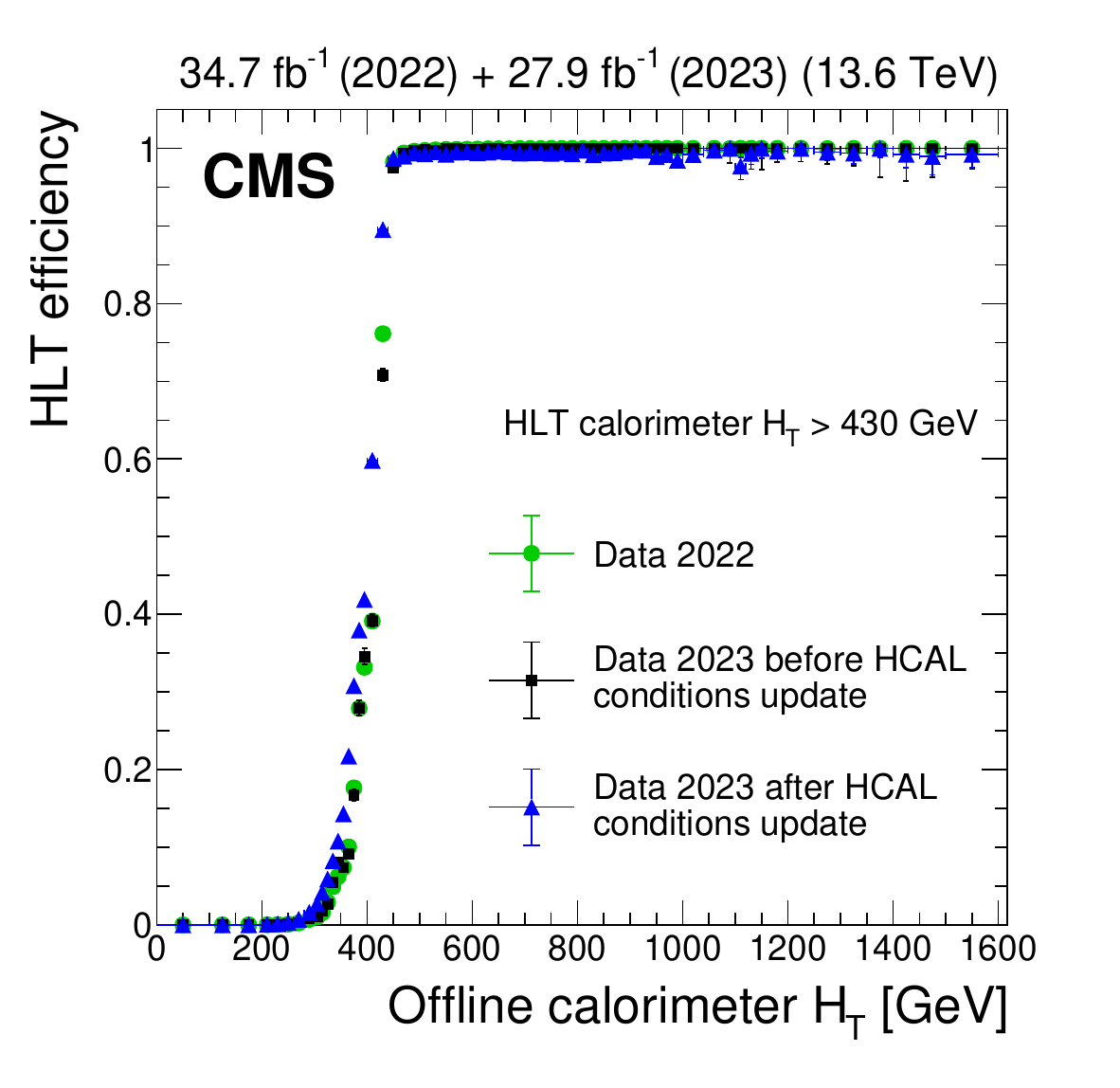}
\includegraphics[width=0.49\textwidth]{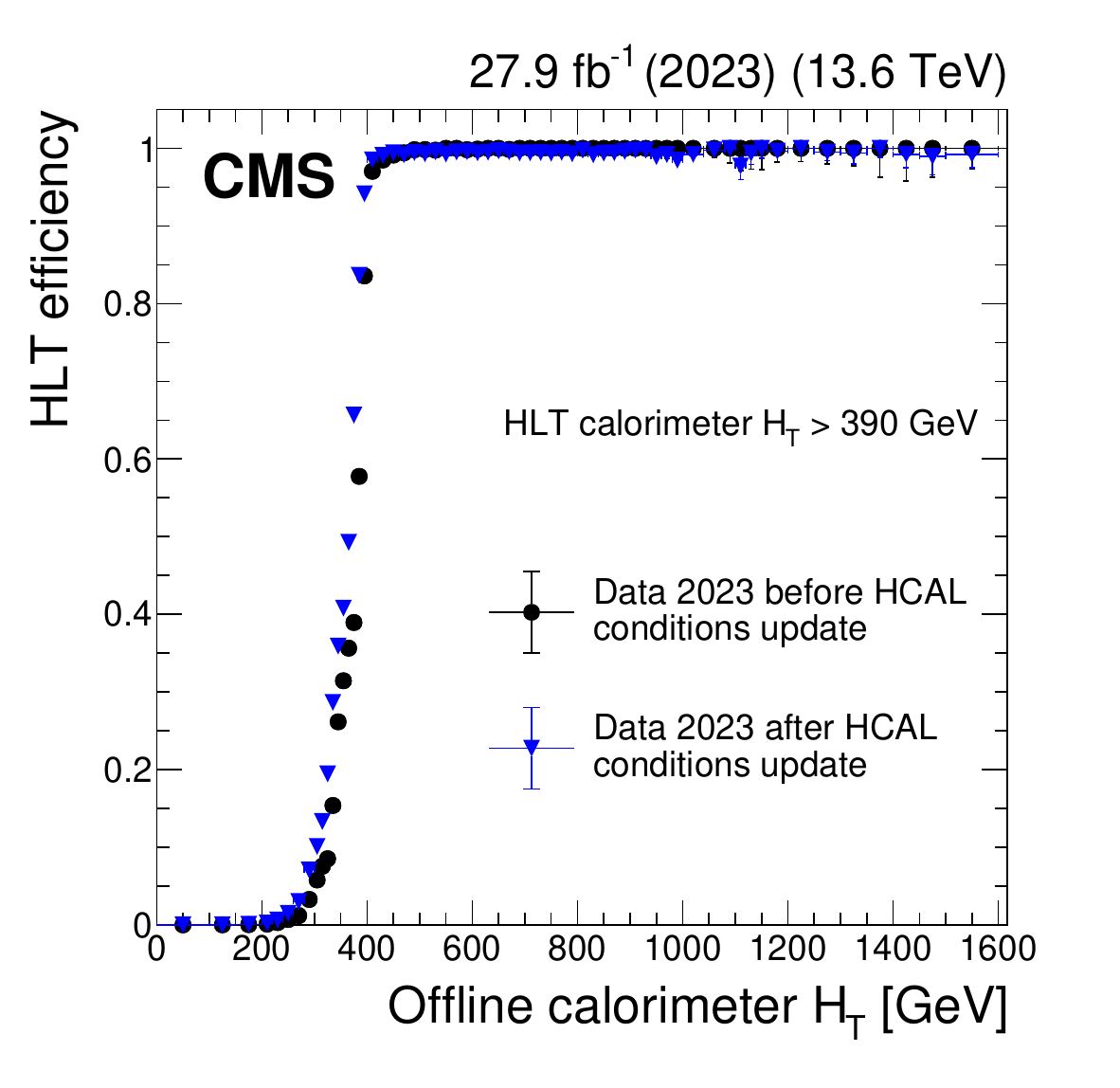}
\caption{The HLT efficiency for a given event passing the main displaced-jet trigger to satisfy HLT calorimeter $\HT>430\GeV$ (\cmsLeft) and $\HT>390\GeV$ (\cmsRight) as a function of the offline calorimeter \HT. For this trigger, the minimum calorimeter \HT threshold was $430\,(390)\GeV$ in 2022 (2023 and later). The measurements are performed with data collected in 2022 (green circles), in 2023 before an update of the HCAL gain values and energy response corrections (black squares), and in 2023 after the update (blue triangles).}
\label{fig:displacedjets_HT}
\end{figure}

\begin{figure}[hbt!]
\centering
\includegraphics[width=0.49\textwidth]{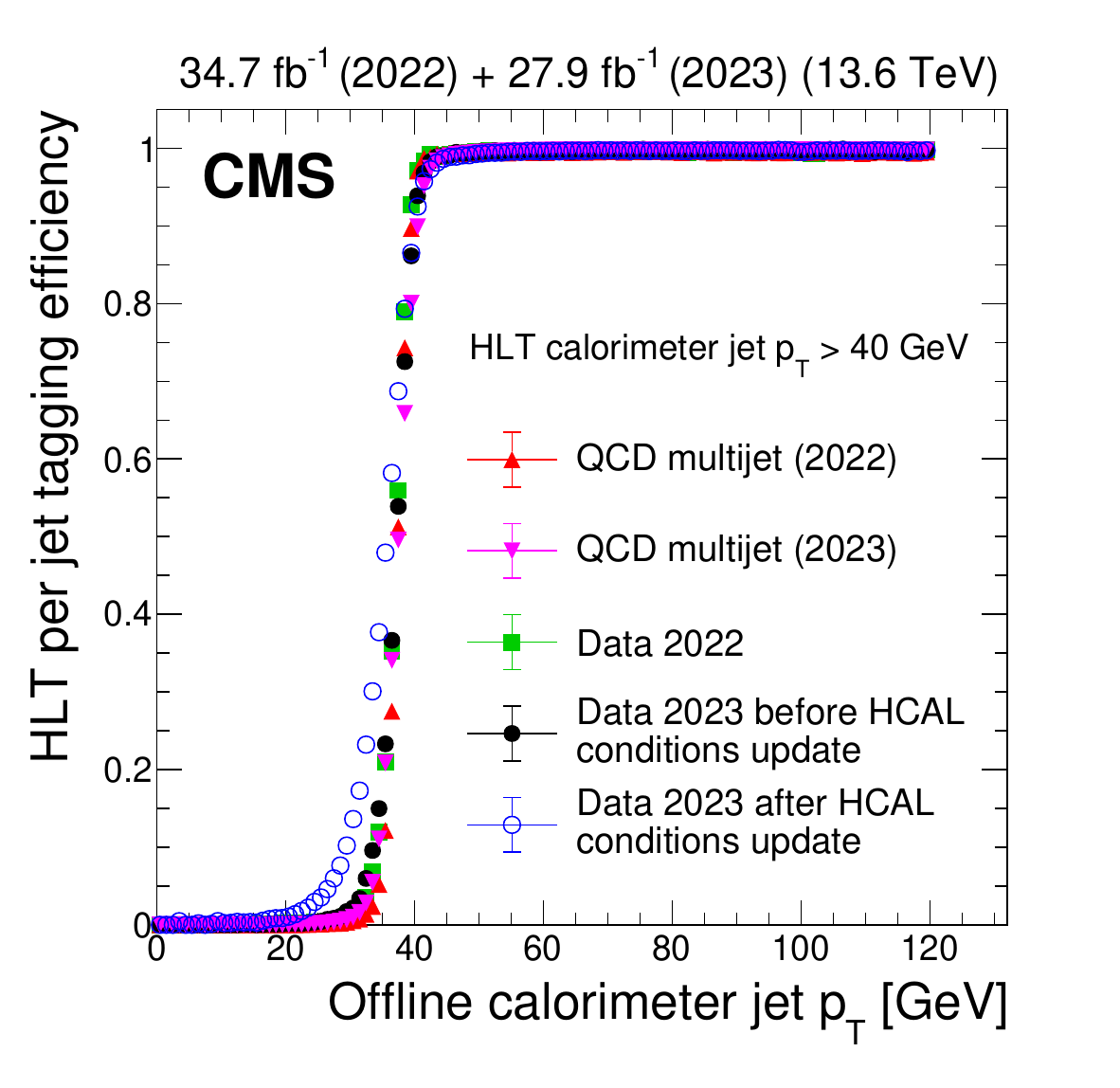}
\includegraphics[width=0.49\textwidth]{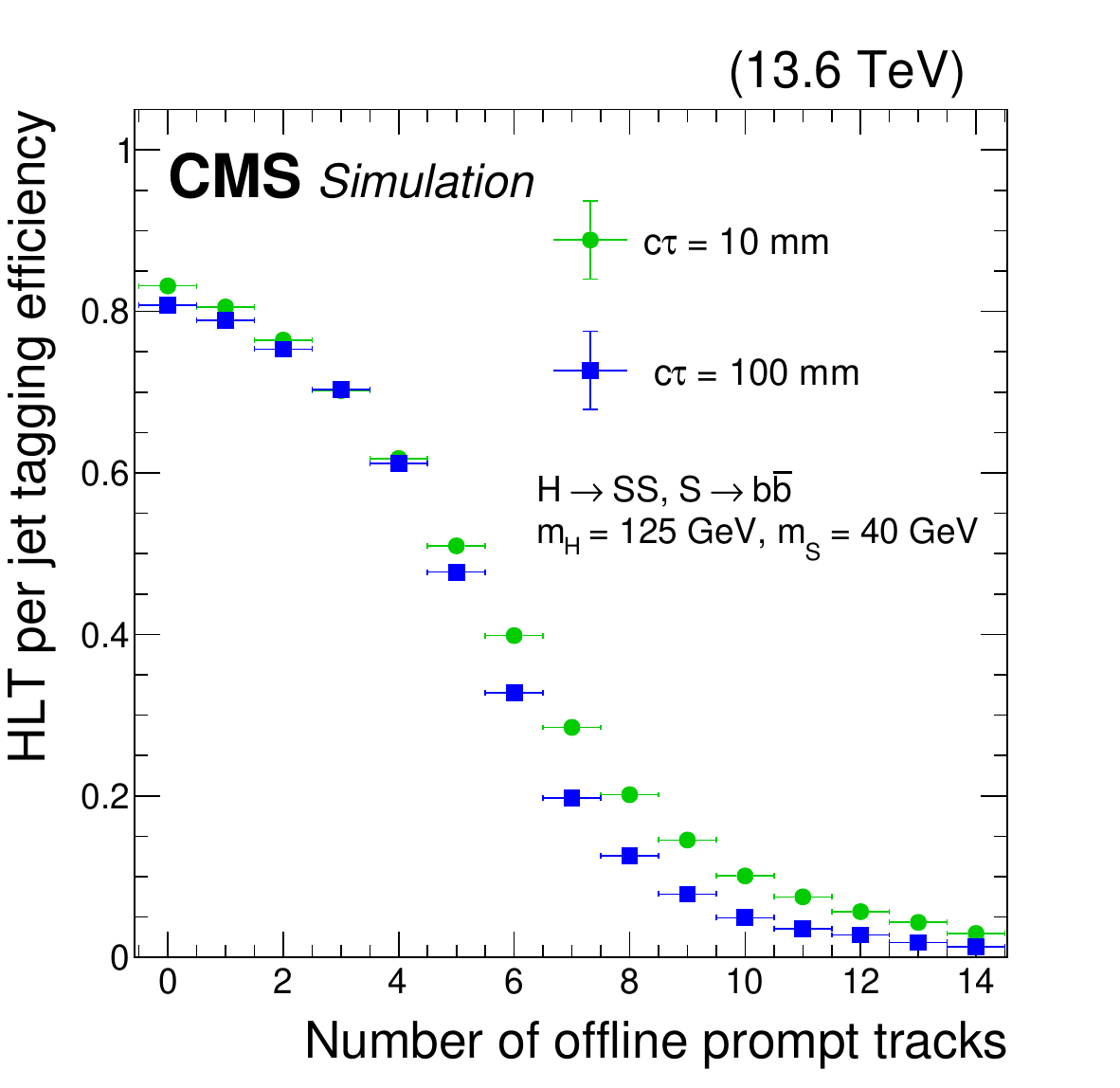}
\caption{The HLT efficiency of the main displaced-jet trigger: Efficiency of an offline calorimeter jet to (\cmsLeft) pass the online \pt requirement in the displaced-jet triggers and (\cmsRight) to have at most one HLT prompt track. In the \cmsLeft plot, the HLT calorimeter jets must have $\pt>40\GeV$. The efficiency is shown for data collected in 2022 (green squares), in 2023 before an update of the HCAL gains and energy response corrections (black filled circles), and in 2023 after the update (blue open circles). The efficiencies measured with QCD multijet simulations are also shown, for 2022 (red triangles) and 2023 (purple triangles) conditions. These measurements are performed using events collected with a prescaled trigger that requires $\HT >425\GeV$ at the HLT. An offline $\HT >450 \GeV$ selection is also applied to ensure the prescaled trigger reaches its plateau. The efficiency is ${>} 96\%$ when the offline jet has $\pt>40\GeV$. The efficiency threshold is lower for the later 2023 data following updates to the HCAL energy corrections and readout gains, although the \pt value at the start of the plateau is unchanged. The efficiency in the \cmsRight plot is shown for 2022 conditions, as a function of the number of offline prompt tracks, in simulated $\PH\to\PS\PS$, $\PS\to\bbbar$ signal events where $\mH=125\GeV$ and $\mS=40\GeV$. Two proper decay lengths of the \PS particle are shown: $\cTau=10\mm$ (green circles) and $\cTau=100\mm$ (blue squares). For jets in signal events, when the number of offline prompt tracks is ${<}4$, the tagging efficiency is larger than 70\%.
\label{fig:displacedjetstracking-turnon} }
\end{figure}

As described in Section~\ref{sec:eventReconstruction}, the CMS experiment employs an iterative process for track reconstruction, and the displaced tracking iterations are usually not used at the HLT because they are CPU-intensive. However, a displaced tracking iteration is used in the displaced-jet triggers. At the same time, similar to the offline tracking, multiple tracking iterations are employed to reconstruct prompt tracks in the displaced-jet triggers, which helps reduce the trigger rates. To meet the stringent HLT CPU timing constraints, only the seeds near jets are considered for track reconstruction in the displaced-jet triggers. Furthermore, the next tracking iteration proceeds only if a given event has two tagged displaced jets based on the tracks from earlier iterations.

\begin{figure}[htb!]
\centering
\includegraphics[width=0.49\textwidth]{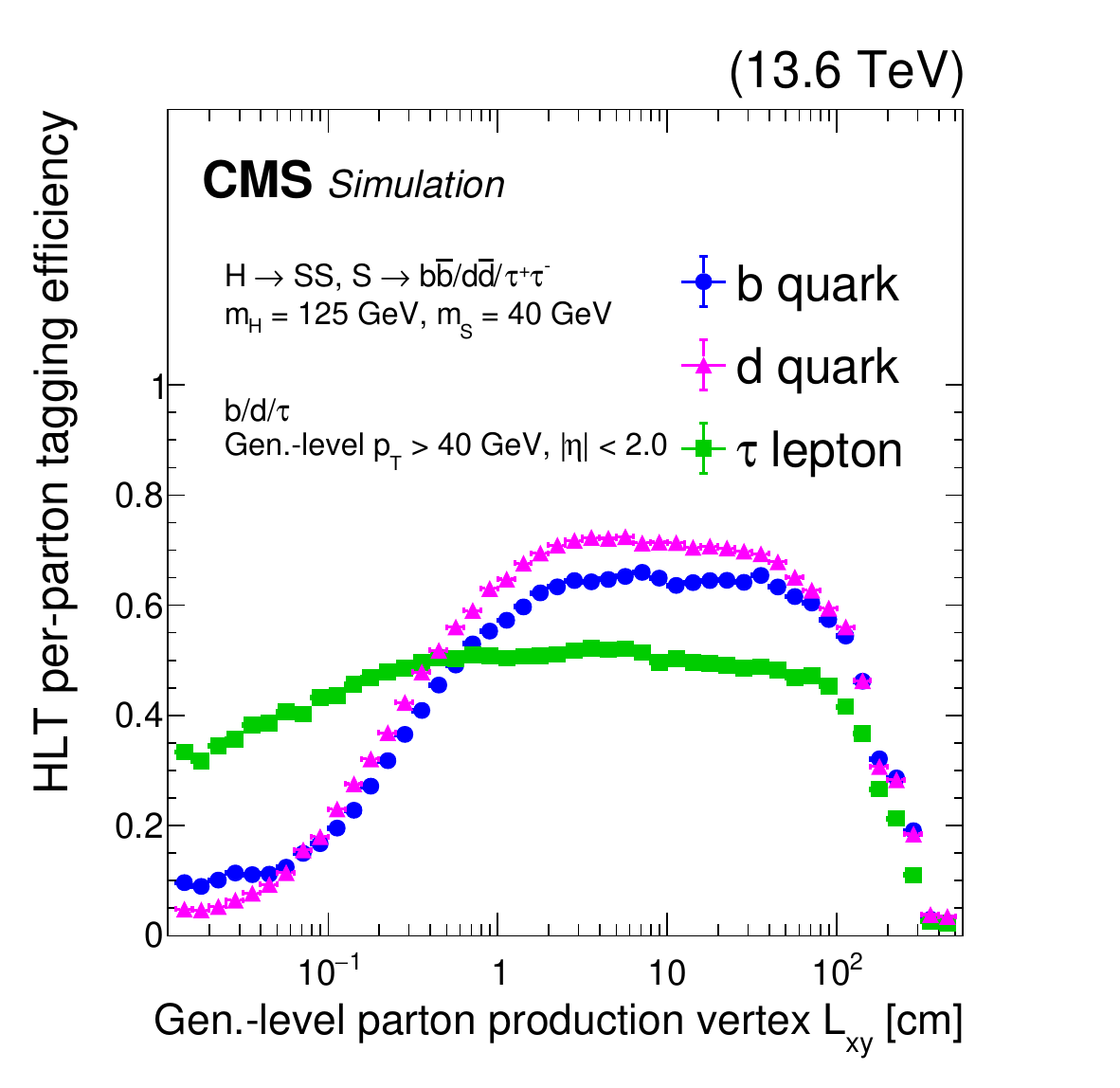}
\caption{The HLT efficiency of the main displaced-jet trigger for 2022 conditions, for $\PH\to\PS\PS$ signal events where $\mH=125\GeV$ and $\mS=40\GeV$. The per-parton (quark or lepton) HLT displaced-jet tagging efficiency as a function of the generator-level \Lxy of the parton is shown for displaced \PQb quarks (blue circles), \PQd quarks (purple triangles), and \PGt leptons (green squares) with $\pt>40\GeV$ and $\abseta <2.0$. Events are required to satisfy the HLT $\HT > 430\GeV$ requirement.}
\label{fig:displacedjetsPerPartonEff}
\end{figure}

The total HLT rate of these displaced-jet paths is $53\,(163)\unit{Hz}$ in the standard (parked) data-taking stream. The efficiency of the main displaced-jet triggers is shown in Figs.~\ref{fig:displacedjets_HT}--\ref{fig:displacedjetsPerPartonEff}. 

Figure~\ref{fig:displacedjets_HT} shows the efficiency for a given event to satisfy the HLT $\HT>430\GeV$ (the 2022 minimum threshold) and $\HT>390\GeV$ (the 2023 and later minimum threshold) requirements measured in data. The measurements are performed using events collected with an isolated single-muon trigger that requires the presence of a muon with $\pt>27\GeV$. The efficiency of the HLT $\HT>390\,(430)\GeV$ requirement is larger than ${\approx}98\%$ when the offline \HT is larger than $400\,(440)\GeV$. Figure~\ref{fig:displacedjetstracking-turnon} (\cmsLeft) shows the efficiency of an offline calo jet to pass the online \pt requirement in displaced-jet triggers as a function of the offline jet \pt. These measurements validate that the HLT \HT and \pt requirements are well modeled in simulation compared with data, especially when the offline thresholds are chosen near the trigger-efficiency plateau. A similarly accurate modeling is also expected for the signal simulation. Figure~\ref{fig:displacedjetstracking-turnon} (\cmsRight) shows the efficiency of an offline calo jet in simulated signal events to pass the online inclusive tagging requirement as a function of the number of offline prompt tracks.

Figures~\ref{fig:displacedjets_HT} and \ref{fig:displacedjetstracking-turnon} show the 2023 data split into two periods: before and after an update of the HCAL energy response corrections (and a simultaneous minor update to the HCAL readout gain values). This update of the conditions was primarily made because of a change in the precision-timing alignment of the HCAL barrel, which increased the energy response. The change in energy response can impact the trigger efficiency, as observed for the tracking-based displaced-jet triggers. The trigger efficiencies are generally higher after the update, while the trigger rates stay around the same.

\begin{figure}[hbt!]
\centering
\includegraphics[width=0.49\textwidth]{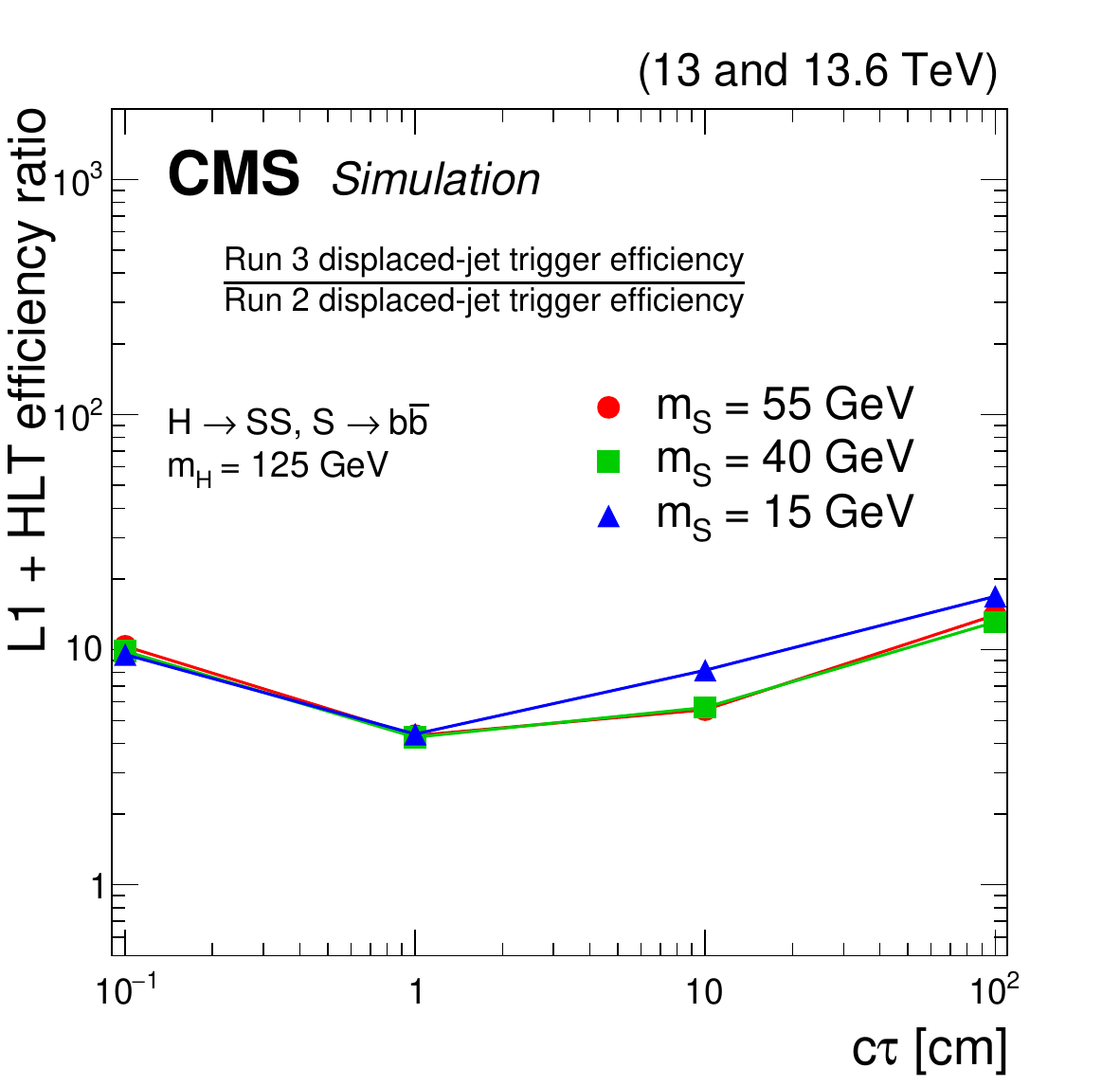}
\caption{The ratio between the Run~3
displaced-jet trigger efficiency and the Run~2 displaced-jet trigger efficiency as a function of LLP \cTau, in simulated $\PH\to\PS\PS$, $\PS\to\bbbar$ signal events where $\mH=125\GeV$ and $\mS=15$ (blue triangles), 40 (green squares), or 55 (red circles)\GeV. The Run~3 displaced-jet trigger efficiencies are measured for 2022 conditions.
\label{fig:displacedjetstracking-Run3vsRun2} }
\end{figure}

Figure~\ref{fig:displacedjetsPerPartonEff} shows the fraction of displaced \PQb quarks, \PQd quarks, and tau leptons that pass the displaced-jet tagging requirement at the HLT as a function of the generator-level parton vertex transverse displacement \Lxy, indicating that the displaced-jet triggers have good performance for a variety of final states across the tracker volume. When $\Lxy\gtrsim 0.3\cm$, the tagging efficiencies for tau leptons are smaller than those for \PQb or \PQd quarks since the branching fraction for tau leptons to decay into leptonic final states is nonzero. When $\Lxy\lesssim0.3\cm$, the tau-lepton tagging efficiencies are higher than those for \PQb or \PQd quarks because there is no hadronization when tau leptons are produced. Additionally, tau lepton decays produce fewer charged particles than \PQb or \PQd quark hadronization, thereby increasing the probability for tau leptons to satisfy the ``inclusive'' tagging requirement. The overall, per-event trigger efficiencies, which are primarily constrained by the L1T efficiencies, are ${\approx}0.4$--1.0\% for $\PH \to \PS\PS \to \qqbar\qqbar$ signatures with $\cTau < 1\unit{m}$. Figure~\ref{fig:displacedjetstracking-Run3vsRun2} shows the gain in trigger efficiency for the displaced-jet triggers in Run~3 compared to Run~2. The Run~3 trigger efficiencies are larger than those of Run~2 by a factor of 4 to 17 for $10<\mS<60\GeV$ and $1<\cTau<1000\mm$. Compared to the standard unprescaled \HT trigger, which requires $\HT>1050\GeV$ at the HLT, the displaced-jet triggers have more than a factor of 20 higher efficiency for the $\PH\to\PS\PS\to\qqbar\qqbar $ signatures with $\cTau\gtrsim 10\mm$.

\begin{sloppypar}The data collected with the displaced-jet triggers in 2022, corresponding to an integrated luminosity of 34.7\fbinv, has been analyzed in a Run~3 displaced-jet search~\cite{displacedJets2022Data}. Despite analyzing a data set corresponding to a much smaller integrated luminosity, this search has achieved up to a factor of 10 increase in sensitivity compared to previous searches~\cite{CMS:2020iwv,ATLAS:2024qoo} for $\PH\to\PS\PS\to\qqbar\qqbar$ signatures that use a Run~2 data set corresponding to an integrated luminosity of 132\fbinv. The full Run~3 data set collected with the displaced-jet triggers is expected to provide further significant improvements in the CMS sensitivity to this signature.\end{sloppypar}

\subsection{Calorimeter-based algorithms}
\label{sec:caloLLPtriggers}

We now describe the LLP trigger algorithms that feature calorimeter information. In this section, we cover displaced-jet triggers that use the HCAL (Section~\ref{sec:displacedjetshcal}), delayed-jet triggers that use ECAL timing (Section~\ref{sec:delayedjetsecal}), delayed-diphoton triggers (Section~\ref{sec:delayeddiphoton}), and displaced-photon + \HT triggers (Section~\ref{sec:displacedphotonandht}). 

\subsubsection{Displaced- and delayed-jet triggers using the HCAL}
\label{sec:displacedjetshcal}

The recent Phase I upgrade to the CMS HCAL~\cite{CMS:TDR-010} introduced both depth segmentation and precision cell timing via a time-to-digital converter (TDC) in the HCAL barrel (HB). This information was made available for the first time for both offline reconstruction and online triggering at the L1T and HLT. Although the primary motivation for these depth and timing enhancements is to maintain high efficiencies for particle identification in increasingly high-PU environments, both aspects of this upgrade also enable the identification of displaced- and delayed-jet signatures. Making full use of the upgraded HCAL requires new algorithms for the L1T. Unlike many other triggers presented in this paper, this required implementing new hardware-level LLP jet trigger algorithms that seed displaced- and delayed-jet HLT paths.

\begin{figure}[!htbp]
\centering
\centering
\includegraphics[width = 0.48\textwidth]{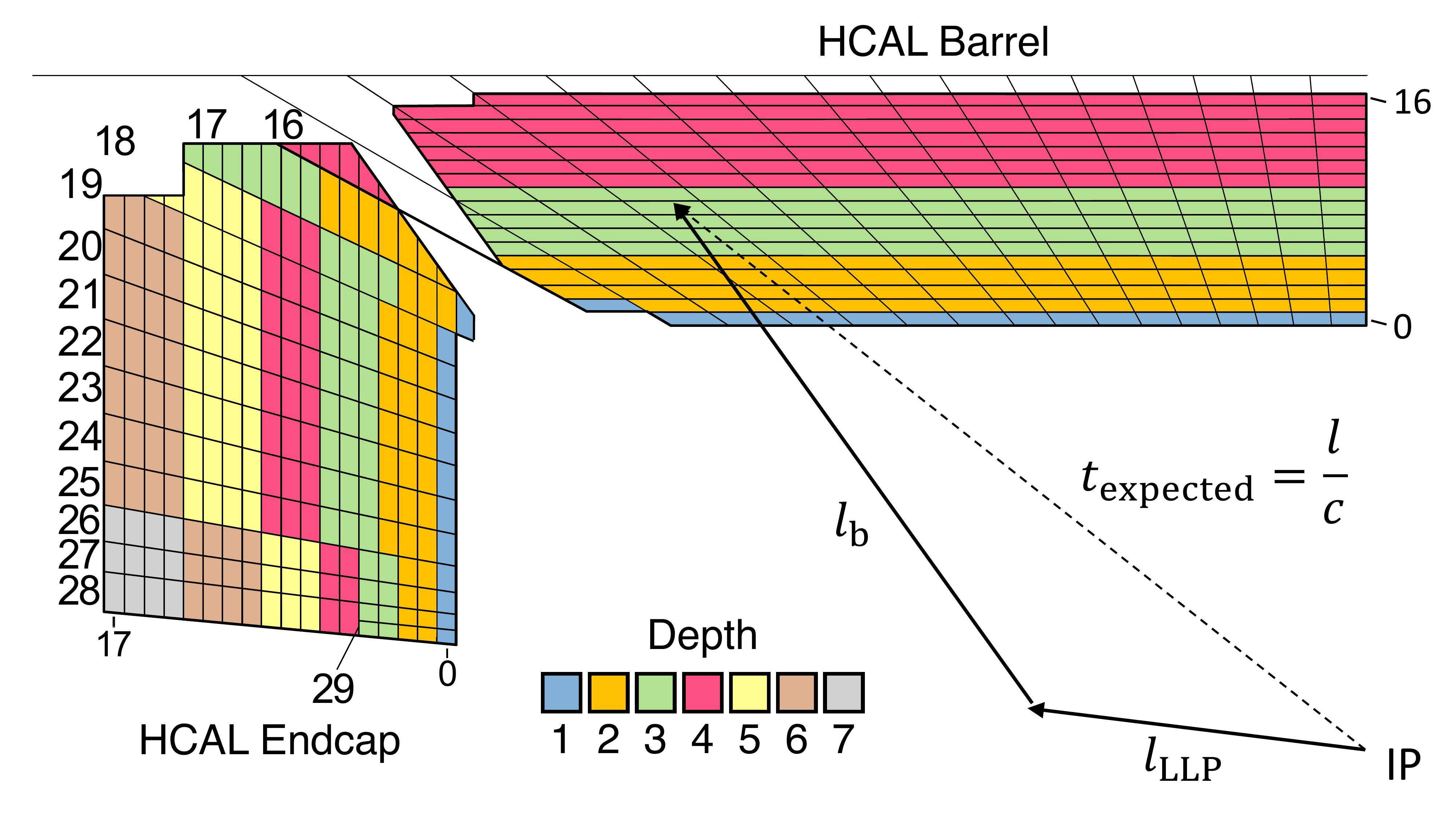}
\hfill
\centering
\includegraphics[width = 0.48\textwidth]{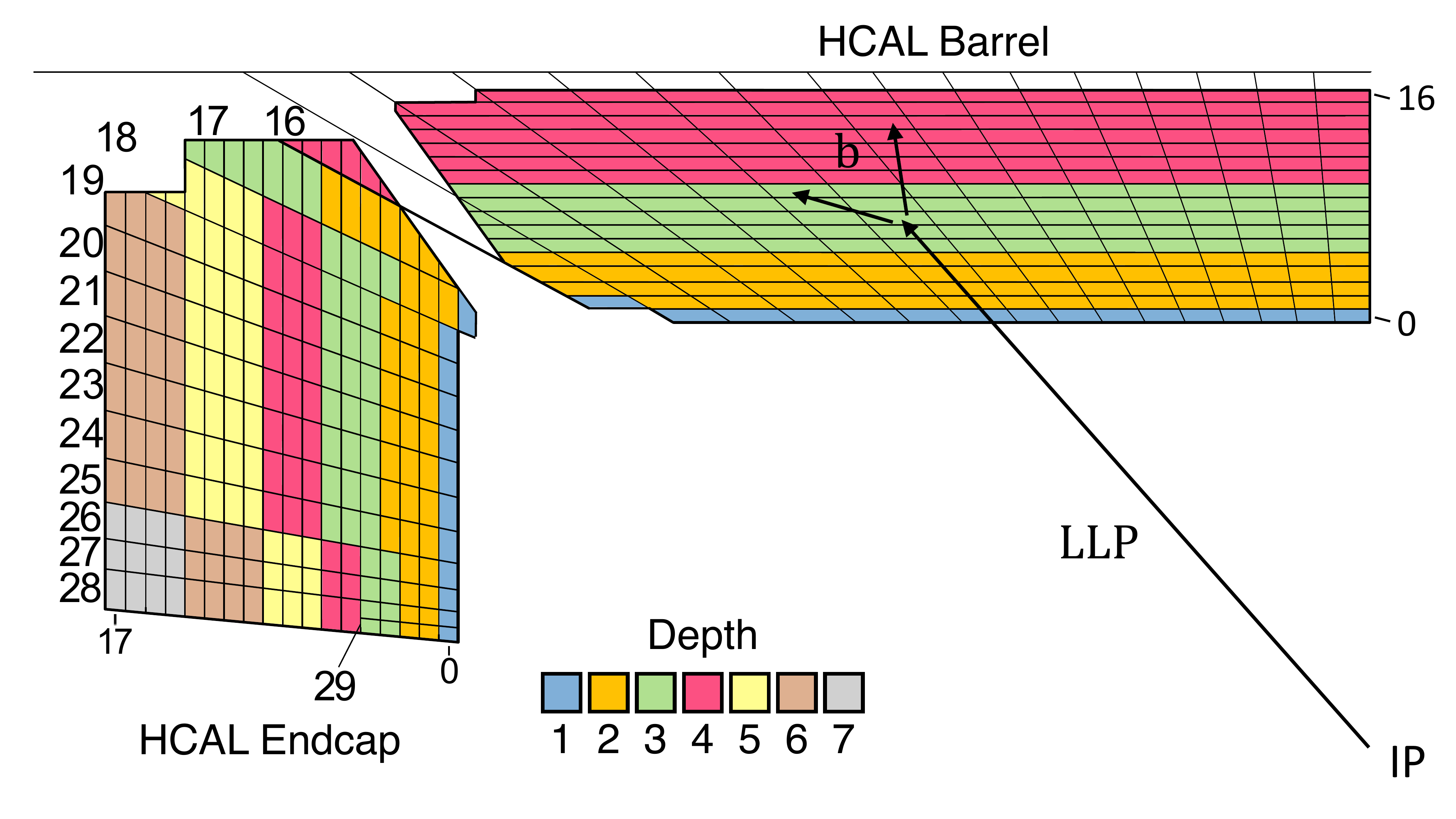}
\caption{Diagram showing the lateral view of a quadrant of the CMS HCAL barrel and endcap in Run~3. The distinct depth segmentation is indicated by the different colors, where the HCAL barrel consists of 4 depths and the HCAL endcap has up to 7 depths. A delayed-jet signature is shown (\cmsLeft), where the time delay results from a combination of the low LLP velocity because of its relatively high mass and the path length difference with respect to a promptly decaying particle. A time delay consists of the path length difference between a particle traveling from the IP directly to the calorimeter ($l$), as compared to the LLP path ($l_{\text{LLP}}$) plus the \PQb quark path ($l_{\cPqb}$). Additionally, the LLP may also have a significantly lower velocity ($v_{\text{LLP}}$), while other particles are assumed in this example to be highly relativistic. Thus, the difference in time of arrival at the calorimeter is $\Delta t = \frac{l_{\text{LLP}}}{v_{\text{LLP}}} + \frac{l_{\cPqb}}{c} - \frac{l}{c}$. A displaced-jet signature is also depicted (\cmsRight), resulting from an LLP decaying within the HCAL volume to produce significant energy deposits deeper in the HCAL with minimal energy deposits in shallower calorimeter layers.}
\label{fig:HBHE_LLPpath}
\end{figure}

This section focuses on a novel L1T targeting LLPs that decay hadronically before or within the HB volume. The targeted signature is illustrated in Fig.~\ref{fig:HBHE_LLPpath}. The signal model used to evaluate the trigger performance is an exotic decay of the Higgs boson (or decay of a high-mass, exotic Higgs boson) that produces two scalar LLPs (\PS), which each decay into two \PQb quarks ($\PH \to \PS\PS \to \bbbar\bbbar$), as shown in Fig.~\ref{fig:HtoXX_feynmanDiagrams}. The scalar \PS can be long-lived, with a \cTau of the order of meters, permitting decays before and within the HCAL volume. The mass of the scalar particle is constrained to be $\mS < \mH/2$.

The L1 HCAL-based LLP triggers significantly increase the LLP acceptance with respect to standard triggers because of their lower jet \pt thresholds. For example, the LLP acceptance is increased by up to a factor of 4 for a signal with $\mH = 125\GeV$, $\mS = 50\GeV$, and $\cTau = 3\unit{m}$~\cite{LLPRun3TriggerDPNote}. These triggers will be used in an upcoming Run~3 search for LLPs, which relies on the upgraded HB information at both trigger and offline analysis levels. This search aims to increase the coverage of LLP models decaying hadronically in the calorimeter region, while also adding sensitivity to lower jet and event energies. 

The TDC measures timing with 6-bit precision for each HCAL channel that is recorded. This information is used in the trigger and offline analysis. Within each 25\unit{ns} bunch crossing, the TDC utilizes 6 bits to encode the pulse arrival time, allowing for fifty valid time bins (0--25\unit{ns}, in half-ns steps). Special codes indicate situations where the pulse started above threshold, never rose above threshold, or there were phase-locked loop errors. Values with any of these invalid or error codes are excluded from use in the timing requirements in the trigger.
Because of bandwidth constraints in the HB detector readout, each 6-bit TDC address is compressed into a 2-bit value corresponding to the following four timing ranges: prompt, slightly delayed, very delayed, and invalid or error. These categories are defined as:
\begin{itemize}
\item Prompt: $t_{\text{pulse}} < 6\unit{ns}$; 
\item Slightly delayed: $6 < t_{\text{pulse}} < 7\unit{ns}$;
\item Very delayed: $7 < t_{\text{pulse}} < 25\unit{ns}$;
\item Invalid pulse: invalid or error code.
\end{itemize}

The full HB 6-bit TDC cannot be recovered at a later stage (\eg, for offline reconstruction) because the 6-to-2-bit compression occurs on the data-encoding FPGA before transmitting data to the off-detector electronics.

In Run~3, the barrel region of HCAL is segmented into four depths, each with time and energy readouts, as illustrated in Fig.~\ref{fig:HBHE_LLPpath}. Each HB depth is made of interleaved layers of scintillator and brass absorber, or stainless steel for the first and last absorbers~\cite{CMS:TDR-010}. To accommodate the different behavior and pulse shapes observed in the innermost depth, two updates have been made for 2024 data taking~\cite{HCAL-DN-2023-022}. First, the prompt range for depth 1 is widened to account for the larger spread of arrival times. Second, the slightly delayed range is extended for all depths, as the wider range provides a larger sample of events for timing evaluation and alignment. The timing ranges are shown in Table~\ref{Table:TDCLUT}.

\begin{table}
    \centering
    \topcaption{The prompt and delayed ranges of the HB TDC timing encoding in 2022--2023 and from 2024 onward. The timing ranges were changed in 2024 to account for the different behavior observed in depth 1.}
\renewcommand{\arraystretch}{1.1}
    \begin{tabular}{lcccc}
        Year & Depth & Prompt [{ns}] & Slightly delayed [{ns}] & Very delayed [{ns}] \\ [\cmsTabTinySkip] \hline &\\[-\cmsTabTinySkipInverse]
        2022--2023 & 1--4 & 0--6 & 6.5--7 & ${>}7.5$ \\ [\cmsTabSkip]
        \multirow{2}{*}{2024, onward} &  1 & 0--9 & 9.5--12 & ${>}12.5$ \\ 
        &  2--4 & 0--6 & 6.5--9 & ${>}9.5$ \\
    \end{tabular}
    \label{Table:TDCLUT}
\end{table}

The complete information from each of the four HB depth segments cannot be sent off-detector to the trigger because of bandwidth constraints. Therefore, quantities are calculated per trigger tower, which is defined as a group of cells across four depths at the same $\eta$ and $\phi$ coordinates. Each tower is assigned a single bit, referred to as the ``LLP flag'', which encodes the following information. A tower will pass the tower LLP flag if it satisfies the criteria for a delayed tower or a displaced tower. A delayed tower includes at least one delayed cell with energy ${>}4\GeV$ and no prompt cells with energy ${>}4\GeV$. A displaced tower includes at least one cell in depths 3 or 4 with energy ${>} 5\GeV$ and has both cells in depths 1 and 2 with energy ${<} 1\GeV$. Any L1 jet containing at least two LLP-flagged towers within the $9{\times}9$ HCAL towers that make up the jet is labeled as an L1 LLP jet. These L1 LLP jets are then used, along with additional kinematic selection criteria, to create single and double L1 LLP jet triggers: the lowest unprescaled single L1 LLP jet trigger requires at least one L1 LLP jet with $\pt > 60\GeV$ and $\HT > 200\GeV$, and the lowest unprescaled double L1 LLP jet trigger requires at least two L1 LLP jets with $\pt > 40\GeV$ and no additional \HT requirement. The L1 \HT calculation includes jets with $\pt > 30\GeV$ and $\abseta <2.5$. The stringent requirement to identify multiple delayed or displaced towers at L1 permits a correspondingly lower minimum \HT requirement, increasing the sensitivity to a range of LLP models~\cite{ArkaniHamed:2004fb,Giudice:2004tc,Hewett:2004nw,ArkaniHamed:2004yi,Gambino:2005eh,Arvanitaki:2012ps,ArkaniHamed:2012gw,GIUDICE1999419,Meade:2008wd,Buican:2008ws,Fayet:1974pd,Farrar:1978xj,Weinberg:1981wj,Hall:1983id,Barbier:2004ez,Csaki:2013jza,Csaki:2015fea}.

Timing scans in the HCAL have been performed at the start of data taking during the LHC intensity ramp-up each year since 2022 to calibrate the HCAL timing. These scans are taken during special runs coordinated with the whole experiment. The scans in 2022, 2023, 2024, and 2025 collected data corresponding to integrated luminosities of 42.2, 418.5, 70, and 195\pbinv, respectively~\cite{HCAL-DN-2023-022}. As these runs involve unique HCAL settings needed to change the calorimeter time, the data from them is marked ``bad'' for physics use, and therefore, the scan time is minimized. However, a relatively large amount of integrated luminosity was dedicated to the scans in 2023 and 2025, as the HCAL barrel (2023) and endcap (2025) precision-timing calibrations, which require a large number of events, were computed from these scans. Subsequent annual HCAL barrel calibrations have required only minor, few-ns shifts and smaller samples, demonstrating the stability of the time alignment throughout the year and from year to year. 

\begin{figure}[htb!]
\centering
\includegraphics[width=0.6\textwidth]{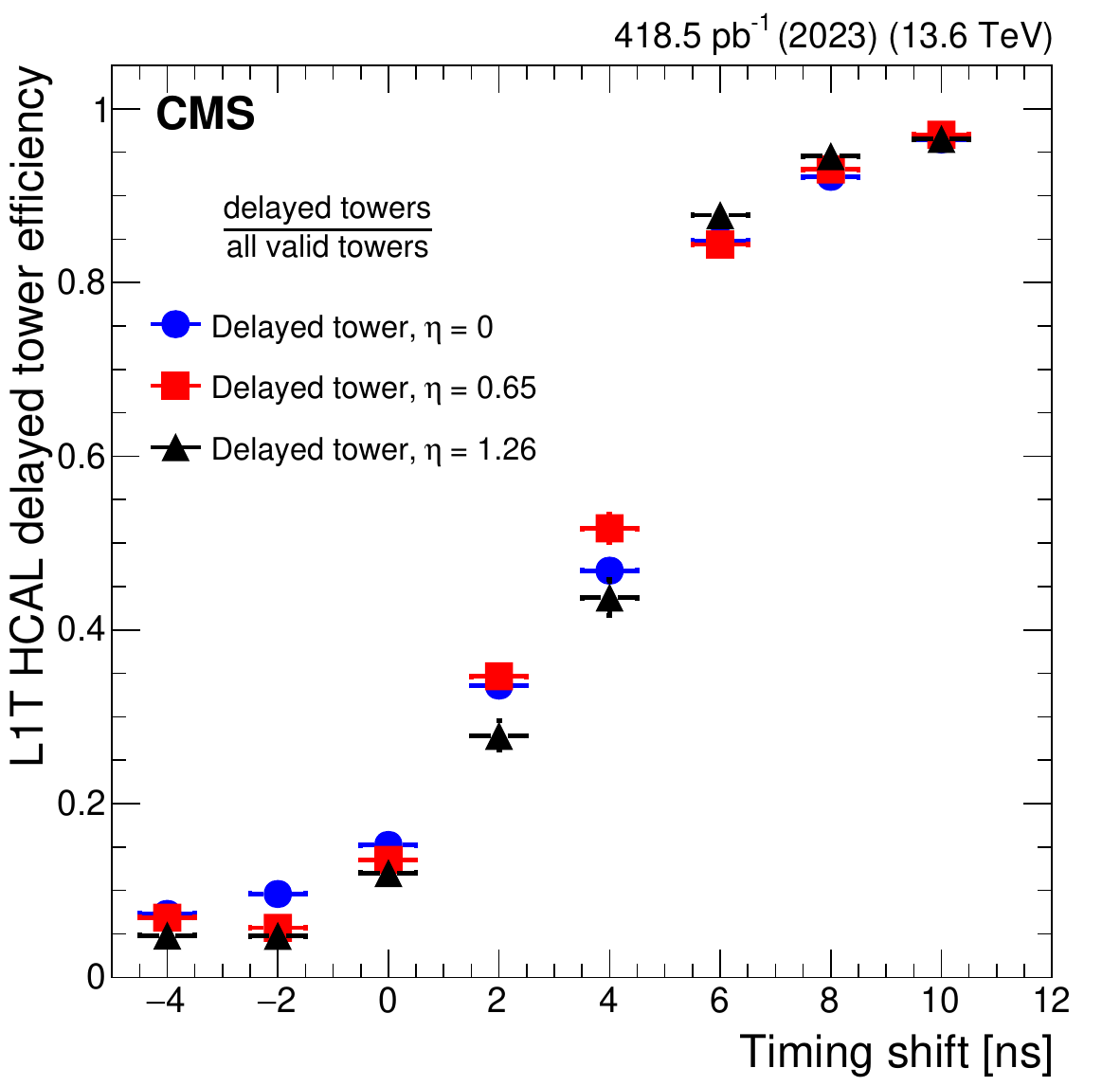} 
\caption{The L1T HCAL trigger tower efficiency of the delayed timing towers in the 2023 HCAL timing-scan data, with efficiencies split by trigger towers centered at $\eta\approx 0$ (blue circles), 0.65 (red squares), 1.26 (black triangles), and with width $\Delta \eta = 0.087$. The sharp rise in efficiency between timing delays of 0--6\unit{ns} is expected, as the prompt timing range includes pulses up to and including those recorded at a 6\unit{ns} arrival time (digitized in half-ns steps by the TDC), demonstrating the timing trigger performance. The delayed timing towers must have at least one delayed cell, no prompt cells, and energy ${>} 4\GeV$. The efficiency is calculated relative to towers with any valid timing code, meaning the tower contains at least one cell with energy ${>} 4\GeV$ and a TDC code of prompt, slightly delayed, or very delayed. Multiple delayed or displaced towers are required for the HCAL-based displaced- and delayed-jet L1T to pass.}
\label{fig:displacedjetshcal-turnon-tower}
\end{figure}

\begin{figure}[htb!]
\centering
\includegraphics[width=0.6\textwidth]{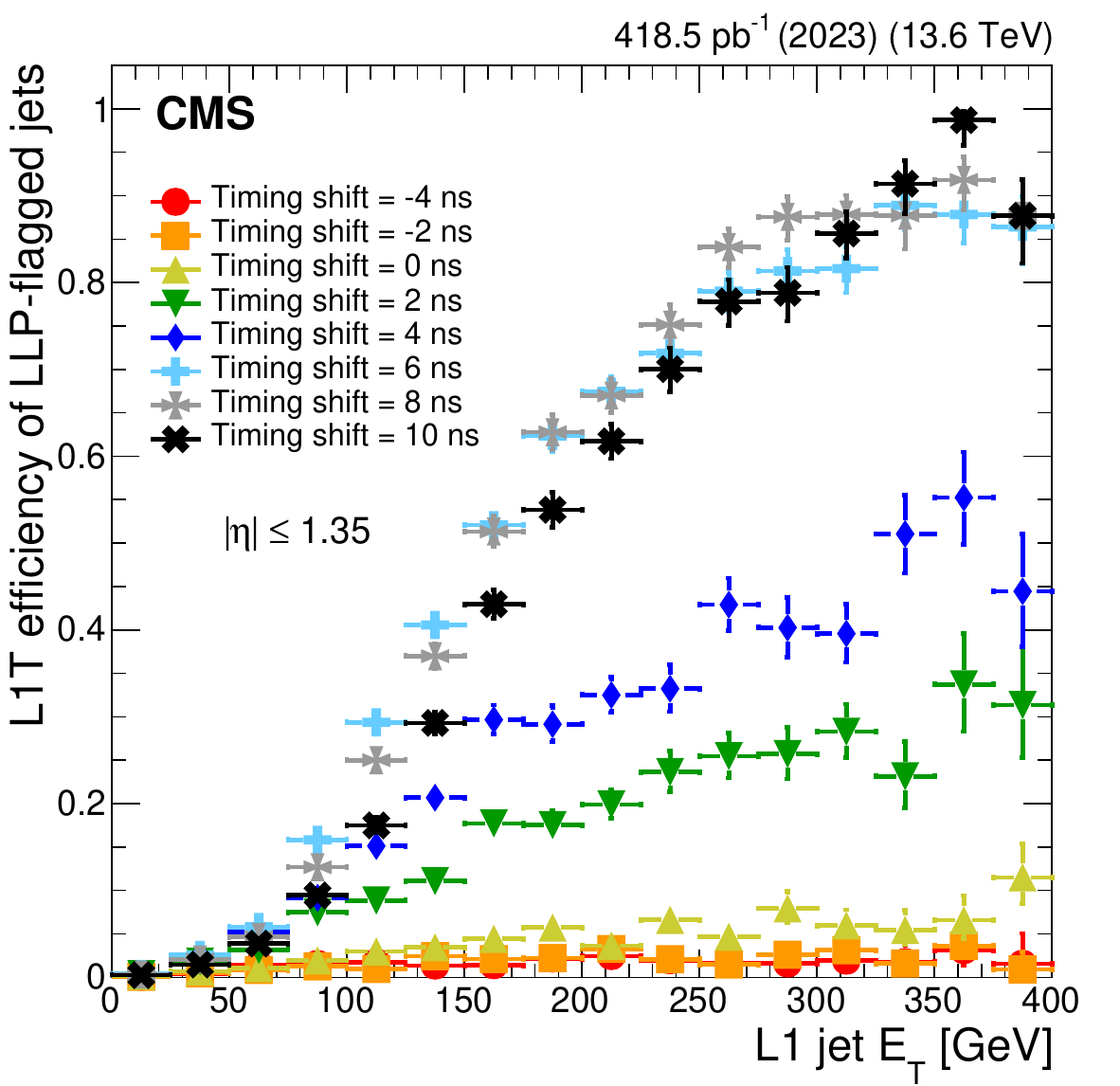} \caption{The L1T efficiency of the LLP jet trigger in 2023 HCAL timing-scan data as a function of L1 jet \ET. The results are inclusive in $\eta$ for the HCAL barrel, corresponding to $\abseta < 1.35$. The fraction of LLP-flagged L1 jets is compared to all L1 jets from a data set of events enriched in jets or \ptmiss. No explicit selection criterion is applied on the jet \ET, though the implicit requirement for a jet to have at least two cells with $\ET > 4\GeV$ shapes the resulting jet trigger efficiency curve. While the trigger configuration is based on the presence of both delayed and displaced towers, the timing scan demonstrates the ability of the trigger to select delayed jets only.}
\label{fig:displacedjetshcal-turnon-jet}
\end{figure}

These timing scans involve shifting the HCAL timing by known values to produce artificially early or delayed jets in collision data, in an approach that is crucial for understanding the detector and trigger performance. Negative timing shifts indicate that the jet arrived earlier than the nominal time. From the timing-scan data, the delayed tower and jet trigger efficiencies are measured, with the timing resolutions of jets in collisions. Figures~\ref{fig:displacedjetshcal-turnon-tower} and~\ref{fig:displacedjetshcal-turnon-jet} show the efficiency of the per-tower LLP flag and the per-jet L1 LLP label, demonstrating the high degree of sensitivity of these algorithms to jet delay. Jets with an arrival time ${>} 6\unit{ns}$ are identified as delayed, with high efficiency. 

\begin{figure}[htb!]
\centering
\centering
\includegraphics[width=0.49\textwidth]{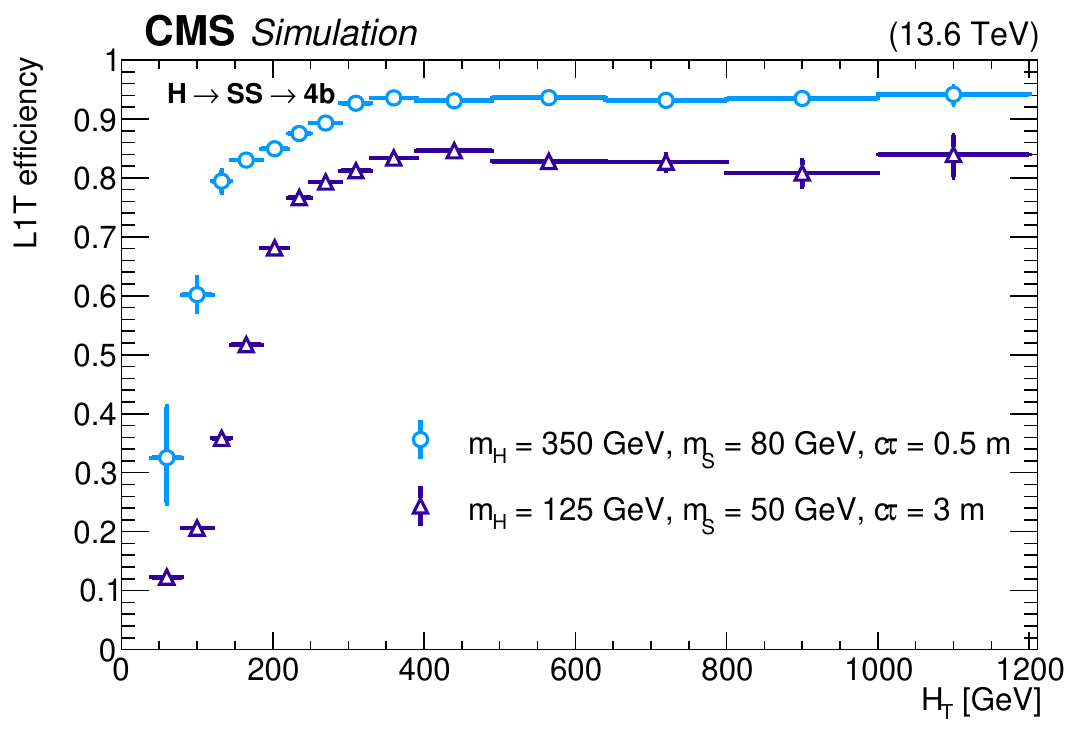}
\hfill
\centering
\includegraphics[width=0.49\textwidth]{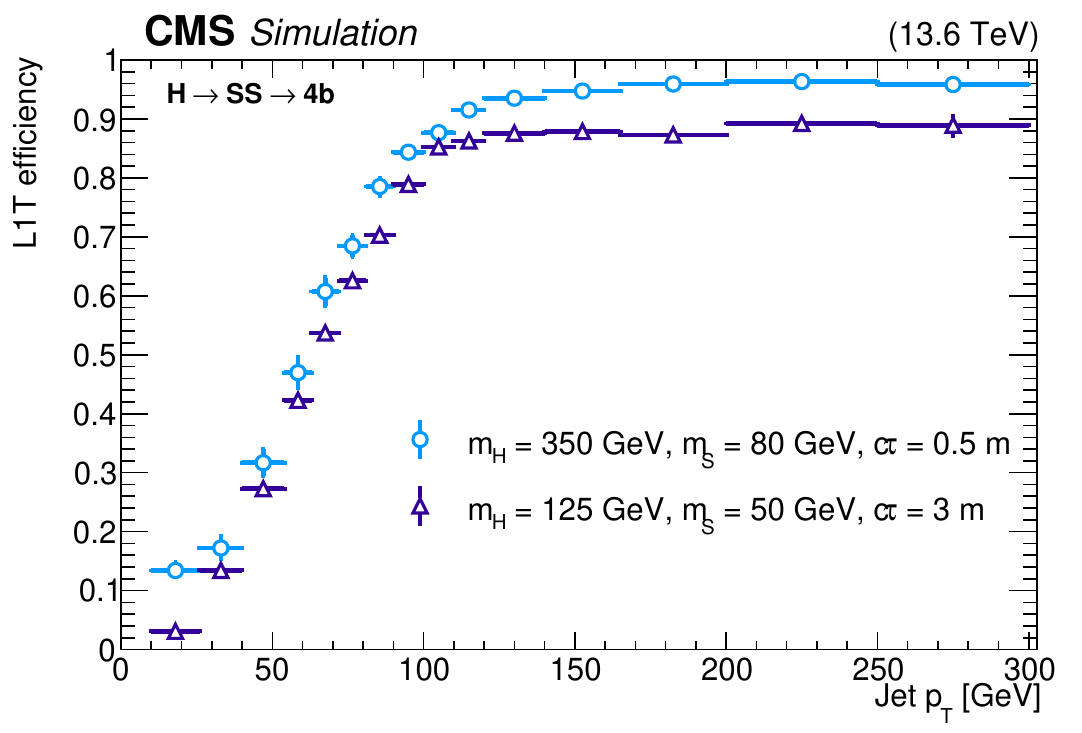}
\caption{The L1T efficiency of the HCAL-based LLP jet triggers, as a function of event \HT (\cmsLeft) and jet \pt (\cmsRight), for $\PH \to \PS\PS \to \bbbar\bbbar$ events with $\mH=350\GeV$, $\mS=80\GeV$, and $\cTauPS=0.5\unit{m}$ (light blue circles) and $\mH=125\GeV$, $\mS=50\GeV$, and $\cTauPS=3\unit{m}$ (purple triangles), for 2023 conditions. The trigger efficiency is evaluated for LLPs decaying in HB depths 3 or 4, corresponding to $214.2< R<295\cm$ and $\abseta< 1.26$. These LLPs are also required to be matched to an offline jet with $\pt > 40 \GeV$ in HB.}
\label{fig:HLT_energy}
\end{figure}

Figure~\ref{fig:HLT_energy} shows the L1T efficiency of the HCAL-based LLP jet trigger as a function of offline \HT and calorimeter jet \pt. The efficiency is shown for jets matched to LLPs decaying in HCAL depths 3 and 4, corresponding to an LLP decay radial position $R = \sqrt{\Delta x^2 + \Delta y^2}$ of $214.2 < R < 295\cm$ and $\abseta < 1.26$. The efficiency plateau is reached at 300\GeV in event \HT and 125\GeV in calorimeter jet \pt. Notably, these dedicated LLP jet triggers enable increased efficiency for relatively low-\pt events that are inaccessible with purely \pt-based triggers. The HCAL-based LLP jet triggers increase the acceptance at L1 for signals with $\mH=125\GeV$, $\mS=50\GeV$, and $\cTauPS=3\unit{m}$ by up to a factor of 4 compared to standard L1 \HT triggers. These paths provide added efficiency for low-\pt and low-\HT LLP events, complementing the displaced-jet HLT paths seeded by L1 \HT triggers discussed in Section~\ref{sec:displacedjetstracking}.

Prior to the implementation of a dedicated L1T HCAL-based LLP trigger, standard L1T \HT or jet triggers were used to seed the displaced-jet paths at the HLT. Before 2023 data taking, the lowest threshold for an unprescaled \HT L1T was 360\GeV, and this trigger reaches 80\% efficiency at an offline \HT of 400\GeV. This L1T was used to seed the tracker-based displaced-jet HLT paths (Section~\ref{sec:displacedjetstracking}) in Run~2 and Run~3 and thus is referenced for comparison. On the other hand, the lowest unprescaled HCAL-based L1T LLP jet path requires $\HT > 200\GeV$, and Fig.~\ref{fig:HLT_energy} (\cmsLeft) demonstrates that this trigger reaches an efficiency of 80\% at an offline \HT of 300\GeV (for the $\mH=125\GeV$ signal point). In Run~3, the lowest \pt threshold for an unprescaled single jet L1T is 180\GeV, and this trigger reaches 80\% efficiency at an offline jet \pt of 200\GeV. By contrast, Fig.~\ref{fig:HLT_energy} (\cmsRight) demonstrates that the HCAL-based L1T LLP triggers achieve an efficiency of 80\% at an offline jet \pt of only 90\GeV. As illustrated in Table~\ref{Table:thresholds}, the dedicated HCAL-based delayed- and displaced-jet L1Ts implemented for Run~3 thus provide access to significantly lower energy final states than those accepted by standard jet or \HT-based triggers, broadening the kinematic phase space accessed by the CMS trigger program.

\begin{table}
    \centering
    \topcaption{The L1T minimum thresholds for \HT and jet \pt, comparing the thresholds of the standard trigger to those of the HCAL-based LLP jet trigger, for unprescaled paths that are used to seed LLP HLTs. The values in parentheses indicate the offline thresholds at which the L1Ts reach 80\% efficiency. The dedicated LLP L1Ts enable significantly lower energy requirements.}
    \cmsTable
{\renewcommand{\arraystretch}{1.1}
    \begin{tabular}{lcc}
        Trigger object & Standard L1T (offline) threshold [{\GeVns}] & HCAL-based LLP jet L1T (offline) threshold [{\GeVns}]\\ [\cmsTabTinySkip] \hline &\\[-\cmsTabTinySkipInverse]
        Jet \pt & 180 (200) & 60 (90)\\ 
        \HT & 360 (400) & 200 (300) \\
    \end{tabular}
    }
    \label{Table:thresholds}
\end{table}

\begin{figure}[htbp!]
\centering
\includegraphics[width=0.6\textwidth]{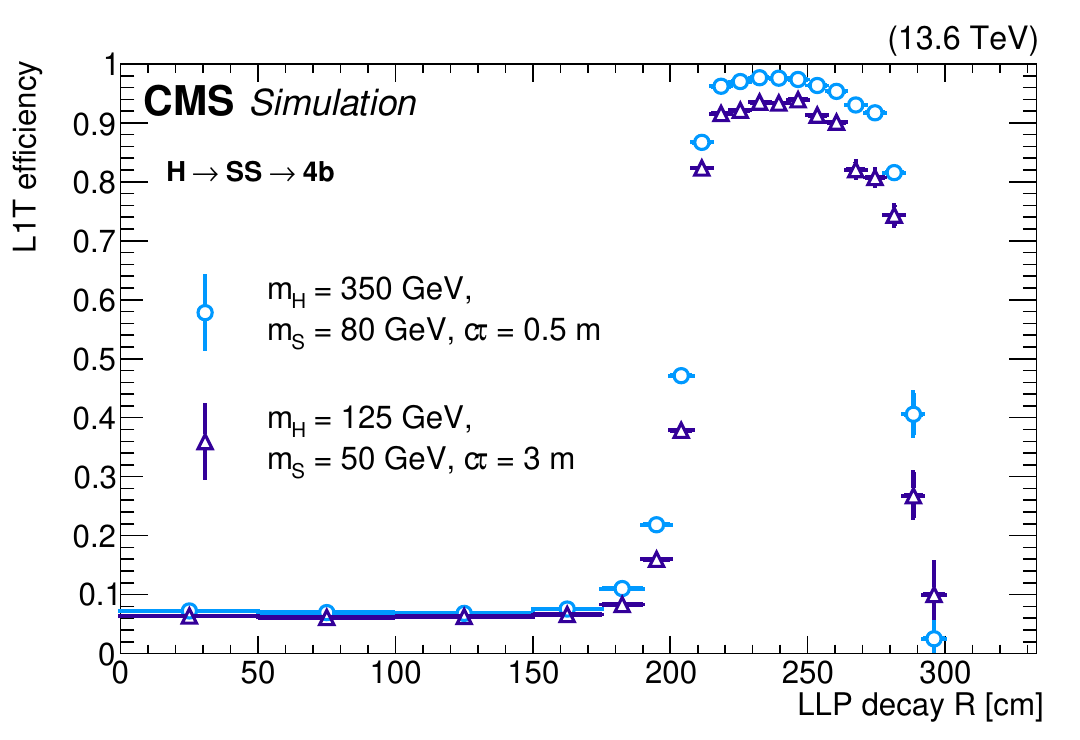} 
\caption{The L1T efficiency of the HCAL-based LLP jet triggers as a function of LLP decay radial position $R$ for $\PH \to \PS\PS \to \bbbar\bbbar$ events with $\mH=350\GeV$, $\mS=80\GeV$, and $c\tau_{\PS}=0.5\unit{m}$ (light blue circles) and $\mH=125\GeV$, $\mS=50\GeV$, and $c\tau_{\PS}=3\unit{m}$ (purple triangles), for 2023 conditions. The trigger efficiency is evaluated for LLPs within $\abseta <1.26$ where either the LLP or its decay products are matched to an offline jet in HB with $\pt>100\GeV$.}
\label{fig:HCAL_L1}
\end{figure}

Figure~\ref{fig:HCAL_L1} demonstrates the efficiency of the HCAL-based LLP jet trigger as a function of the LLP decay $R$. The LLP is required to decay within $\abseta <1.26$ and be matched to an offline jet with $\pt > 100\GeV$. The L1T has an efficiency of more than 90\% for LLP decays within HCAL depths 3 and 4, the region targeted by the depth criteria in the trigger to identify displaced LLP jets.

In general, the high-mass signal model reaches a higher efficiency at L1, as heavier LLPs are more likely to produce jets that pass the stringent calorimeter energy deposit requirements (multiple cells with over 5\GeV energy deposited in an individual depth) for the depth-based trigger. Triggering on low-mass signals is more difficult since they tend to have lower \HT and therefore are less likely to satisfy energy-based selections. The high-mass signal also reaches higher L1T efficiencies in jet \pt and event \HT because of the decay kinematic behavior. In this signal, the LLP is more boosted, leading to more collimated decay products. This is beneficial as the trigger requires multiple nearby depth-flagged trigger towers. The LLP proper decay length primarily impacts the signal acceptance rather than the efficiency.

A suite of displaced- and delayed-jet HLT paths is seeded by the HCAL-based LLP L1 jet triggers. The total rate of these HLT paths is about 35\unit{Hz}. Since the start of Run~3, a number of these paths follow the HLT approach detailed in Sections~\ref{sec:displacedjetstracking} and~\ref{sec:delayedjetsecal}. As the displaced- and delayed-jet HLT paths are seeded with dedicated LLP jet L1 seeds, and not standard \HT-based L1Ts, low \HT thresholds can be used at the HLT, while keeping the rates acceptable. The lowest-threshold unprescaled HCAL-based displaced-jet HLT path requires calo jet $\pt > 40\GeV$ and $\HT>170\GeV$, which is 230\GeV lower than the minimum \HT threshold used by similar standard paths. This path benefits from the unprescaled L1 double-jet seed, which has no \HT requirement at L1. 

\begin{figure}[htb!]
\centering
\includegraphics[width=0.48\textwidth]{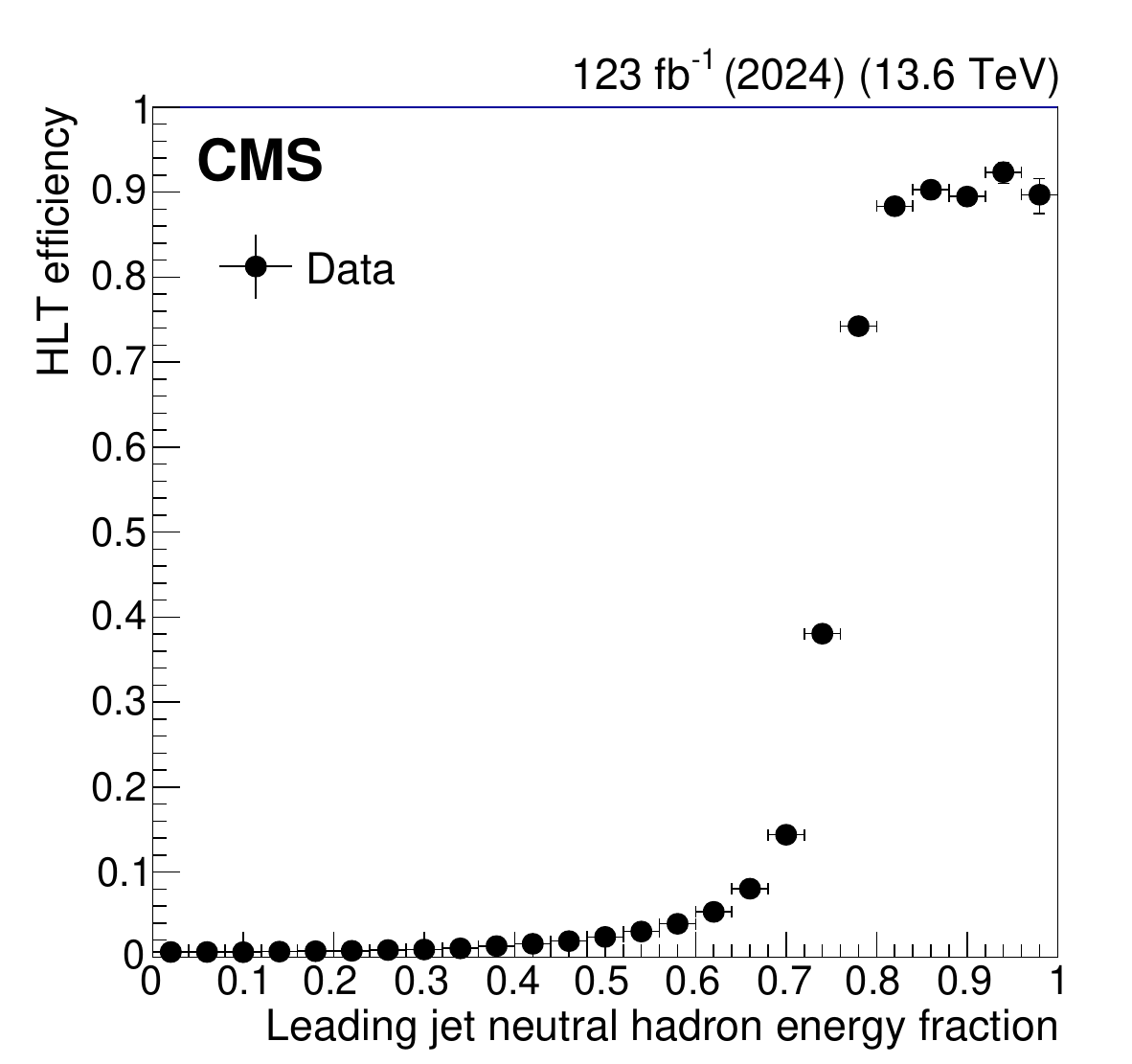}
\hfill
\centering
\includegraphics[width=0.48\textwidth]{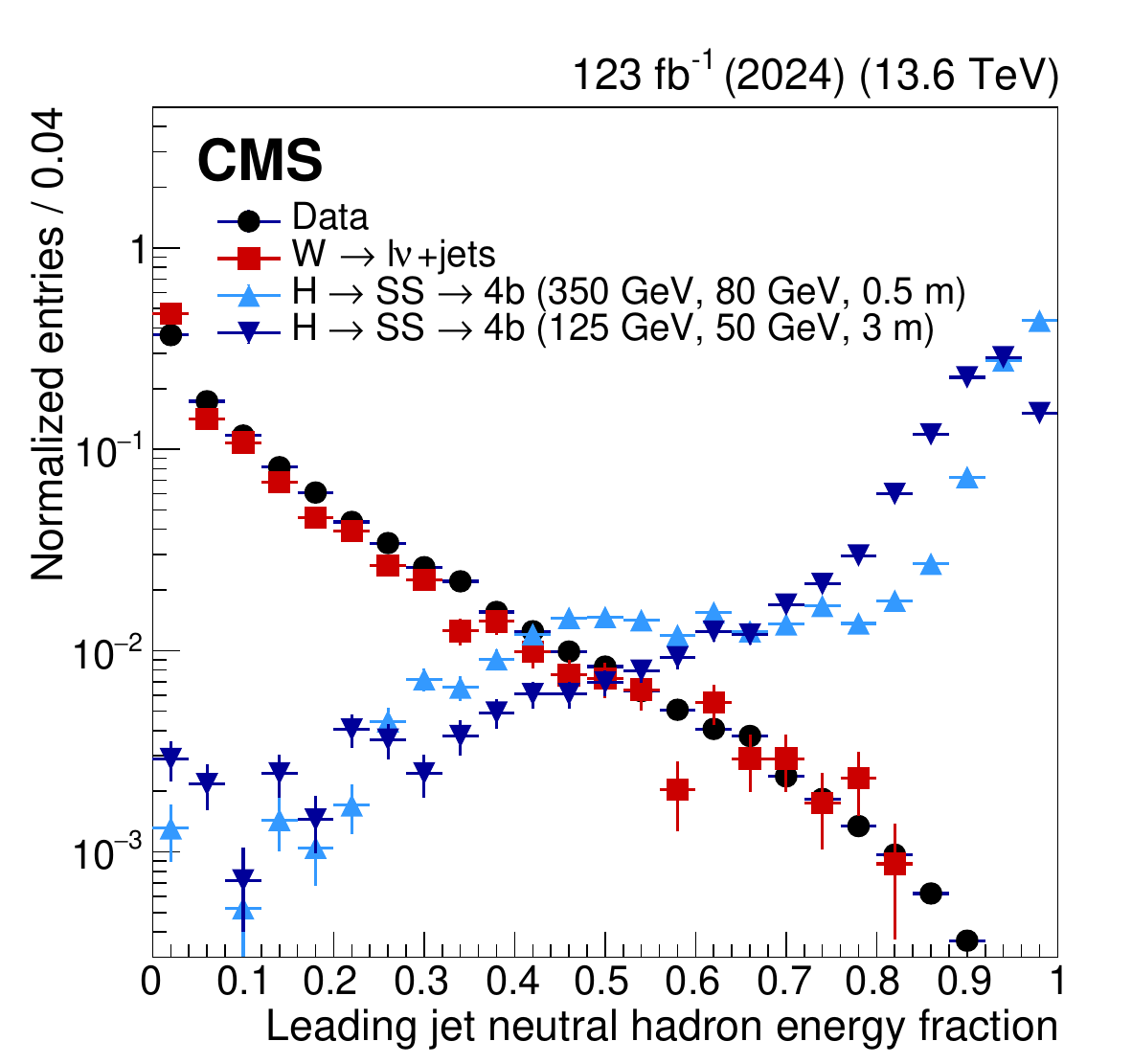}
\caption{The HLT efficiency of the CalRatio trigger as a function of the leading PF jet NHEF in 2024 data, measured with respect to a logical OR of the HCAL-based LLP L1 jet triggers (\cmsLeft). Distribution of the leading PF jet NHEF (\cmsRight) in 2024 data (black circles), \WtoLNu background simulation for 2024 conditions (red squares), and $\PH \to \PS\PS \to \bbbar\bbbar$ signal simulation for 2023 conditions (light blue and purple triangles). Events are required to have $\HT > 200\GeV$ and the leading jet is required to have $\pt > 60\GeV$ and $\abseta < 1.5$; these requirements are equivalent to the respective HLT jet object selections. The signal distributions additionally require the leading jet to be matched to an LLP decaying anywhere inside the barrel calorimeter volume ($129 < R < 295\cm$). The clear separation between the displaced signal and the prompt background in the \cmsRight plot motivates the development of the CalRatio trigger.}
\label{fig:HLT_CalRatio}
\end{figure}

In 2024, an additional HLT path was added to target highly displaced jets arising from LLPs decaying hadronically inside the calorimeters, complementing the depth-based L1 seed. This HLT path follows a similar strategy to the calorimeter-ratio trigger used by ATLAS for LLP decays in the hadronic calorimeter~\cite{ATLAS:2019qrr,ATLAS:2024ocv} and is therefore referred to here as the ``CalRatio'' trigger. This trigger selects jets that have a large neutral hadron energy fraction (NHEF). The composition of these objects may not actually be enriched in neutral hadrons. Instead, this selection is associated with a large fraction of energy deposited in the HCAL with minimal ECAL and tracker activity. In other words, this trigger selects for a high ratio of calorimeter energies, comparing the HCAL to the ECAL---a defining feature of displaced jets produced inside the hadronic calorimeter. This trigger selects events with $\HT>200\GeV$ containing at least one calo jet with $\pt>60\GeV$, $\abseta<1.5$, and hadronic energy fraction~$>0.7$; since calo jets do not contain tracking information, neutral and charged contributions cannot be distinguished. To suppress background and limit the rate, the trigger additionally requires at least one PF jet with the same kinematic requirements and $\text{NHEF}>0.7$. The HLT efficiency of the path, together with characteristic distributions of this variable in data, signal, and background, is shown in Fig.~\ref{fig:HLT_CalRatio} as a function of the NHEF of the highest \pt (leading) PF jet. The efficiency plateaus at around 90\% for $\text{NHEF}>0.8$; it does not reach 100\% largely because the additional offline kinematic selections (\eg, \HT and jet \pt), which are equivalent to their online quantities, are below their respective efficiency plateaus.

\subsubsection{Delayed-jet triggers using ECAL timing}
\label{sec:delayedjetsecal}

\begin{figure}[htb!]
\centering    \includegraphics[width=0.65\textwidth]{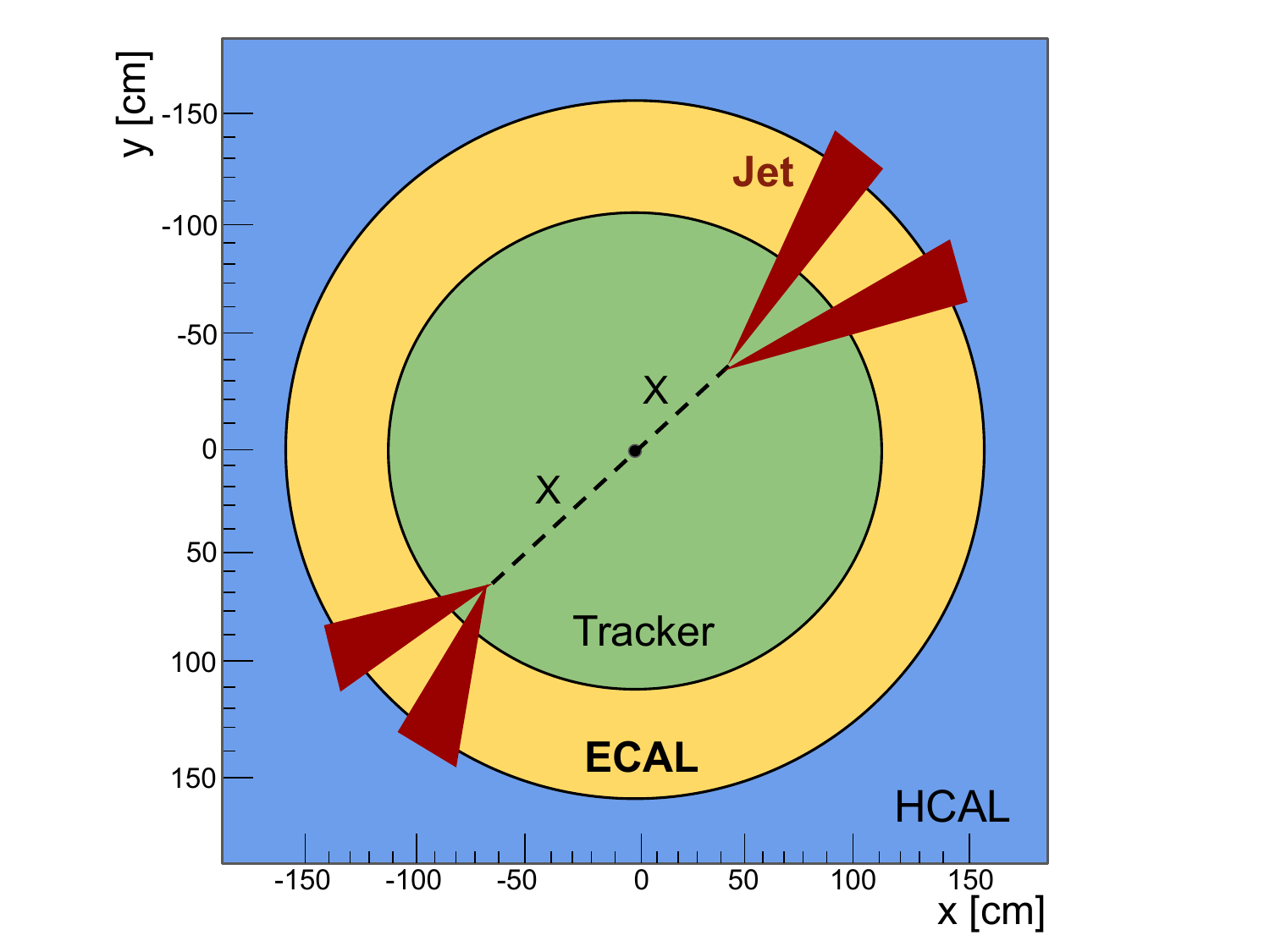}
\caption{\label{fig:EXO-19-001_signalDiagram} Diagram of a typical delayed-jet signal event.}
\end{figure}

Dedicated triggers targeting delayed-jet signatures were introduced at the HLT during Run~3 to improve the performance for nonprompt jets~\cite{CMS:2019qjk}. These triggers are highly efficient for LLPs that decay into jets within the tracker or the ECAL volume, as shown in Fig.~\ref{fig:EXO-19-001_signalDiagram}. Many extensions of the SM predict the existence of LLPs that decay into jet final states, including GMSB models~\cite{GIUDICE1999419} and hidden valley models~\cite{Strassler_2007}. The delayed-jet triggers utilize the timing capabilities of the ECAL~\cite{CERN-LHCC-97-033} to search for such signatures.

\begin{figure}[htbp!]
\centering
\includegraphics[width=0.6\textwidth]{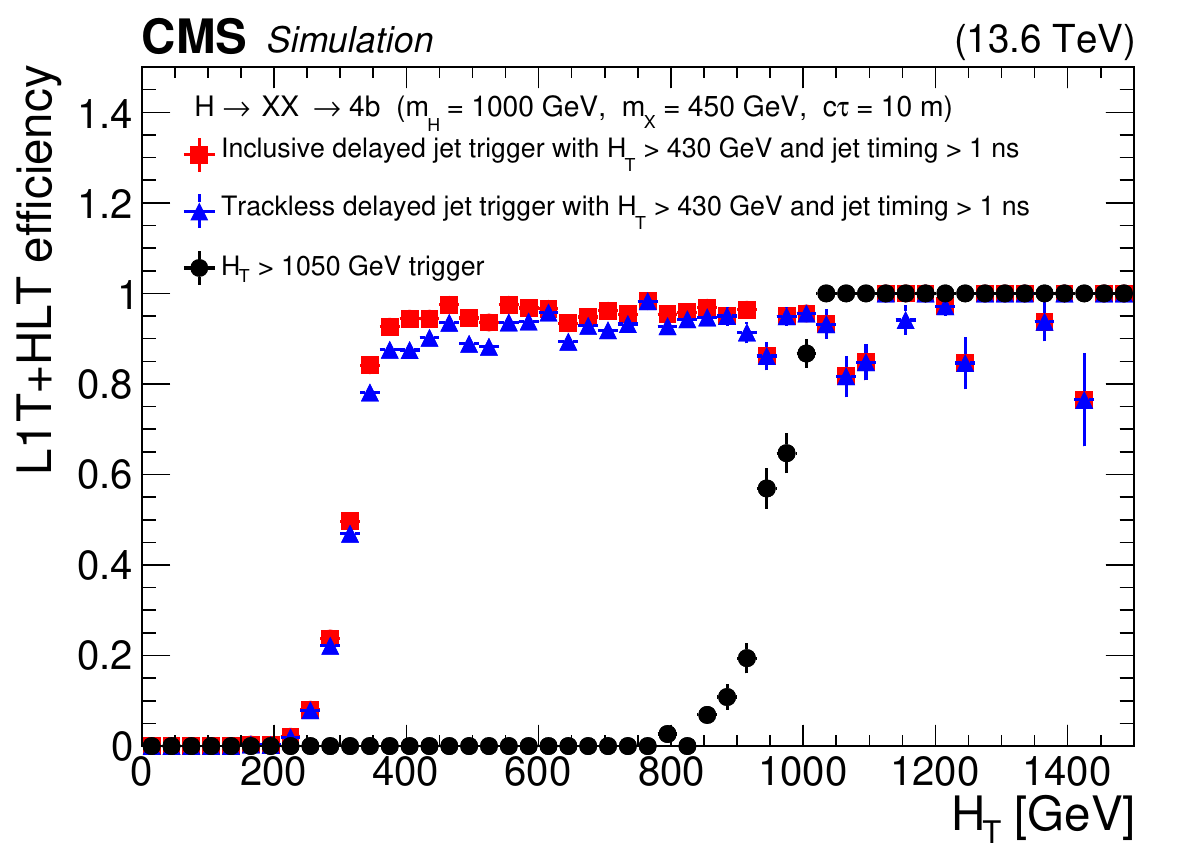}
\caption{The L1T+HLT efficiency of the inclusive (red squares) and trackless (blue triangles) delayed-jet triggers introduced in Run~3, for 2022 conditions, and the \HT trigger (black circles), which was the most appropriate path available in Run~2, for a $\PH \to \PX\PX \to 4\PQb$ signal with $\mH=1000\GeV$, $\mX=450\GeV$, and $\cTau=10\unit{m}$. The addition of these delayed-jet triggers results in a significant improvement in the efficiency of the signal for $430 < \HT < 1050\GeV$. These plots include events with jets with $\pt > 40\GeV$, number of ECAL cells $> 5$, barrel region with $\abseta<1.48$, and jet timing $> 1\unit{ns}$.}
\label{fig:delayedjetsecal-run2comparison}
\end{figure}

These HLT paths are seeded by L1Ts that require either $\HT > 350\GeV$, followed by an
additional $\HT > 430\GeV$ requirement at the HLT, or a \tauh candidate with $\pt > 120\GeV$. Jet timing requirements are applied at the HLT to single or double inclusive and trackless calo jets. A trackless jet is defined as a calo jet with at most two associated prompt tracks, while the inclusive jets do not have this additional requirement. Single (double) delayed-jet triggers refer to a requirement of at least one (two) delayed jets satisfying the trigger conditions. The triggers at the HLT include a variety of timing thresholds to balance the trigger rates against the signal efficiency. For example, the timing thresholds for the single jet triggers at the HLT are ${>}1\unit{ns}$, and the timing resolution is around 0.4\unit{ns}. The delayed-jet triggers are designed to be model-independent and more efficient for lower-\HT final states than the triggers used in the Run~2 analysis, which targeted final states with $\ptmiss>300\GeV$~\cite{CMS:2019qjk}. Figure~\ref{fig:delayedjetsecal-run2comparison} shows the improvement in the trigger efficiency in Run~3 with the dedicated delayed-jet triggers, compared to Run~2, where a standard trigger that requires $\HT>1050\GeV$ was available. The improvement is particularly noticeable at low \HT, for a signal model where very heavy BSM Higgs bosons decay into generic LLPs ($\PX$) and subsequently into 4 \PQb jets, $\PH \to \PX\PX \to \bbbar\bbbar$, the diagram for which is shown in Fig.~\ref{fig:HtoXX_feynmanDiagrams}. Some trigger inefficiencies are introduced because \HT is calculated offline using jets with $\pt>40\GeV$ and differs from the \HT calculated at the HLT. The trackless delayed-jet trigger has lower efficiency compared to the inclusive one (red squares) since it requires at most two prompt tracks associated with the delayed jet. However, the trackless trigger adds to the signal acceptance for certain signal parameter space regions, and its lower rate enables reducing the timing thresholds compared to the inclusive trigger. The delayed-jet trigger suite is a combination of inclusive triggers and trackless triggers with lower timing thresholds, both of which contribute to the overall signal acceptance.

\begin{figure}[htbp!]
\centering
\includegraphics[width=0.49\textwidth]{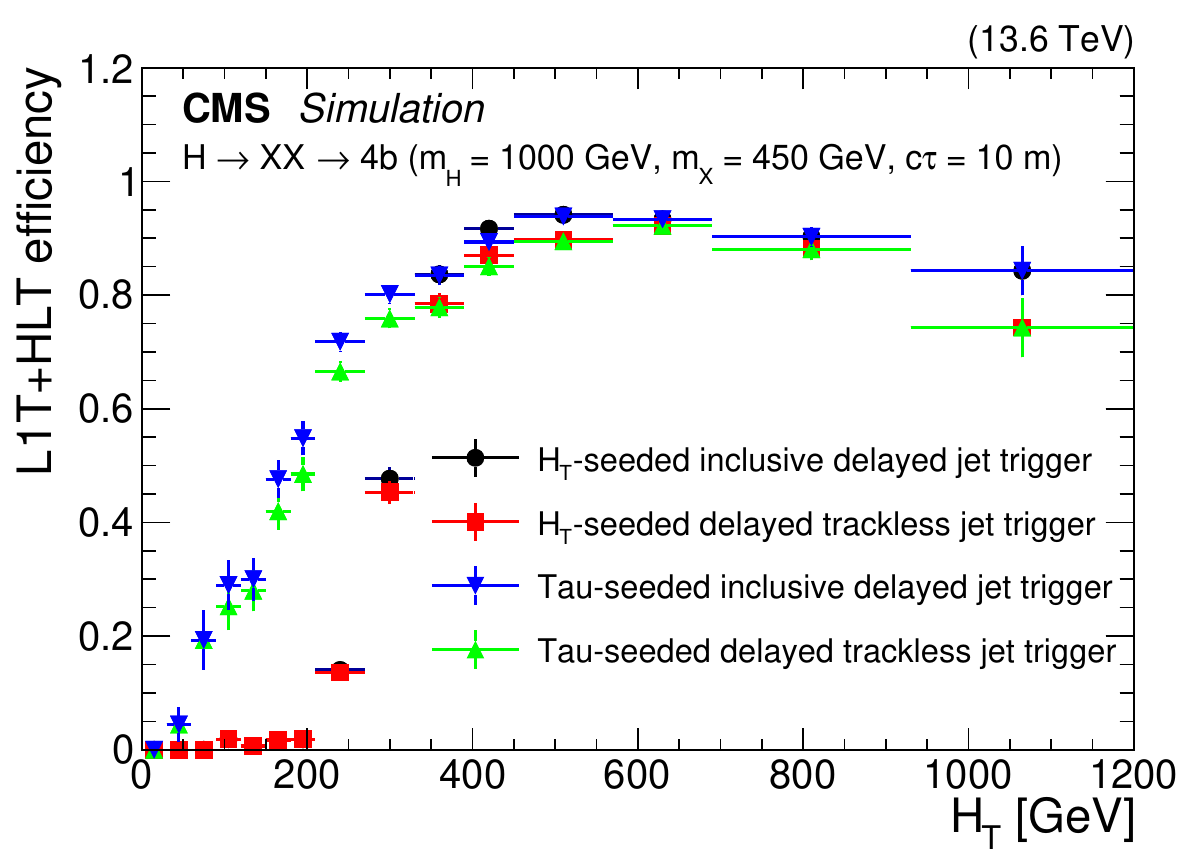}
\includegraphics[width=0.49\textwidth]{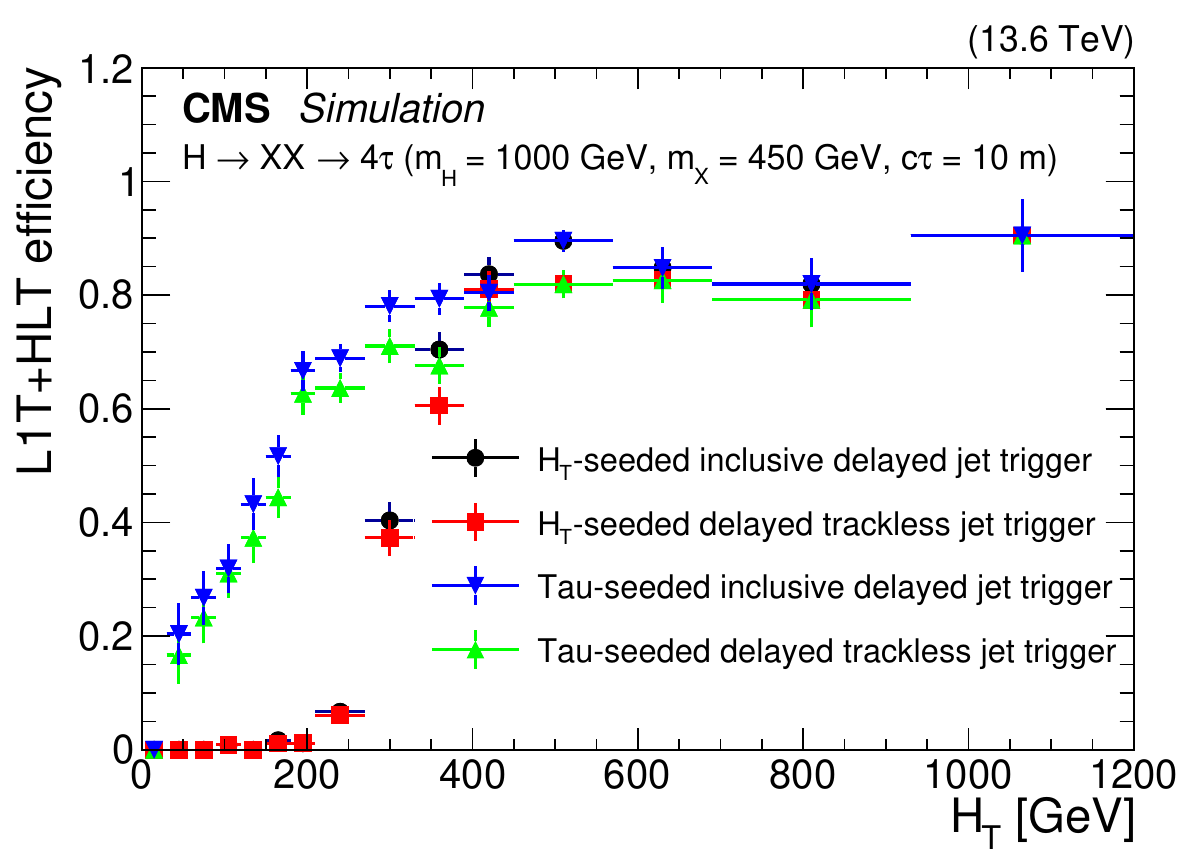}
\caption{The L1T+HLT efficiency of the delayed-jet triggers for signal models with $\PH \to \PX\PX \to 4\PQb$ (\cmsLeft) and $\PH \to \PX\PX \to 4\PGt$ (\cmsRight), with $\mH=1000\GeV$, $\mX=450\GeV$ and $\cTau=10\unit{m}$, for 2022 conditions. The improvement from the tau triggers (blue and green triangles) is evident in the $\HT < 430\GeV$ region compared to the \HT-seeded triggers (black circles and red squares). These plots include events with jets with $\pt > 40\GeV$, number of ECAL cells $> 5$, barrel region with $\abseta<1.48$, and jet timing $> 2\unit{ns}$.}
\label{fig:delayedjetsecal-tau}
\end{figure}

The inclusion of the tau candidate seed in the L1T improves the analysis sensitivity to low-\HT signal models, targeting LLPs that decay into tau leptons in addition to jets. Figure~\ref{fig:delayedjetsecal-tau} demonstrates the improvement in signal efficiency because of the L1 tau seed addition, particularly for the low-\HT final states for $\PH \to \PX\PX \to 4\PQb$ and $\PH \to \PX\PX \to 4\PGt$ decays. The decrease in efficiency at large \HT values is attributed to the large discrepancy between online and offline \HT for LLP signals with a large \cTau of 10\unit{m}.

\begin{figure}[htb!]
\centering
\includegraphics[width=0.49\textwidth]
{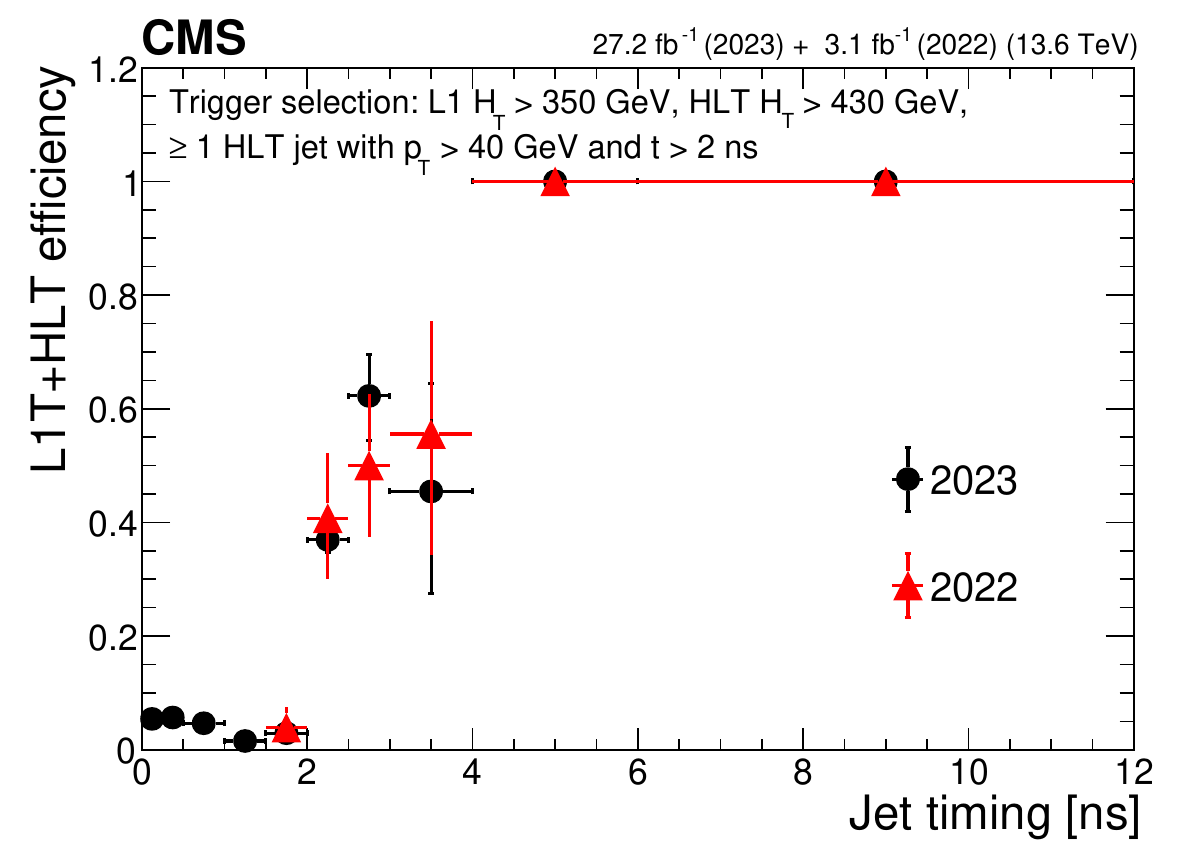}
\includegraphics[width=0.49\textwidth]{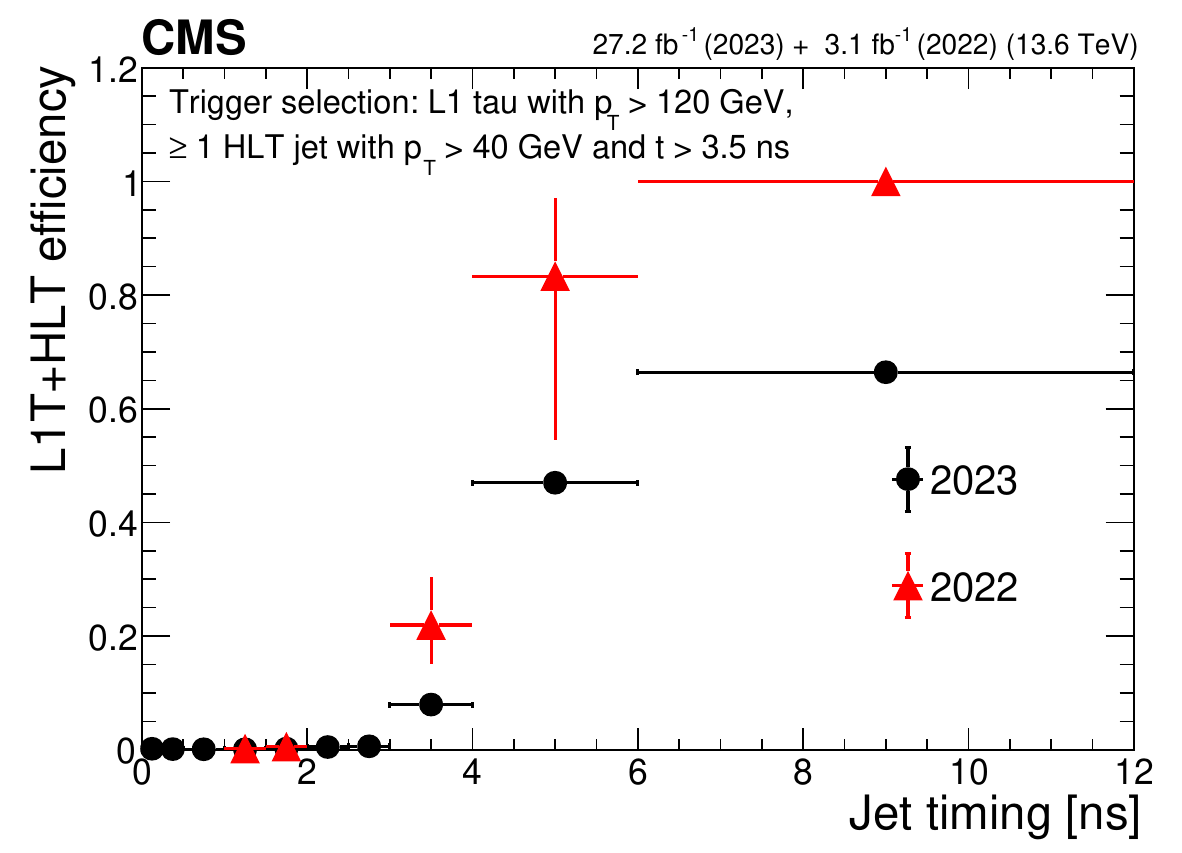}
\caption{The L1T+HLT efficiency of the delayed-jet triggers as a function of jet timing for 2022 (red triangles) and 2023 (black circles) data-taking periods for the \HT-seeded trigger (\cmsLeft) and the \PGt-seeded trigger (\cmsRight). A clear rise in efficiency is evident around the threshold values. The plots include events that pass the $\ptmiss>200\GeV$ trigger and have at least one barrel jet with $\pt>50\GeV$, number of ECAL cells $>8$, and ECAL energy $>25\GeV$. The \HT is calculated using the scalar \pt sum of jets with offline $\pt>40\GeV$, and this is different from the \HT calculation used at the HLT level, contributing to the trigger inefficiencies. The maximum jet time accepted by the trigger is 12.5\unit{ns}.}
\label{fig:delayedjetsecal-turnon}
\end{figure}

The total trigger rate of the ECAL-based delayed-jet triggers at the HLT is 37\unit{Hz} in standard data taking. Figure~\ref{fig:delayedjetsecal-turnon} shows the trigger performance in Run~3 data collected in 2022 and 2023 as a function of jet timing. A clear rise in efficiency is evident at the trigger timing threshold. The lower efficiency plateau observed for the \PGt-seeded path in 2023 with respect to that of 2022 is related to an update of the offline jet-timing reconstruction, which removes isolated out-of-time ECAL cells that are inconsistent with the timing of neighboring cells and are likely to arise from noise or mismeasurement.

To increase the analysis sensitivity, triggers with lower timing thresholds are introduced in the parked data with an upper bound on timing to ensure minimal overlap with the triggers in standard data taking~\cite{CMS:2024zhe}. These triggers take advantage of the higher rate allocation for the parked data-taking stream and have a lower timing threshold, thus providing higher signal acceptance. The parking triggers have a signal acceptance that is more than 30\% higher than their counterparts in the standard data stream. The total trigger rate for these triggers in the parked data-taking stream is 77\unit{Hz}.

\subsubsection{Delayed-diphoton triggers}
\label{sec:delayeddiphoton}

\begin{figure}[htb!]
\centering    
\includegraphics[width=0.64\textwidth]{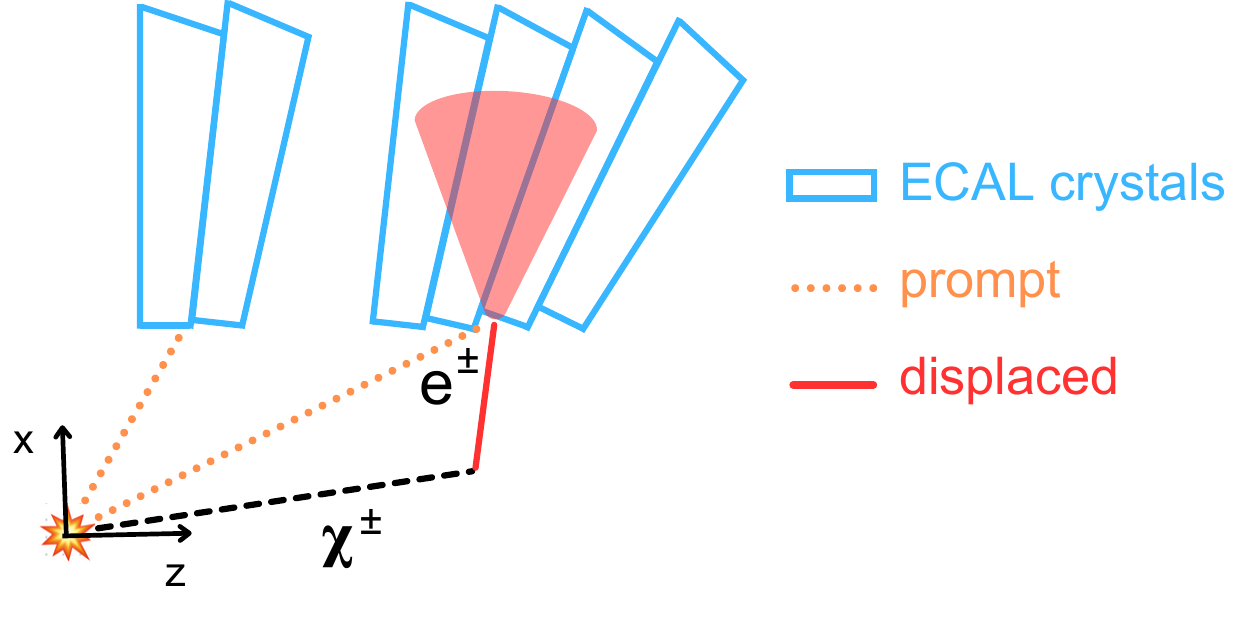}
\caption{\label{fig:llp_ecal_concept} Diagram of a prompt (orange dotted line) and displaced (red solid line) electron, which comes from the decay of a chargino, reaching the ECAL in the $x$-$z$ plane. The time measurement is shifted so that all prompt electron signals register the same time, no matter where they are located in the ECAL. An electron produced at a displaced vertex thus arrives delayed compared to a prompt particle.}
\end{figure}

Electrons and photons from the decay of an LLP whose mean path length is of the order of the CMS tracker dimensions will produce a distinct signature in the ECAL. The peak signal time of such an electromagnetic signature will be delayed compared to an electron or photon created from the proton collision vertex at the center of the detector. This is conceptually illustrated in Fig.~\ref{fig:llp_ecal_concept}. The time delay at the seed crystal of the ECAL supercluster, called $\Pe/\PGg$ seed time, is chosen as a representative value of the time delay of the delayed ECAL signature. For a displaced vertex in the inner pixel tracker region, the expected delay in the peak ECAL electromagnetic signature is of the order of nanoseconds, which is significantly larger than the ECAL timing resolution of ${<}0.2\unit{ns}$ for energies ${>}10\GeV$ in the barrel, as mentioned in Section~\ref{sec:detector}.

\begin{figure}[htb!]
\centering
\includegraphics[width=0.49\textwidth]{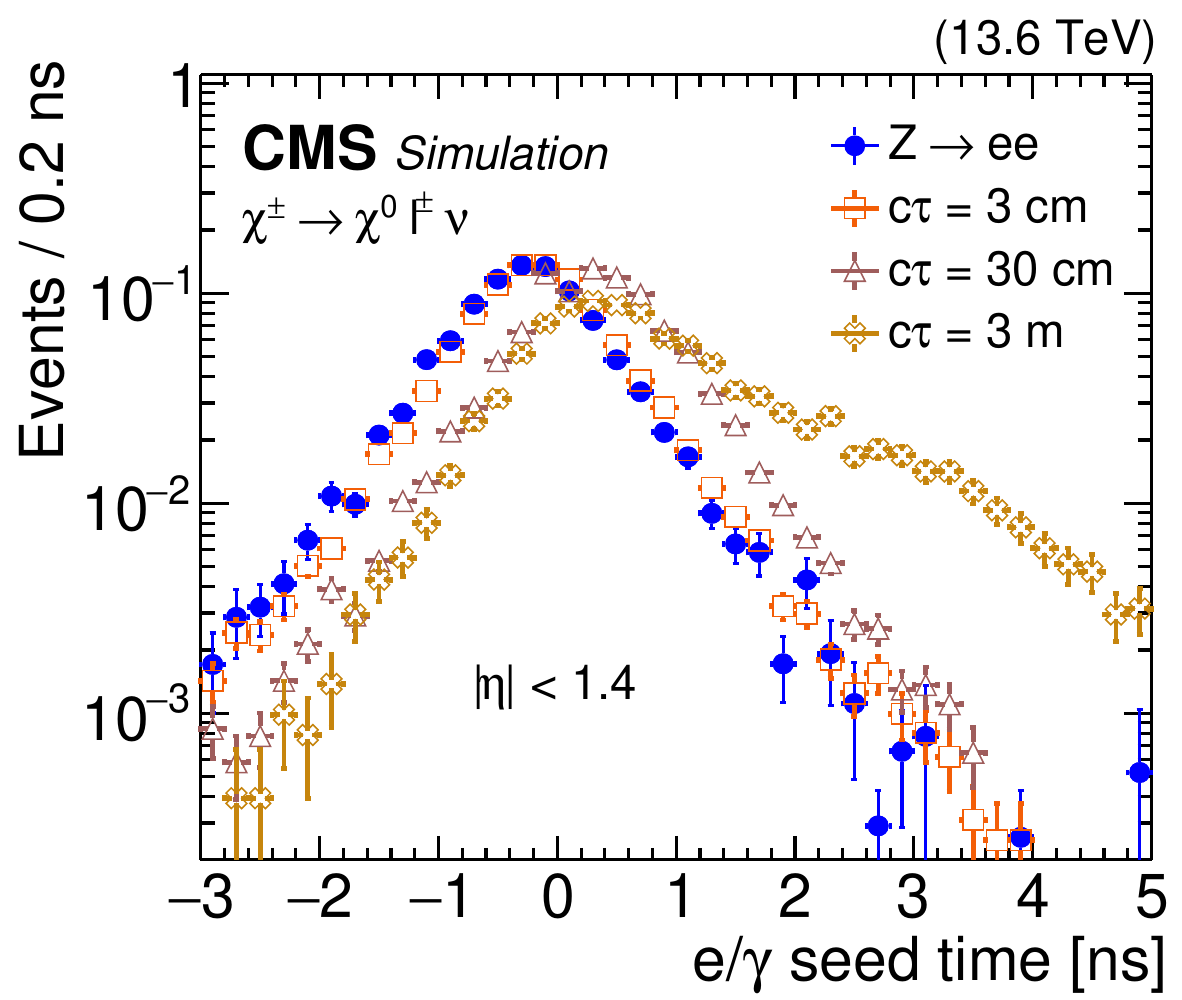}
\includegraphics[width=0.49\textwidth]{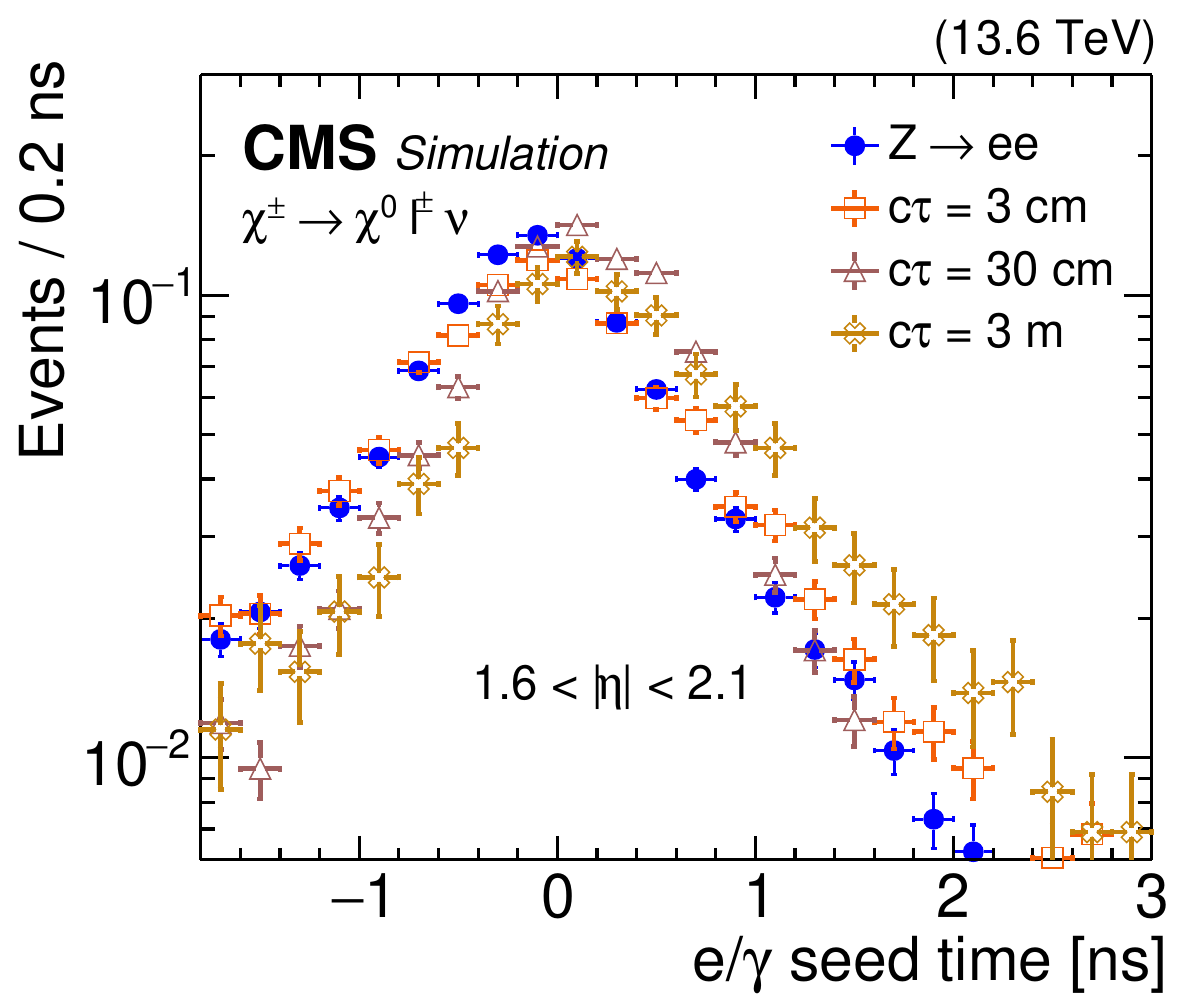}
\caption{\label{fig:llp_ecal_time}The ECAL time delay of the $\Pe/\PGg$ L1 seeds in the barrel (\cmsLeft) and endcap (\cmsRight). The distributions are shown for $\PZ \to \Pe\Pe$ simulation (filled blue circles) and \PARchipm \cTau values of 3\cm (open red squares), 30\cm (open purple triangles) and 3\unit{m} (open yellow crosses), assuming the singlet-triplet Higgs dark portal model ($\chi^{\pm} \to \chi^{0} \Pell \PGn$, where the \PARchipm has a mass of 220\GeV and the \PARchizero has a mass of 200\GeV), for 2023 conditions. The distributions are normalized to unity.}
\end{figure}

Figure~\ref{fig:llp_ecal_time} shows the expected ECAL time delay for simulated LLP signals with different \cTau values and for $\PZ \to \Pe \Pe$ simulation. An ECAL supercluster reconstructed with the online reconstruction algorithm is matched by angle to a generator-level electron. The seed time distribution for electrons from the \PZ boson decay shows the spread in the distribution of the prompt $\Pe/\PGg$ objects~\cite{CMS:2009onm}. The LLP decay appears in the ECAL $\mathcal{O}(ns)$ later than the prompt $\PZ \to \Pe\Pe$ simulation. The ECAL timing measurement is affected by radiation damage to the ECAL crystals, which reduces the efficiency to trigger on delayed electrons and photons. The radiation damage is larger in the ECAL endcap, which results in worse timing resolution than the barrel and thereby smaller apparent differences in the seed time distribution between LLP signals with different \cTau values. 

A new trigger was introduced in Run~3 to improve the CMS sensitivity to delayed electrons and photons arising in the singlet-triplet Higgs portal dark matter model discussed in Ref.~\cite{Blekman:2020hwr}. This trigger uses single and double $\Pe/\PGg$ seeds at the L1T, and it selects both delayed electrons and photons with the same HLT path by applying requirements only to ECAL superclusters. No electron track requirement is applied to create specific trigger paths for electrons because the typical electron track reconstruction algorithm is not efficient for displaced electrons. Although this trigger is highly efficient for both delayed electrons and photons, we call it ``the delayed-diphoton trigger'' in what follows, for simplicity. 

\begin{figure}[ht!]
\centering    
\includegraphics[width=0.46\textwidth]{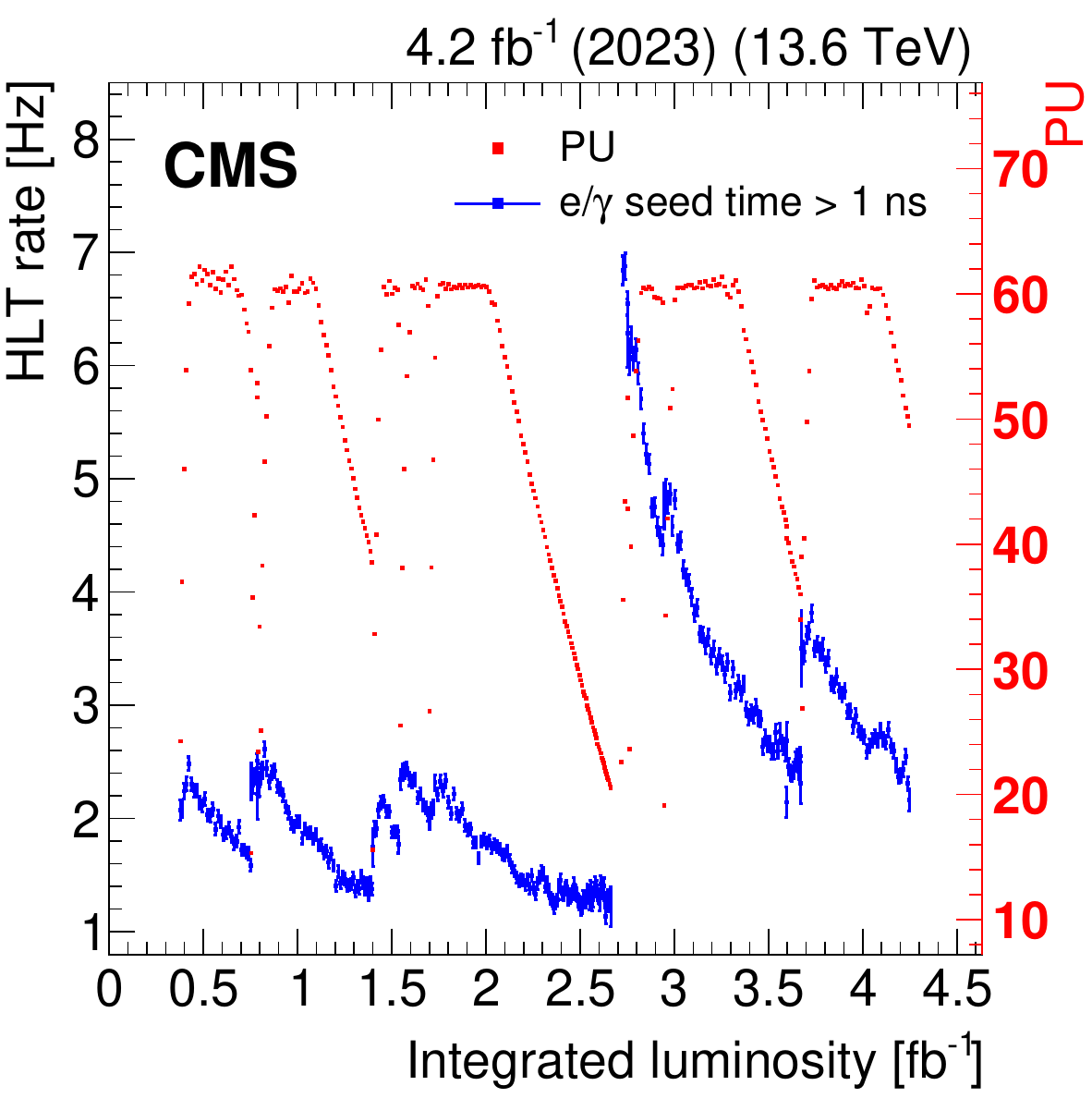}
\includegraphics[width=0.49\textwidth]{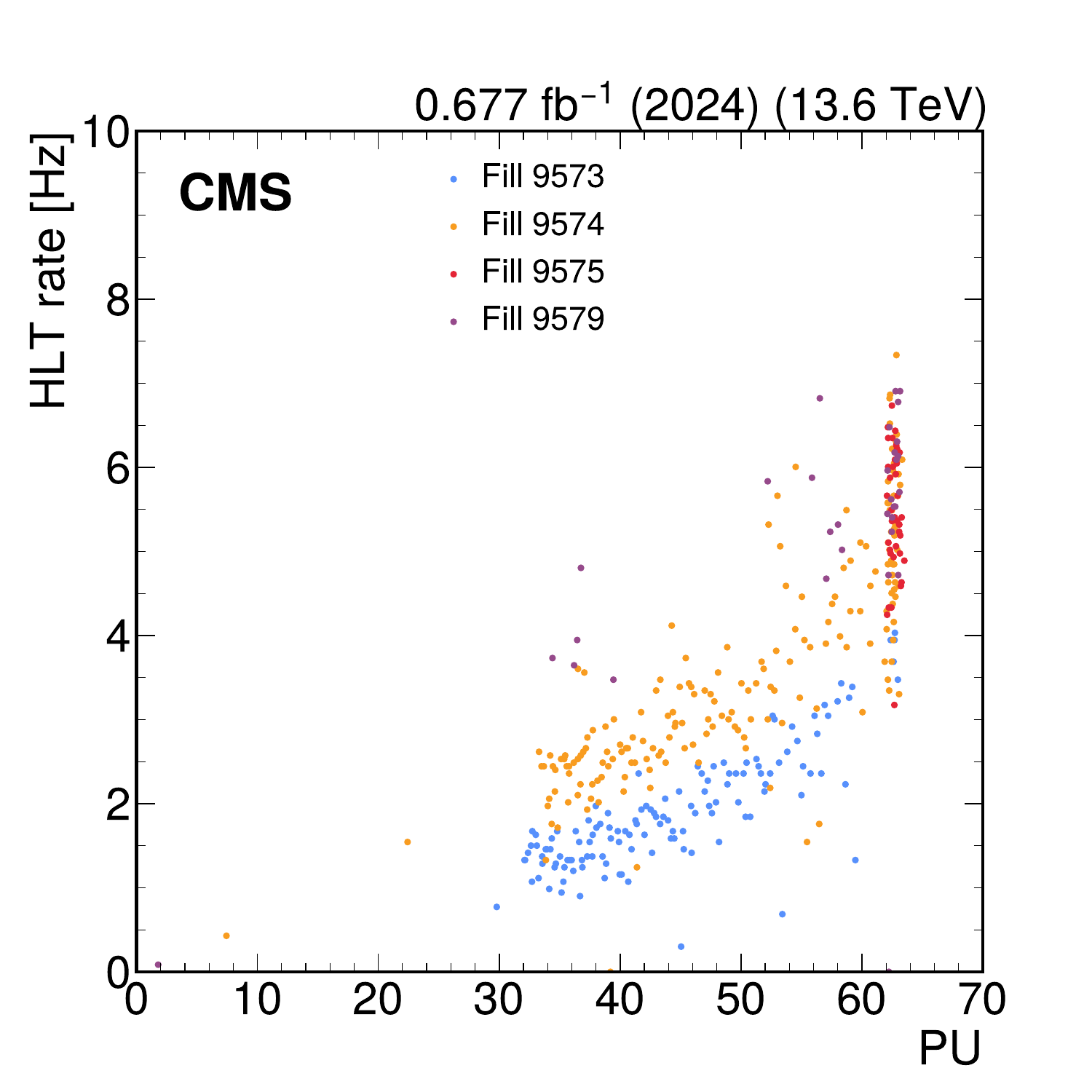}
\caption{\label{fig:llp_ecal_rate}The HLT rate (blue points) of the delayed-diphoton trigger for a few representative runs in the first data collected in 2023, corresponding to an integrated luminosity of 4.2\fbinv, compared with the PU during the same data-taking period (red points), as a function of integrated luminosity (\cmsLeft). The rate decreases nonlinearly during a single fill as a result of the increasing crystal opacity. It recovers by the start of the next fill with ${<}1\%$ reduction in rate between the fills. The rate generally increased throughout the year because of periodic online calibrations to mitigate the loss in trigger efficiency, which is a result of the ECAL crystal radiation damage. The delayed-diphoton trigger rate is shown as a function of PU for selected fills in 2024 data, at an instantaneous luminosity of approximately \instL{1.8} (\cmsRight). The trigger rate displays a linear dependency on PU.}
\end{figure}

The delayed-diphoton trigger requires at least two ECAL superclusters with $\abseta<2.1$ and $\ET > 22\,(12)\GeV$ for the leading (subleading) supercluster, where the ``subleading'' supercluster has the second-highest \ET. This asymmetric minimum energy threshold is reduced to a symmetric $\ET > 10\GeV$ threshold on both superclusters if the event has PF $\HT > 300\GeV$. The time delay for the crystal with the largest energy is required to be larger than 1\unit{ns} for both ECAL superclusters.

\begin{figure}[ht!]
\centering
\includegraphics[width=0.49\textwidth]{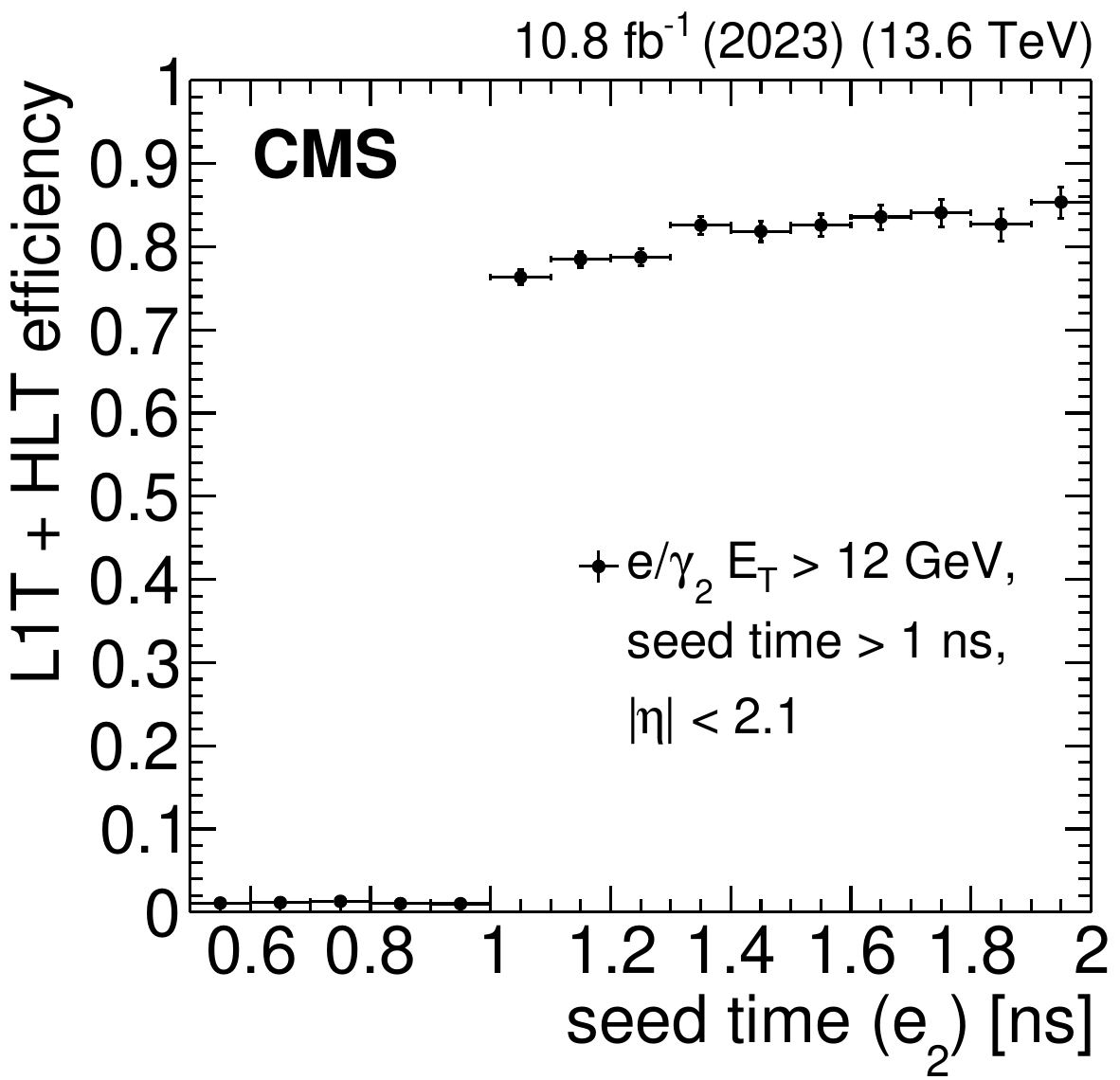}
\caption{\label{fig:llp_ecal_time_eff}The L1T+HLT efficiency of the delayed-diphoton trigger as a function of the subleading probe electron ($\Pe_2$) supercluster seed time, measured with data collected in 2023. At the HLT, the subleading $\Pe/\PGg$ supercluster ($\Pe/\PGg_2$) is required to have $\ET>12\GeV$, $\abseta<2.1$, and a seed time ${>} 1\unit{ns}$. The trigger reaches maximum efficiency above 1\unit{ns}.}
\end{figure}

\begin{figure}[hb!]
\centering
\includegraphics[width=0.49\textwidth]{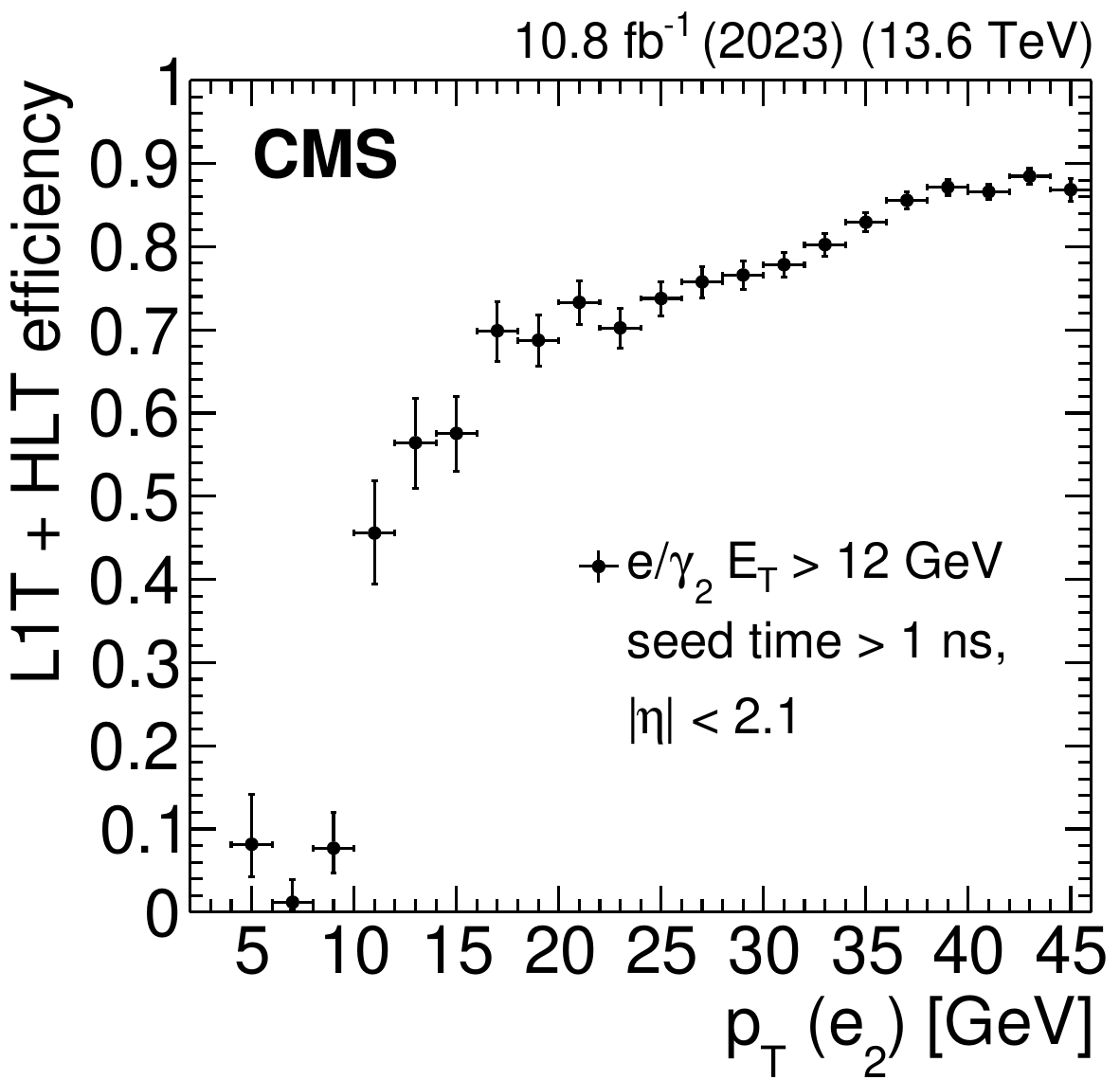}
\includegraphics[width=0.49\textwidth]{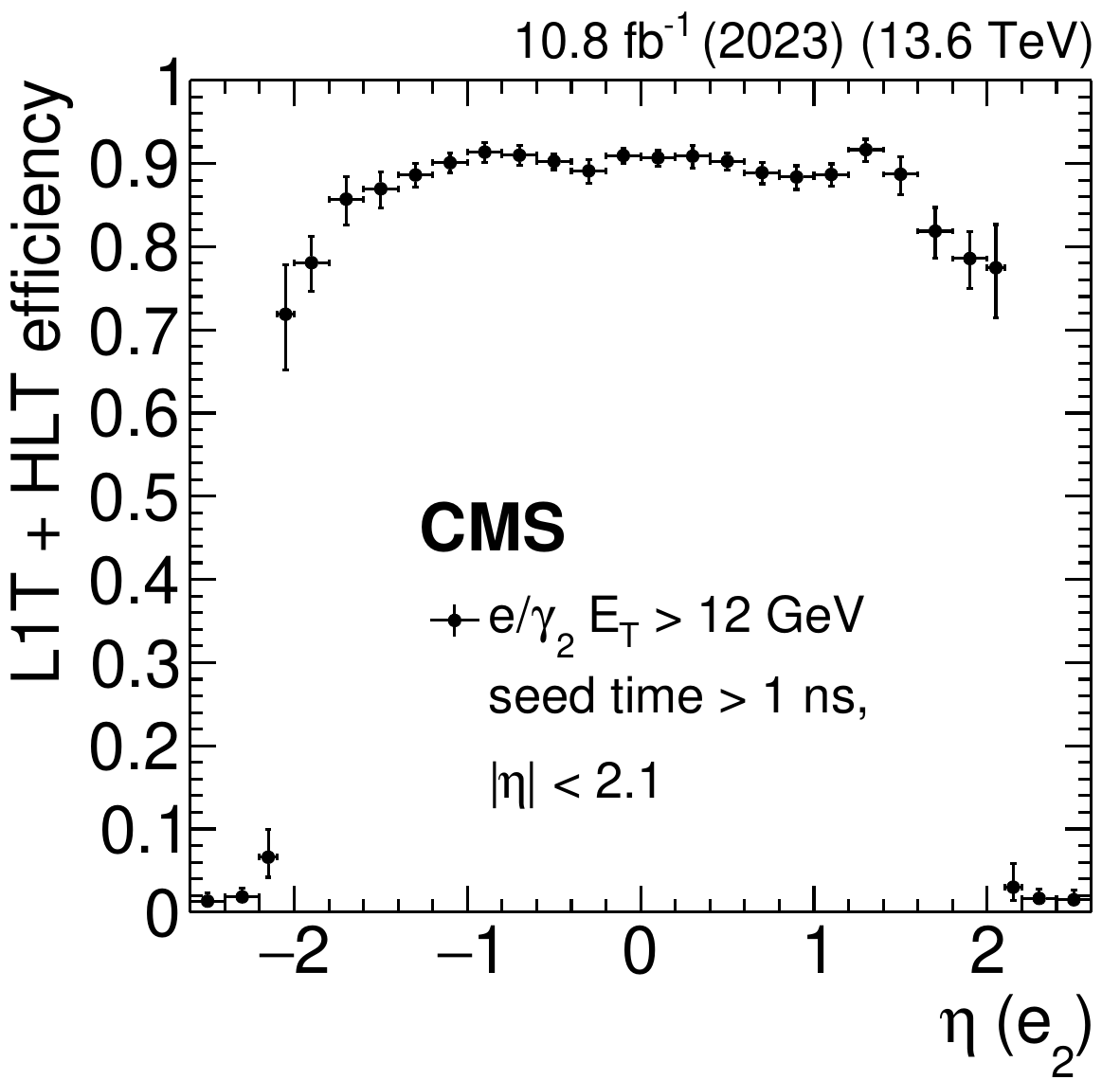}
\caption{\label{fig:llp_ecal_kin_eff}The L1T+HLT efficiency of the delayed-diphoton trigger as a function of subleading probe electron ($\Pe_2$) \pt (\cmsLeft) and $\eta$ (\cmsRight), measured with data collected in 2023. At the HLT, the subleading $\Pe/\PGg$ supercluster ($\Pe/\PGg_2$) is required to have $\ET>12\GeV$, $\abseta<2.1$, and a seed time ${>}1\unit{ns}$. The efficiency rises sharply for $\pt > 12\GeV$ and plateaus for $\pt > 35\GeV$. The slow rise in between is from additional L1 \HT requirements. The trigger is efficient in the region $\abseta < 2.1$.}
\end{figure}

Figure~\ref{fig:llp_ecal_rate} (\cmsLeft) shows the dependence of the delayed-diphoton trigger rate on integrated luminosity and PU in 2024. Within a fill, even while the instantaneous luminosity is kept constant, the rate decreases nonlinearly as a result of the increasing ECAL crystal opacity, which reduces the light yield and shifts the energy spectrum to smaller values and the timing distribution towards negative values. Interfill energy calibrations restore efficiency, as shown at integrated luminosities of 0.7 and 1.4\fbinv, indicating minimal permanent damage on short timescales. Over extended running periods, however, radiation damage can significantly degrade sensitivity, necessitating periodic online timing calibrations to correct negative shifts. These calibration updates maintain high efficiency for LLP signals with a $1\pm0.2\unit{ns}$ delay, though the calibration is intentionally overestimated by 0.2\unit{ns} to account for radiation effects between long shutdowns of the LHC. At an integrated luminosity of 2.7\fbinv, this produces a temporary ``spike'' in rate, which diminishes over subsequent fills as a result of the nonlinear timing dependence, leaving the average rate higher than before. Additionally, progressive timing resolution degradation throughout the year necessitates a higher trigger rate. In 2024, the rate scales linearly with PU, as seen in Fig.~\ref{fig:llp_ecal_rate} (\cmsRight). The spread in rate for a single PU value is caused by the reduction in rate during the period of constant instantaneous luminosity, as described above.

The total rate of these HLT paths is about 15\unit{Hz}. As explained at the beginning of Section~\ref{sec:LLPtriggers}, this rate is calculated from data taken on October 4, 2024, soon after an energy calibration update was applied. As a result, this rate is about a factor of 3 larger than the rates reported in Fig.~\ref{fig:llp_ecal_rate}.

\begin{sloppypar}Data collected in the later part of 2023 and corresponding to an integrated luminosity of 10.8\fbinv are used to measure the trigger efficiency of the delayed-diphoton HLT path, using the ``tag-and-probe'' method with $\PZ \to \Pe\Pe$ decays~\cite{CMS:2011aa}. The limited resolution of the online calibration results in a set of \PZ boson decays with a prompt signature that can appear delayed by more than $1\unit{ns}$, which is large enough to facilitate the trigger efficiency measurement. The tag electron is required to have $\pt > 20\GeV$, $\abseta < 2.1$, cluster seed time ${>} 1\unit{ns}$, and to pass the tight identification~\cite{CMS:2020uim}. Probe electrons are only considered if the tag-and-probe electron pair has two electrons of opposite charges and the dielectron invariant mass is within a window of $84 < m_{\Pe\Pe} < 96\GeV$, consistent with the mass of the \PZ boson. The probe electron is required to have $\pt > 4\GeV$ and $\abseta < 2.5$. The trigger efficiency is defined as the ratio of the probe electrons that pass the HLT path to all the probe electrons that pass the selection criteria. Figure~\ref{fig:llp_ecal_time_eff} shows the trigger efficiency as a function of the subleading probe electron seed time, which sharply rises at a seed time of $1\unit{ns}$ and plateaus thereafter. Figure~\ref{fig:llp_ecal_kin_eff} shows the efficiency as a function of the subleading probe electron \pt and $\eta$, for events with a seed time ${>}1\unit{ns}$.\end{sloppypar}

\subsubsection{Displaced-photon + {\texorpdfstring{\HT}{HT}} triggers}
\label{sec:displacedphotonandht}

\begin{figure}[htb!]
\centering
\includegraphics[width=0.95\textwidth]{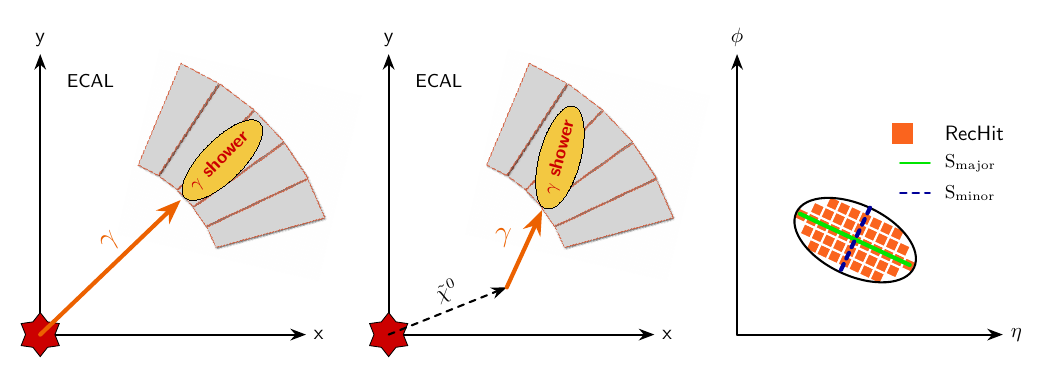}
\caption{Diagrams of a promptly decaying photon in the $x$-$y$ plane showering in the ECAL (\cmsLeft), a delayed photon produced from a long-lived neutralino \PSGcz in the $x$-$y$ plane showering later in the ECAL (middle), and an elliptical shower in the $\eta$-$\phi$ plane produced from a delayed photon (\cmsRight). The lengths of the major ($S_{\text{major}}$) and minor ($S_{\text{minor}}$) axes of the shower and the reconstructed hits (RecHits) that compose the shower are labeled.}
\label{fig:displacedPhotonDiagram}
\end{figure}

A trigger designed to collect events containing a displaced photon in addition to hadronic activity was originally deployed online in 2017, during Run~2. This displaced-photon + \HT trigger was designed for the search for delayed photons~\cite{CMS:2019zxa}, which targets the decays of long-lived neutralinos into displaced and delayed photons in a GMSB model~\cite{Allanach:2002nj}. The delayed photon signature is shown in Fig.~\ref{fig:displacedPhotonDiagram}. At the HLT, this trigger requires a photon with $\pt>60\GeV$; a set of loose, calorimeter-based identification requirements; loose isolation; and a loose set of requirements on the calorimeter shower shape, such that the shower is compatible with a displaced and delayed photon. The ECAL hit pattern from displaced photons is more elliptical in the $\eta$-$\phi$ plane than that of prompt photons, and so the length $S_{\text{major}}$ ($S_{\text{minor}}$) of the major (minor) axis of the shower is required to be less than $1.5\,(0.4)$. In addition, in Run~3, the trigger requires PF $\HT>350\GeV$, where \HT is the scalar sum of \pt for all jets satisfying $\pt>30\GeV$ and $\abseta<2.5$, to keep the rate around 12\unit{Hz} at the HLT. The L1 seeds of the HLT path require at least one $\Pe/\PGg$ object, tau lepton, or jet.

\begin{figure}[htb!]
\centering
\includegraphics[width=0.49\textwidth]{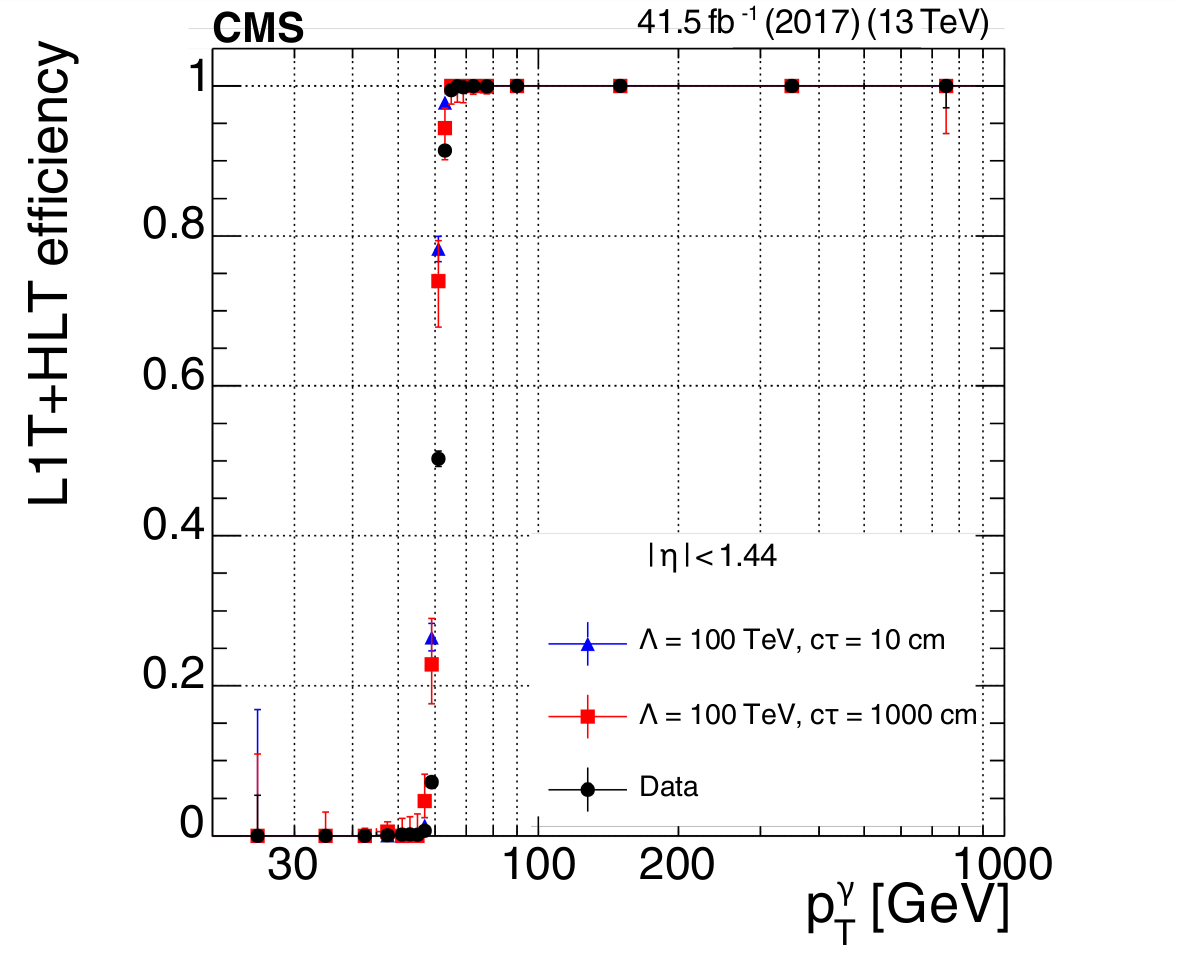}
\includegraphics[width=0.49\textwidth]{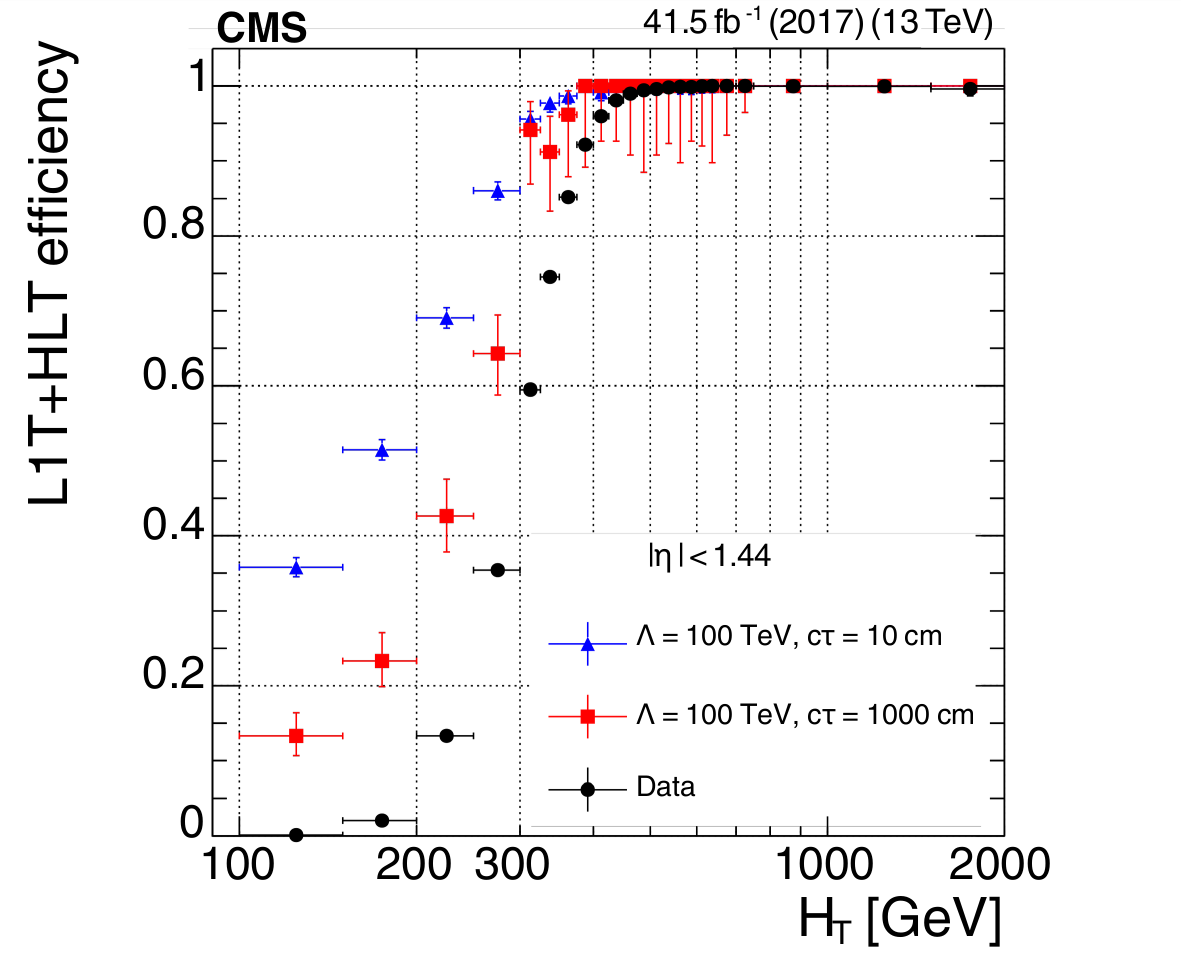}
\caption{The L1T+HLT efficiency of the displaced-photon + \HT trigger as a function of photon \pt (\cmsLeft) and event \HT (\cmsRight), for 2017 data (black circles) and GMSB signals with $\Lambda = 100\TeV$ and $\cTau=10\cm$ (blue triangles) or $\cTau=1000\cm$ (red squares).}
\label{fig:displacedPhotonPlusHTEffVsPhotonPtAndHT}
\end{figure}

The efficiency of the displaced-photon + \HT trigger is shown in Fig.~\ref{fig:displacedPhotonPlusHTEffVsPhotonPtAndHT} as a function of the offline photon \pt (\cmsLeft) and the event \HT (\cmsRight). Events in 2017 data (Run~2) and a GMSB signal benchmark with $\Lambda = 100\TeV$ and $\cTau=10$ or 1000\cm for 2017 conditions are selected that pass a trigger requiring at least one isolated muon with $\pt>27\GeV$ and at least one offline photon with $\abseta<1.44$ and passing the displaced-photon identification criteria as described in Ref.~\cite{CMS:2019zxa}. Only the leading photon satisfying these requirements is considered in the trigger efficiency measurements. These plots indicate that the trigger is fully efficient for photon $\pt > 70\GeV$ and $\HT > 400\GeV$. In 2017, the HLT path required $\HT > 350\GeV$ for jets satisfying $\pt>15\GeV$ and $\abseta<3$. The combined efficiency of the standard-photon identification and the displaced-photon identification in the displaced-photon$+\HT$ trigger is over 95\%. For a typical high-PU run in 2017, the displaced-photon identification requirements reduced the trigger rate by 20\%.

\begin{figure}[htb!]
\centering
\includegraphics[width=0.49\textwidth]{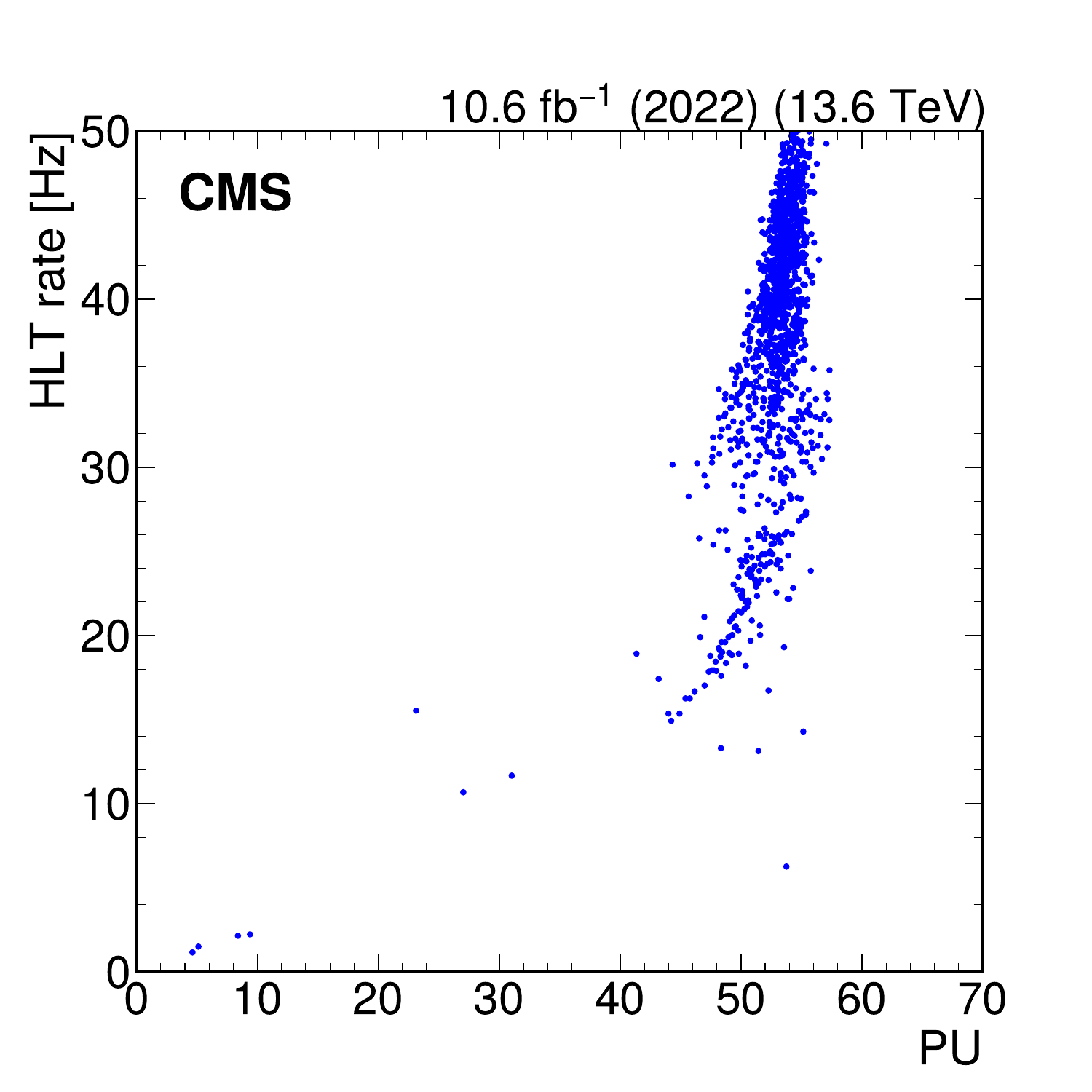}
\includegraphics[width=0.49\textwidth]{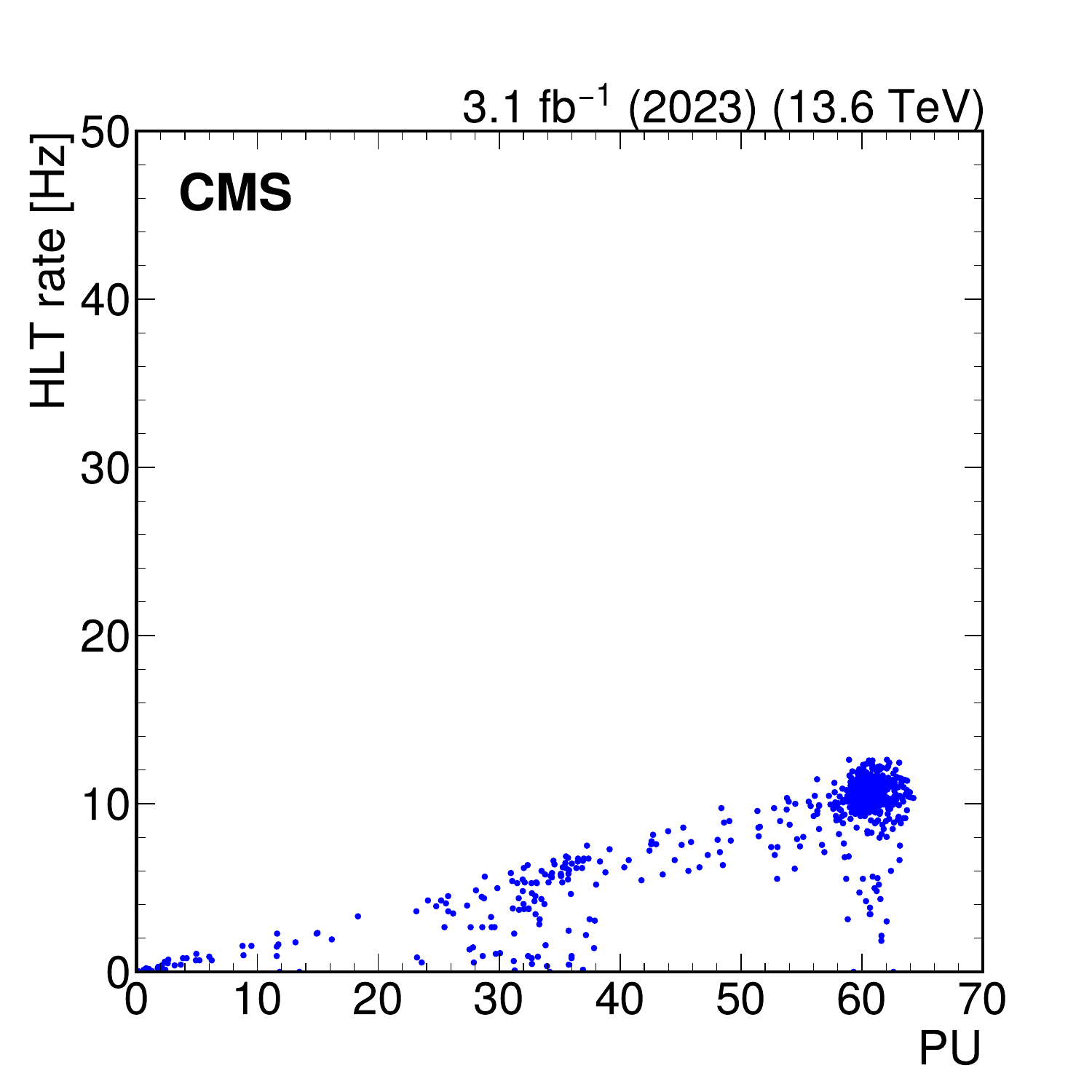}
\caption{Total rate of the displaced-photon + \HT HLT path for a few representative runs in 2022 data (\cmsLeft), at an instantaneous luminosity of approximately \instL{1.8}, and 2023 data (\cmsRight), at an instantaneous luminosity of approximately \instL{2.0}, as a function of PU. The rate vs PU behavior was nonlinear in 2022, and the jet \pt was increased to restore a linear dependence for 2023 data taking.}
\label{fig:displacedPhotonPlusHTRates}
\end{figure}

In 2022, the rate of this trigger increased more than initially expected as a result of the higher PU. The HLT rate as a function of PU was also nonlinear and thus difficult to maintain. The trigger was modified for 2023 data taking by increasing the minimum \pt requirement on the jets included in the PF \HT calculation at the HLT. After raising the minimum jet \pt threshold from 15 to 30\GeV, the total HLT rate was reduced to 12\unit{Hz} (about 4\unit{Hz} pure), and the behavior with respect to PU was again linear. Figure~\ref{fig:displacedPhotonPlusHTRates} shows the rate of this trigger in 2022 and 2023 data as a function of PU.

\subsection{Muon spectrometer-based algorithms}
\label{sec:muonLLPtriggers}

This section describes trigger algorithms designed to select LLP decays in the muon system. These include displaced-dimuon triggers (Section~\ref{sec:displaceddimuon}), double displaced L3 muon triggers (Section~\ref{sec:displacedL3muon}), displaced muon + photon triggers (Section~\ref{sec:displacedL3muonandphoton}), dimuon scouting triggers (Section~\ref{sec:dimuonscouting}), and MDS triggers with the CSCs (Section~\ref{sec:mdsshowers}) and with the DTs (Section~\ref{sec:dtcluster}).

\subsubsection{Displaced single-muon and dimuon triggers}
\label{sec:displaceddimuon}

\begin{figure}[htb!]
\centering
\includegraphics[width=0.8\textwidth]{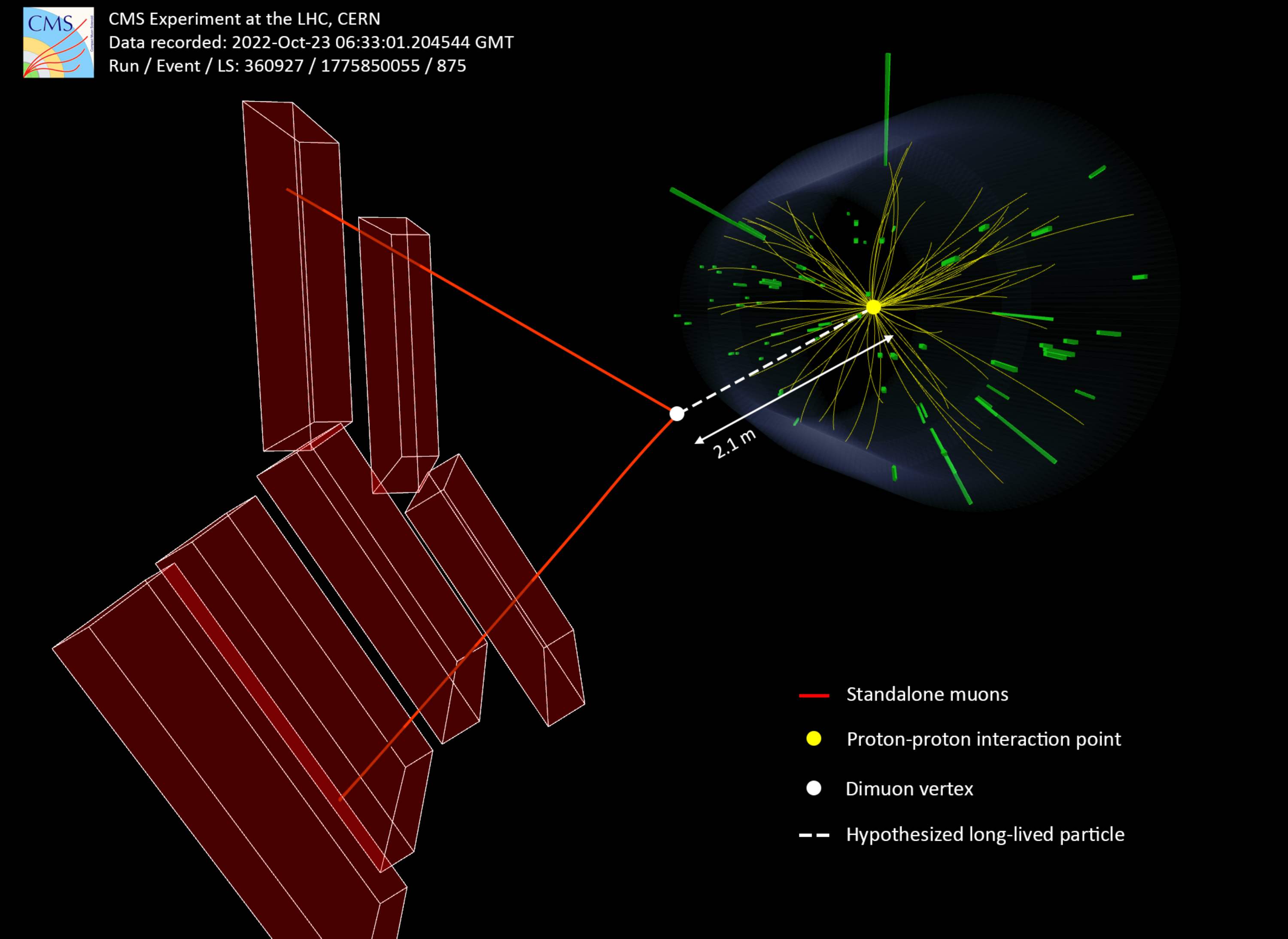}
\caption{A Run~3 data event containing a candidate LLP decay into a pair of muons away from the IP, reconstructed in the CMS detector. The red lines correspond to the two muons, which are detected only in the muon system. The muon tracks are used to calculate a dimuon vertex, indicated by the white circle, where the LLP is hypothesized to have decayed.
}
\label{fig:displacedDimuons_eventDisplay}
\end{figure}

An LLP produced in a hard interaction of the colliding protons can traverse a considerable distance in the detector before decaying into muons. When muons are produced well within the silicon tracker, they can be reconstructed by both the tracker and the muon system. However, if the muons are produced with larger displacements, they can only be reconstructed by the muon system. A pair of reconstructed muon tracks is fitted to a common secondary vertex forming a ``displaced dimuon''. All reconstructed dimuon events are classified into two mutually exclusive categories: (1) where both muons are global muons and are reconstructed using both the tracker and the muon system (TMS-TMS category), or (2) where both muons are standalone muons (STA-STA category). The TMS-STA category is not considered because it provides only a small fraction of events for the considered signals. Owing to the excellent resolution of the tracker, the position and momentum resolution of global and tracker muons are far superior to that of their standalone muon counterparts. This enhanced position resolution leads to an improvement in the resolution of the TMS-TMS dimuon vertex by a factor of 100 to 1000 for simulated HAHM events~\cite{Curtin:2014cca}, depending on the lifetime of the LLP and the masses of the exotic Higgs boson and the LLP. On the other hand, the STA-STA category provides sensitivity to LLPs that decay outside the tracker volume. Figure~\ref{fig:displacedDimuons_eventDisplay} shows an event display from Run~3 data that illustrates the displaced dimuon signature.

Since Run~2, a set of dedicated dimuon triggers has existed that relies on muons reconstructed only in the muon system (the STA-STA category) using algorithms that are similar to those used in the standalone offline muon reconstruction. These triggers were deployed in 2016 and 2018. As discussed in Section~\ref{sec:muons}, in the L1T muon \pt assignment, a beam-axis constraint was used: the muon candidate was treated as originating from the nominal beam line, or equivalently as having a negligible \dzero with respect to the beam axis. On the other hand, the muon track fits in the HLT reconstruction did not impose this beam-axis constraint for these paths. This L1T beam-axis requirement caused a large reduction in displaced-muon reconstruction efficiency for large LLP lifetimes. Each HLT muon was required to be within the region $\abseta < 2.0$ and to have $\pt > 28\,(23)\GeV$ for 2016 (2018) data taking. The CMS Collaboration performed a Run~2 search for LLPs decaying to a pair of muons that used these triggers~\cite{EXO-21-006}. During Run~3, these triggers (labeled ``Run~2 (2018)'') are also used, maintaining the \pt threshold of 23\GeV from 2018.  Displaced-muon signatures have also motivated dedicated trigger and reconstruction developments in ATLAS, including searches using displaced muon tracks and displaced vertices, as well as Run~3 displaced-muon trigger developments~\cite{ATLAS:2020xyo,Rousso:2026srg}.

To enhance the signal efficiency, a novel set of displaced-dimuon triggers, both at L1 and the HLT, was introduced in CMS at the beginning of Run~3. These improvements made an early search~\cite{EXO-23-014} possible, and its results are competitive with those of Run~2, using only 2022 Run~3 data with less than half of the Run~2 integrated luminosity.

At L1, two additional sets of triggers were introduced. One set comprises traditional beam axis-constrained double-muon triggers with either no or a very low (4.5\GeV) minimum muon-\pt threshold. The triggers in this set have the following requirements: the L1 muon candidates have opposite charges; segments in at least three different muon stations; an angular separation $\DR$ between the two muons that does not exceed a maximum threshold that varies between 1.2 and 1.4; and $\abseta< 1.5$. The other set takes advantage of a new track-finding procedure for the barrel section of the L1 muon trigger in the barrel muon track finder (BMTF) algorithm. This novel technique enables the reconstruction of L1 muon candidates and the determination of their \pt without using a beam-axis constraint~\cite{CMS:2023gfb}. These triggers require $\pt > 15$ and 7\GeV for the leading and subleading L1 muon, respectively. The minimum \pt thresholds are lowered to 6 and 4\GeV when $\dzero > 25\cm$ for both L1 muons.

In the barrel/endcap overlap and the endcap sections of the L1 muon trigger, the determination of muon \pt without a beam-axis constraint was implemented for 2024 data taking, extending the coverage of L1 displaced-muon triggers to the full $\eta$ range of CMS. 

In the overlap section, to allow for the beam axis-unconstrained \pt measurement, the overlap muon track finder (OMTF) algorithm first extrapolates the direction of a reference hit, selected from hits in the two innermost muon stations, to other muon stations. Then, the differences between $\phi$ values of hits in these other stations and extrapolated $\phi$ values from the reference hit are used by the OMTF algorithm~\cite{omtfAlgo} to reconstruct the muons and determine both beam axis-constrained and beam axis-unconstrained \pt. 

In the endcap section, the endcap muon track finder (EMTF) algorithm uses a neural network (NN) to determine the beam axis-unconstrained muon \pt and \dzero, using 29 features extracted from the muon track. In parallel to the NN, the original EMTF algorithm, based on a boosted decision tree~\cite{emtfBDT}, is used to determine the beam axis-constrained \pt.

\begin{figure}[htb!]
\centering
\includegraphics[width=0.49\textwidth]{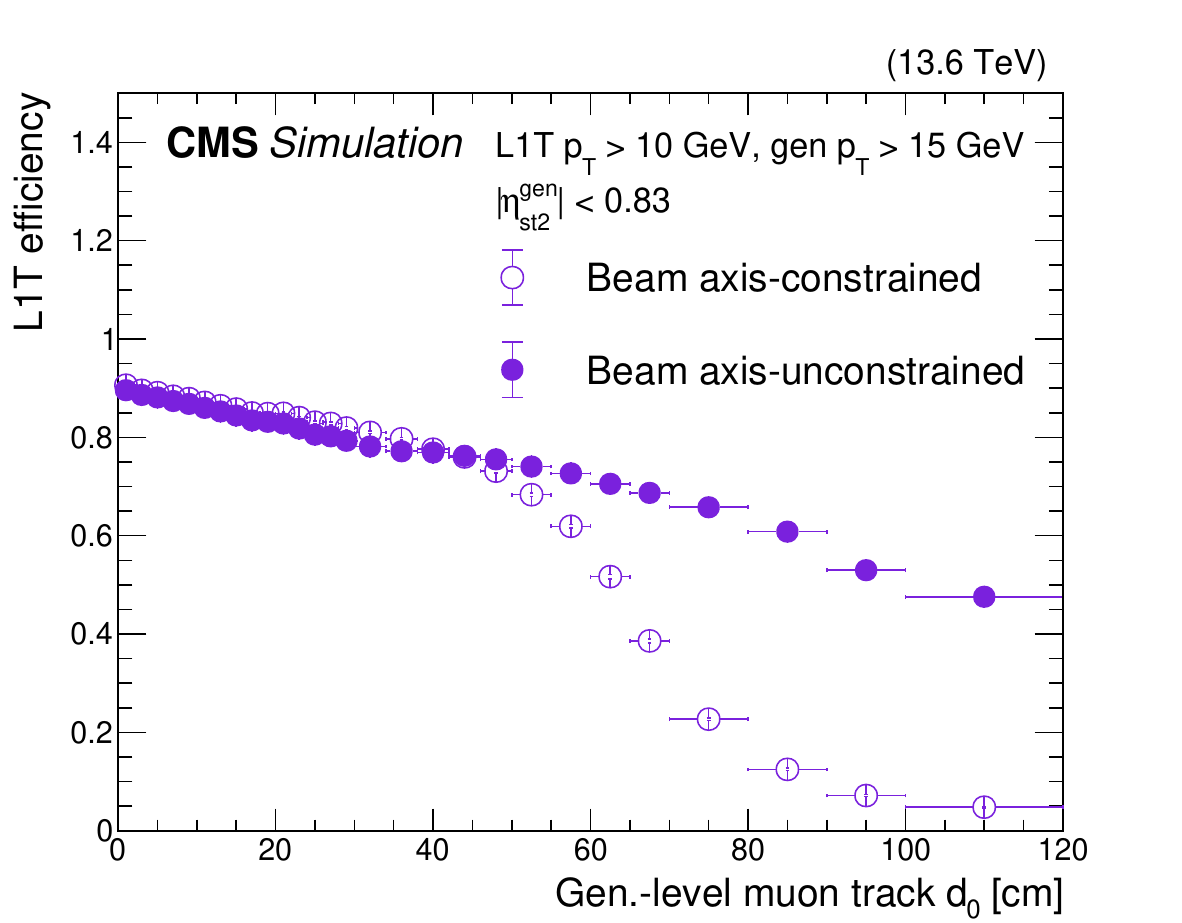}
\includegraphics[width=0.49\textwidth]{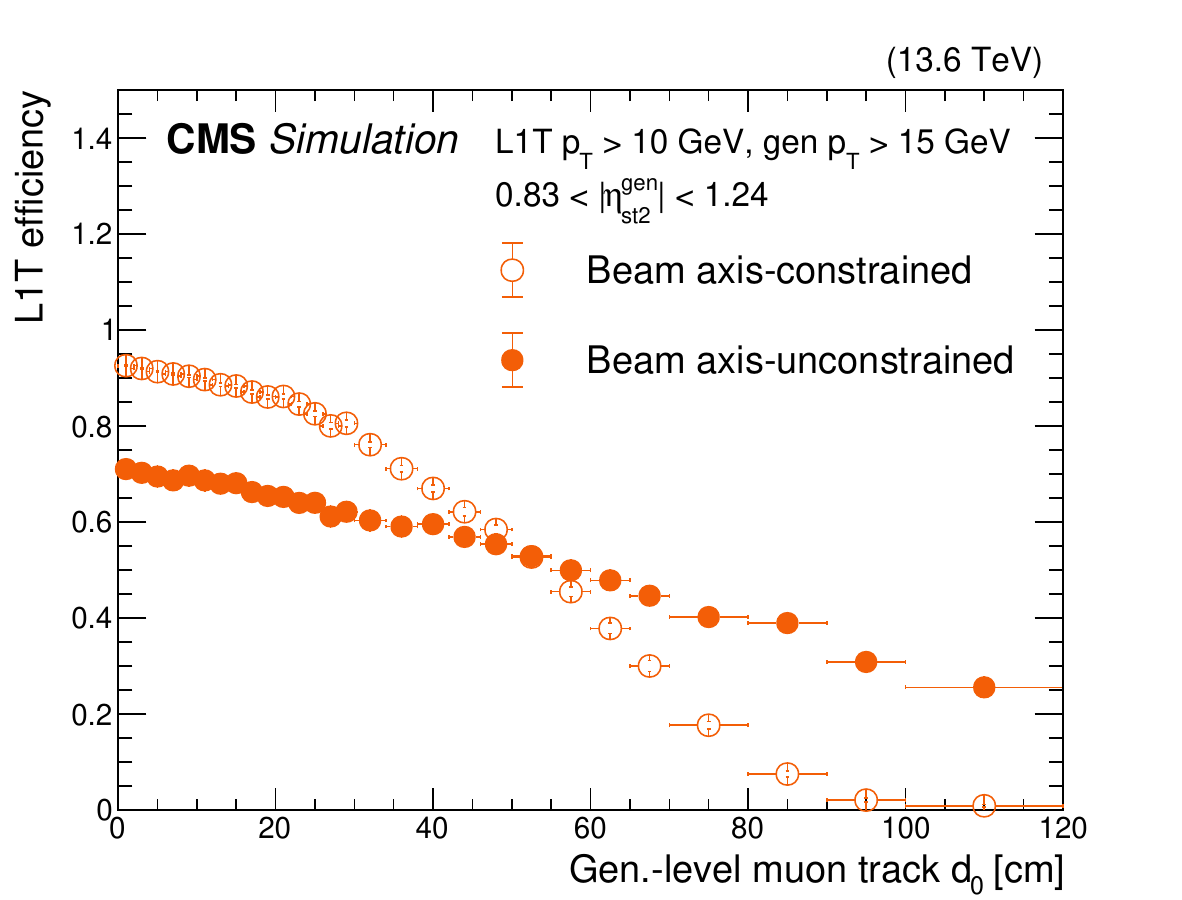}
\includegraphics[width=0.49\textwidth]{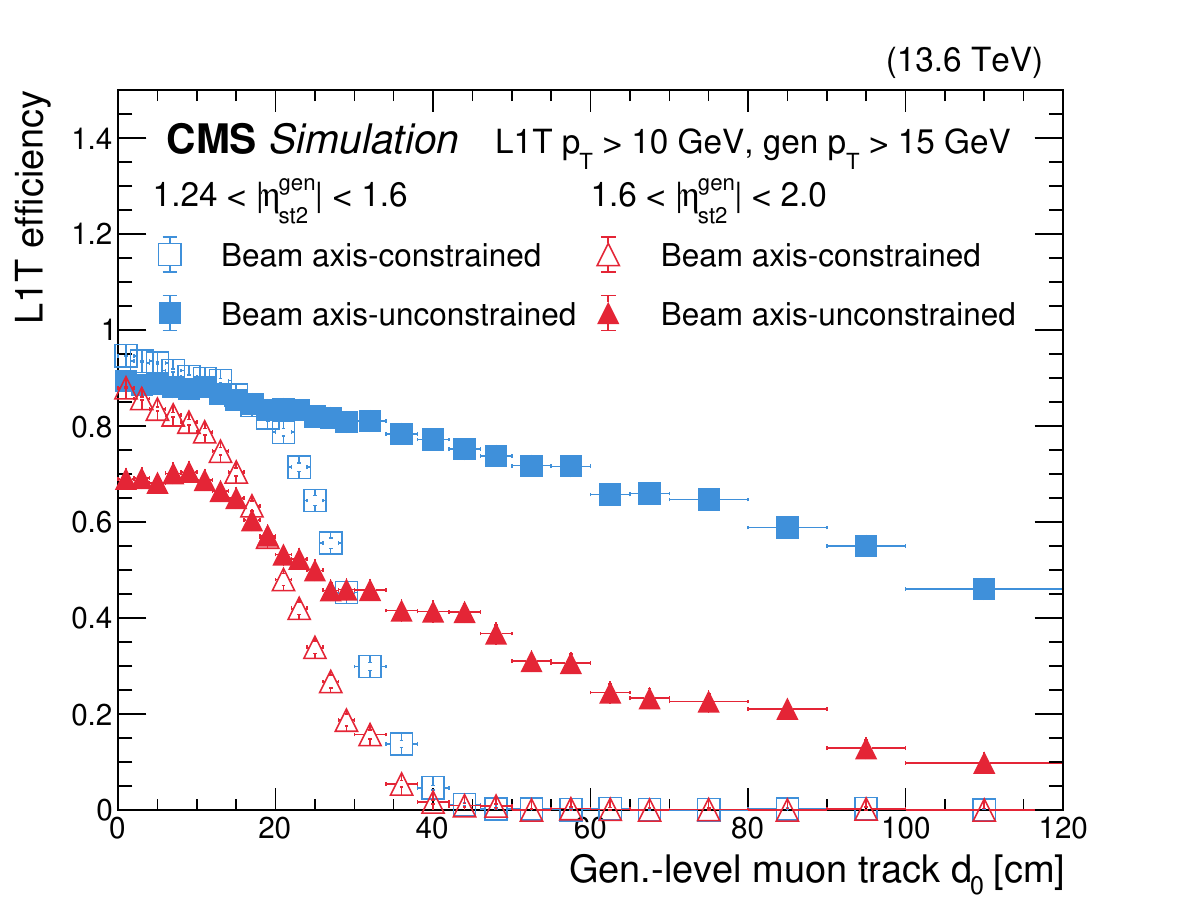}
\caption{The BMTF (upper \cmsLeft), OMTF (upper \cmsRight), and EMTF (lower) L1T efficiencies for beam axis-constrained (open circles, squares, and triangles) and beam axis-unconstrained (filled circles, squares, and triangles) \pt assignment algorithms for L1T $\pt > 10\GeV$ with respect to generator-level muon track \dzero, obtained using a sample which produces LLPs that decay to dimuons. The L1T algorithms and data-taking conditions correspond to 2024. A selection on the generator-level muon track $\pt > 15\GeV$ is applied to show the performance at the efficiency plateau. The generator-level muon tracks are extrapolated to the second muon station to determine the $\eta^{\text{gen}}_{\text{st2}}$ values that are used in the plot. Both the new vertex-unconstrained algorithm performance and the default beam axis-constrained algorithm performance are shown. In the EMTF plot, the different colors and symbols show different $\abseta$ regions: $1.24 < \eta^{\text{gen}}_{\text{st2}} < 1.6$ (blue squares), $1.6 < \eta^{\text{gen}}_{\text{st2}} < 2.0$ (red triangles).
}
\label{fig:displacedDimuonsL1}
\end{figure}

Figure~\ref{fig:displacedDimuonsL1} shows the efficiency of the BMTF, OMTF, and EMTF algorithms for both beam axis-constrained and beam axis-unconstrained \pt, using a simulated sample of LLPs that decay into a pair of muons. 

At the L1 global trigger, an additional displaced-dimuon trigger algorithm was introduced in 2024. This trigger uses the outputs of BMTF, OMTF, and EMTF and requires two high-quality muons with beam axis-unconstrained $\pt > 6\GeV$. It covers the range $\abseta < 2$, thus extending the $\eta$ range of the displaced dimuon triggers used previously.

At the HLT, two new displaced-dimuon paths were introduced. The new triggers are characterized by minimum \pt thresholds that depend on the muon \dzero. These two triggers are similar in that they both perform the L3 muon reconstruction, but the way the L3 muons are used in the triggers is different. In the first path, the L3 muon is used to reject events, while in the second path, the L3 muon is used to select events.

The first algorithm, labeled ``Run~3 (2022, L2)'', attempts to reconstruct the muon candidates at the L3 stage (similar to global muons), taking advantage of their superior tracker resolution for determining muon \dzero, instead of stopping the online reconstruction at the L2 stage (muon system alone), as was done in the Run~2 trigger. If either of the two L2 muon candidates is reconstructed as an L3 muon with $\dzero < 1\cm$, the event is discarded, since such an L2 muon candidate is likely to originate from the background processes. The resulting trigger operates with minimum L2 muon \pt thresholds of 10\GeV and significantly improves the signal efficiency while adding only ${\approx}10\unit{Hz}$ to the HLT rate.

The second new HLT algorithm, labeled ``Run~3 (2022, L3)'', relies entirely on the online L3 muon reconstruction, in contrast to the ``Run~3 (2022, L2)'' trigger. A minimum threshold on the impact parameter of each muon, $\dzero > 0.01\cm$, allows the minimum \pt thresholds to be low: $16\,(10)\GeV$ for the leading (subleading) L3 muon. The resulting trigger greatly improves the signal efficiency at smaller \cTau compared to the Run~3 (2022, L2) trigger, while adding only 10\unit{Hz} to the HLT rate.

An equivalent trigger to the Run~3 (2022, L3), labeled ``Run~3 (2022, L3 dTks)'', is also included for data taking. This trigger uses a newly developed L3 muon reconstruction algorithm that closely resembles the algorithm used offline. The main improvements are the inclusion of the displaced tracker muons and the optimization of the seeding stage of the muon trajectory building, which is made more efficient for displaced muons by removing the \dzero requirement. The new trigger, currently in the commissioning phase, has the same signal efficiency but with a rate that is ${\approx}15\%$ less than that of the Run~3 (2022, L3) trigger.

\begin{figure}[htbp!]
  \centering
  \includegraphics[width=0.6\textwidth]{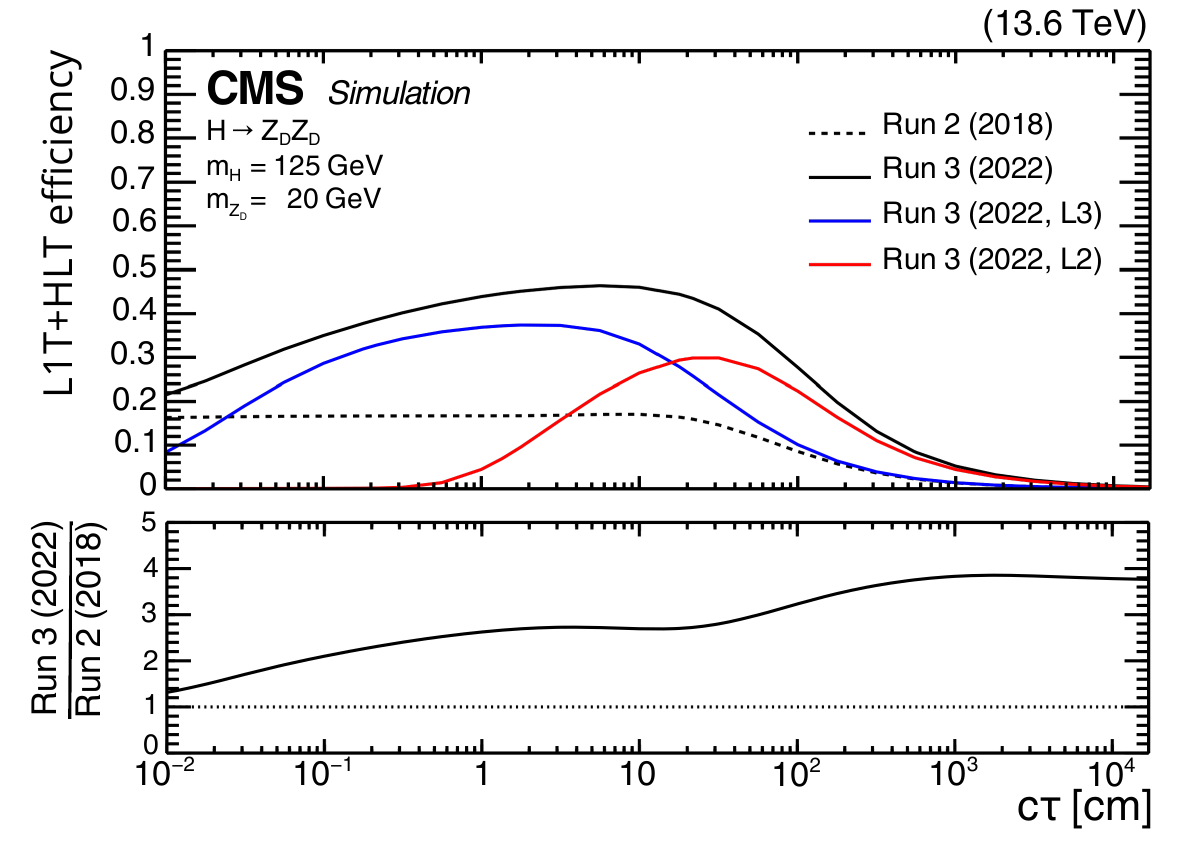}
  \caption{The L1T+HLT efficiencies of the various displaced-dimuon triggers and their logical OR as a function of \cTau for the HAHM signal events with $\mH=125\GeV$ and $\mZD = 20\GeV$, for 2022 conditions.  The efficiency is defined as the fraction of simulated events that satisfy the detector acceptance and the requirements of the following sets of triggers: the Run~2 (2018) triggers (dashed black); the Run~3 (2022, L3) triggers (blue); the Run~3 (2022, L2) triggers (red); and the logical OR of all these triggers (Run~3 (2022), solid black). The lower panel shows the ratio of the overall Run~3 (2022) efficiency to the Run~2 (2018) efficiency. Figure adapted from Ref.~\cite{EXO-23-014}.}
  \label{fig:HLT_efficiency_split_by_path}
\end{figure}

The combined L1T+HLT efficiency of the various displaced-dimuon triggers and their logical OR as a function of \cTau is shown in Fig.~\ref{fig:HLT_efficiency_split_by_path} for simulated HAHM signal events, where the dark Higgs boson (\Hdark) mixes with the SM Higgs boson (\PH) and decays into a pair of long-lived dark photons (\PZD), each with a mass $\mZD = 20\GeV$, as shown in Fig.~\ref{fig:H_darkPhotons_feynmanDiagrams}. The addition of new triggers at L1 and the HLT improves the efficiency for $\PZD$ with $\mZD > 10\GeV$ and $\cTau \gtrsim 0.1\cm$ by a factor of 2 to 4, depending on \cTau and mass.

\begin{figure}[htb!]
  \centering
  \includegraphics[width=0.49\textwidth]{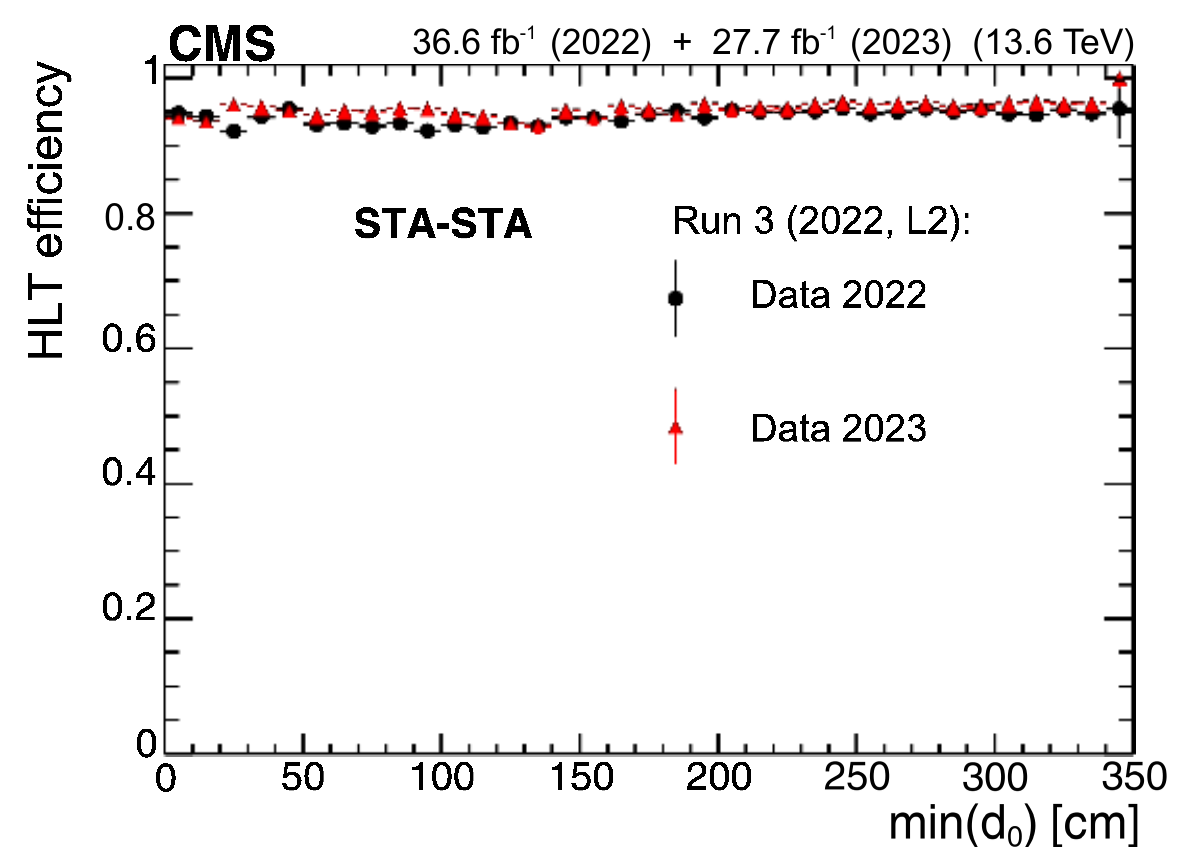}
  \includegraphics[width=0.49\textwidth]{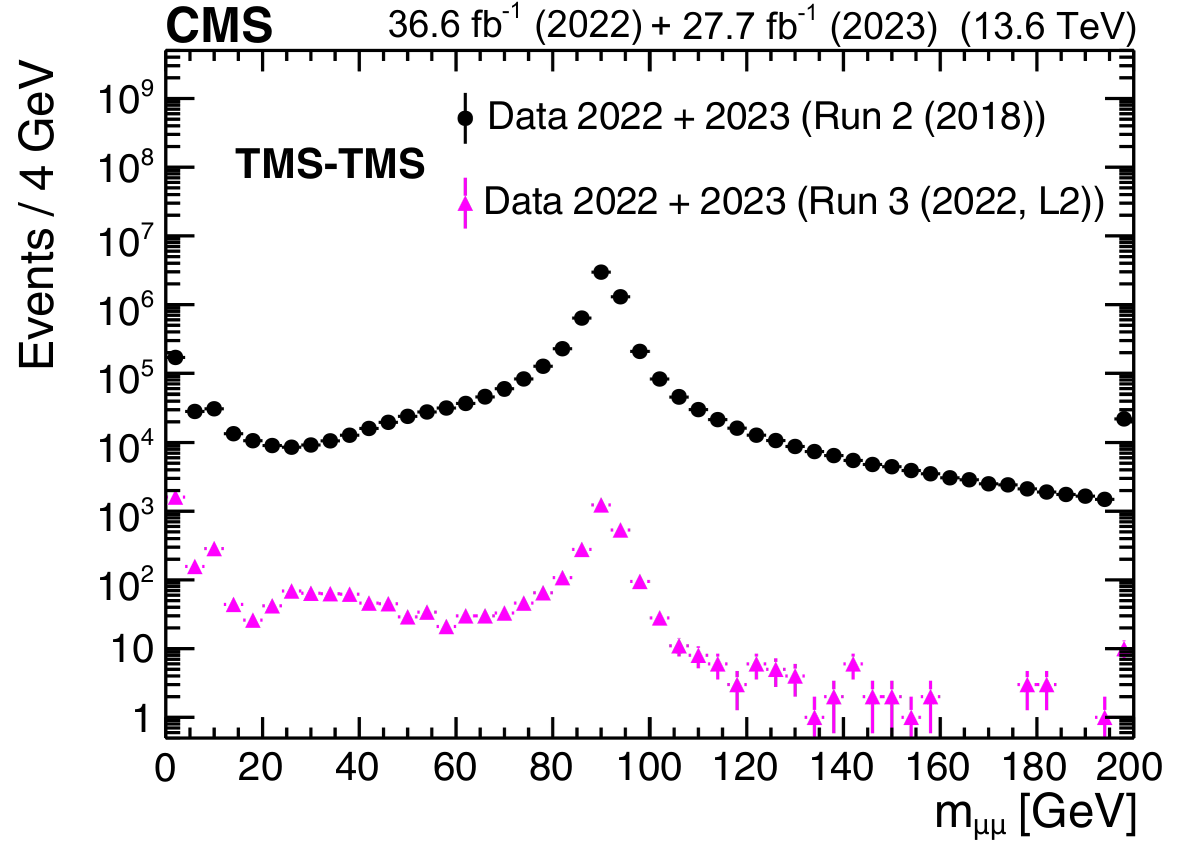} 
  \caption{The HLT efficiency (\cmsLeft), defined as the fraction of events recorded by the Run~2 (2018) triggers that also satisfy the requirements of the Run~3 (2022, L2) triggers, as a function of offline-reconstructed \mindzero of the two muons forming STA-STA dimuons in events enriched in cosmic ray muons. The black circles represent efficiencies during the 2022 data-taking period, and the red triangles represent the 2023 period. For displaced muons, the efficiency of the online \mindzero requirement is larger than 95\% in all data-taking periods. The invariant mass distribution for TMS-TMS dimuons (\cmsRight) in events recorded by the Run~2 (2018) triggers in the combined 2022 and 2023 data set (black circles), and in the subset of events also selected by the Run~3 (2022, L2) triggers (pink triangles), illustrating the prompt muon rejection of the Run~3 (2022, L2) triggers.}
  \label{fig:trigger_eff_cosmic_L2HLT}
\end{figure}

We use STA-STA dimuons that have a three-dimensional (3D) opening angle compatible with a cosmic ray muon traversing CMS recorded with the Run~2 (2018) trigger to evaluate the efficiency of the Run~3 (2022, L2) triggers. The efficiency as a function of \mindzero is shown in Fig.~\ref{fig:trigger_eff_cosmic_L2HLT} (\cmsLeft). It is greater than 95\% for dimuons with \mindzero up to 350\cm in both 2022 and 2023. Figure~\ref{fig:trigger_eff_cosmic_L2HLT} (\cmsRight) shows the invariant mass distribution for TMS-TMS dimuons in events recorded by the Run~2 (2018) triggers, and the subset of events also selected by the Run~3 (2022, L2).

\begin{figure}[hbp!]
  \centering
  \includegraphics[width=0.49\textwidth]{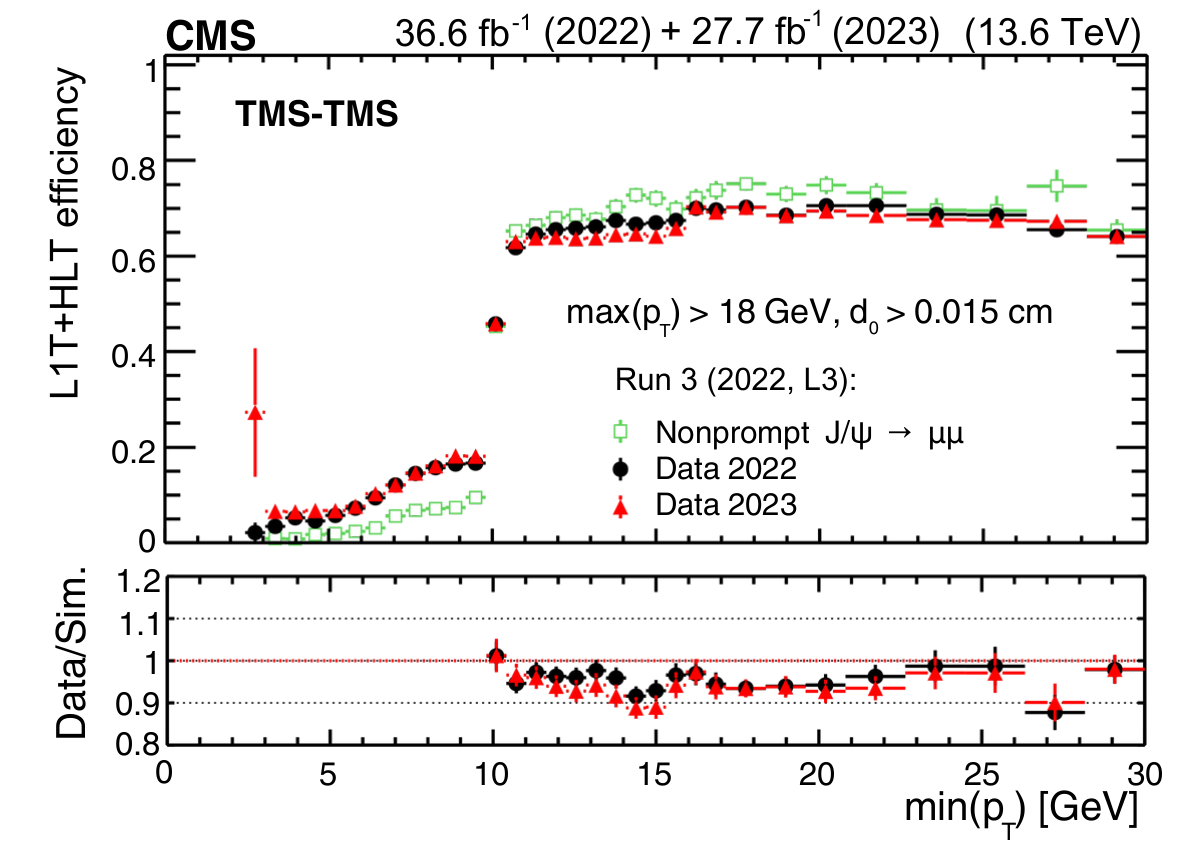}
  \includegraphics[width=0.49\textwidth]{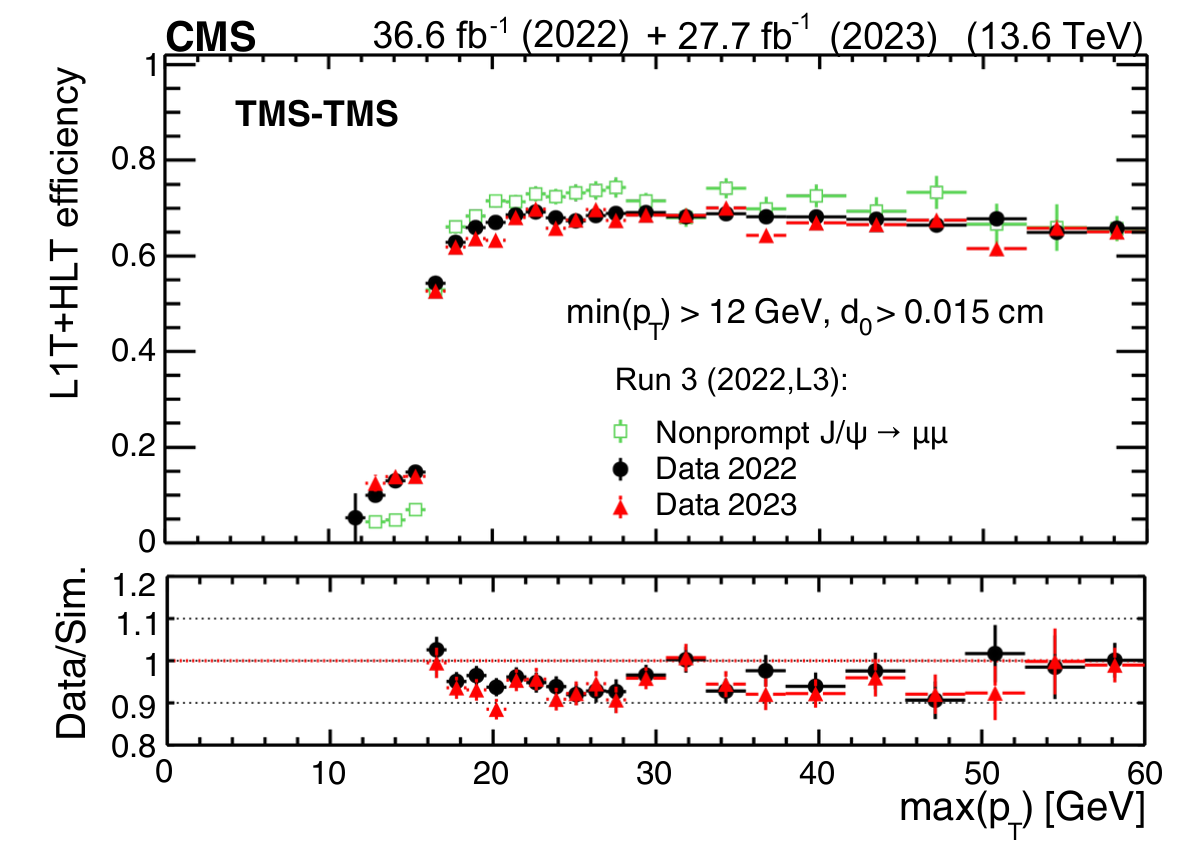}
  \includegraphics[width=0.49\textwidth]{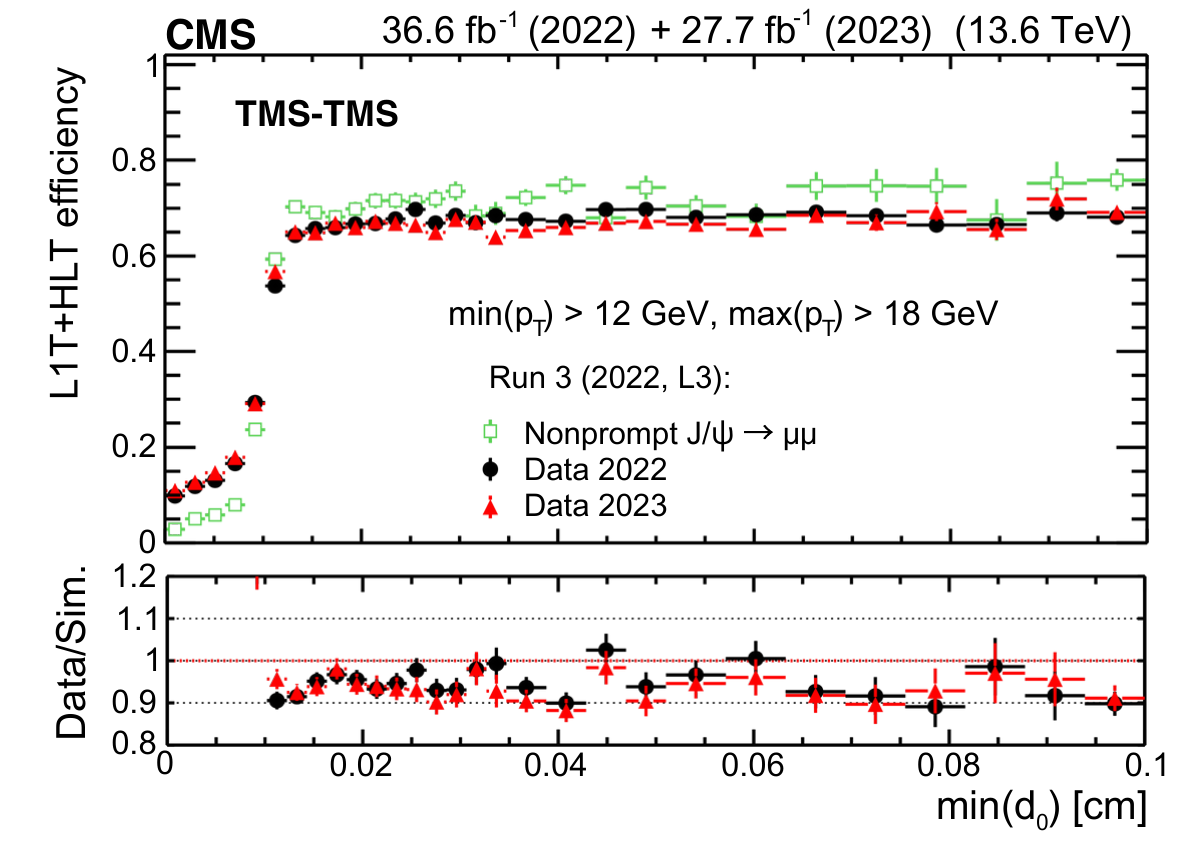}
  \caption{The L1T+HLT efficiency of the Run~3 (2022, L3) triggers in 2022 data (filled black circles), 2023 data (filled red triangles), and simulation (open green squares) as a function of \minpt (upper \cmsLeft), \maxpt (upper \cmsRight), and \mindzero (lower) of the two muons forming TMS-TMS dimuons in events enriched in $\PJGy\to\PGm\PGm$ events. The efficiency in data is the fraction of $\PJGy\to\PGm\PGm$ events recorded by the triggers based on the information from jets and \ptmiss that also satisfy the requirements of the Run~3 (2022, L3) triggers. It is compared to the efficiency of the Run~3 (2022, L3) triggers in a combination of simulated samples of $\PJGy\to\PGm\PGm$ events produced in various \PQb hadron decays. The lower panels show the ratio of the data to simulated events.}
  \label{fig:trigger_eff_orthogonal_jpsi_L3HLT}
\end{figure}

An efficiency measurement is performed for the Run~3 (2022, L3) trigger based on $\PJGy\to\PGm\PGm$ events, recorded using jets and \ptmiss triggers with 2022 and 2023 data. This data set corresponds to an integrated luminosity of 64.3\fbinv, which is ${\approx}4\%$ more than is used in other sections of this paper that focus on 2022 and 2023 data, since in this case, only the subdetectors needed for muon reconstruction are required to provide good-quality data. The efficiencies are defined as the subset of these events that also pass the Run~3 (2022, L3) path. The measurement is compared with the efficiency in simulation, defined as the fraction of events that pass the same trigger, obtained by combining various simulated samples of \PQb hadrons that decay into $\PJGy\to\PGm\PGm$. Figure~\ref{fig:trigger_eff_orthogonal_jpsi_L3HLT} shows trigger efficiencies as a function of the minimum \pt of the two muons in the event, \minpt; the maximum \pt of the two muons, \maxpt; and the minimum \dzero of the two muons, \mindzero. For each measurement, the other two variables are required to be at the trigger efficiency plateau. The efficiency plateau, where the efficiencies are close to 65\% in data and 70\% in simulation, is reached for $\minpt > 12\GeV$, $\maxpt> 18\GeV$, and $\mindzero > 0.015\cm$. The main loss of efficiency at the plateau comes from inefficiencies in the double-muon L1Ts to select two muons with small spatial separation. This inefficiency is particularly prominent in $\PJGy\to\PGm\PGm$ events, as the two muons typically have a separation of $0.05<\DR <0.15$. There is some remaining efficiency of around 15\% at \pt values less than the trigger \pt thresholds. This is because of differences between the online and offline L3 muon reconstructions, which can provide different estimations of the \pt and \dzero parameters.

\begin{figure}[htb!]
  \centering
  \includegraphics[width=0.49\textwidth]{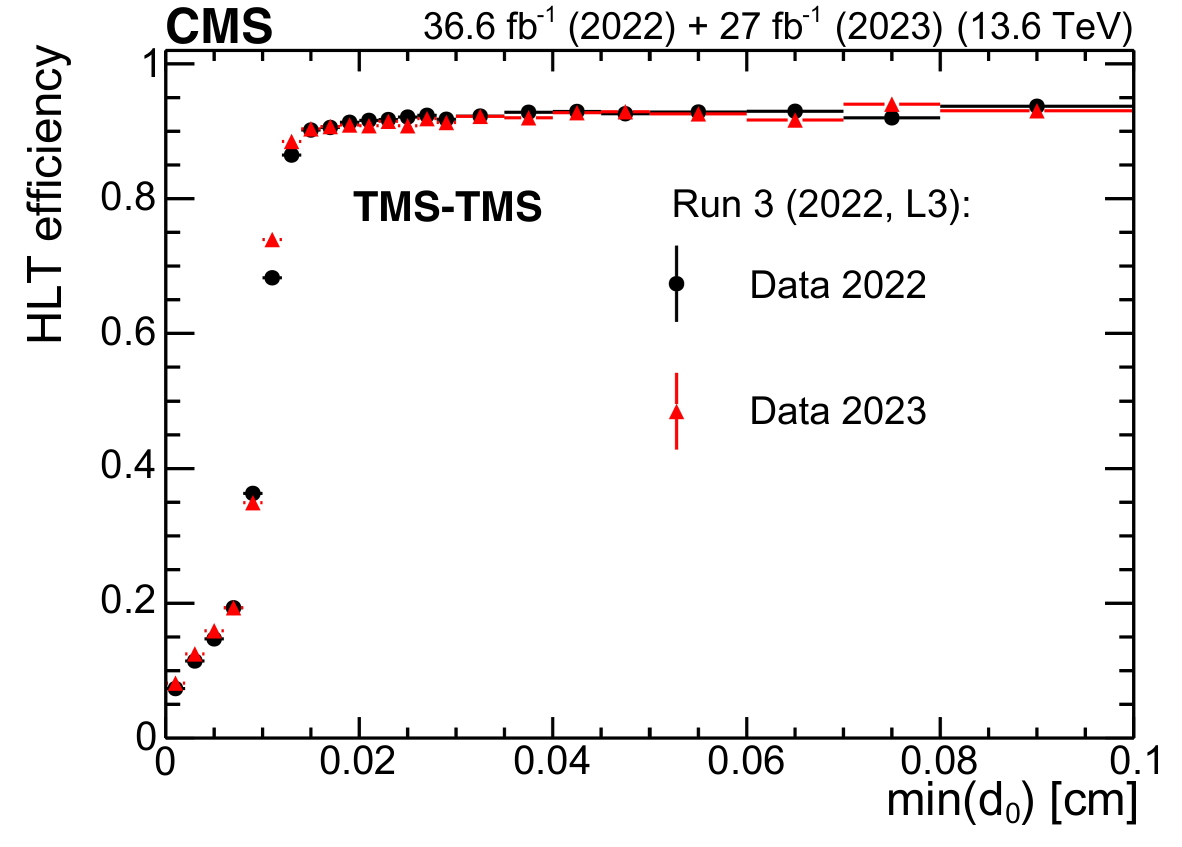} 
  \includegraphics[width=0.49\textwidth]{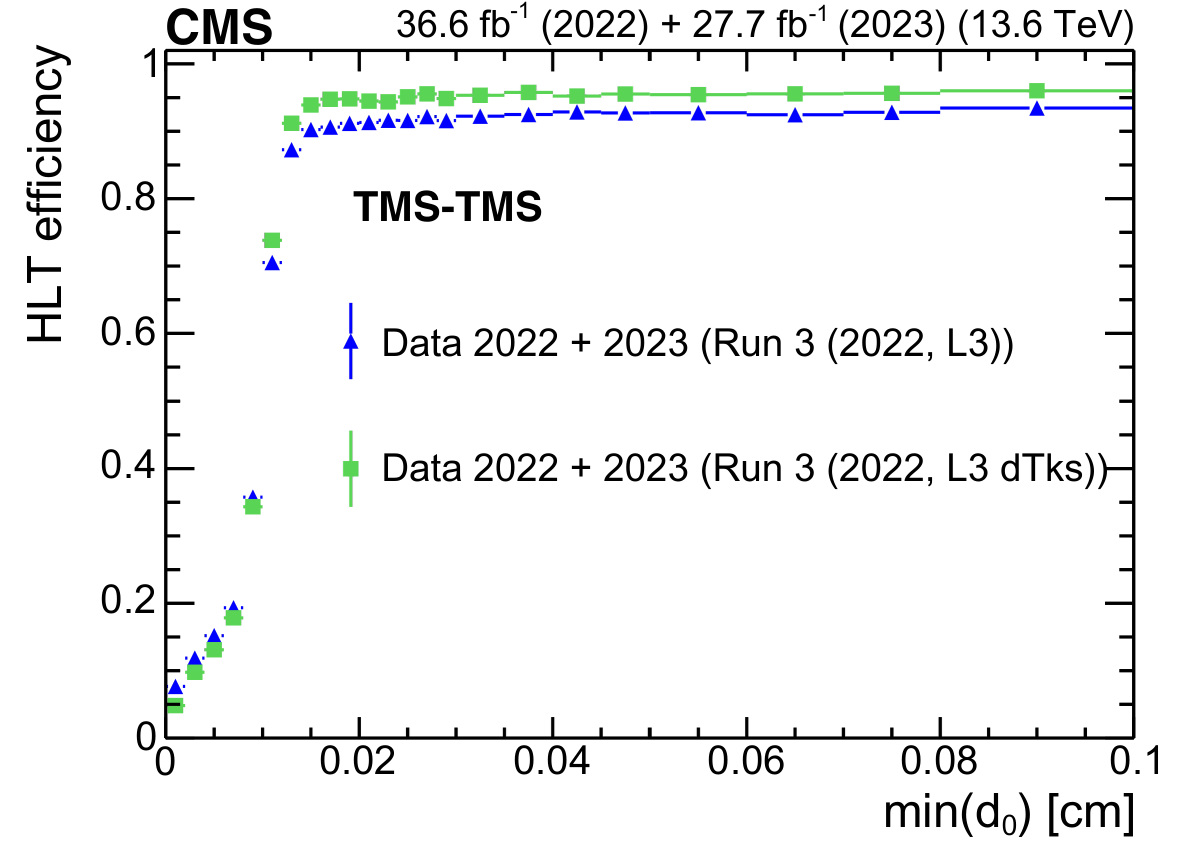}
  \includegraphics[width=0.49\textwidth]{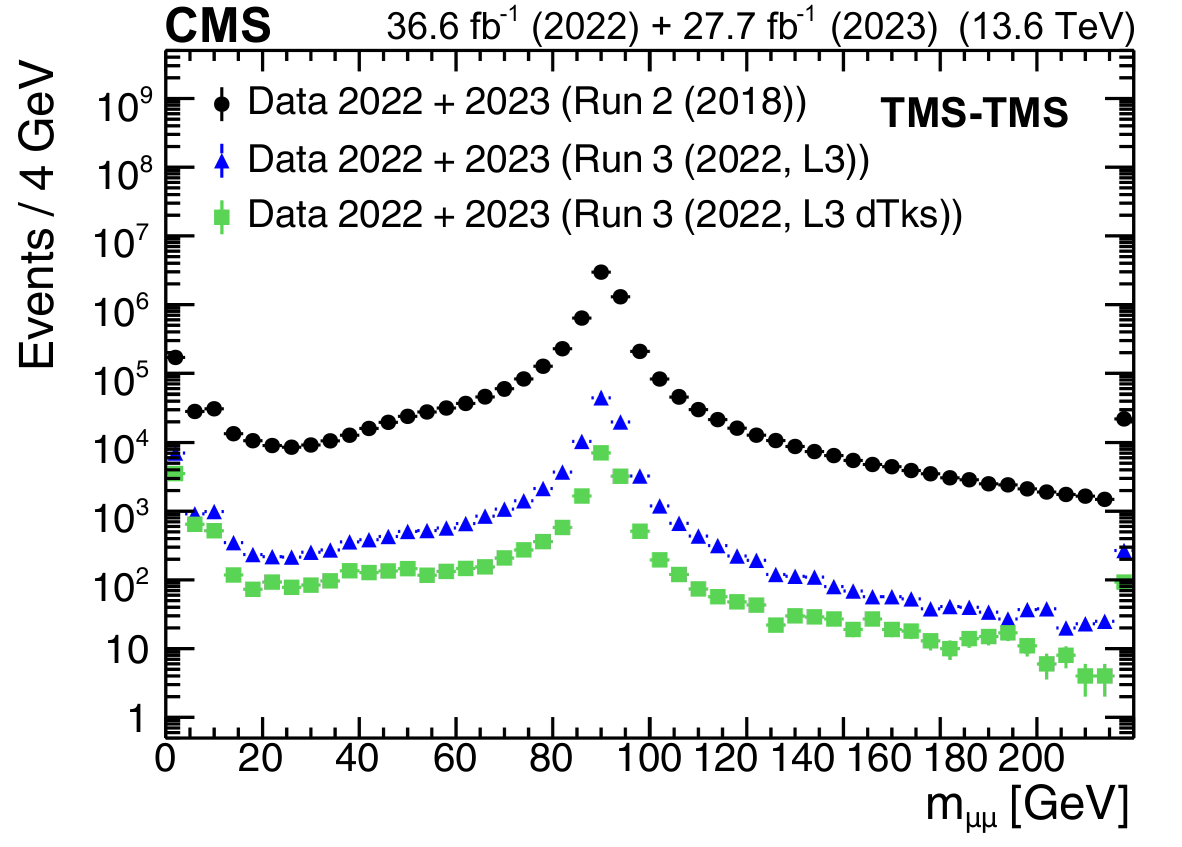}
  \caption{The HLT efficiency (upper \cmsLeft), defined as the fraction of events recorded by the Run~2 (2018) triggers that also satisfy the requirements of the Run~3 (2022, L3) triggers, as a function of offline-reconstructed \mindzero of the two muons forming TMS-TMS dimuons in events enriched in $\PJGy\to\PGm\PGm$, for the 2022 and 2023 data-taking periods. The black circles represent efficiencies during the 2022 data-taking period, and the red triangles represent the 2023 period. For dimuons with offline $\mindzero > 0.012\cm$, the combined efficiency of the L3 muon reconstruction and the online \mindzero requirement is larger than 90\% in all data-taking periods. The HLT efficiency of the Run~3 (2022, L3) triggers (upper \cmsRight), shown with blue triangles, and the Run~3 (2022, L3 dTks) triggers, shown with green squares, for $\PJGy\to\PGm\PGm$ events in the 2022 and 2023 data sets. Invariant mass distribution for TMS-TMS dimuons (lower) in events recorded by the Run~2 (2018) triggers in the combined 2022 and 2023 data set (black circles), and in the subset of events also selected by the Run~3 (2022, L3) trigger (blue triangles) and Run~3 (2022, L3 dTks) trigger (green squares), illustrating the prompt muon rejection of the L3 triggers.}
  \label{fig:trigger_eff_jpsi_L3HLT}
\end{figure}

We also use a sample of $\PJGy\to\PGm\PGm$ events, recorded with the Run~2 (2018) trigger and containing TMS-TMS dimuon candidates with an invariant mass of $3.0<\mMuMu<3.2$\GeV, to evaluate the efficiency of the Run~3 (2022, L3) and Run~3 (2022, L3 dTks) triggers. Out of this sample of events recorded with the Run~2 (2018) trigger, we compute how many events also pass the Run~3 (2022, L3) and Run~3 (2022, L3 dTks) triggers. This efficiency is shown in Fig.~\ref{fig:trigger_eff_jpsi_L3HLT} (upper \cmsLeft) for the Run~3 (2022, L3) trigger, split in the 2022 and 2023 years, with an efficiency higher than 90\% for dimuons with min$(\dzero)>0.012\cm$; and in Fig.~\ref{fig:trigger_eff_jpsi_L3HLT} (upper \cmsRight) for the combined 2022 and 2023 years and both triggers. The observation that the Run~3 (2022, L3) and Run~3 (2022, L3 dTks) triggers have a lower efficiency than the Run~2 (2018) trigger is expected, since a veto is applied in the Run~3 triggers. As stated earlier, the Run~2 (2018) trigger continues to be used during Run~3.

\begin{sloppypar}Figure~\ref{fig:trigger_eff_jpsi_L3HLT} (lower) shows the invariant mass distribution for TMS-TMS dimuons in events recorded by the Run~2 (2018) triggers, and the subset of events also selected by the Run~3 (2022, L3) or Run~3 (2022, L3 dTks) triggers. This last trigger provides larger signal efficiency and background rejection than the Run~3 (2022, L3) trigger.
\end{sloppypar}

In 2024, a few displaced single-muon triggers were added. At L1, a single-muon BMTF trigger requires a high-purity muon with $\pt > 15\GeV$, calculated without applying a beam-axis constraint to the track. This trigger seeds two paths at the HLT: one path that requires at least one L2 muon with $\pt > 50\GeV$ and rejects events with at least one L3 muon with $\dzero < 1\cm$, and another path that requires at least one L3 muon with $\pt>30\GeV$ and $\dzero > 0.01\cm$. The first path aims to select events with L2 muons but reject prompt L3 muons. The second path aims to select events with L3 muons that are at least slightly displaced.

The motivation for including these displaced single-muon triggers is twofold. First, these triggers provide the opportunity to collect displaced dimuon events when the subleading muon is difficult to reconstruct at the L1T or HLT, thus providing additional signal efficiency. Second, the displaced single-muon triggers also expand the CMS sensitivity to additional models, such as those containing leptonic decays of displaced tau leptons~\cite{Buican:2008ws,Meade:2008wd,GIUDICE1999419} or models with a single displaced muon, such as those produced from the decay of a supersymmetric muon~\cite{Alwall:2008ag}.

The total rates of the displaced single-muon and dimuon triggers described here are 67 and 102\unit{Hz}, respectively.

\subsubsection{Double displaced L3 muon triggers}
\label{sec:displacedL3muon}

\begin{figure}[htb!]
\centering
\includegraphics[width=0.45\textwidth]{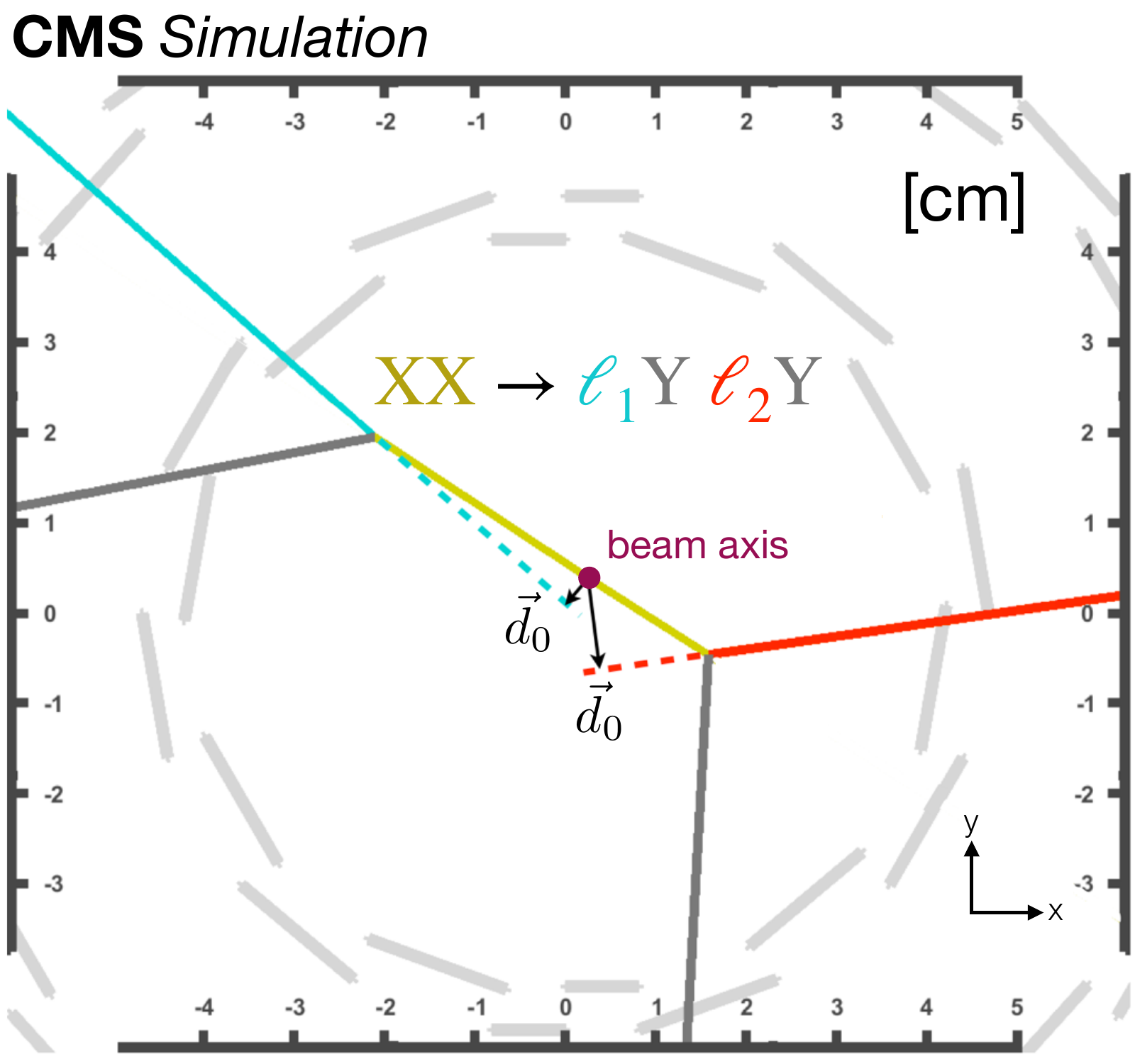}
\caption{Diagram of a simulated signal event in the inclusive displaced-lepton search~\cite{EXO-18-003}, from a transverse view of the IP. The two yellow lines indicate the trajectories of the two long-lived \PX particles. Each \PX decays into a displaced lepton (denoted $\Pell_{1}$ and $\Pell_{2}$) and another displaced SM particle (denoted \PY). The gray solid lines indicate the trajectories of the two \PY particles. The cyan and red solid lines indicate the trajectories of the two leptons, and the two dashed lines indicate the extrapolation of the trajectories towards their point of closest approach to the beam axis. The black arrows, which point from the beam axis and are orthogonal to the cyan and red dotted lines, indicate the lepton \dzero vectors. Figure adapted from Ref.~\cite{CMS:2024zqs}.
}
\label{fig:EXO_18_003_signal}
\end{figure}

Another displaced-dimuon trigger has also been implemented since Run~2. This trigger is designed for the displaced-lepton analysis, described in Ref.~\cite{EXO-18-003}, where the two leptons are not required to come from a common displaced vertex. This signature is shown with a simulated signal event in Fig.~\ref{fig:EXO_18_003_signal} and can arise, \eg, in an $R$-parity violating SUSY model with long-lived top squarks that decay into displaced quarks and leptons~\cite{Graham:2012th}, a GMSB model with long-lived sleptons that decay into a gravitino and a lepton~\cite{Evans:2016zau}, and a model with BSM Higgs bosons that decay into two long-lived scalar particles, which each decay into two displaced leptons~\cite{Strassler:2006ri}. At the HLT, this trigger requires at least two L3 muons, each with $\pt>43\GeV$. The fits to the muon tracks are not constrained to the PV, thus providing the trigger with a higher acceptance for displaced muons. This HLT path is seeded by L1 dimuon triggers. The total rate of this HLT path is about 2\unit{Hz}.

\begin{figure}[htb!]
\centering
\includegraphics[width=0.65\textwidth]{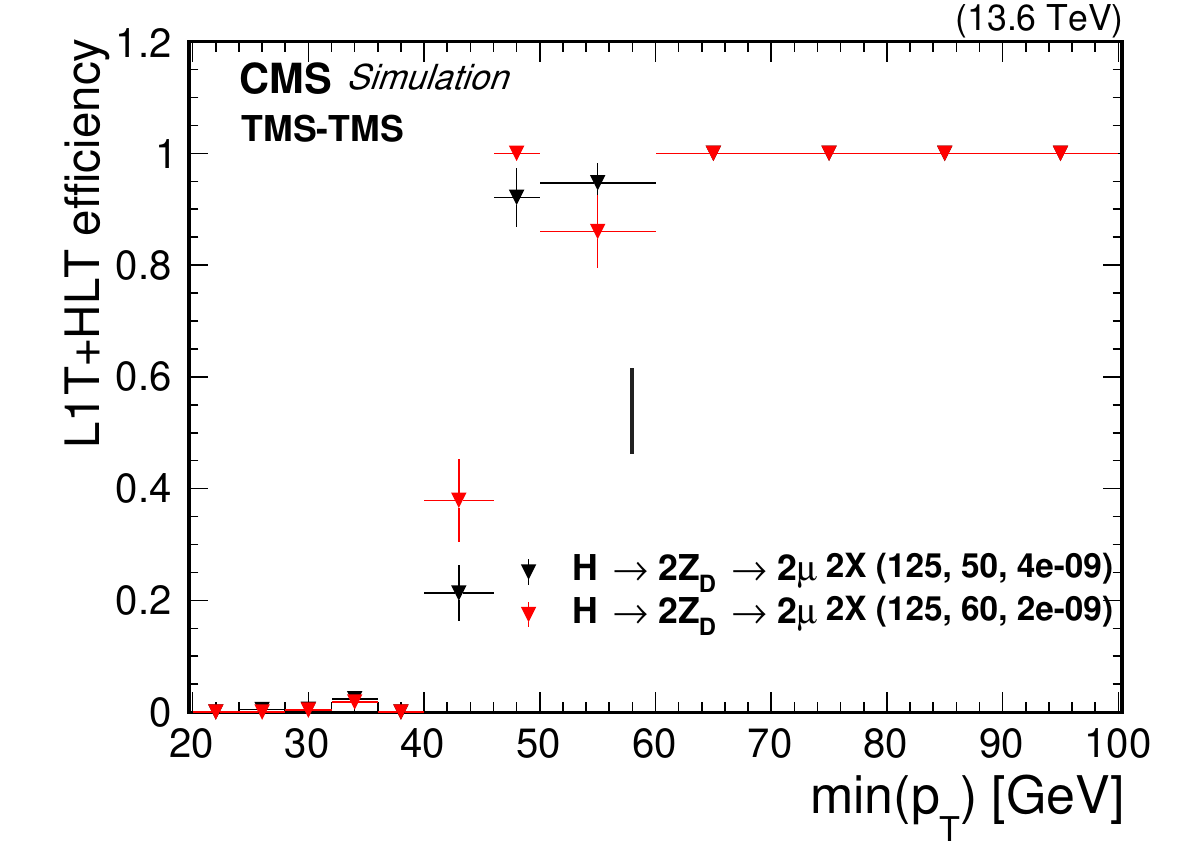}
\caption{The L1T+HLT efficiency of the double displaced L3 muon trigger as a function of \minpt of the two global or tracker muons in the event. The efficiency is plotted for HAHM signal events for 2022 conditions with $\mZD = 50\GeV$ and kinetic mixing \mbox{parameter $\epsilon = 4 \times 10^{-9}$} (black triangles), $\mZD = 60\GeV$ and $\epsilon = 2 \times 10^{-9}$ (red triangles), and $\mH=125\GeV$ in both cases. The events are required to have at least two good global or tracker muons with $\pt>23\GeV$.}
\label{fig:doubleDisplacedL3Muon_Hto2ZDto2mu2X}
\end{figure}

Figure~\ref{fig:doubleDisplacedL3Muon_Hto2ZDto2mu2X} shows the efficiency of this trigger for HAHM signal events, as a function of the \minpt of the two global or tracker muons in the event. The HAHM signal is used here, rather than one of the signals mentioned above that are targeted by this trigger, to facilitate comparison with the efficiency of other displaced-muon triggers. The trigger efficiency is shown as a function of muon \pt, illustrating that the double displaced L3 muon trigger is fully efficient for offline muons with $\minpt > 48\GeV$. Furthermore, this trigger has 100\% efficiency in the HAHM signal benchmark models, for \mindzero values smaller than the inner radius of the tracker. The HLT path does not exhibit the typical rise in efficiency at small \mindzero values because no displacement requirement is imposed by this trigger.

\begin{figure}[htb!]
\centering
\includegraphics[width=0.49\textwidth]{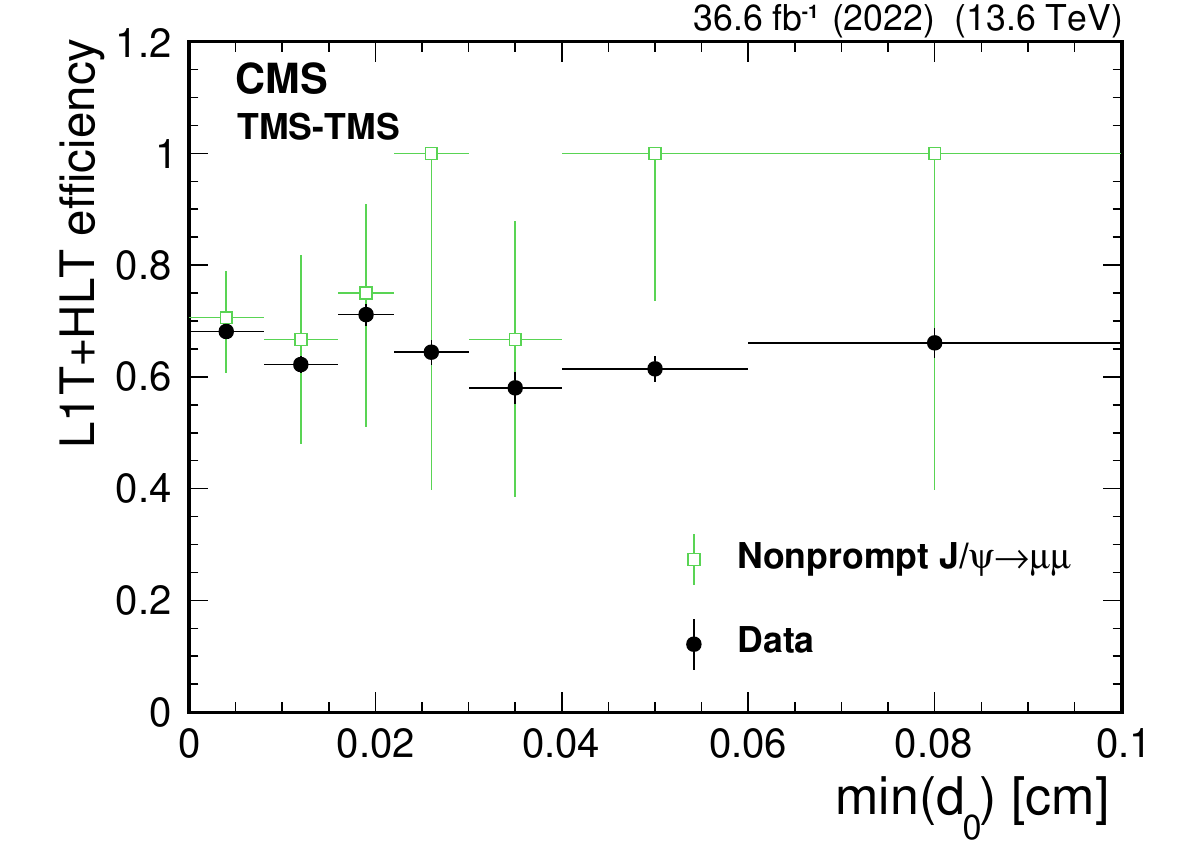}
\includegraphics[width=0.49\textwidth]{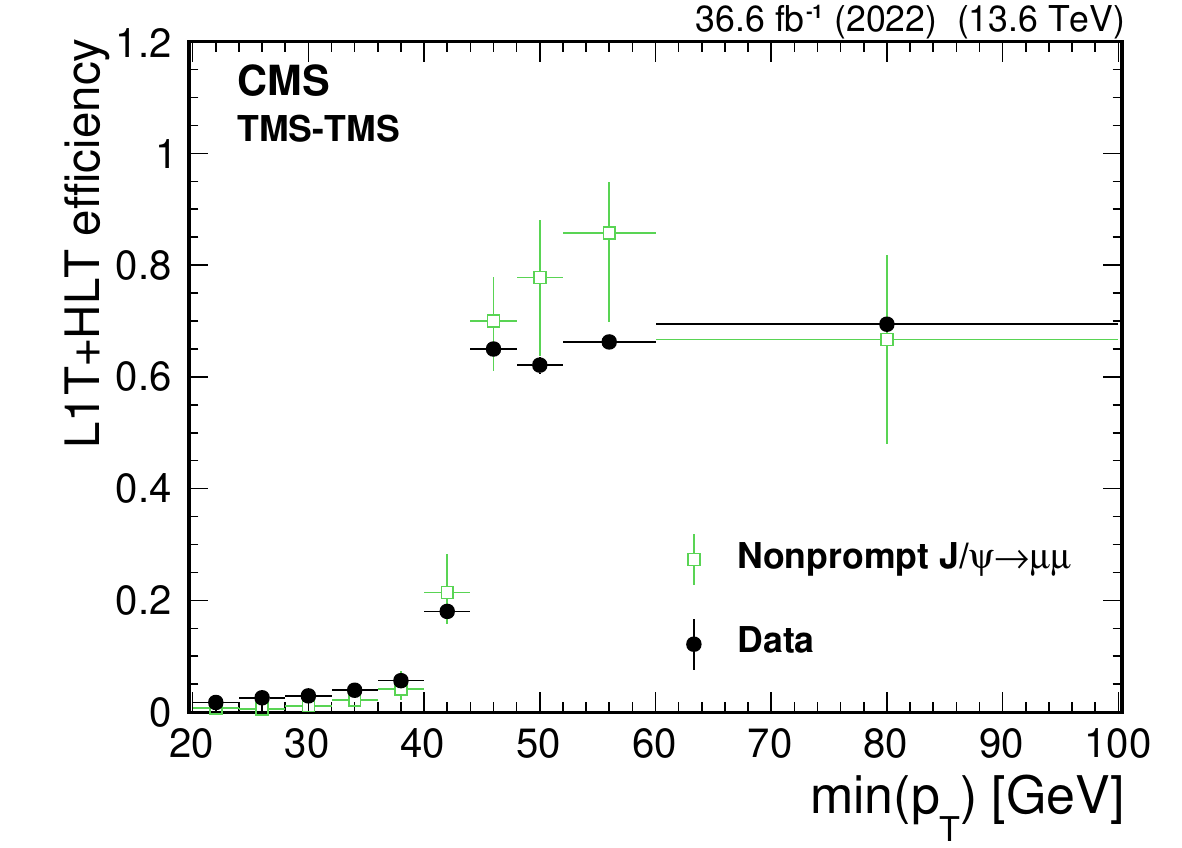}
\caption{The L1T+HLT efficiency of the double displaced L3 muon trigger in 2022, as a function of \mindzero (\cmsLeft) and \minpt (\cmsRight) of the two global or tracker muons in the event. The efficiency is plotted for simulated $\PJGy\to\PGm\PGm$ events produced in various \PQb hadron decays (open green squares) and data enriched in $\PJGy\to\PGm\PGm$ events recorded by jet- and \ptmiss-based triggers (filled black circles). The events are required to have at least two good global or tracker muons compatible with the \PJGy meson mass and with $\pt>45\GeV$ (\cmsLeft) or $\pt>23\GeV$ (\cmsRight). 
}
\label{fig:doubleDisplacedL3Muon_JPsiDataAndBkgMC}
\end{figure}

Figure~\ref{fig:doubleDisplacedL3Muon_JPsiDataAndBkgMC} shows the efficiency of the double displaced L3 muon trigger in \PJGy-enriched events in 2022 data and simulation (the same \PJGy-enriched samples used in Section~\ref{sec:displaceddimuon}), as a function of the \mindzero (\cmsLeft) and \minpt (\cmsRight) of the two global or tracker muons in the event. Data and simulation agree within the uncertainties. In the displaced-lepton analysis~\cite{EXO-18-003}, trigger scale factors are applied to the simulation to improve the data-to-simulation agreement. 

\subsubsection{Displaced L3 muon + photon triggers}
\label{sec:displacedL3muonandphoton}

Several mixed-object triggers, which select at least one photon and at least one displaced muon at the HLT, have been used in CMS since Run~2. These triggers are designed for the analysis described in Ref.~\cite{EXO-18-003} that searches for two LLPs that are produced in $\Pp\Pp$ collisions and decay in the detector. One of the LLPs decays into a displaced muon and other final state particles, and the other LLP decays into a displaced electron and other final state particles. This signature, shown in Fig.~\ref{fig:EXO_18_003_signal}, can arise in the same theoretical scenarios discussed in Section~\ref{sec:displacedL3muon}. The displaced electron and muon need not necessarily come from the same displaced vertex. In this way, the HLT photon is a proxy for a displaced electron. To be sensitive to larger displacements and to limit bias towards lower displacements, the displaced electron track is not reconstructed at the HLT, but only in the offline analysis.

At the HLT, there is a muon-photon mixed-object trigger that requires an L3 muon, whose track fit is not constrained to the PV, with $\pt>43\GeV$. This trigger also requires a photon at the HLT with $\pt>43\GeV$ as well as a set of loose, calorimeter-based identification selections. No displacement requirement is imposed in this trigger. In addition, there is another muon-photon mixed-object trigger available at the HLT, which makes similar requirements except it lowers the minimum \pt requirements on both the HLT photon and the muon to 38\GeV, and to compensate for the rate increase, it requires the muon to have $\dzero>1\cm$. In this way, the efficiency to trigger on lower-\pt signals is increased. Both muon-photon mixed-object HLT paths are seeded by muon-$\Pe/\PGg$ L1Ts.

The total rate of these HLT paths is about 5\unit{Hz}. The performance of the displaced L3 muon leg of this trigger can be seen from the performance of double displaced L3 muon triggers, as described in Section~\ref{sec:displacedL3muon} and shown in Figs.~\ref{fig:doubleDisplacedL3Muon_Hto2ZDto2mu2X} and \ref{fig:doubleDisplacedL3Muon_JPsiDataAndBkgMC}. The performance of the photon leg of the trigger, which is a standard photon reconstructed at the HLT, is described in Ref.~\cite{CMS:2024psu}.

\subsubsection{Dimuon scouting triggers}
\label{sec:dimuonscouting}

Dimuon scouting triggers, which are fully documented in Ref.~\cite{CMS:2024zhe}, have been available since 2017. Their purpose is to provide substantially lower muon minimum \pt thresholds compared to standard triggers. In particular, the dimuon scouting triggers require both muons to have $\pt > 3\GeV$ at the HLT, as opposed to standard dimuon triggers, which historically required $\pt > 17$ and 8\GeV for the leading and subleading \pt muons, respectively. Since Run~3, an inclusive dimuon trigger strategy has been implemented to collect events in the low-mass region (lower mass than the \PgU peak) using the parked data, with \pt thresholds as low as those used for scouting. However, for displaced dimuon signatures, a relevant distinction is the online reconstruction used in the trigger selection. The dimuon scouting triggers use HLT reconstruction algorithms that lack any constraint on the muon trajectories to the PV. This enables scouting searches for resonances that have nonzero displacement from the PV, such as the search described in Ref.~\cite{CMS:2021sch}. This search targets a displaced dimuon signature similar to the one described in Section~\ref{sec:displaceddimuon}, but focused on signals with lower dimuon masses, such as a scenario with an SM-like Higgs boson that decays to four leptons via two intermediate dark photons or through the Higgs portal via a dark Higgs boson~\cite{Curtin:2014cca}. Figure~\ref{fig:displacedDimuons_eventDisplay}, which is shown above in Section~\ref{sec:displaceddimuon}, illustrates an example signal event that is also applicable for the search that uses dimuon scouting triggers.

In Run~2, the dimuon scouting HLT paths required that muon tracks leave hits in at least two layers of the pixel detector. This requirement limited the maximum muon \dzero to about 11\cm. At the beginning of Run~3, this pixel detector hit requirement was removed from the scouting muon reconstruction algorithm, which increased the trigger efficiency for displaced muons, as they may be sufficiently displaced to leave no hits in the pixel detector. As a result, the efficiency for reconstructing dimuon events containing at least one displaced-vertex candidate increases from 29 to 50\%, corresponding to an improvement of a factor of approximately 1.7. This pixel detector hit requirement change occurred at the same time as the L1T BMTF improvement mentioned in Section~\ref{sec:displaceddimuon}, which also improved the displaced-muon trigger performance, as well as other Run~3 improvements to the scouting program, including the transition from calorimeter-based to PF scouting and substantial upgrades of the HLT reconstruction enabled by GPU-accelerated tracking. All of these changes collectively improved the dimuon scouting trigger performance, significantly increasing the efficiency for displaced-muon signatures while maintaining a trigger rate comparable to that of Run~2.

\begin{figure}[htb!]
\centering
\includegraphics[width=0.7\textwidth]{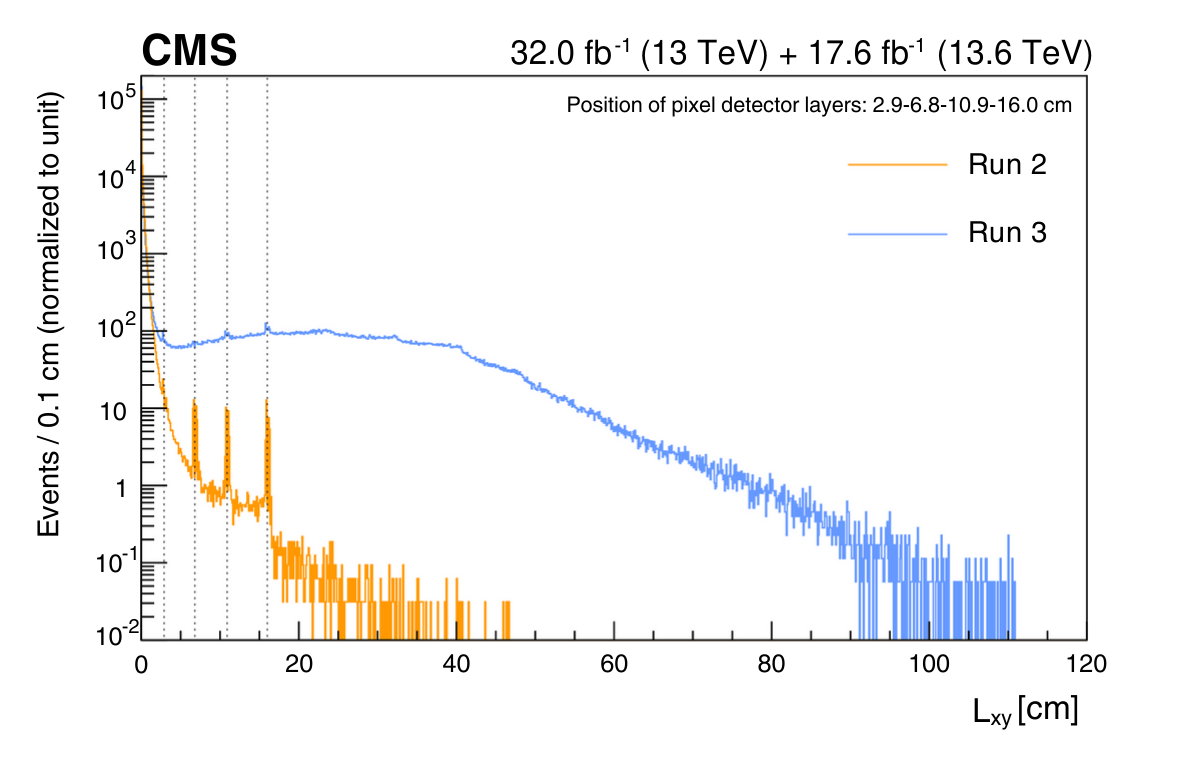}
\caption{Comparison between the \Lxy distribution for dimuon scouting events for Run~2 (orange) and Run~3 (blue) events in data that contain dimuon pairs with a common displaced vertex and a minimal selection on the vertex quality. The blue Run~3 line is obtained from the vertex-unconstrained muon reconstruction algorithm. The dashed vertical lines, at radii of 2.9, 6.8, 10.9, and 16.0\cm, correspond to the positions of the pixel detector layers where photons undergo conversion processes in the material, causing the observed peaks in the \Lxy distribution. Figure taken from Ref.~\cite{CMS:2024zhe}.}
\label{fig:dimuonScoutingRun3vsRun2lxy}
\end{figure}

\begin{figure}[hbt!]
\centering
\includegraphics[width=0.73\textwidth]{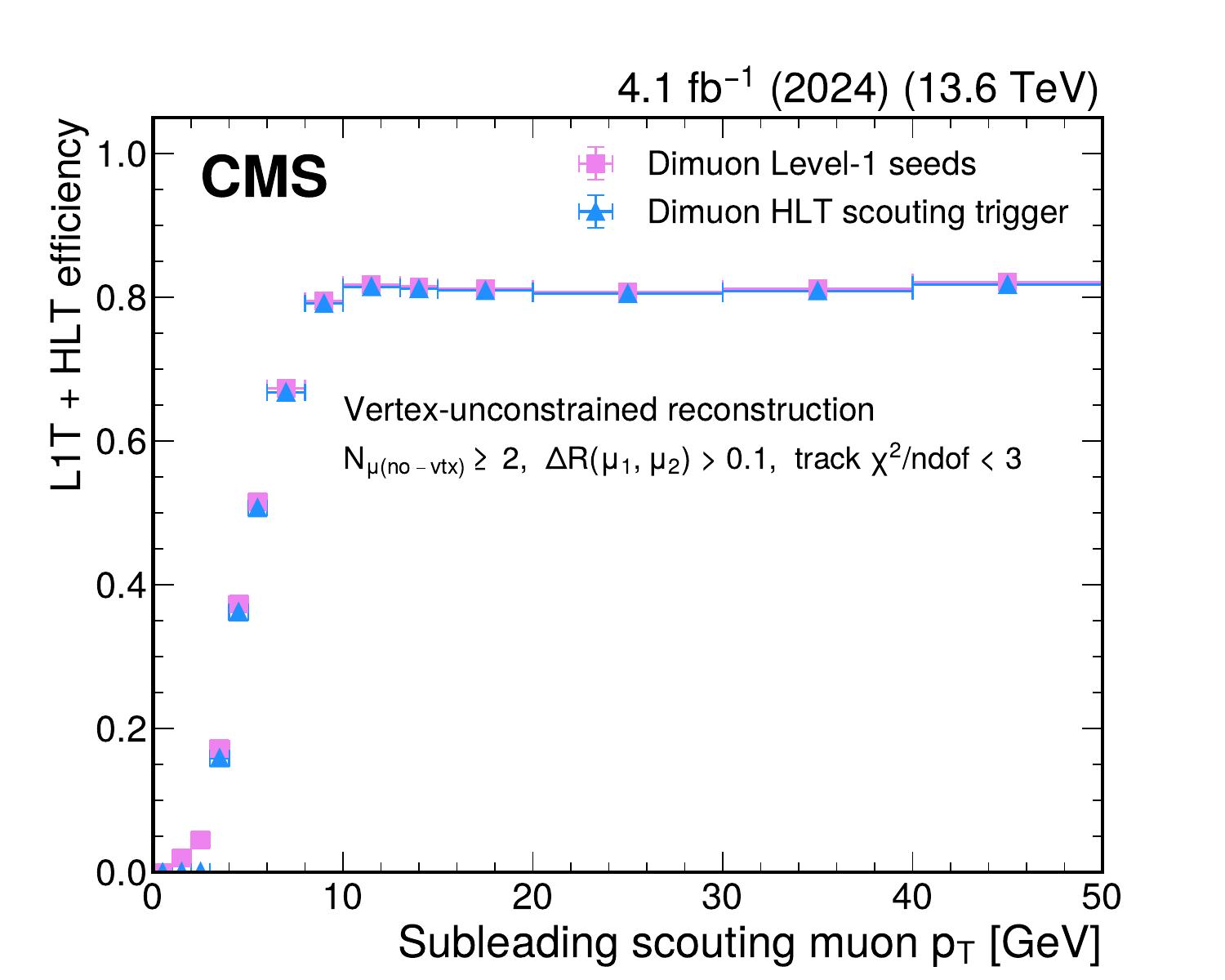}
\caption{The L1T+HLT efficiency of the dimuon scouting trigger as a function of the subleading muon \pt, for 2024 data. The efficiency of the L1T dimuon seeds (pink squares) and the HLT dimuon scouting trigger with the vertex-unconstrained reconstruction algorithm (blue triangles) is shown. The events in the denominator are required to have at least two vertex-unconstrained muons ($N_{\PGm(\text{no-vtx})} \geq 2$) and additionally to have $\chi^2/\Ndof < 3$ and $\DR>0.1$.}
\label{fig:dimuonScoutingEffVsPtData}
\end{figure}

Figure~\ref{fig:dimuonScoutingRun3vsRun2lxy} shows the distribution of the dimuon vertex transverse displacement \Lxy, for events that contain at least one pair of opposite-sign muons associated with a selected secondary vertex. Events are collected with dimuon displacements up to about 100\cm, a radial distance that corresponds to the effective end of the region of the tracker available for track reconstruction. At the positions of the pixel detector layers, with radii of 29, 68, 109, and 160\mm, photons undergoing conversion processes in the material lead to peaks in the \Lxy distribution. These peaks are less pronounced in the Run~3 distribution because of the removal of the pixel detector hit requirement, which leads to higher efficiency, but also lower purity, if no additional analysis-specific quality criteria are required, as is the case here. While the removal of this requirement increases the background yield, it recovers potential displaced signal vertices that would otherwise be lost. This strategy, combined with appropriate background rejection techniques, ultimately improves the CMS sensitivity to exotic displaced decays.

A combination of L1T algorithms requiring one of two options is employed. These options are to require either at least two muons with minimum \pt thresholds of 15 and 7\GeV for the leading and subleading muon, respectively, or two opposite-sign muons with no explicit \pt thresholds (other than requiring the muon to satisfy the minimum \pt restrictions of the L1T muon reconstruction algorithms, as discussed in Section~\ref{sec:muons}) and an additional selection on the invariant mass or angular separation. The proportional rate of these triggers at L1 amounts to about 10\unit{kHz} at the highest luminosity for an average PU of 60. Events selected by at least one of the L1 seeds are then reconstructed at HLT with a minimum muon \pt requirement of 3\GeV, reducing the HLT rate of the dimuon scouting trigger to about 4.2\unit{kHz}.

\begin{figure}[htb!]
\centering
\includegraphics[width=0.73\textwidth]{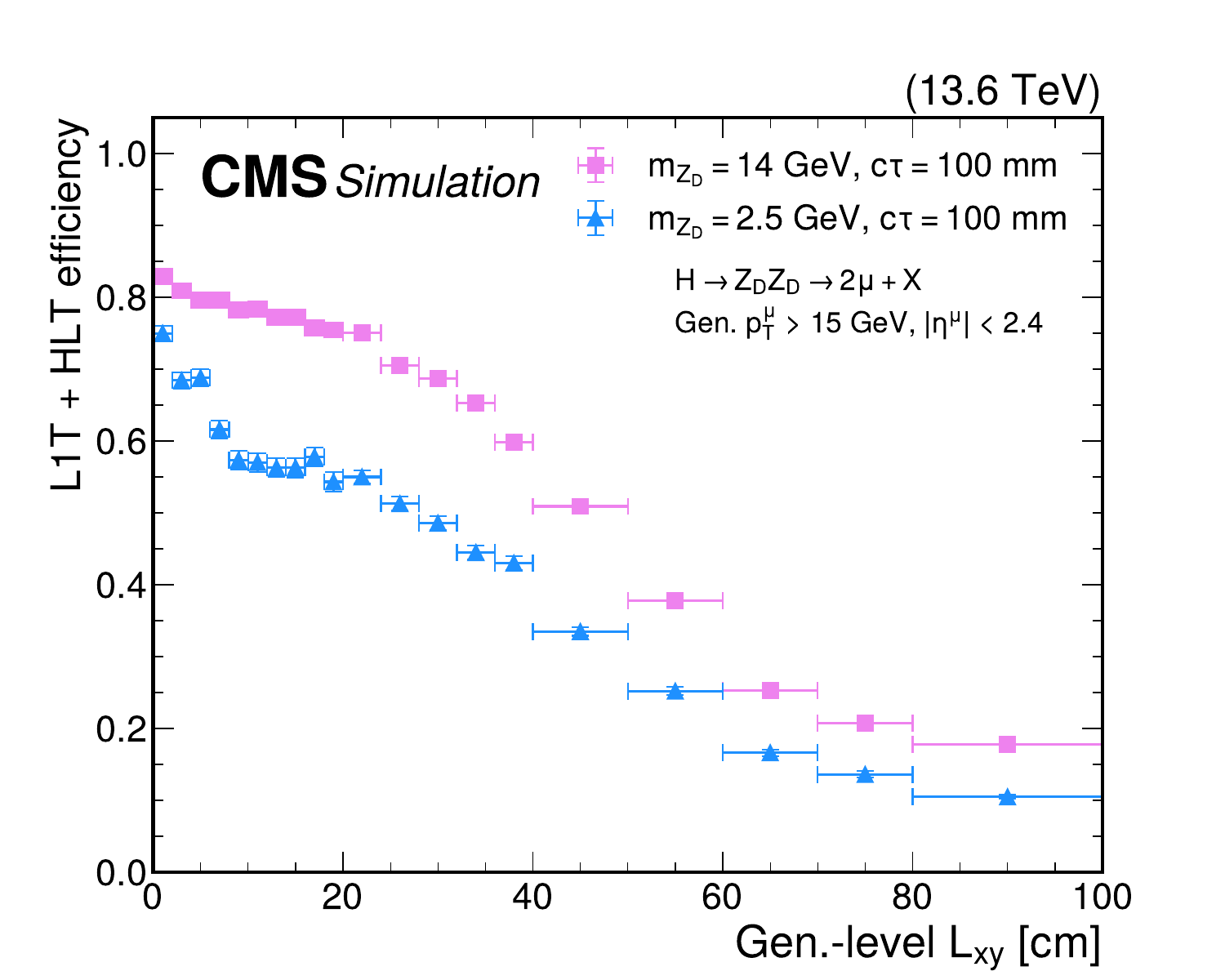}
\caption{The L1T+HLT efficiency of the dimuon scouting trigger as a function of the generator-level \Lxy, for HAHM signal events, for 2024 conditions. The efficiency is shown for $\mZD = 14\GeV$ and $\cTau=100\mm$ (pink squares) and $\mZD = 2.5\GeV$ and $\cTau=100\mm$ (blue triangles). The muons are required to have $\pt>15\GeV$ and $\abseta<2.4$ at the generator level.}
\label{fig:dimuonScoutingEffVsLxySignal}
\end{figure}

\begin{figure}[hbt!]
\centering
\includegraphics[width=0.49\textwidth]{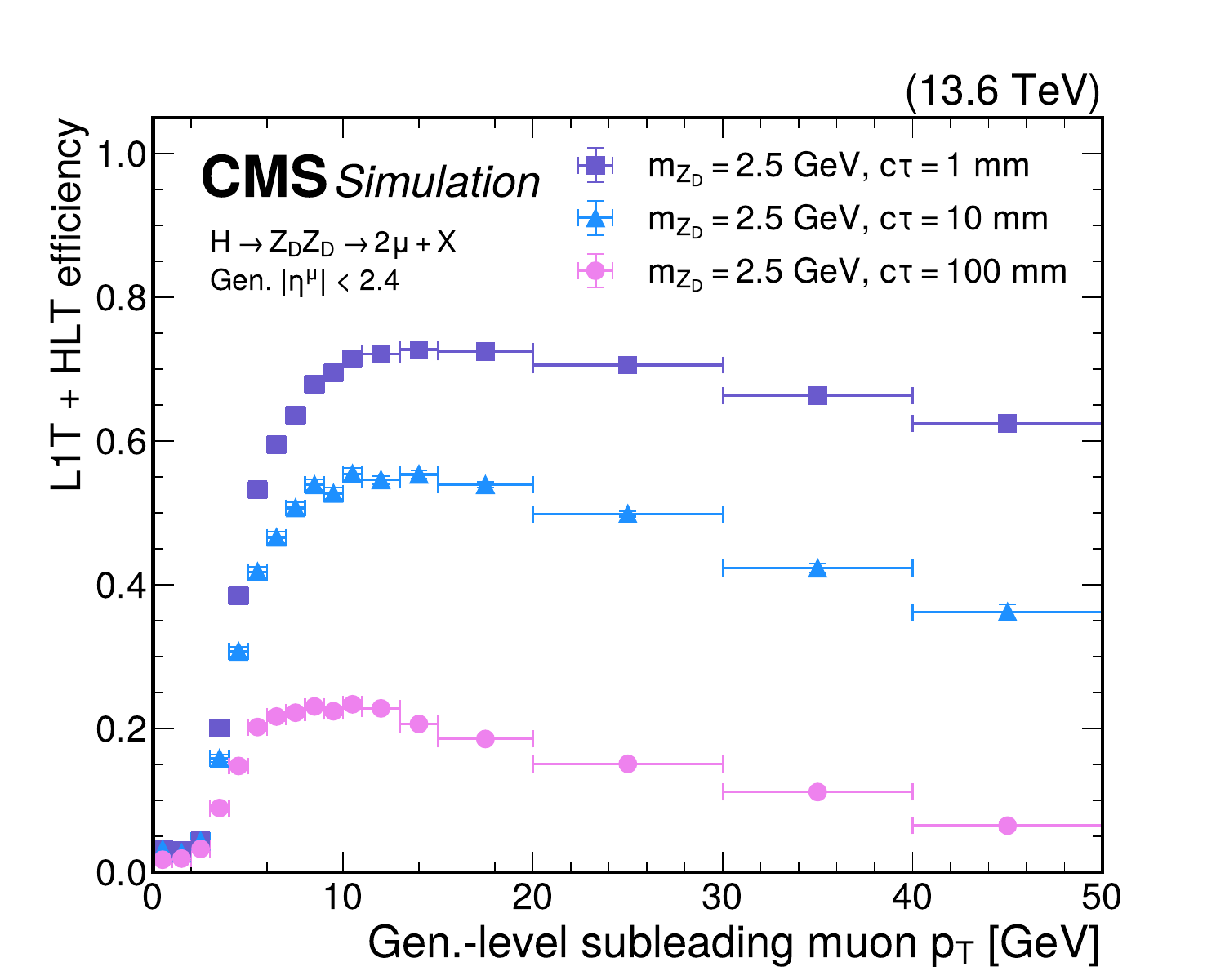}
\includegraphics[width=0.49\textwidth]{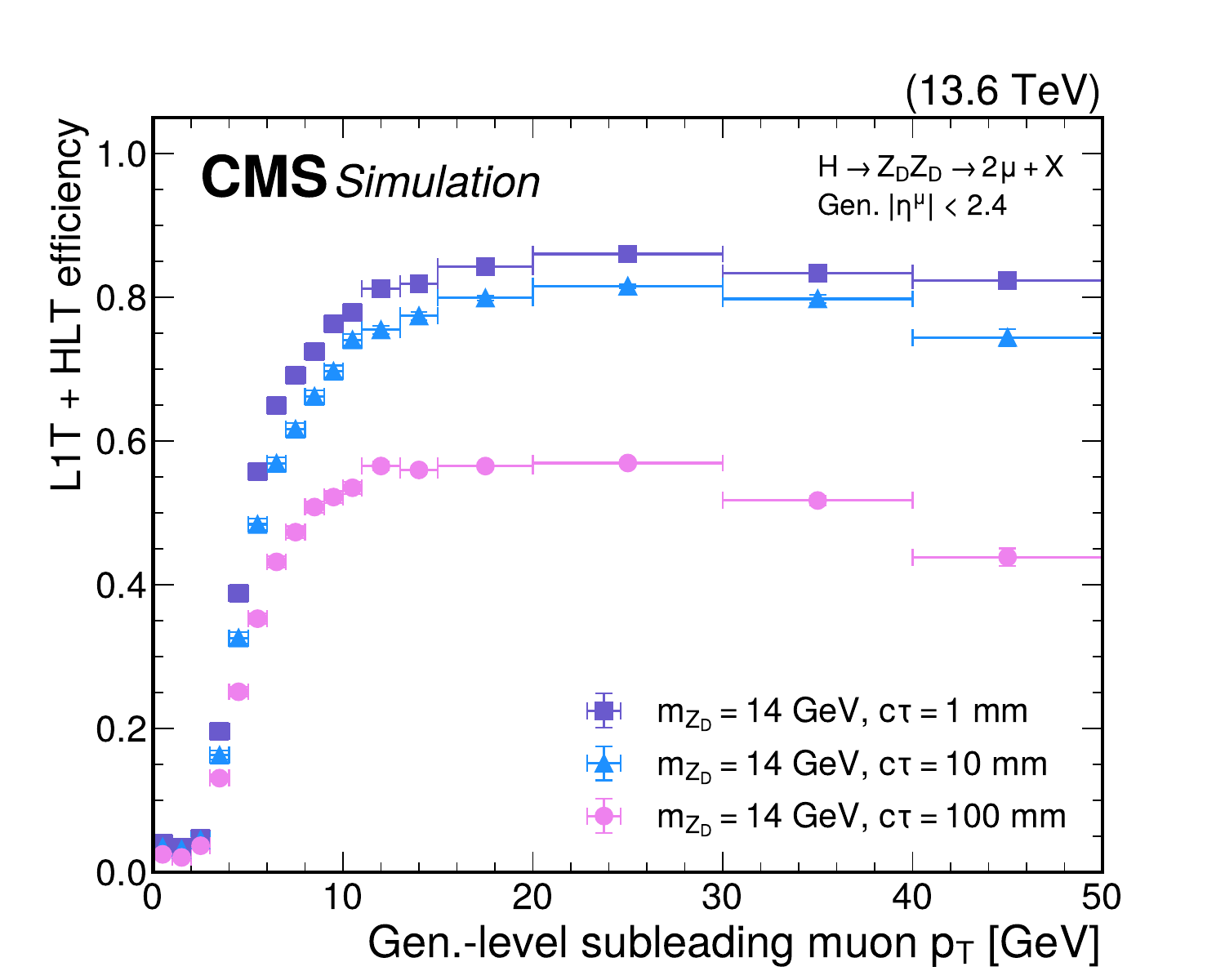}
\caption{The L1T+HLT efficiency of the dimuon scouting trigger as a function of the generator-level subleading muon \pt, for HAHM signal events for 2024 conditions. The efficiency is shown for \PZD masses of 2.5 (\cmsLeft) and 14\GeV (\cmsRight), and \cTau values of 1 (purple squares), 10 (blue triangles), and 100\mm (pink circles). The muons are required to have $\abseta<2.4$ at the generator level.}
\label{fig:dimuonScoutingEffVsPtSignal}
\end{figure}

The efficiency of the dimuon scouting trigger in 2024 data is shown in Fig.~\ref{fig:dimuonScoutingEffVsPtData} and in simulated HAHM signal events for two illustrative masses of $\mZD = 2.5$ and 14\GeV in Figs.~\ref{fig:dimuonScoutingEffVsLxySignal} and \ref{fig:dimuonScoutingEffVsPtSignal}. Figure~\ref{fig:dimuonScoutingEffVsPtData} shows that the dimuon scouting trigger efficiency is driven by the L1T, which requires high-quality muons. Figures~\ref{fig:dimuonScoutingEffVsLxySignal} and \ref{fig:dimuonScoutingEffVsPtSignal} demonstrate that the dimuon scouting triggers are highly efficient for HAHM events, especially for sufficiently large \mZD and small boost. Furthermore, a significant fraction of the muons are reconstructed even for the longest considered lifetimes. High efficiency is maintained for \Lxy values up to about 40\cm, demonstrating the improved efficiency of the Run~3 trigger configuration for displaced signals. These signal efficiencies also reflect the behavior observed in boosted decays, where the muons are very close to each other, as occurs for very low masses in the HAHM model. This effect is particularly significant when $\mZD = 2.5\GeV$, where muons with larger \pt values but also small angular separation exhibit a reduced trigger efficiency as the \pt increases.

\begin{figure}[htb!]
\centering
\includegraphics[width=0.49\textwidth]{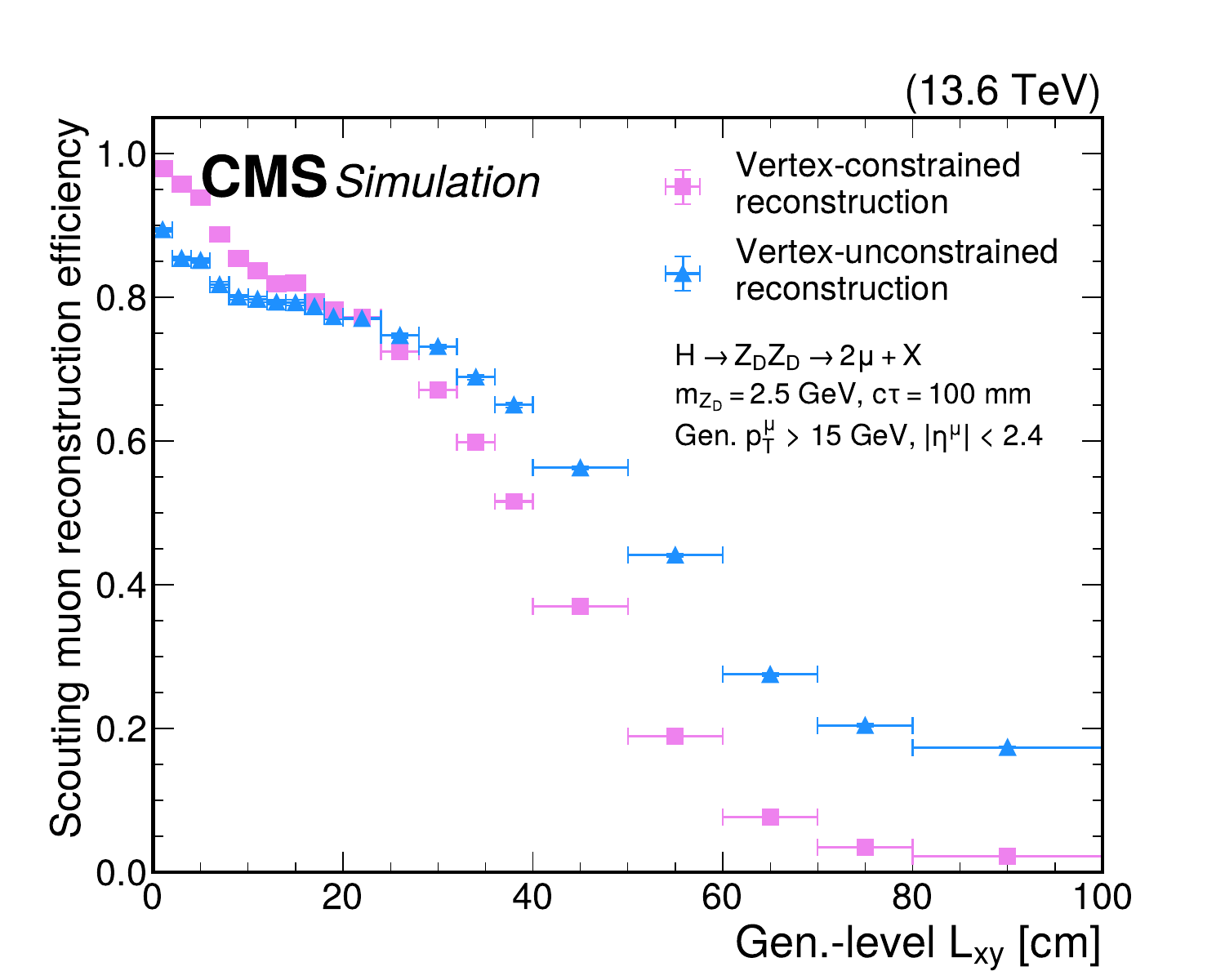}
\includegraphics[width=0.49\textwidth]{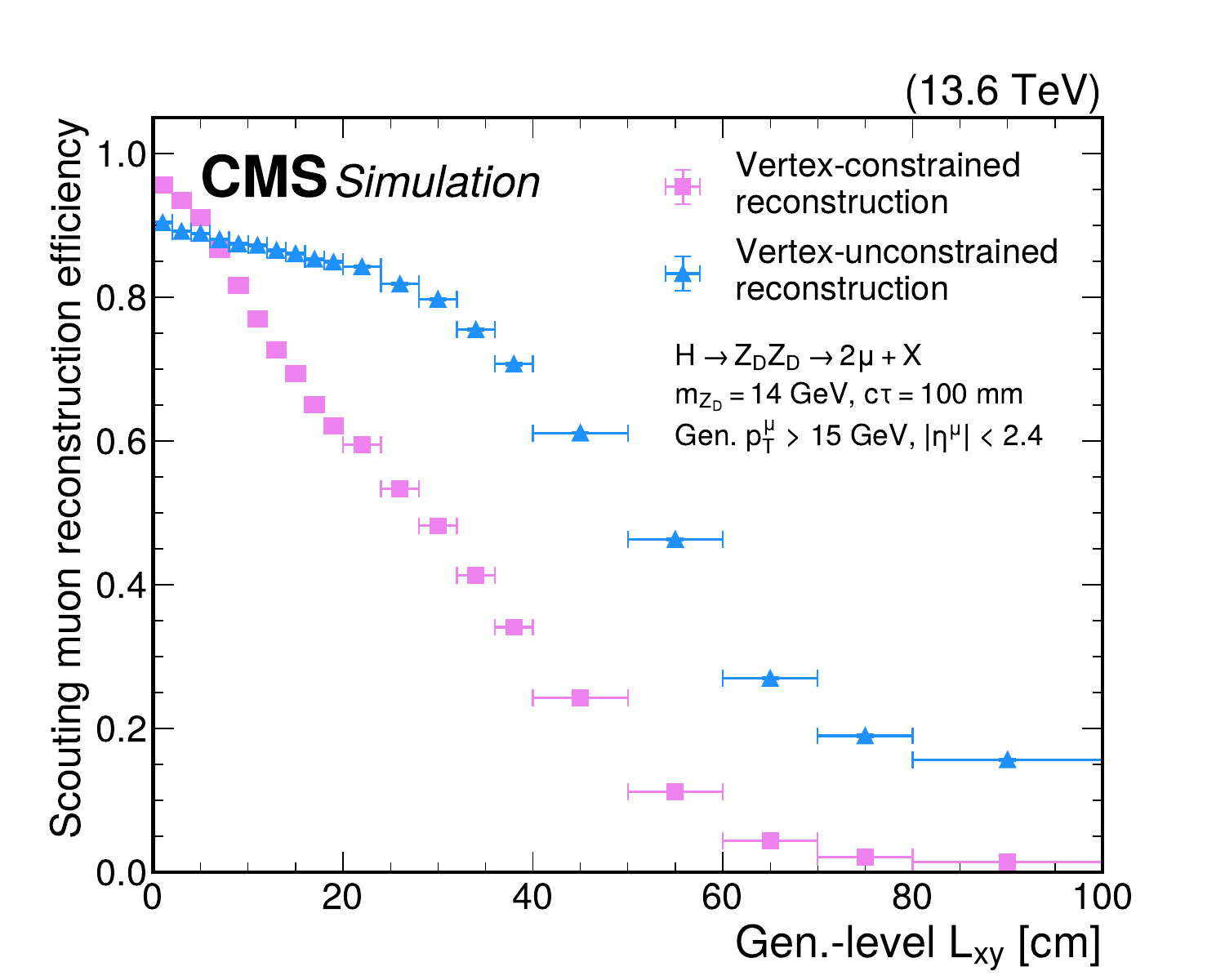}
\caption{Scouting muon reconstruction efficiency of the vertex-constrained (pink squares) and vertex-unconstrained (blue triangles) algorithms as a function of the generator-level \Lxy, for HAHM signal events for 2024 conditions. This efficiency is representative of the reconstruction efficiency of the L2 and L3 HLT muon reconstruction employed for scouting data. The efficiency is shown for $\mZD = 2.5\GeV$ and $\cTau=100\mm$ (\cmsLeft) and $\mZD = 14\GeV$ and $\cTau=100\mm$ (\cmsRight). The muons are required to have $\pt>15\GeV$ and $\abseta<2.4$ at the generator level.}
\label{fig:dimuonScoutingRecoEffVsLxySignal}
\end{figure}

\begin{figure}[hbt!]
\centering
\includegraphics[width=0.7\textwidth]{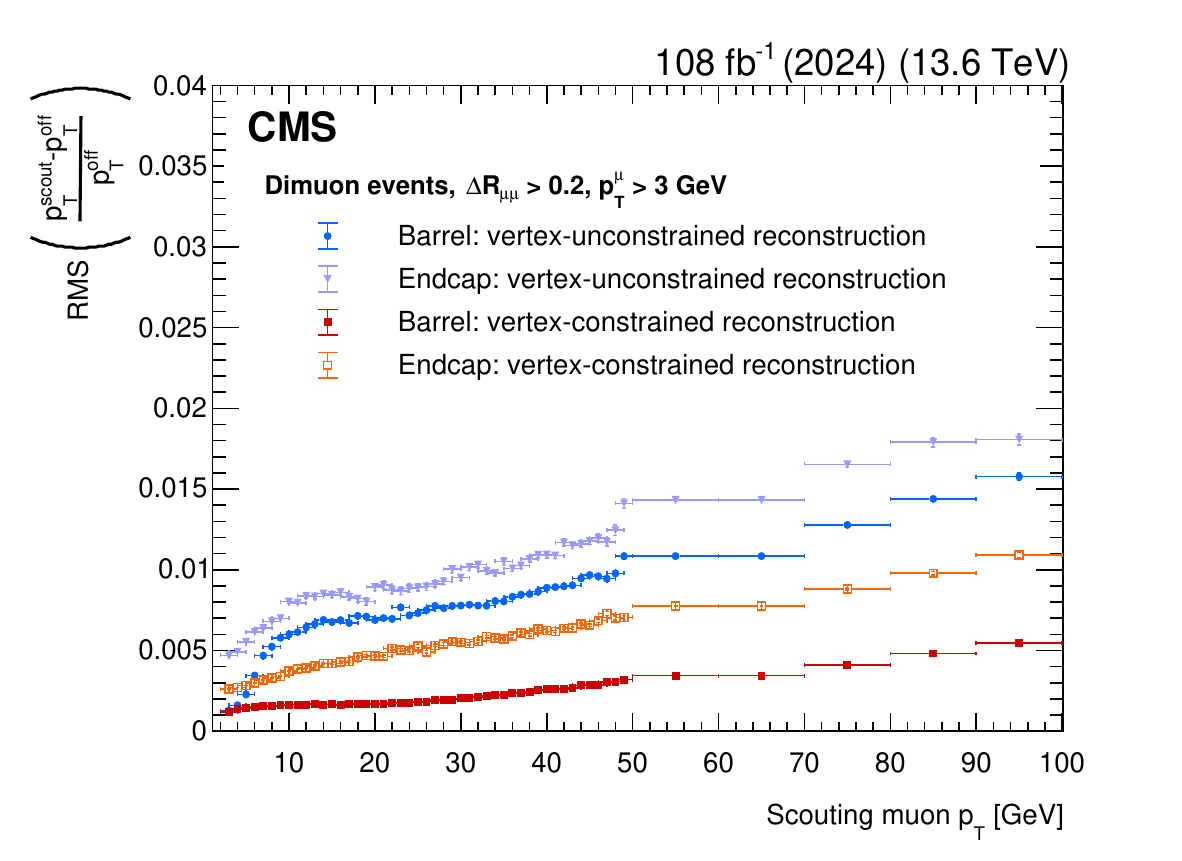}
\caption{The \pt resolution of scouting muons with respect to offline muons, as a function of the scouting muon \pt, for 2024 data events. The root mean square (RMS) of the difference of the scouting muon \pt and the offline muon \pt, divided by the offline muon \pt, is plotted. The dimuon \DR is required to be greater than 0.2, and the scouting muon \pt is required to be greater than 3\GeV. The resolution is shown for muons in the barrel (filled blue circles) and the endcaps (filled purple triangles) that are reconstructed with the vertex-unconstrained reconstruction algorithm, as well as for muons in the barrel (filled red squares) and the endcaps (open orange squares) that are reconstructed with the vertex-constrained reconstruction algorithm. A special monitoring data set is used that collects events triggered by a mixture of HLT paths (both scouting and standard triggers) with a very high prescaling factor, in which all information about the muon objects is stored from the offline and scouting reconstruction.
}
\label{fig:dimuonScoutingPtResolutionVsOffline}
\end{figure}

Two different scouting muon reconstruction algorithms were developed: one constrained to the PV and another unconstrained. The vertex-constrained algorithm, which was enabled at the start of data taking in 2024, is appropriate for prompt muons, while the vertex-unconstrained algorithm, which has been enabled since 2022, provides higher efficiency for displaced muons. The reconstruction efficiency of both algorithms as a function of \Lxy is compared in Fig.~\ref{fig:dimuonScoutingRecoEffVsLxySignal} for HAHM signal events with the same two illustrative masses. Figure~\ref{fig:dimuonScoutingPtResolutionVsOffline} shows the \pt resolution of scouting muons with respect to offline muons, as a function of the scouting muon \pt, for 2024 data events. The vertex-unconstrained algorithm provides \pt resolutions ranging from 0.5 to 1\% for $\pt< 50\GeV$ and increasing to 2\% at 100\GeV, for both barrel and endcap muons. The vertex-constrained algorithm provides \pt resolutions of less than 0.5\% for the barrel and between 0.5 and 1\% in the endcaps. This object-level comparison is shown in terms of \pt because the dimuon scouting trigger requirements and the corresponding analyses are formulated directly using the online muon \pt; the resolution of the inverse HLT muon \pt is shown separately in Section~\ref{sec:trigPerf}.

\subsubsection{Muon detector shower triggers with the CSCs}
\label{sec:mdsshowers}

In CMS, LLPs with sufficiently long lifetime, produced at the PV in $\Pp\Pp$ collisions, could reach the muon system before decaying, creating an MDS signature. The defining feature of LLP decays within the CMS muon system is the multiplicity of hits created by particle showers. This starkly contrasts with what a typical muon would deposit in the detector, which is several hits in each detection layer that form a trajectory pointing to the IP. Some Run~2 analyses have used this MDS signature offline, as described in Refs.~\cite{CMS:2021juv,CMS:2023arc}, relying on other objects to provide the trigger. To improve the CMS sensitivity to this striking signature, several novel MDS triggers have been commissioned for Run~3 data taking. The ATLAS experiment uses muon-region-of-interest cluster triggers to select LLP decays producing localized activity in its muon spectrometer~\cite{ATLAS:2018tup,ATLAS:2022gbw}; the CMS MDS triggers described here target a different detector signature, namely high-multiplicity hit clusters produced by showers in the CMS muon system and steel return yoke.

\begin{figure}[htb!]
  \centering
  \includegraphics[width=0.49\textwidth]{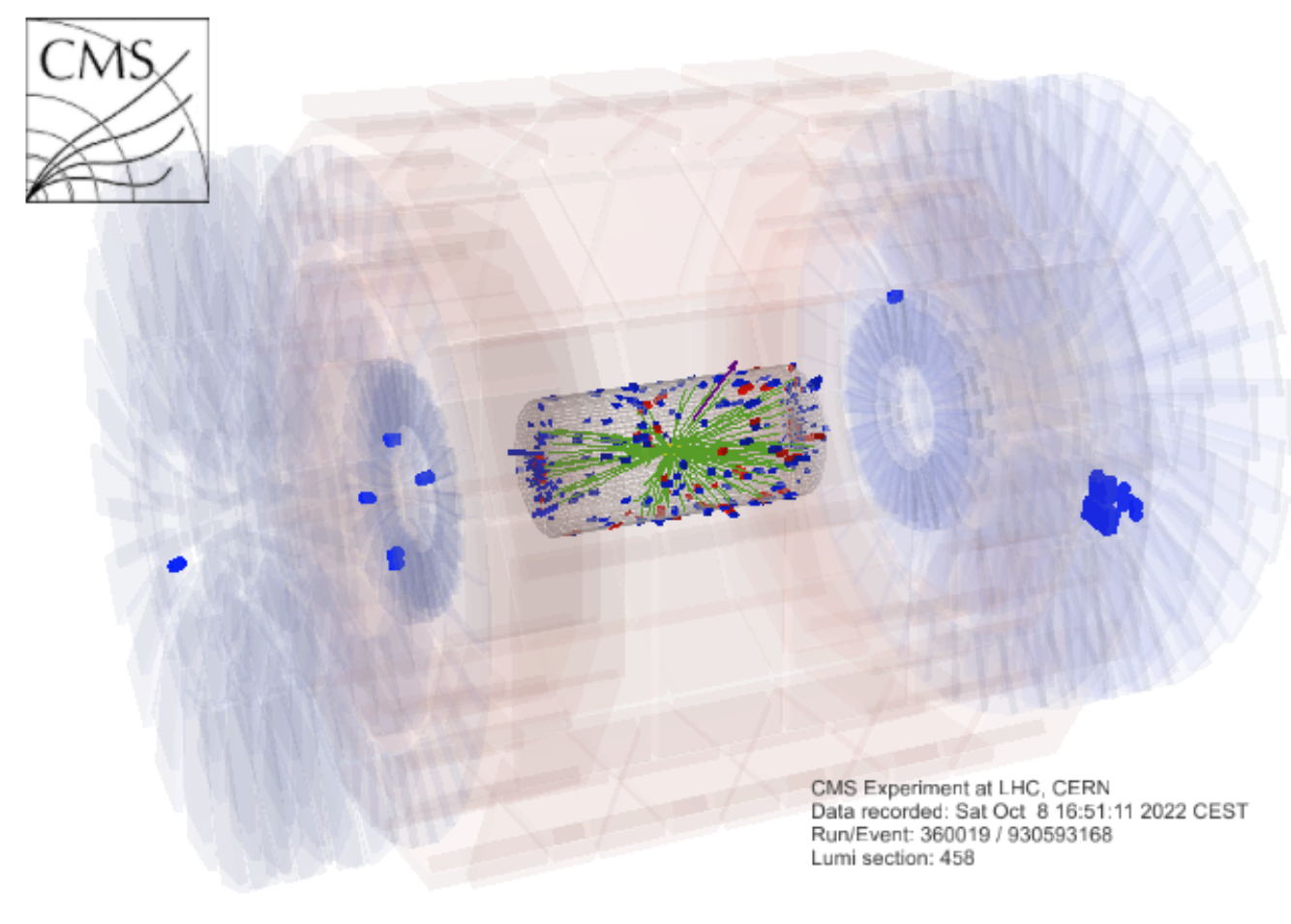}
  \includegraphics[width=0.49\textwidth]{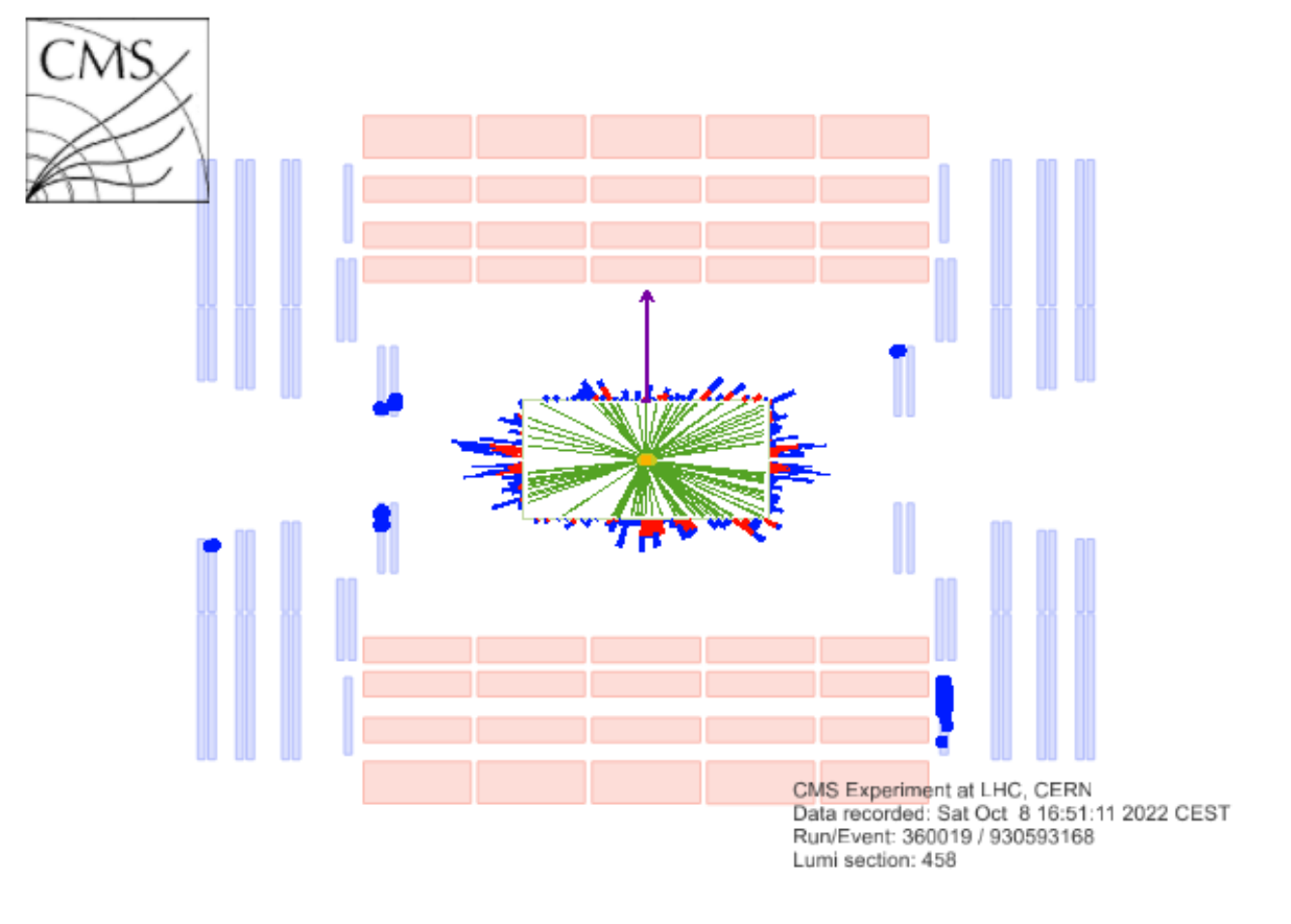}
  \caption{Event display of a collision triggered by the CSC MDS trigger in Run~3 data. The CSC reconstructed hits are represented by blue dots in the muon endcap region. This event features a CSC cluster of 210 hits in the ME1/3 ring. The event was recorded on October 8th, 2022.}
  \label{fig:MDS_event_display}
\end{figure}

The overall strategy at the L1T is first to identify the LLP showers by counting the number of hits in individual CSC chambers (\Nhits). Because the capacity of each stage of the trigger is limited, the information is progressively compressed as it moves through the system. The first stage reduces the chamber-level hit information to a small number of bits indicating whether a shower is present. Subsequent stages then aggregate these chamber-level flags. The processor checks whether a chamber shows a shower and encodes this in only a few bits. Finally, an algorithm in the global muon trigger determines whether a shower is present anywhere in the full detector, again encoding this result in a compact form and sending it to the global trigger for the final decision. In essence, each stage performs a logical OR-type reduction and transmits an increasingly condensed representation of the shower activity through the off-detector electronics. Hits per bunch crossing are counted in each CSC chamber by the CSC local trigger, which has both anode wire and cathode strip readout. Optimized thresholds are set on each station individually to evaluate whether \Nhits is above or below such thresholds. The greater available resources of the optical trigger motherboard allow the trigger to detect the coincidence of anode and cathode counts above thresholds for chambers in ring 1, while an anode-only logic is used for chambers in rings 2 and 3.  This information is encoded in several bits called ``loose'', ``nominal'', and ``tight'', each with a different threshold, and then transferred to the EMTF processors. Likewise, the EMTF transmits information to the global muon trigger about the existence or absence of any chamber passing the counting criteria. Finally, the global muon trigger transfers the information to the global trigger, and the L1 physics seeds (One-Nominal, One-Tight, and Two-Loose) are created. In the case of the Two-Loose strategy, which targets pair-produced LLPs, the global muon trigger evaluates if there are at least two different stations with any chamber above relaxed thresholds, which vary depending on the chamber. A display of one CMS event selected by this trigger in 2022 data is shown in Fig.~\ref{fig:MDS_event_display}.

The efficiency of the trigger to select LLPs is evaluated with a sample of $\PH \to \PS\PS \to \bbbar\bbbar$ events, the diagram for which is shown in Fig.~\ref{fig:HtoXX_feynmanDiagrams}. Signal efficiencies are shown in Table~\ref{table:signal_eff_LLP} for simulated events with one and two LLPs decaying in the CSC detector acceptance. A larger mass splitting between \mH and \mS, which results in LLPs with higher energy, leads to greater trigger efficiency. The addition of the Two-Loose L1 seed also yields up to a 10\% increase in the signal efficiency, compared to having only the One-Nominal seed, when both LLPs decay within the CSC detector acceptance.

\begin{table}[htbp]
  \centering
    \topcaption{The L1T CSC MDS trigger efficiency $\epsilon$ in $\PH \to \PS\PS \to \bbbar\bbbar$ events, for different choices of \mH, \mS, and \cTauPS, for simulated events with at least 1 (middle columns) and exactly 2 (right columns) LLPs decaying within the CSC detector acceptance. The $\epsilon_{\text{acc}}$ refers to the percentage of events within the CSC detector acceptance. The $\epsilon_{\text{1-N}}$, $\epsilon_{\text{1-T}}$, and $\epsilon_{\text{2-L}}$ denote the efficiencies for the One-Nominal, One-Tight, and Two-Loose algorithms, respectively, using events that pass the CSC detector acceptance ($\epsilon_{\text{acc}}$) as the denominator.} \label{table:signal_eff_LLP}
    \begin{tabular}{@{}%
        >{\centering\arraybackslash}p{1.7cm}%
        >{\centering\arraybackslash}p{1.7cm}%
        >{\centering\arraybackslash}p{1.7cm}
        cccccc@{}}
        \multirow{2}{*}{\mH [{\GeVns}]} & 
        \multirow{2}{*}{\mS [{\GeVns}]} & 
        \multirow{2}{*}{\cTauPS [{mm}]} 
        & \multicolumn{3}{c}{$\geq 1$ LLP decay in CSCs} 
        & \multicolumn{3}{c}{2 LLP decays in CSCs} \\
         &  &  
        & $\epsilon_{\text{acc}}$ & $\epsilon_{\text{1-N}}$ & $\epsilon_{\text{1-T}}$ 
        & $\epsilon_{\text{acc}}$ & $\epsilon_{\text{1-N}}$ & 
        $\epsilon_{\text{2-L}}$ OR $\epsilon_{\text{1-N}}$\\[\cmsTabTinySkip]
        \hline & \\[-\cmsTabTinySkipInverse]
        125 & 12  & 900   & 16.2\% & 43.1\% & 39.2\% & 0.7\% & 69.6\% & 75.0\% \\
        125 & 25  & 1500  & 15.9\% & 43.7\% & 40.0\% & 0.7\% & 75.1\% & 79.7\% \\
        250 & 60  & 10000 & 10.3\% & 55.5\% & 52.1\% & 0.3\% & 74.0\% & 78.2\% \\
        250 & 120 & 10000 & 15.4\% & 29.9\% & 27.7\% & 1.0\% & 44.4\% & 49.6\% \\
        350 & 160 & 10000 & 17.9\% & 39.1\% & 36.7\% & 1.1\% & 63.4\% & 69.6\% \\
    \end{tabular}
\end{table}

At the HLT, hits from all the CSC chambers are available and can be used to reconstruct cluster objects. This allows the HLT to capture the properties of signal showers across multiple CSC chambers, thus rejecting more backgrounds. The CSC 3D spatial points (hits) are clustered using the Cambridge--Aachen algorithm~\cite{Dokshitzer:1997in, Wobisch:1998wt} in $\eta$-$\phi$ space with a distance parameter of 0.4. A cluster object is required to have at least 50 hits, a threshold approximately two times higher than a typical muon would create in the muon system.

The additional cluster properties available at the HLT allow for the design of various trigger paths to target different LLP production modes, as shown in Table~\ref{tab:triggerlist}. The single-MDS path, which was introduced in time for 2022 data taking, is the most generic, whereas the double-MDS path (introduced in 2023) targets pair-produced LLP models more efficiently because of its looser cluster selections. For LLPs produced together with a lepton or photon, additional mixed-object trigger paths were introduced in time for 2024 data taking, where the trigger thresholds for both the cluster and the lepton/photon selections are lowered.

\begin{figure}[htb!]
  \centering
  \includegraphics[width=0.49\textwidth]{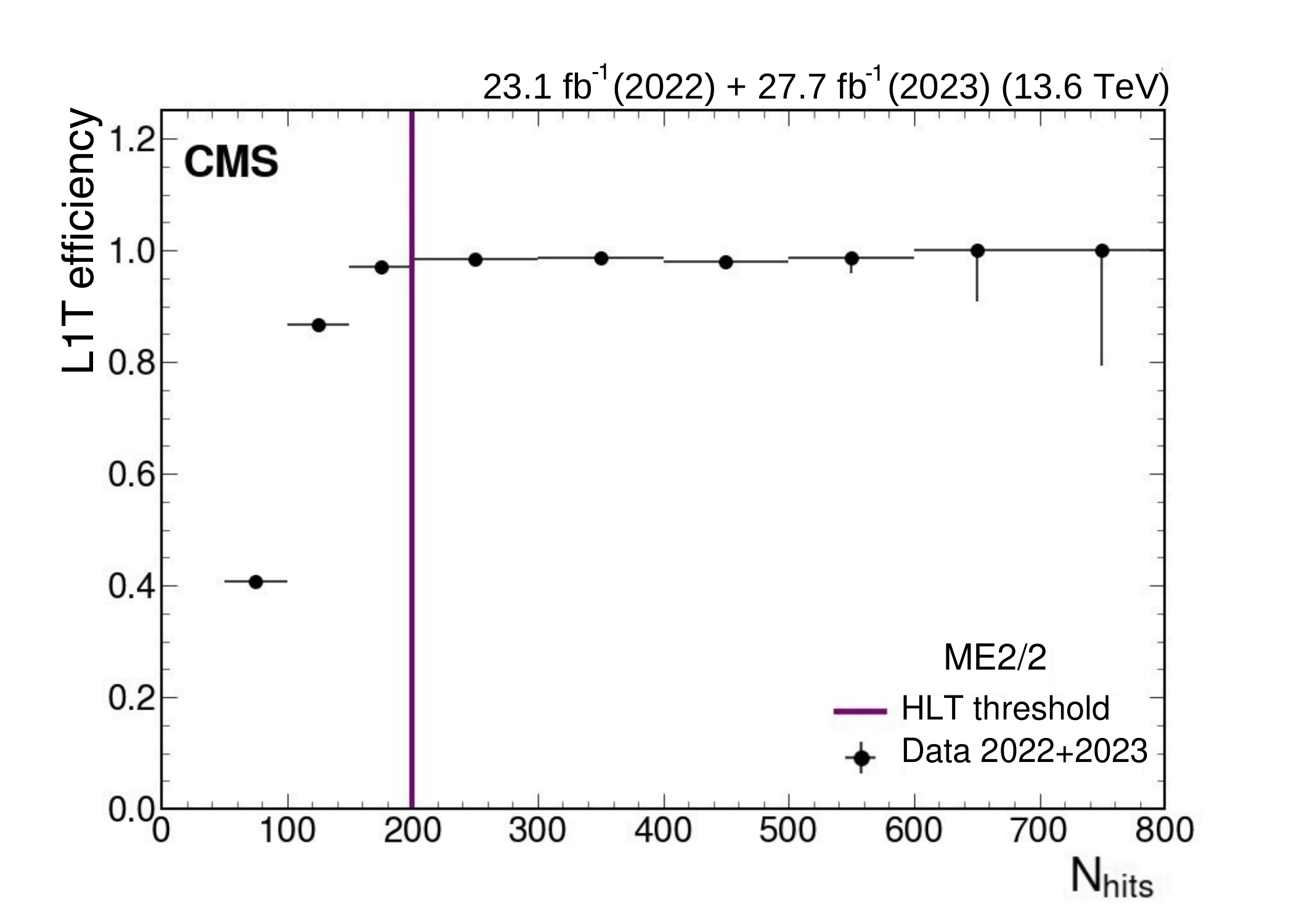}
  \includegraphics[width=0.49\textwidth]{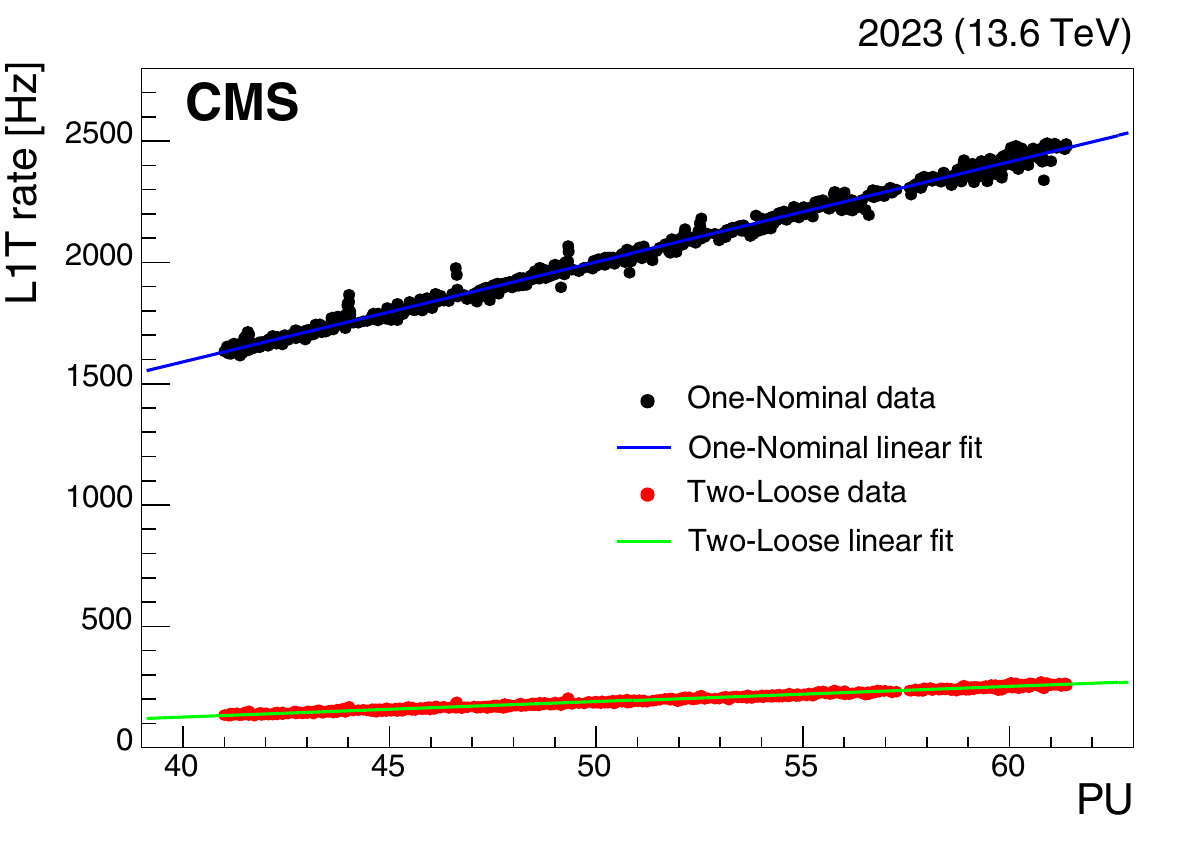}
  \caption{The L1T efficiency of the One-Nominal CSC MDS trigger in a single CSC ring (ME2/2, indicated in Fig.~\ref{fig:cms}) as a function of the number of hits in the cluster, measured with muon bremsstrahlung-induced electromagnetic showers using data taken in 2022 and 2023 (\cmsLeft). The efficiency is defined as the fraction of events that pass the One-Nominal CSC MDS trigger, given that an uncorrelated trigger selects the data and that a muon is matched to a CSC cluster. The L1T rate as a function of PU for the One-Nominal and Two-Loose seed for a 2023 data-taking run (\cmsRight). The rate dependence on PU is extracted by a linear fit.}
  \label{fig:MDS_eff_rate}
\end{figure}

The L1T efficiency as a function of \Nhits for one of the CSC rings and the L1T rates for One-Nominal and Two-Loose as a function of PU are shown in Fig.~\ref{fig:MDS_eff_rate}. The efficiencies for other CSC rings are shown in Ref.~\cite{HMTnote22To23}. To evaluate the efficiency of individual CSC rings, CSC clusters created by muon bremsstrahlung-induced electromagnetic showers are used as proxies for LLP signal clusters in data. The efficiency is defined as the fraction of events that pass the CSC MDS One-Nominal L1 seed, given that an uncorrelated trigger selects the data and that a muon is matched to a CSC cluster. Because the L1 thresholds of different CSC rings are different, the efficiency is measured separately for each CSC ring. Thus, the centroid position of the cluster is restricted to be within a single CSC ring (\eg, ME2/2), and at least 90\% of the hits in the clusters must be contained in the same CSC ring. 

Figure~\ref{fig:MDS_eff_rate} also shows that the L1T rates of both the One-Nominal and Two-Loose seeds are proportional to the average PU. The cluster selections at the HLT further reduce the L1T rates by a factor of 170, such that the single- and double-MDS paths have a combined total rate of 14\unit{Hz}. The CSC MDS mixed-object triggers provide an additional 14\unit{Hz} total rate at the HLT at the same instantaneous luminosity.

\subsubsection{Muon detector shower triggers with the DTs}
\label{sec:dtcluster}

Similar to the MDS signature in the CSCs described in Section~\ref{sec:mdsshowers}, an LLP decaying in the DTs in the muon system can produce showers from the resulting decay products. These showers can be identified via clusters of hits in the DTs, using this subsystem as a sampling calorimeter. The DTs provide complementary coverage to that of the CSC MDS triggers described in Section~\ref{sec:mdsshowers}. In Run~2, CMS performed a search for LLPs using the DT cluster showers with competitive limits on various models, including those with Higgs boson decays to LLPs, and produced the first LHC limit on a strong dark-sector coupling to the Higgs boson~\cite{CMS:2023arc}.

\begin{figure}[htb!]
  \centering
  \includegraphics[width=0.55\textwidth]{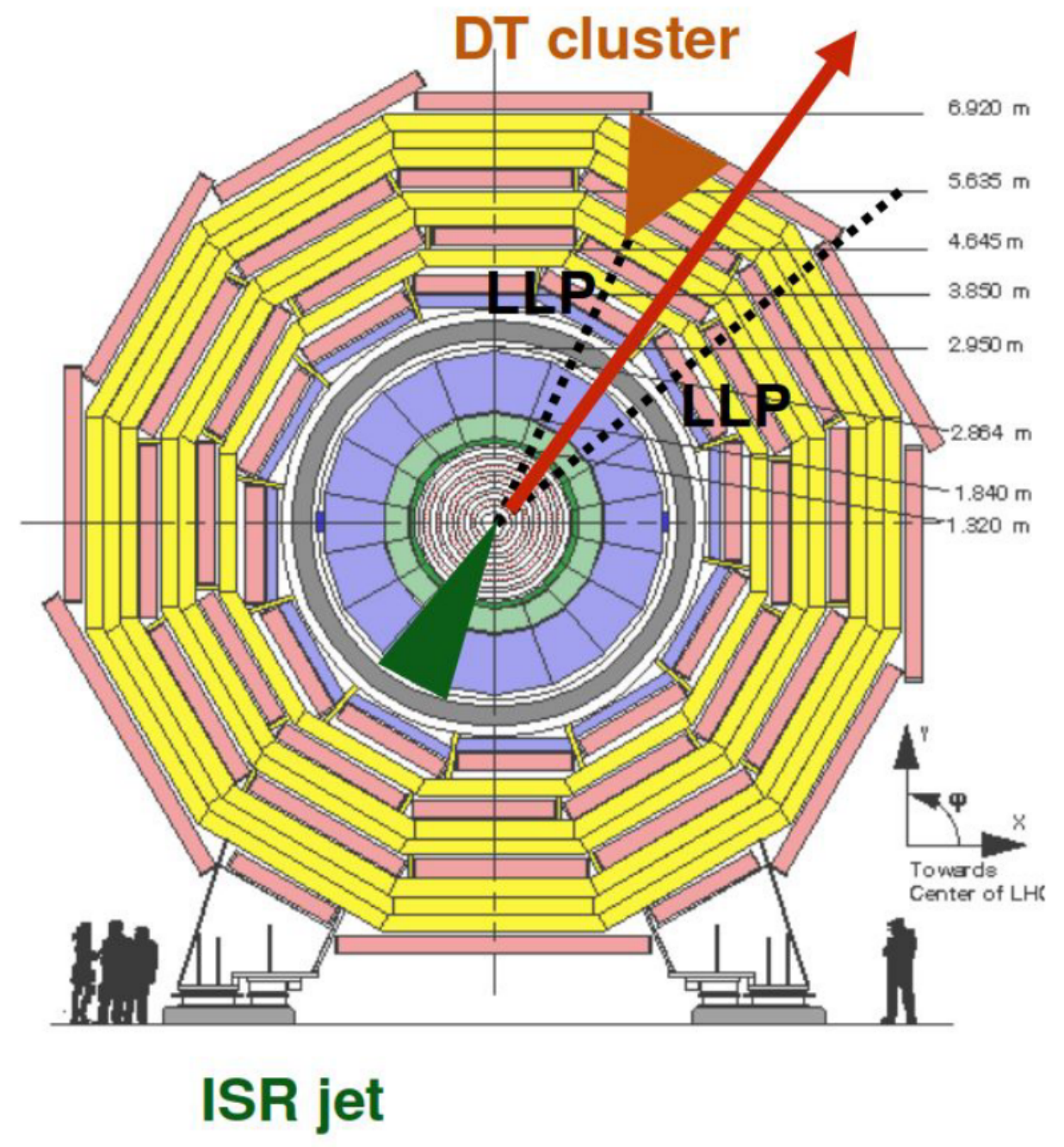}
  \caption{Diagram of the MDS signature in the DTs. The LLP decays in the DTs, opposite a jet from ISR.}
  \label{fig:dtcluster}
\end{figure}

In Run~3, dedicated triggers have been introduced that target a cluster of hits in the DTs opposite a jet from ISR or final-state radiation. Figure~\ref{fig:dtcluster} provides a visual representation of this signature. This trigger enables a more model-independent search for LLPs decaying in the DTs. The trigger uses \ptmiss with a minimum threshold of 150\GeV as the L1 seed; implementing a dedicated DT MDS trigger at the L1T was impossible because of the lack of processing capacity. At the HLT, DT clusters are reconstructed similarly to the CSCs, and $\Nhits > 50$ is required. Hits from the first DT station, MB1, are excluded from the clusters, as it has the least amount of material between it and the IP and thus higher backgrounds. Furthermore, another HLT path was implemented that requires the One-Nominal MDS CSC cluster at the L1T and at least one DT cluster at the HLT. This path recovers some efficiency for the case where multiple clusters are created in the transition region between the barrel and the endcaps. The total trigger rate of the DT MDS paths at the HLT is 9\unit{Hz}.

\begin{figure}[htb!]
  \centering
  \includegraphics[width=0.49\textwidth]{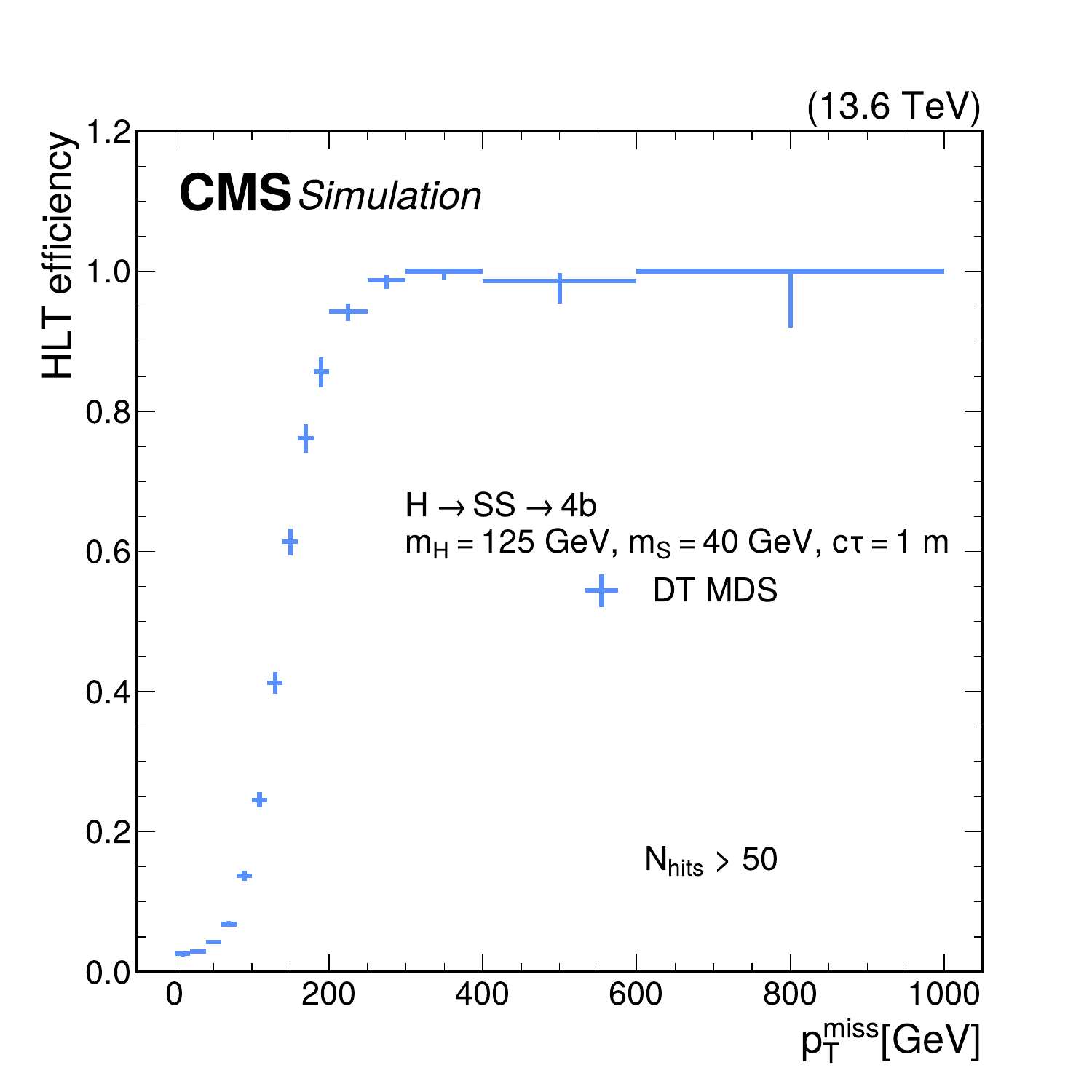}
    \includegraphics[width=0.49\textwidth]{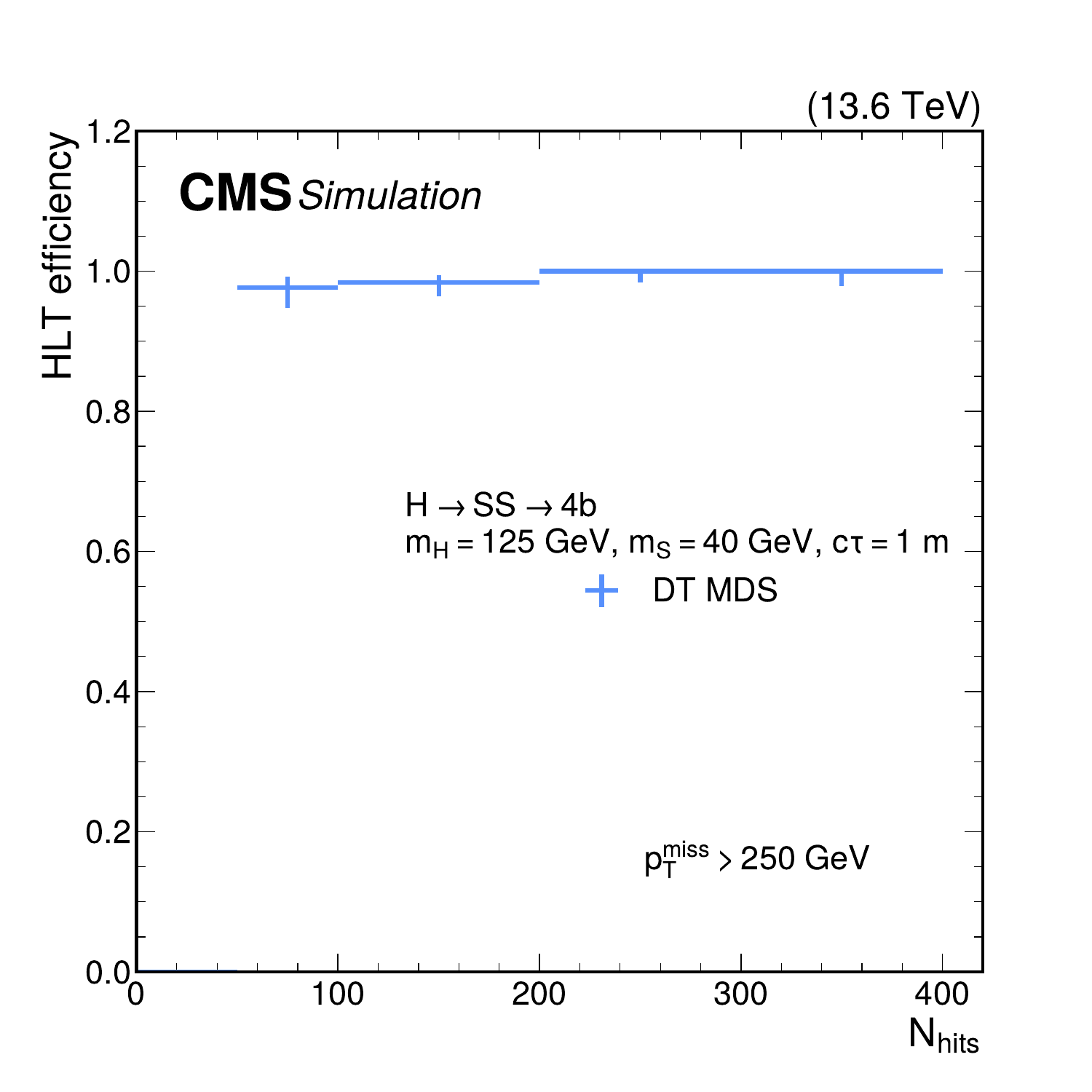}
  \caption{The HLT efficiency of the DT MDS triggers as a function of \ptmiss~(\cmsLeft) and the number of hits in the cluster~(\cmsRight), for simulated $\PH \to \PS\PS \to \bbbar\bbbar$ events with $\mH= 125\GeV$, $\mS = 40\GeV$, and $\cTauPS=1\unit{m}$, for 2023 conditions. Events are required to have at least one cluster with more than 50 hits~(\cmsLeft) and $\ptmiss>250\GeV$ (\cmsRight).}
  \label{fig:DT_MDS_eff_v_MET_cls}
\end{figure}

Figure~\ref{fig:DT_MDS_eff_v_MET_cls} shows the trigger efficiencies with respect to the offline selections, for simulated signal events. The cluster used at the HLT is identical to the offline one, which leads to close to 100\% efficiency. Events with fewer than 50 hits are not allowed by the trigger algorithm, and so there are no measurement points in Fig.~\ref{fig:DT_MDS_eff_v_MET_cls} with $\Nhits < 50$.

\subsection{The No-BPTX algorithms}
\label{sec:NoBPTXtriggers}

\begin{sloppypar}The No-BPTX trigger algorithms are unique in that they are active only when no proton bunches are colliding in the detector. They are designed to trigger on LLPs that have especially long lifetimes, are slow-moving, and come to rest in the detector, decaying at a later time. These triggers collect data when the beam pickup timing device (BPTX) does not detect the proton beam, and hence they are called the ``No-BPTX'' triggers.\end{sloppypar}

Each LHC beam consists of proton bunches arranged in an irregular pattern of ``trains''~\cite{Bailey:691782}. Within a train, the proton bunches are nominally spaced 25\unit{ns} apart, with larger spacings between trains to allow for the injection process. The LHC timing structure is defined by the 400\unit{MHz} radio frequency system, which generates 2.5\unit{ns} buckets in which protons can be captured. In standard operation, proton bunches are nominally loaded into only every tenth radio frequency bucket, such that collisions can occur at 25\unit{ns} intervals. The CMS trigger and readout are organized in terms of these 25\unit{ns} bunch crossings (BXs). One LHC orbit contains 3564 BXs. Many BXs remain empty, and the trains may be spaced such that multiple empty BXs lie between filled BXs. The maximum occupancy of the LHC in Run~3 was about 2400 colliding bunches, depending on the year. At the end of each LHC orbit, there is a 3.15\mus long ``abort gap'', which is kept free to allow the beam to be dumped cleanly when required.

To search for LLP decays during these empty BXs, the main No-BPTX triggers select events at least two BXs away from any proton bunches. Thus, these triggers are live only during these specific time windows. This distance of two BXs is chosen so that we maximize the search time window while suppressing most of the events from secondary $\Pp\Pp$ interactions and from ``beam halo'', arising from muons traveling outside the LHC beam, which are produced by LHC beam-collimator scattering. As a result, these triggers collect most of their data during the LHC abort gap. The No-BPTX algorithm is present in dedicated L1 seeds with jet and muon variants. Higher \pt thresholds for the jets and muons are employed at the HLT to suppress backgrounds and lower the rate to a few Hz. The jet and muon No-BPTX triggers are discussed in more detail below, in Sections~\ref{sec:nobptxjets} and \ref{sec:nobptxmuons}, respectively.

\subsubsection{Jet No-BPTX triggers}
\label{sec:nobptxjets}

\begin{figure}[htb!]
\centering
\includegraphics[width=0.6\textwidth]{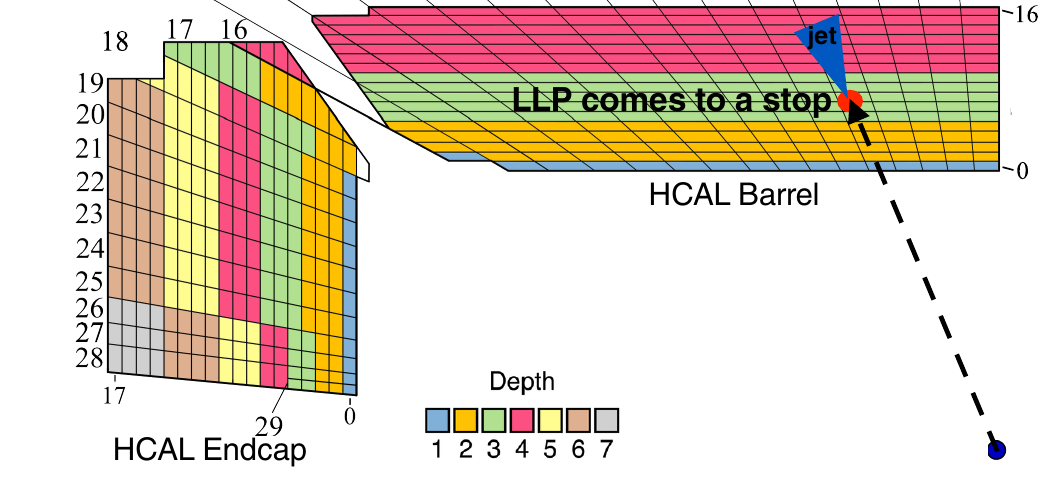}
\caption{Diagram of a stopped-particle event, which can be selected by the jet No-BPTX triggers. The dotted black arrow indicates the LLP, such as a gluino, that travels through the detector before coming to a stop in the dense material of the HCAL. After some time, this stopped particle decays hadronically, producing a significant energy deposit in the HCAL. The distinct depth segmentation is indicated by the different colors, where the HCAL barrel consists of 4 depths and the HCAL endcap has up to 7 depths.
}
\label{fig:EXO_16_004_jetSignal}
\end{figure}

The jet No-BPTX triggers have been used since Run~1 to trigger on hadronic decays of LLPs that become stopped in the dense inactive material in the CMS calorimeters~\cite{CMS:2015mmz,CMS:2017kku}. Such stopped LLPs arise, \eg, in BSM scenarios with long-lived gluinos~\cite{Giudice:2004tc,ArkaniHamed:2004yi} or top squarks~\cite{Abercrombie:2015wmb,Carpenter:2016thc,Covi:2014fba}. This signature is illustrated in Fig.~\ref{fig:EXO_16_004_jetSignal}.

The jet No-BPTX triggers are designed to trigger on extremely out-of-time calo jets at the HLT with energy greater than 60\GeV, $\abseta < 3$, and at least two BXs separation from $\Pp\Pp$ collisions. A requirement is placed on the total energy of the jets instead of the transverse energy because the stopped particles will decay isotropically, not necessarily predominantly in the transverse plane. These triggers primarily select events with beam halo and noise in the calorimeters; thus, their rates are largely independent of PU. The total rate of these HLT paths is less than 2\unit{Hz} at the maximum number of colliding proton bunches.

\subsubsection{Muon No-BPTX triggers}
\label{sec:nobptxmuons}

The muon No-BPTX triggers, like the jet No-BPTX triggers, select events when there are no colliding bunches of protons in the detector, \ie, empty BXs. The muon versions of these triggers select out-of-time L2 muons and have taken data since Run~1. The main trigger selects events at least two BXs away from the $\Pp\Pp$ collision time with at least one muon reconstructed in the muon system with $\pt>40\GeV$. This trigger primarily selects cosmic ray muons that have produced showers of particles in the CMS detector or the rock above it. Thus, the rate of this trigger depends on the number of filled bunch crossings but is largely constant as a function of PU. In addition, about 20--30\% of the rate, depending on the instantaneous luminosity, is caused by beam-related backgrounds.

\begin{figure}[htb!]
  \centering
  \includegraphics[width=0.6\textwidth]{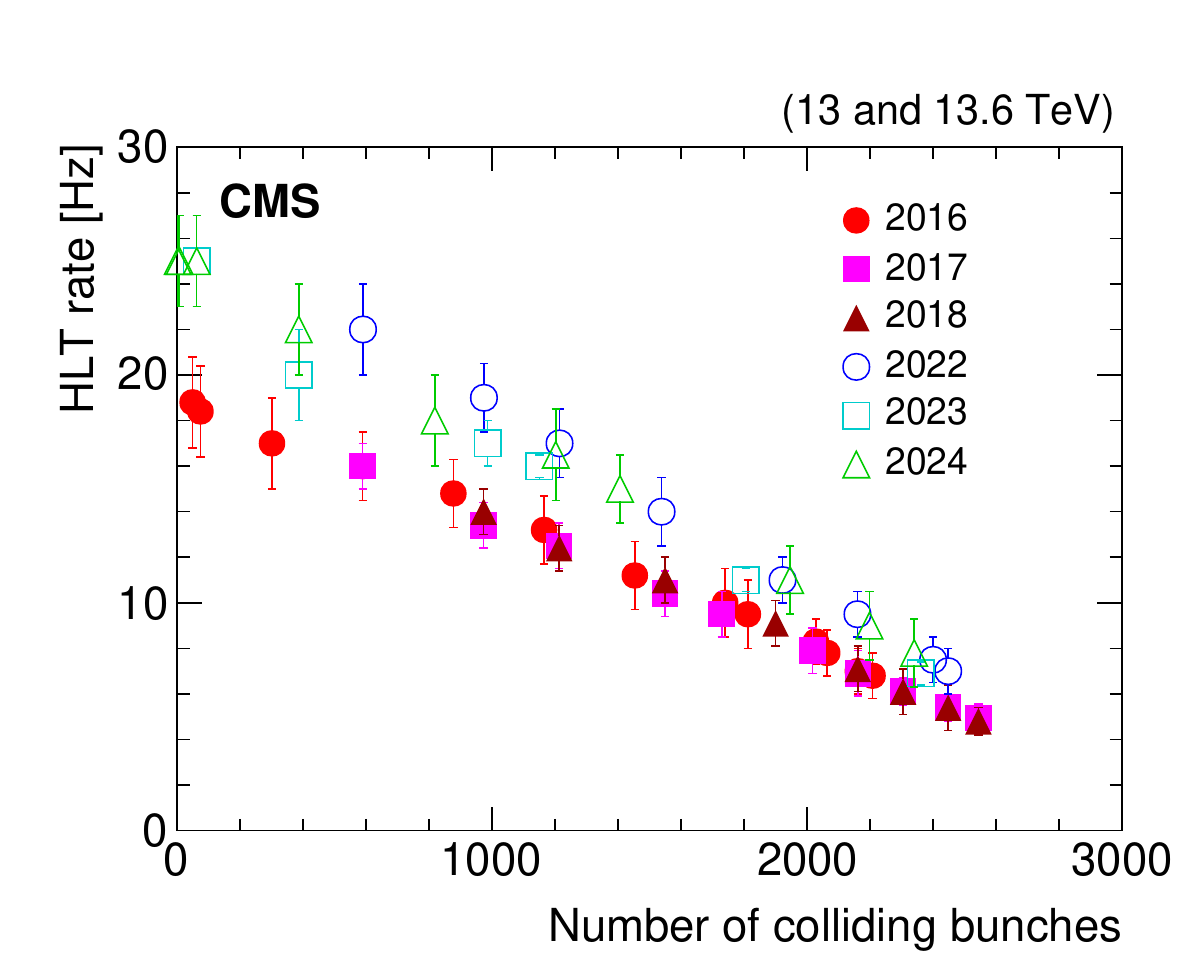}
  \caption{Rate of the main muon No-BPTX HLT path as a function of the number of colliding bunches, for different years in Run~2 and Run~3.}
  \label{fig:MuonNoBPTXRate}
\end{figure}

The rate of the main muon No-BPTX HLT path as a function of the number of colliding bunches is shown in Fig.~\ref{fig:MuonNoBPTXRate}. The rate decreases linearly as the number of proton bunches increases and as fewer empty BXs are available for the No-BPTX trigger. The rate increase from Run~2 to Run~3 is expected, as a result of the increase in rate from the backgrounds in the EMTF, and the increase in rate of the LHC backgrounds overall from the increase in center-of-mass energy. The total rate of these HLT paths is about 7\unit{Hz} at the maximum number of colliding proton bunches.

\begin{figure}[htb!]
\centering
\includegraphics[width=0.6\textwidth]{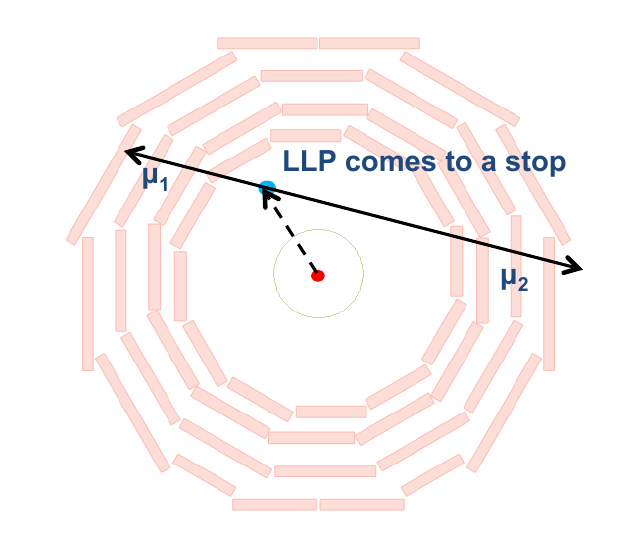}
\caption{Diagram of a stopped-particle event, which can be selected by the muon No-BPTX triggers. The dotted black arrow indicates the LLP that travels through the detector before coming to a stop in the steel yoke in the muon system. After some time, this stopped particle decays to two back-to-back muons.
}
\label{fig:EXO_16_004_muonSignal}
\end{figure}

\begin{sloppypar}The muon No-BPTX triggers are used to trigger on the muonic decays of stopped LLPs~\cite{CMS:2017kku}, that arise, for example, in models with long-lived gluinos~\cite{Giudice:2004tc,ArkaniHamed:2004yi} or multiply charged particles~\cite{Langacker:2011db}. This signature of a stopped particle decaying into two back-to-back muons is illustrated in Fig.~\ref{fig:EXO_16_004_muonSignal} and was probed in the latest stopped-particle analysis~\cite{CMS:2017kku}. It should be noted that the muon No-BPTX triggers are also efficient for other signatures, as the triggers simply require at least one out-of-time muon.\end{sloppypar}

\section{Long-lived particle trigger acceptances}
\label{sec:trigPerf}

Different types of LLP triggers (Section~\ref{sec:LLPtriggers}) often have significant acceptance for a given benchmark LLP signal model (Section~\ref{sec:dataMC}). By comparing the trigger acceptance as a function of the LLP \cTau or decay position, we demonstrate the complementarity of the diverse trigger strategies and their contributions to the CMS physics program. The trigger acceptances for models that predict neutral LLPs are shown first for hadronic and leptonic signatures, followed by the acceptances for models that predict charged LLPs.

\begin{figure}[htbp!]
\centering
\includegraphics[width=0.95\textwidth]{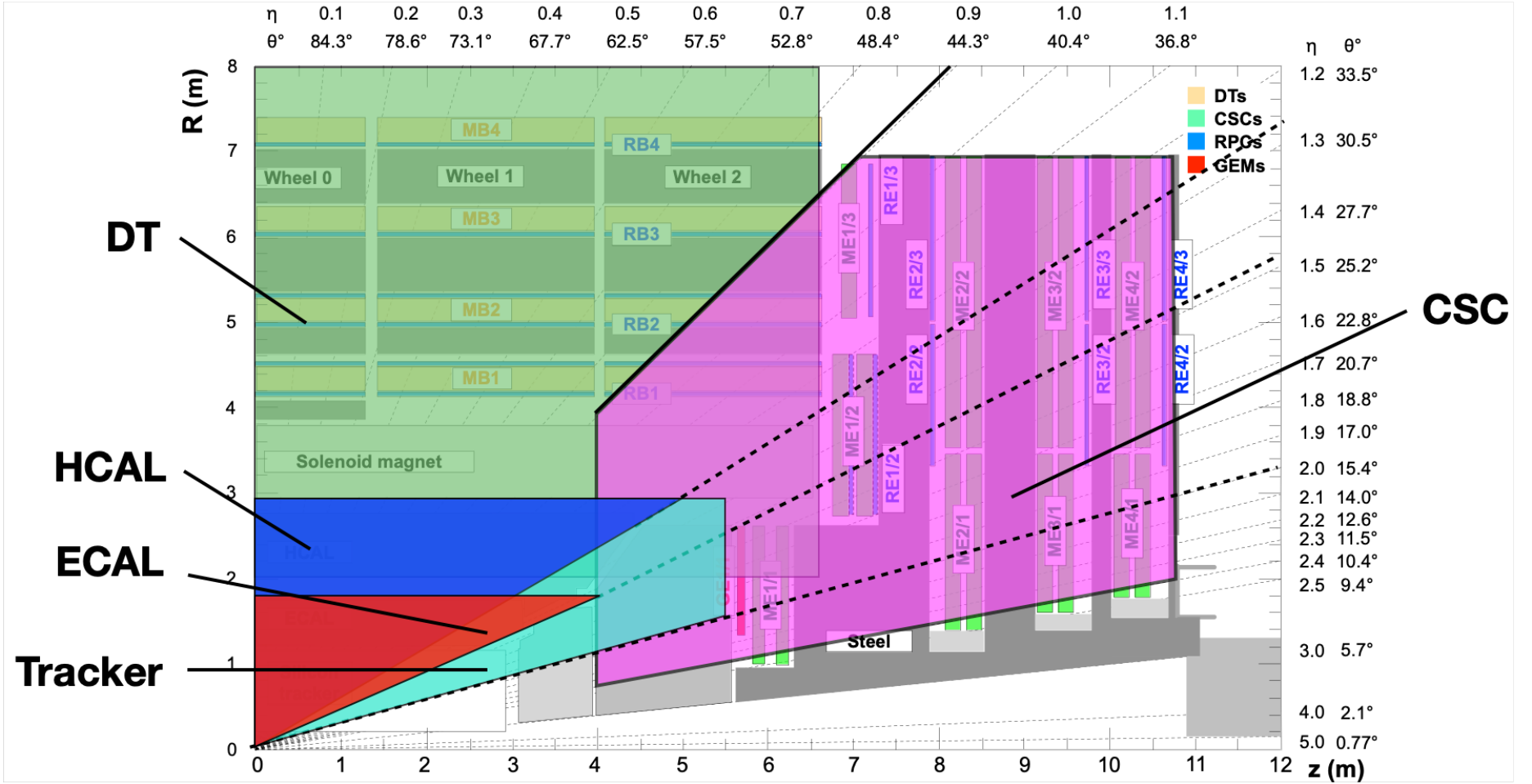}
\caption{The fiducial regions defined for the calculation of each trigger acceptance in Figs.~\ref{fig:jetTriggers_HtoXXto4b_triggerAcceptance} and \ref{fig:jetTriggers_HtoXXto4b_triggerAcceptance_z}. The ``Tracker'' region labeled in this figure refers to the fiducial region of the displaced-jet triggers using the tracker, which is larger than the physical volume of the tracker detector. The text contains the exact definitions of each region.}
\label{fig:jetTriggers_fiducialVolume}
\end{figure}

The signal acceptance is defined as the ratio of the number of LLPs passing the respective HLT requirements, given that the LLPs decay in the corresponding fiducial region of the detector shown in Fig.~\ref{fig:jetTriggers_fiducialVolume}. For example, the signal acceptance for MDS triggers using CSCs is computed with respect to the decay of LLPs within the fiducial region of the CSCs. The fiducial region used in the tracker-based displaced-jet triggers is defined as $R<300\cm$, $\abs{z}<560\cm$, and $\abseta<2.0$. Here, the fiducial region extends from the tracker into the calorimeters because some of the tracking-based displaced-jet algorithms rely on calo jets with a prompt track veto, enabling sensitivity to displaced LLP decays beyond the tracker volume itself. For the ECAL-based (HCAL-based) delayed-jet triggers, the fiducial region is defined as $R<152\,(300)\cm$ and $\abseta<1.48\,(1.26)$, which corresponds to the barrel region of ECAL (the barrel region of HCAL with full segmentation information). For the MDS triggers using the DT and CSCs, the fiducial region is defined as $200<R<800\cm$ and $\abs{z}<661\cm$ for the DTs, and $R<695\cm$, $400<\abs{z}<1100\cm$, and $0.9<\abseta<2.4$ for the CSCs. For the CSC MDS triggers, the fiducial region is extended in front of the muon system to include the LLPs that initiate hadronic showers in the solenoid and HCAL.

The performance of the different LLP triggers is shown here for the benchmark $\PH \to \PX\PX \to \bbbar\bbbar$ process, the diagram for which was shown above in Fig.~\ref{fig:HtoXX_feynmanDiagrams}.

\begin{figure}[htbp!]
\centering
\includegraphics[width=0.499\textwidth]{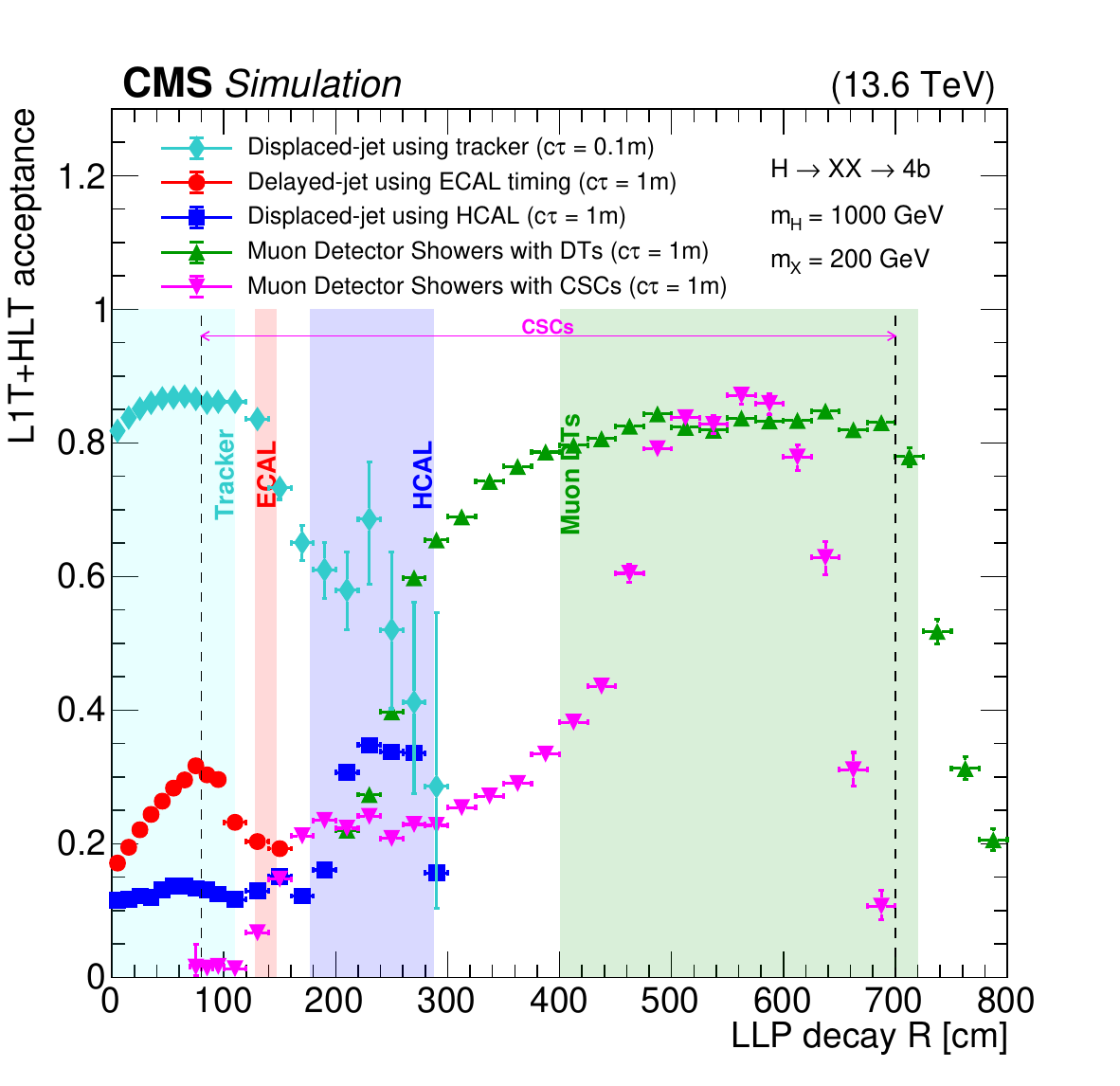}\hfill
\includegraphics[width=0.499\textwidth]{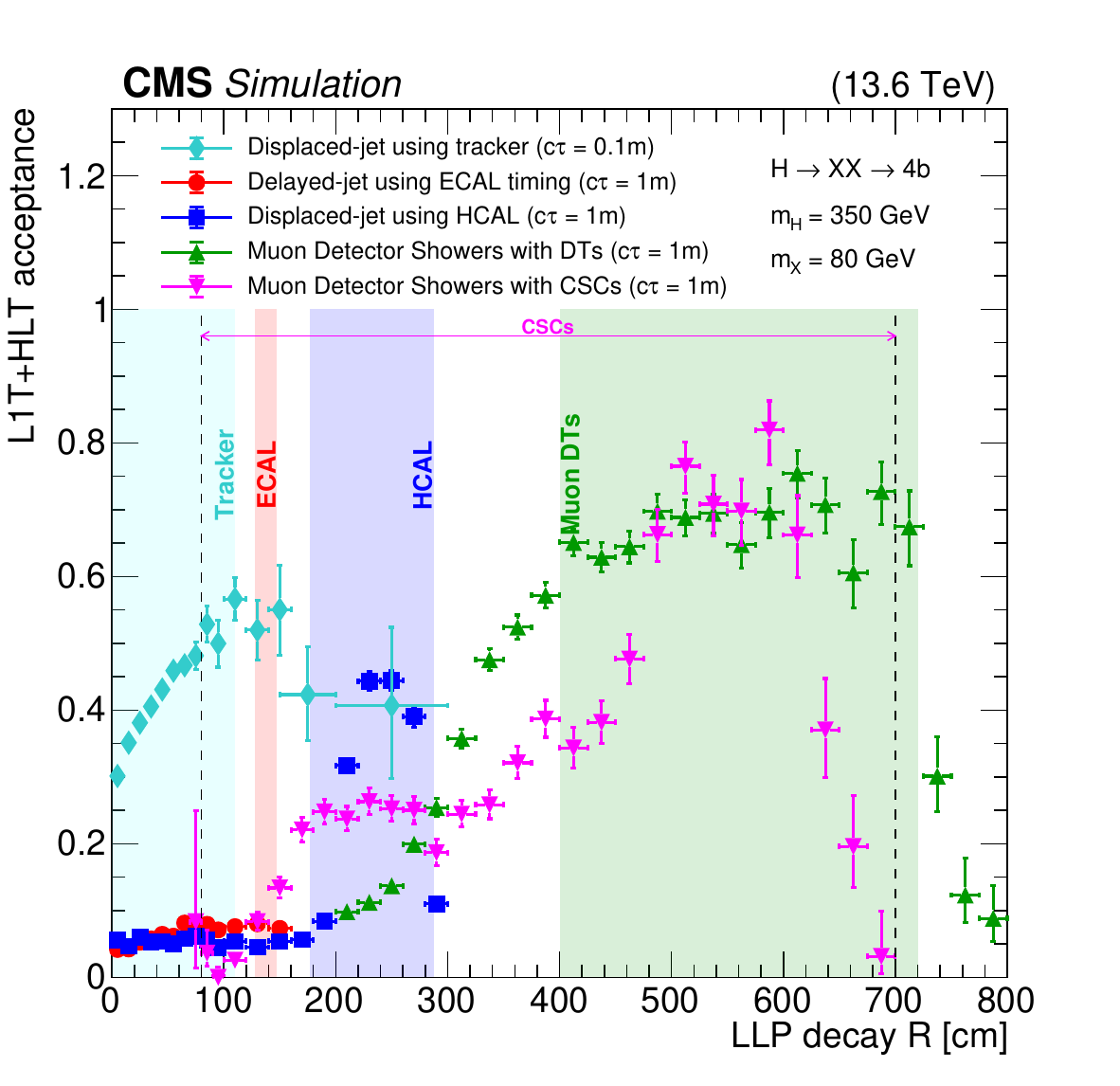}
\includegraphics[width=0.499\textwidth]{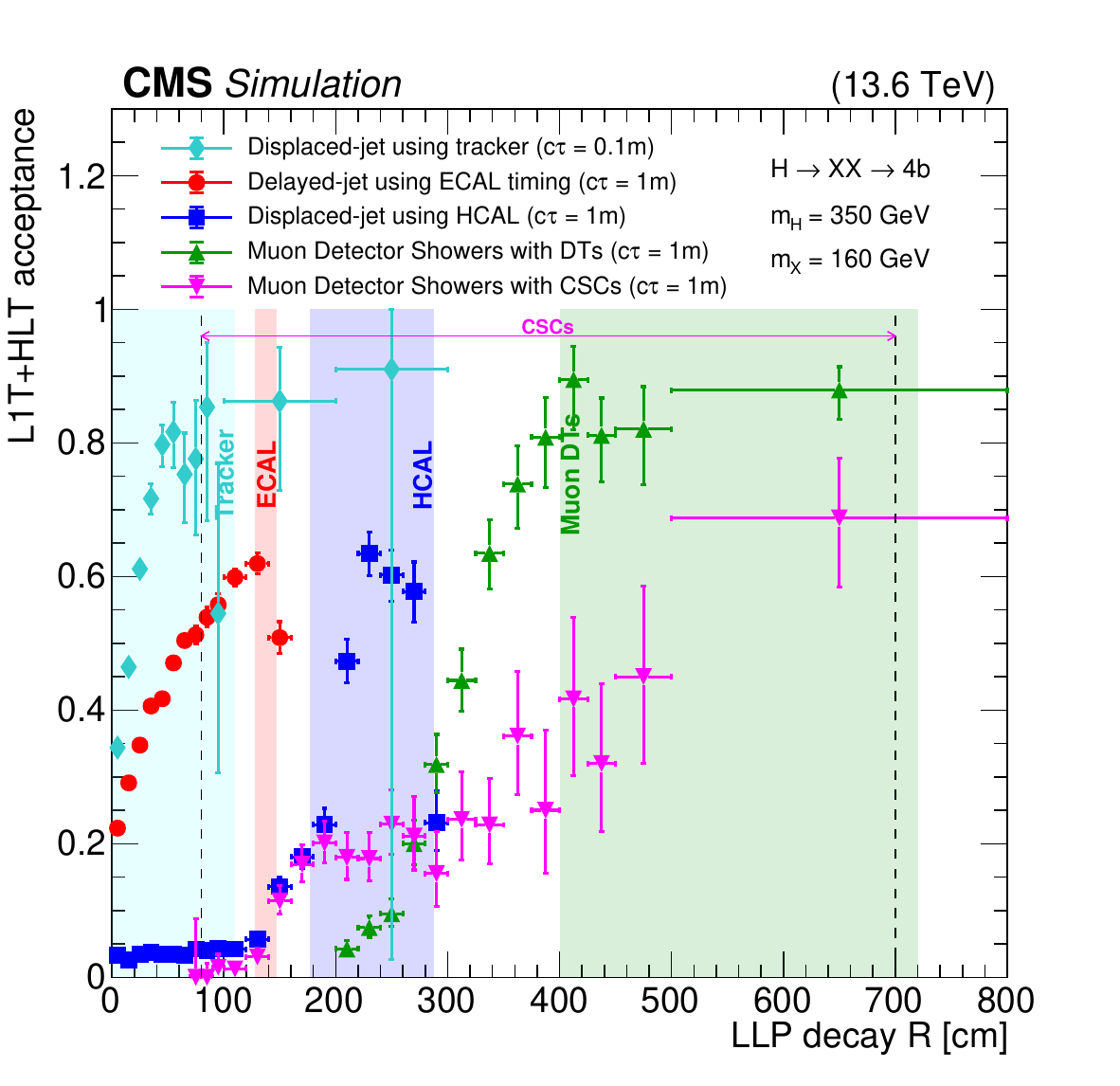}\hfill
\includegraphics[width=0.499\textwidth]{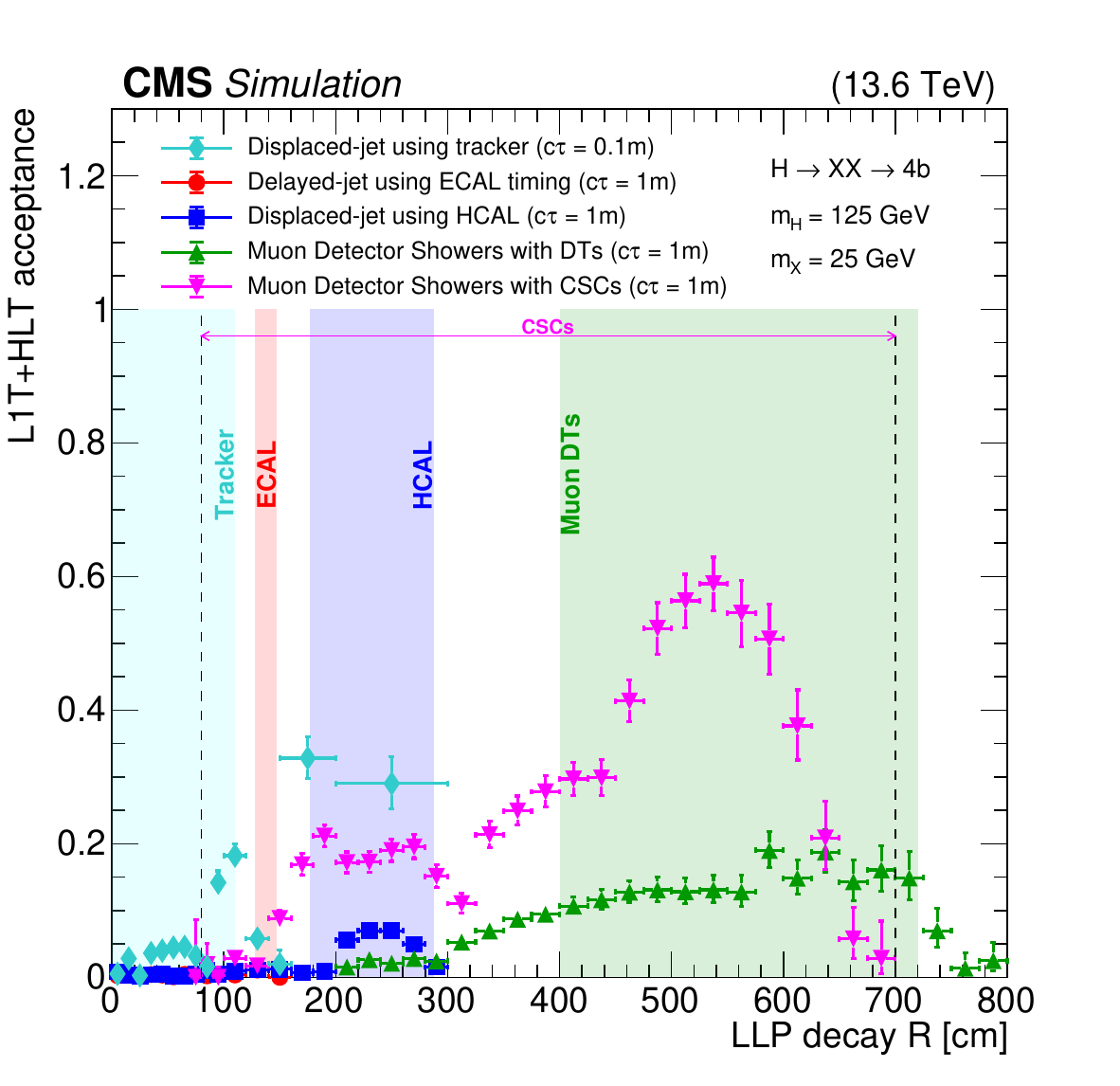}
\caption{The L1T+HLT acceptances for various LLP triggers using different subdetectors, as functions of the LLP decay radial position, for $\PH \to \PX\PX \to \bbbar\bbbar$ events for 2023 conditions. The plots correspond to different signal models with $\mH=1000\GeV$ and $\mX=200\GeV$ on the upper \cmsLeft; $\mH=350\GeV$ and $\mX=80\GeV$ on the upper \cmsRight; $\mH=350\GeV$ and $\mX=160\GeV$ on the lower \cmsLeft; and $\mH=125\GeV$ and $\mX=25\GeV$ on the lower \cmsRight. In each of these plots, the \cTau is 0.1\unit{m} for the displaced-jet triggers using the tracker and 1\unit{m} for the other triggers. The acceptance is shown for the displaced-jet triggers using the tracker (cyan diamonds), for the delayed-jet triggers using ECAL timing (red circles), for the displaced- and delayed-jet triggers using the HCAL (blue squares), for the MDS triggers with the DTs (green triangles), and for the MDS triggers with the CSCs (pink triangles). The boundaries of the tracker, ECAL, HCAL, DTs, and CSCs are also shown in the figures.}
\label{fig:jetTriggers_HtoXXto4b_triggerAcceptance}
\end{figure}

\begin{figure}[htbp!]
\centering
\includegraphics[width=0.499\textwidth]{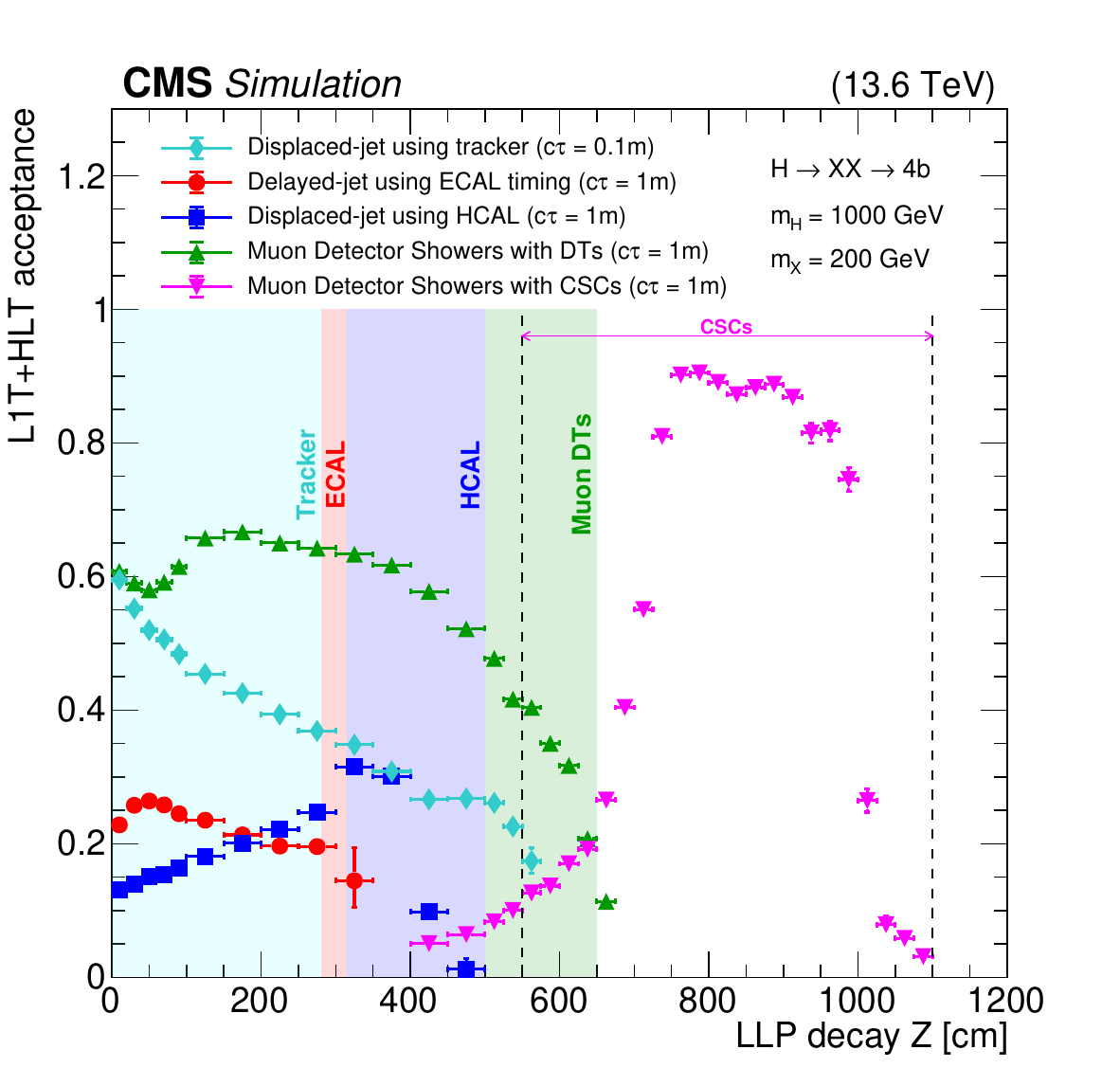}\hfill
\includegraphics[width=0.499\textwidth]{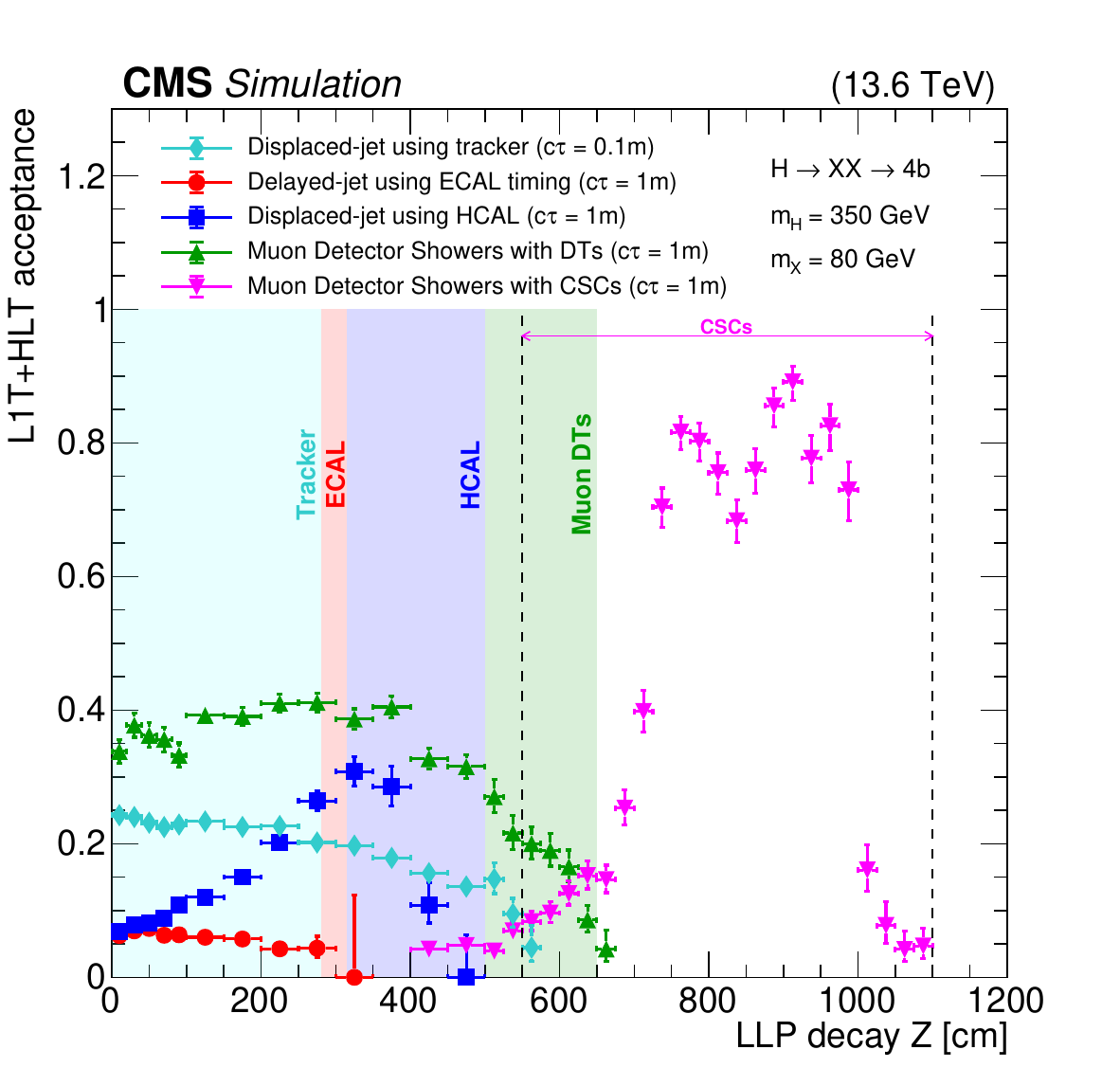}
\includegraphics[width=0.499\textwidth]{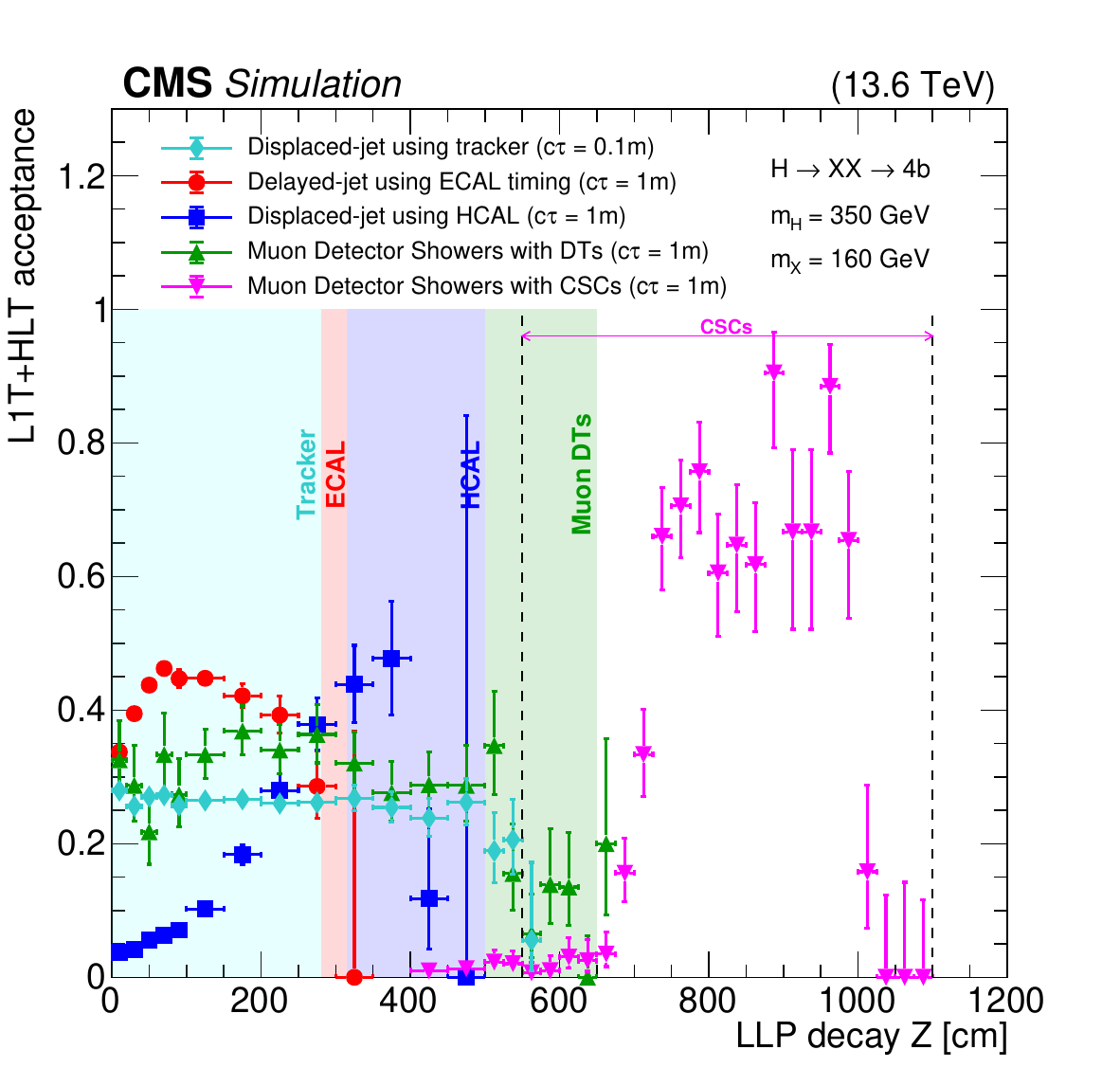}\hfill
\includegraphics[width=0.499\textwidth]{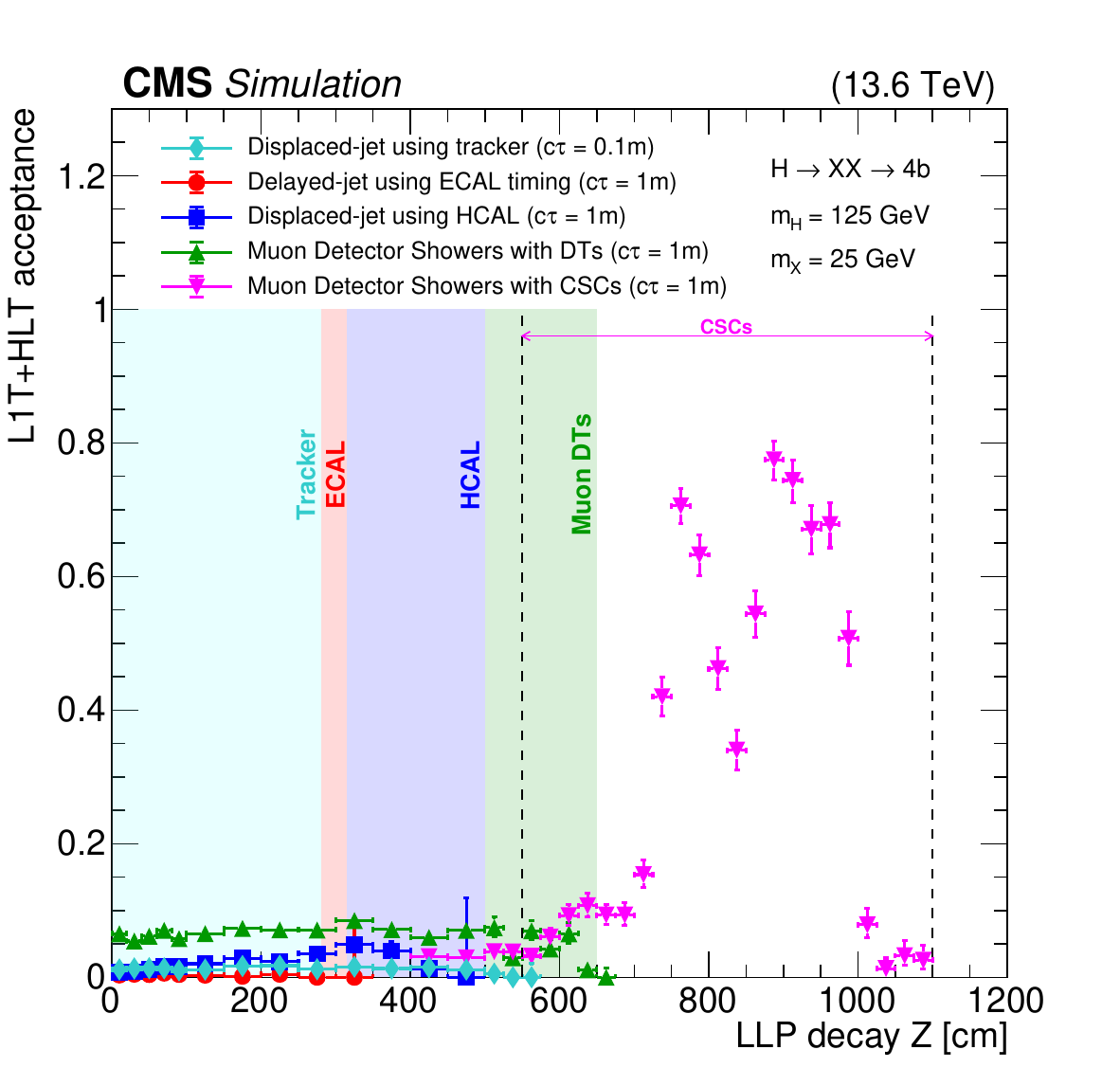}
\caption{The L1T+HLT acceptances for various LLP triggers using different subdetectors, as functions of the $z$ position of the LLP decay, for $\PH \to \PX\PX \to \bbbar\bbbar$ events for 2023 conditions. The plots correspond to different signal points with $\mH=1000\GeV$ and $\mX=200\GeV$ on the upper \cmsLeft; $\mH=350\GeV$ and $\mX=80\GeV$ on the upper \cmsRight; $\mH=350\GeV$ and $\mX=160\GeV$ on the lower \cmsLeft; and $\mH=125\GeV$ and $\mX=25\GeV$ on the lower \cmsRight. In each of these plots, the \cTau is 0.1\unit{m} for the displaced-jet triggers using the tracker and 1\unit{m} for the other triggers. The acceptance is shown for the displaced-jet triggers using the tracker (cyan diamonds), for the delayed-jet triggers using ECAL timing (red circles), for the displaced- and delayed-jet triggers using the HCAL (blue squares), for the MDS triggers with the DTs (green triangles), and for the MDS triggers with the CSCs (pink triangles). The boundaries of the tracker, ECAL, HCAL, DTs, and CSCs are also shown in the figures.}
\label{fig:jetTriggers_HtoXXto4b_triggerAcceptance_z}
\end{figure}

Figures~\ref{fig:jetTriggers_HtoXXto4b_triggerAcceptance} and \ref{fig:jetTriggers_HtoXXto4b_triggerAcceptance_z} show the trigger acceptances at the HLT for various hadronic LLP triggers, as functions of the LLP decay radial and $z$ positions, respectively. The plots correspond to different choices of \mH and \mX.

Figure~\ref{fig:jetTriggers_HtoXXto4b_triggerAcceptance} demonstrates the complementarity of the LLP triggers, as the displaced-jet triggers using the tracker have high acceptance for the smallest decay vertex radial distances, the delayed-jet triggers using ECAL timing provide higher acceptance up until the start of the ECAL, the displaced- and delayed-jet triggers using the HCAL provide acceptance throughout the HCAL, and the MDS triggers with the DTs and CSCs provide acceptance inside the muon system. Figure~\ref{fig:jetTriggers_HtoXXto4b_triggerAcceptance_z} demonstrates the complementarity of these triggers as a function of LLP decay $z$ position using both the barrel and endcap of various subdetector systems.

In Fig.~\ref{fig:jetTriggers_HtoXXto4b_triggerAcceptance_z}, for $z$ positions between 700 and 1000\cm, the acceptance of the MDS triggers with the CSCs (pink points) varies in the 30--80\% range, for $\mH=125\GeV$ and $\mX=25\GeV$ (lower \cmsRight plot). This variation reflects the positions of the CSCs. Depending on the LLP decay position, the decay products pass through different thickness of steel, which needs to be thick enough to initiate the hadronic shower but not too thick to attenuate the particles before reaching a CSC. This effect is more prominent when the LLP energy is smaller, hence the structure in the lower \cmsRight plot is more obvious.

\begin{figure}[htb!]
  \centering
  \includegraphics[width=0.49\textwidth]{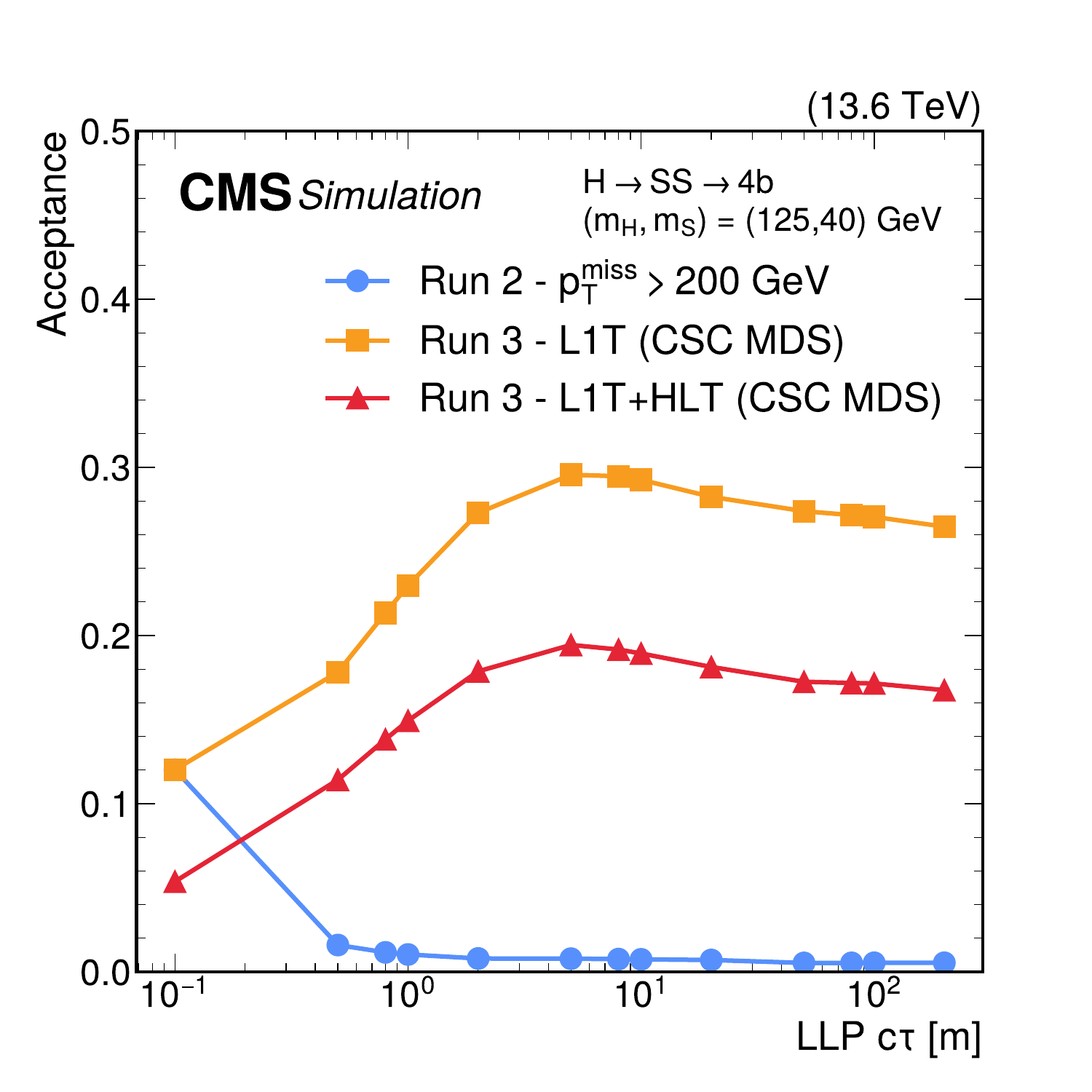}
  \includegraphics[width=0.49\textwidth]{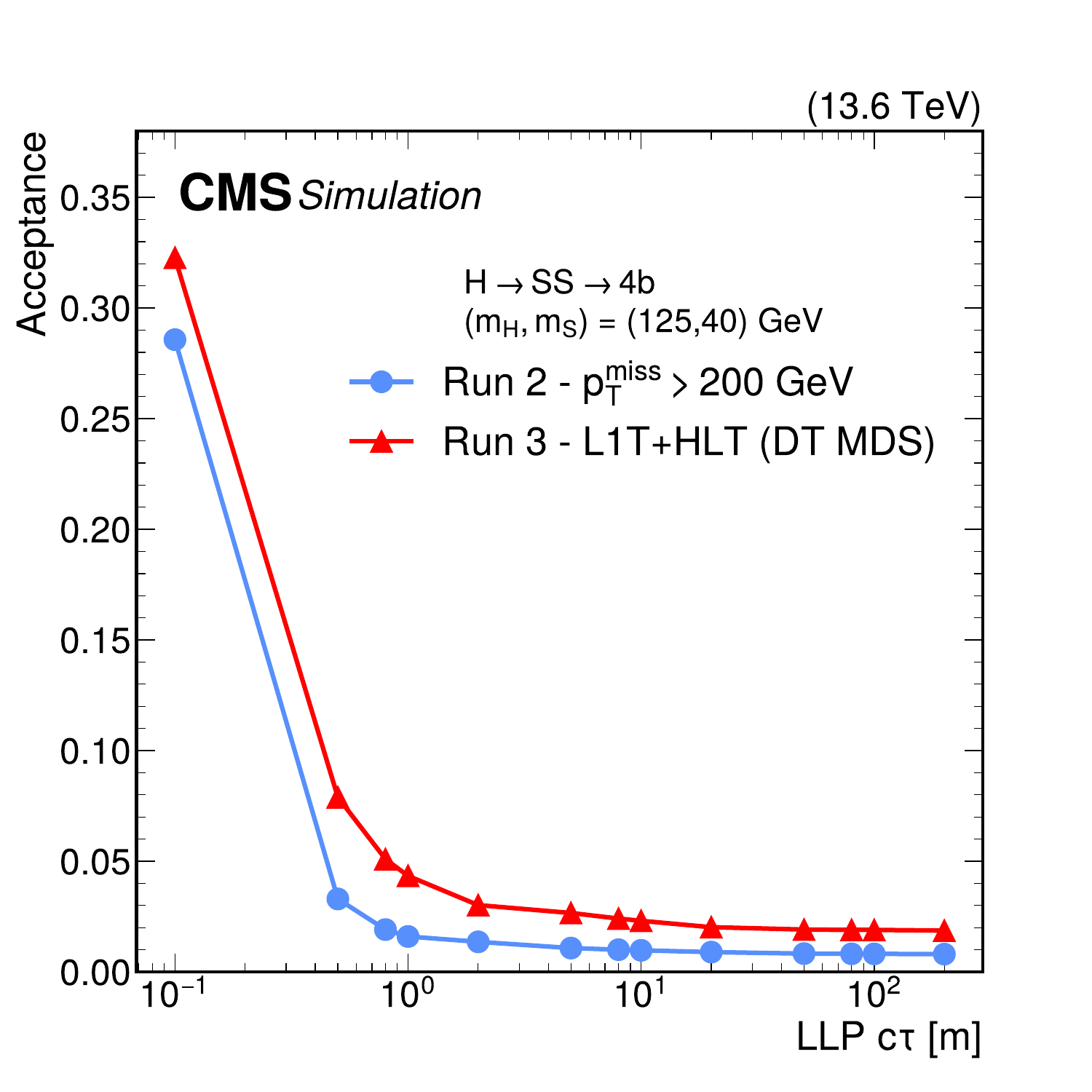}
  \caption{Comparison of the acceptances in Run~2 and Run~3 for the CSC (\cmsLeft) and DT (\cmsRight) MDS triggers at the L1T and HLT as functions of the LLP lifetime, for $\PH \to \PS\PS \to \bbbar\bbbar$ events with $\mH= 125\GeV$ and $\mS = 40\GeV$, for 2023 conditions. The acceptance is defined as the fraction of events that pass the specified selection, given an LLP decay in the fiducial region of the CSCs (\cmsLeft) or the DTs (\cmsRight). The \cmsLeft plot compares the acceptance of the Run~2 strategy of triggering on \ptmiss (blue circles), which corresponds to an offline requirement of $\ptmiss>200\GeV$, with that of the Run~3 strategy of triggering on the MDS signature in the CSCs, where the L1T (L1T+HLT) acceptance is shown with orange squares (red triangles). The \cmsRight plot compares the acceptance of the Run~2 strategy of triggering on \ptmiss (blue circles) with the Run~3 strategy of triggering on the MDS signature in the DTs (red triangles), for L1T+HLT.}
  \label{fig:MDS_acc_v_ctau}
\end{figure}

Figure~\ref{fig:MDS_acc_v_ctau} shows the trigger acceptance for the CSC and DT MDS triggers as a function of LLP \cTau. Figure~\ref{fig:MDS_acc_v_ctau} (\cmsLeft) illustrates the improvement of the dedicated CSC MDS triggers in Run~3 with respect to the Run~2 trigger strategy, which relies on triggering on \ptmiss. Figure~\ref{fig:MDS_acc_v_ctau} (\cmsRight) illustrates the improved performance of the DT MDS triggers in Run~3 over that of the Run~2 trigger strategy, namely, triggering on \ptmiss at the L1T and HLT. Since the Run~3 MDS triggers in the DTs still rely on \ptmiss in the L1T because of rate constraints, the improvement is less significant than that of the CSC trigger paths. 

\begin{figure}[!htb]
\centering
\includegraphics[width = 0.6\textwidth]{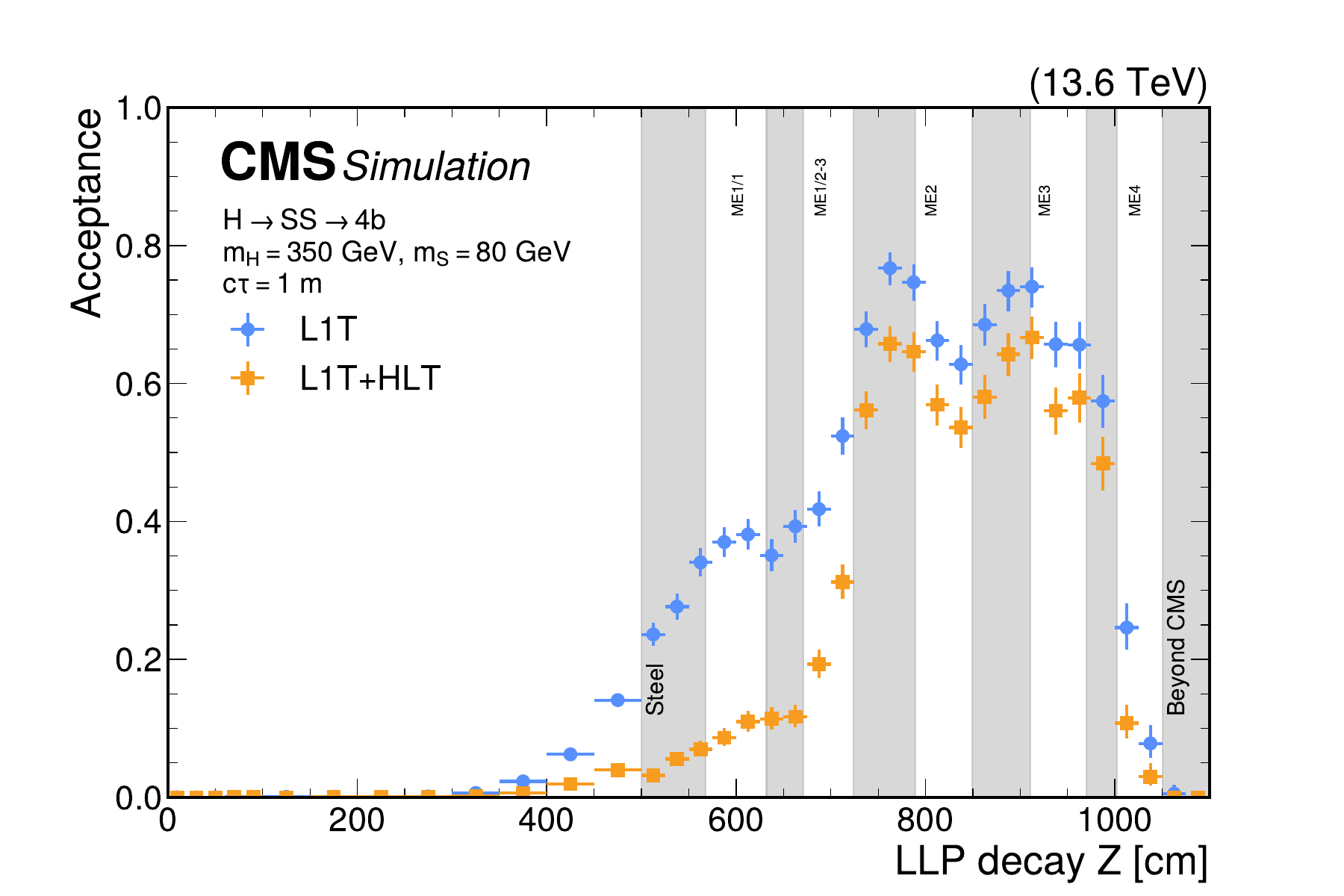}
\caption{The L1T (blue circles) and L1T+HLT (orange squares) acceptances for the CSC MDS trigger as functions of the LLP decay position in the $z$-direction, for $\PH \to \PS\PS \to \bbbar\bbbar$ events with $\mH=350\GeV$, $\mS=80\GeV$, and $c\tau_{\PS}=1\unit{m}$, for 2023 conditions.}
\label{fig:MDS_CSC_1D_acceptance}
\end{figure}

\begin{figure}[!htb]
\centering
\includegraphics[width = 0.6\textwidth]{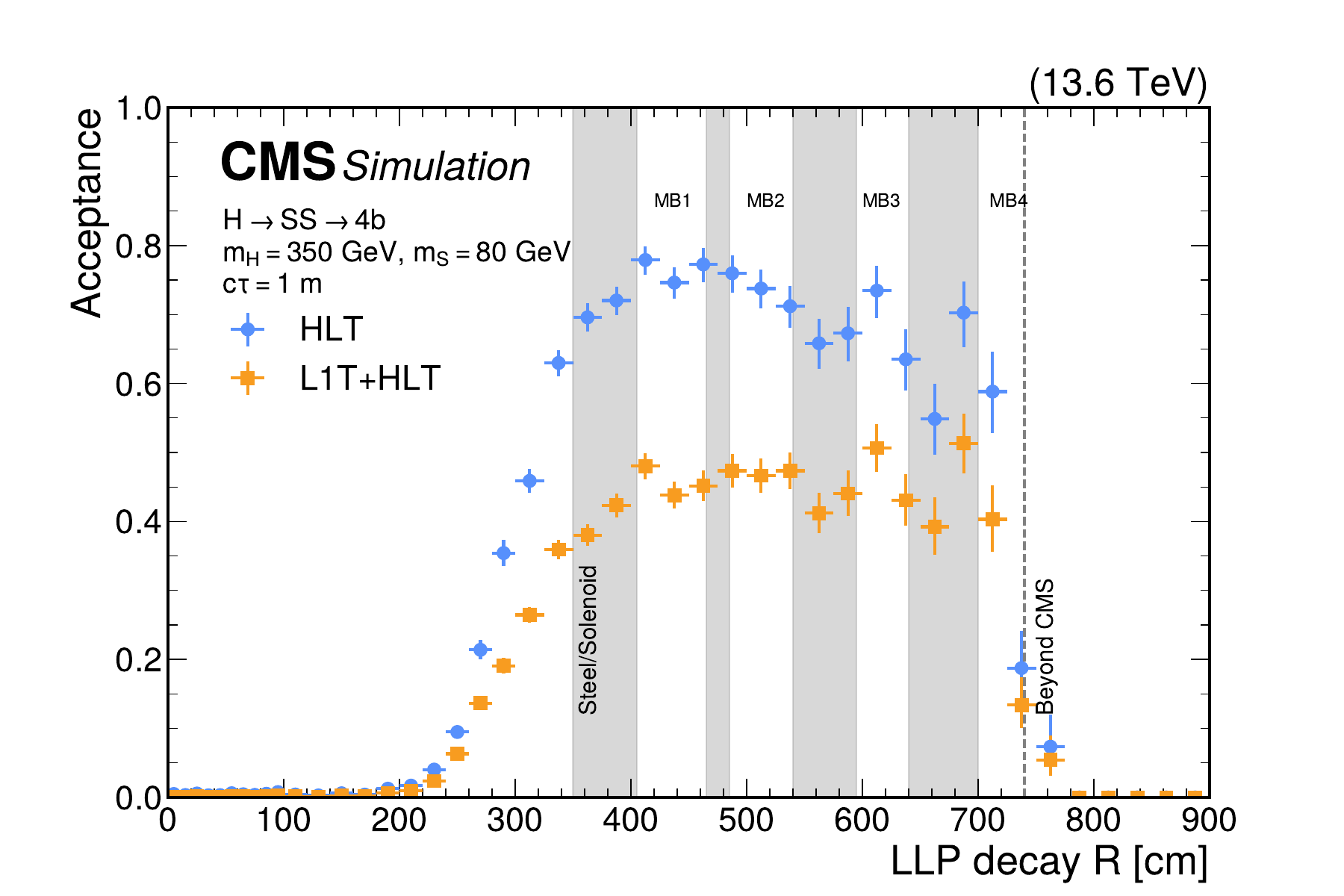}
\caption{The HLT (blue circles) and L1T+HLT (orange squares) acceptances for the DT MDS trigger as functions of the LLP decay position in the radial direction, for $\PH \to \PS\PS \to \bbbar\bbbar$ events with $\mH=350\GeV$, $\mS=80\GeV$, and $c\tau_{\PS}=1\unit{m}$, for 2023 conditions.}
\label{fig:MDS_DT_1D_acceptance}
\end{figure}

Figure~\ref{fig:MDS_CSC_1D_acceptance} shows the L1T and HLT acceptances for the CSC MDS trigger as a function of the LLP decay position along the $z$-axis. The L1T (HLT) acceptance is approximately $40\,(10)\%$ for showers in the ME1/1 region because high thresholds are needed in this CSC station to suppress hadronic punchthrough. However, the acceptance increases to around $70\,(60)\%$ in the other stations. The shaded regions represent steel within the CSC system, and the structure observed in the acceptance distribution partly reflects variations in steel thickness, which influence shower development. The corresponding acceptance distributions for the DT MDS trigger are shown in Fig.~\ref{fig:MDS_DT_1D_acceptance}. In this case, there is no dedicated L1 seed, so acceptance is shown both for all events and for those that pass the L1 \ptmiss trigger. Hadronic punchthrough has a smaller impact on the DT MDS trigger, owing to the lower particle flux in the barrel region and the presence of the solenoid, which results in higher HLT acceptance in the first station compared to that of the CSC MDS trigger.

\begin{figure}[!htb]
\centering
\includegraphics[width = 0.49\textwidth]{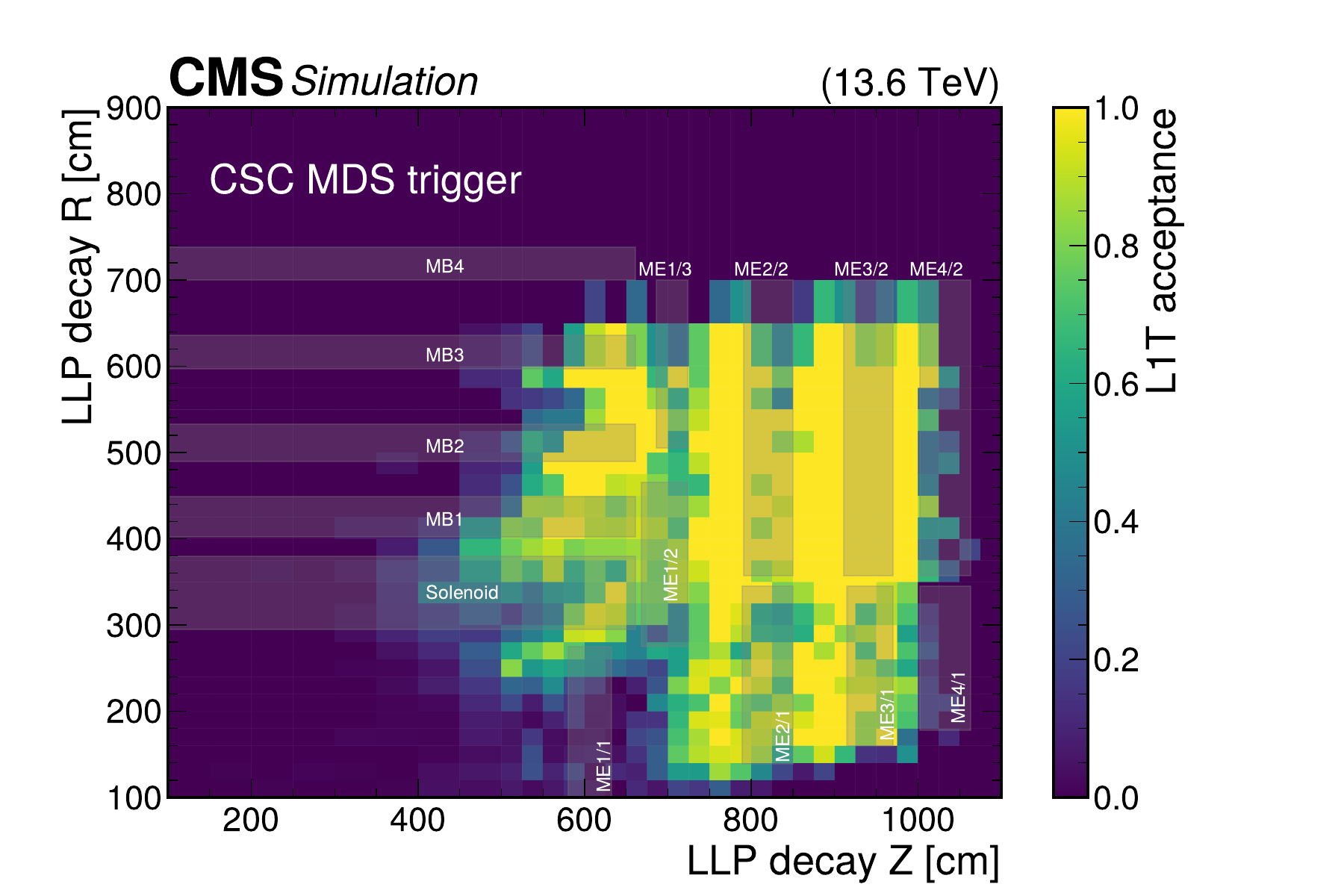}
\includegraphics[width = 0.49\textwidth]{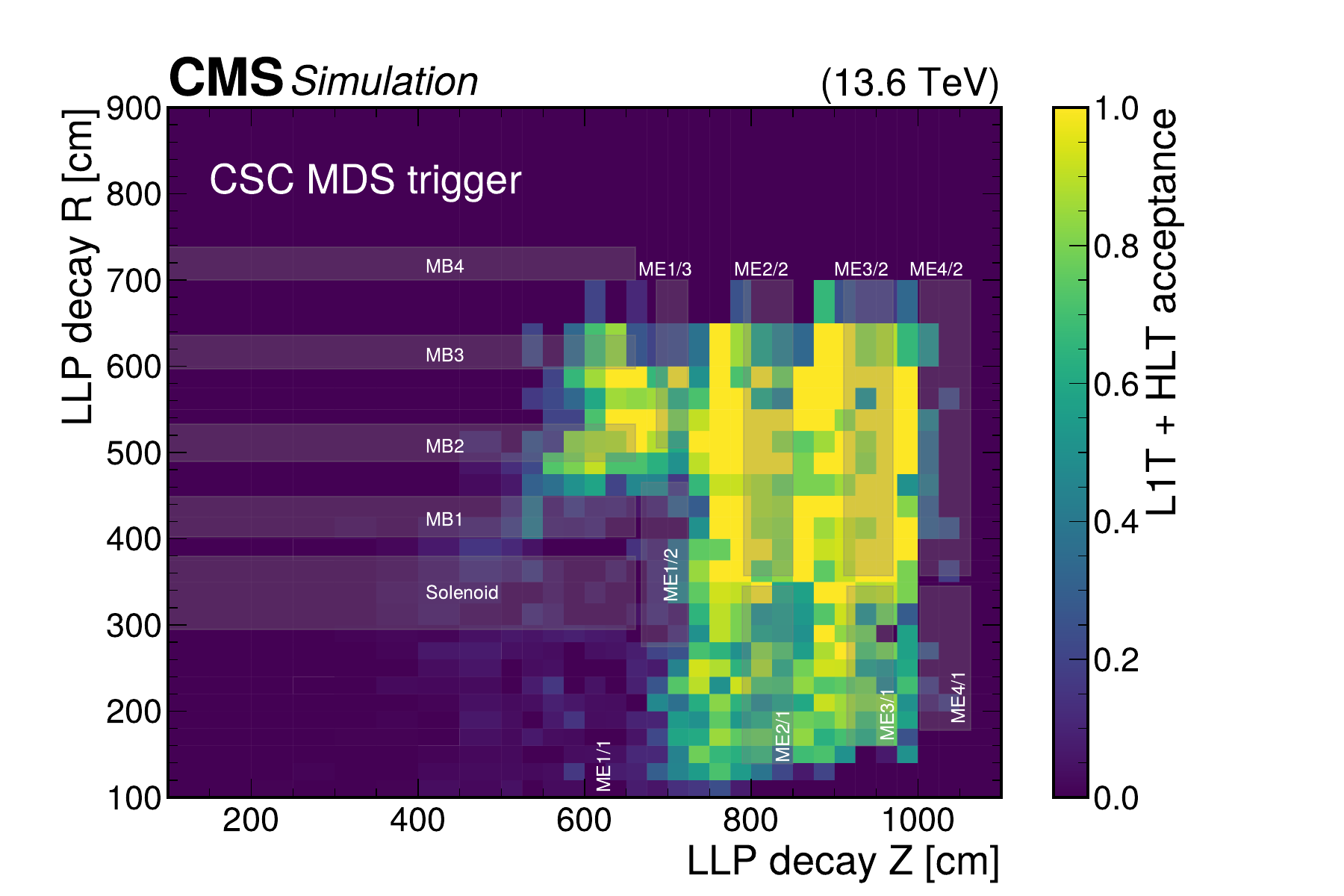}
\caption{The L1T (\cmsLeft) and L1T+HLT (\cmsRight) acceptances for the CSC MDS trigger as functions of the LLP decay position, for $\PH \to \PS\PS \to \bbbar\bbbar$ events with $\mH=350\GeV$, $\mS=80\GeV$, and $c\tau_{\PS}=1\unit{m}$, for 2023 conditions.}
\label{fig:MDS_CSC_2D_acceptance}
\end{figure}

\begin{figure}[!htb]
\centering
\includegraphics[width = 0.49\textwidth]{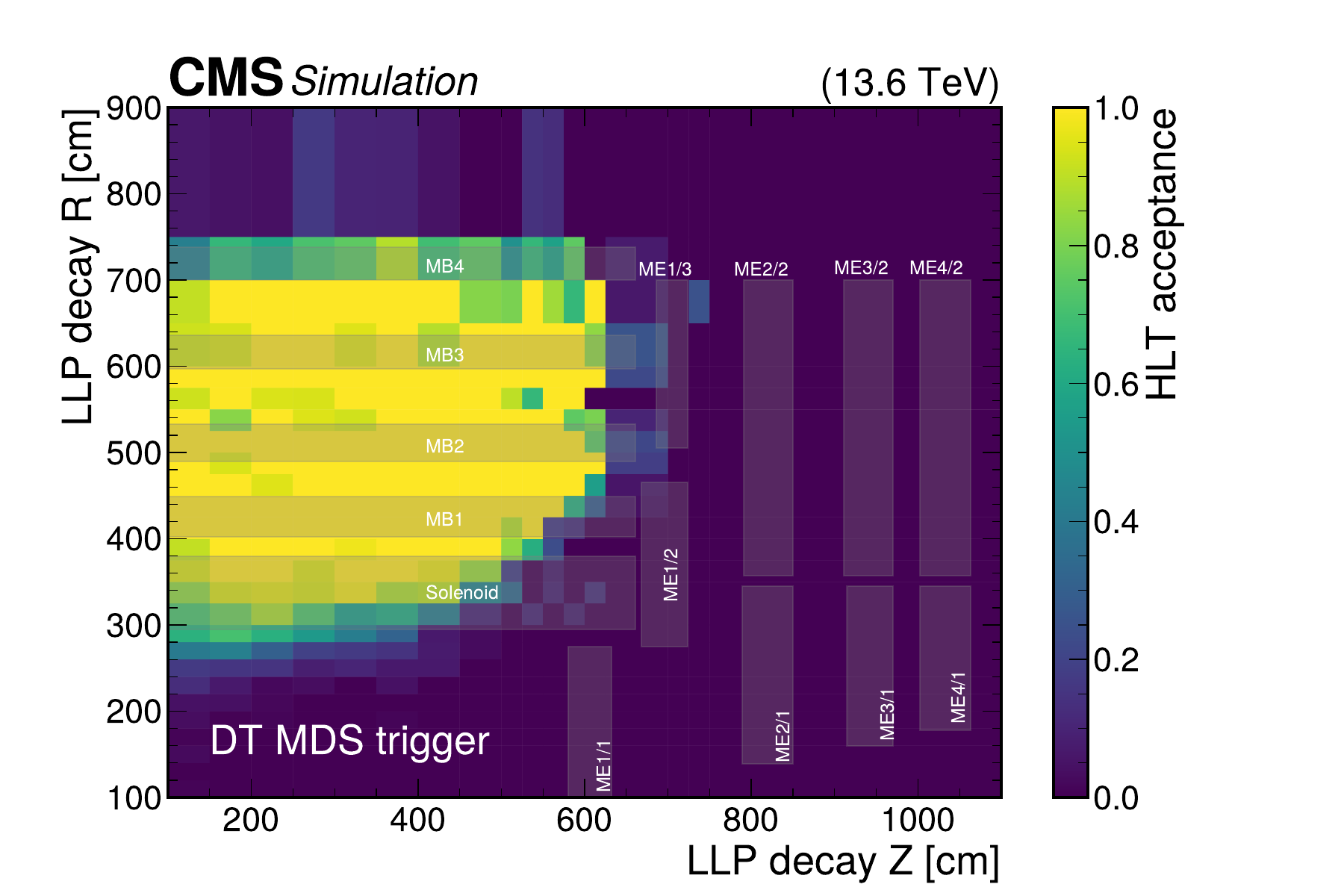}
\caption{The HLT acceptance for the DT MDS trigger as a function of the LLP decay position, for $\PH \to \PS\PS \to \bbbar\bbbar$ events with $\mH=350\GeV$, $\mS=80\GeV$, and $c\tau_{\PS}=1\unit{m}$, for 2023 conditions. The L1T acceptance that is based on the \ptmiss trigger is not included.}
\label{fig:MDS_DT_2D_acceptance}
\end{figure}

Figures~\ref{fig:MDS_CSC_2D_acceptance} and \ref{fig:MDS_DT_2D_acceptance} show the trigger acceptance for the CSC and DT MDS triggers, respectively, as a function of the LLP decay radial and $z$ positions. These figures illustrate that the MDS trigger program is comprehensive and allows CMS to trigger on the MDS signature throughout the entire muon system.

\begin{figure}[htbp!]
\centering
\includegraphics[width=0.45\textwidth]{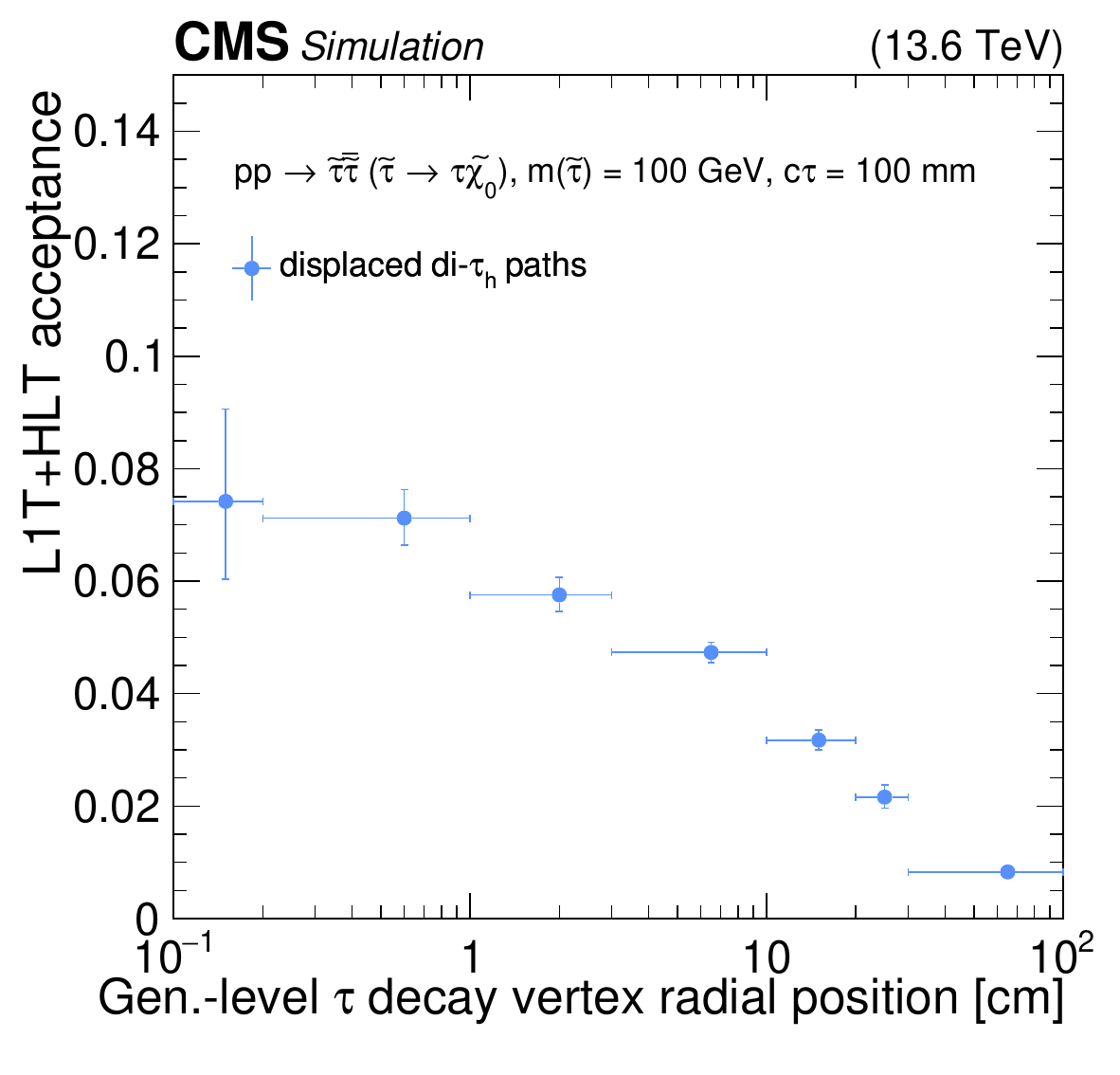}
\caption{The L1T+HLT acceptance of the displaced \tauh trigger, for simulated $\Pp\Pp \to \PSGt\PASGt (\PSGt \to \PGt\PSGczDo)$ events, where each \PGt decays hadronically and the \PSGt has a simulated \cTau of 10\cm. The acceptance is shown for the displaced di-\tauh trigger path for 2022 data-taking conditions and is plotted with respect to the generator-level \PGt lepton decay vertex radial position. Selections on the visible component of the generator-level \PGt lepton \pt ($\pt(\PGt) > 30\GeV$), its $\eta$ ($\abs{\eta(\PGt)} < 2.1$), and its decay vertex radial position ($R < 115\cm$) are applied. 
\label{fig:displacedtau_mc_acceptance} 
}
\end{figure}

The acceptance of the displaced \tauh trigger in $\Pp\Pp \to \PSGt\PASGt\,(\PSGt \to \PGt\PSGczDo)$ events, where each \PGt lepton decays hadronically, is shown in Fig.~\ref{fig:displacedtau_mc_acceptance}.

\begin{figure}[htbp!]
\centering
\includegraphics[width=0.80\textwidth]{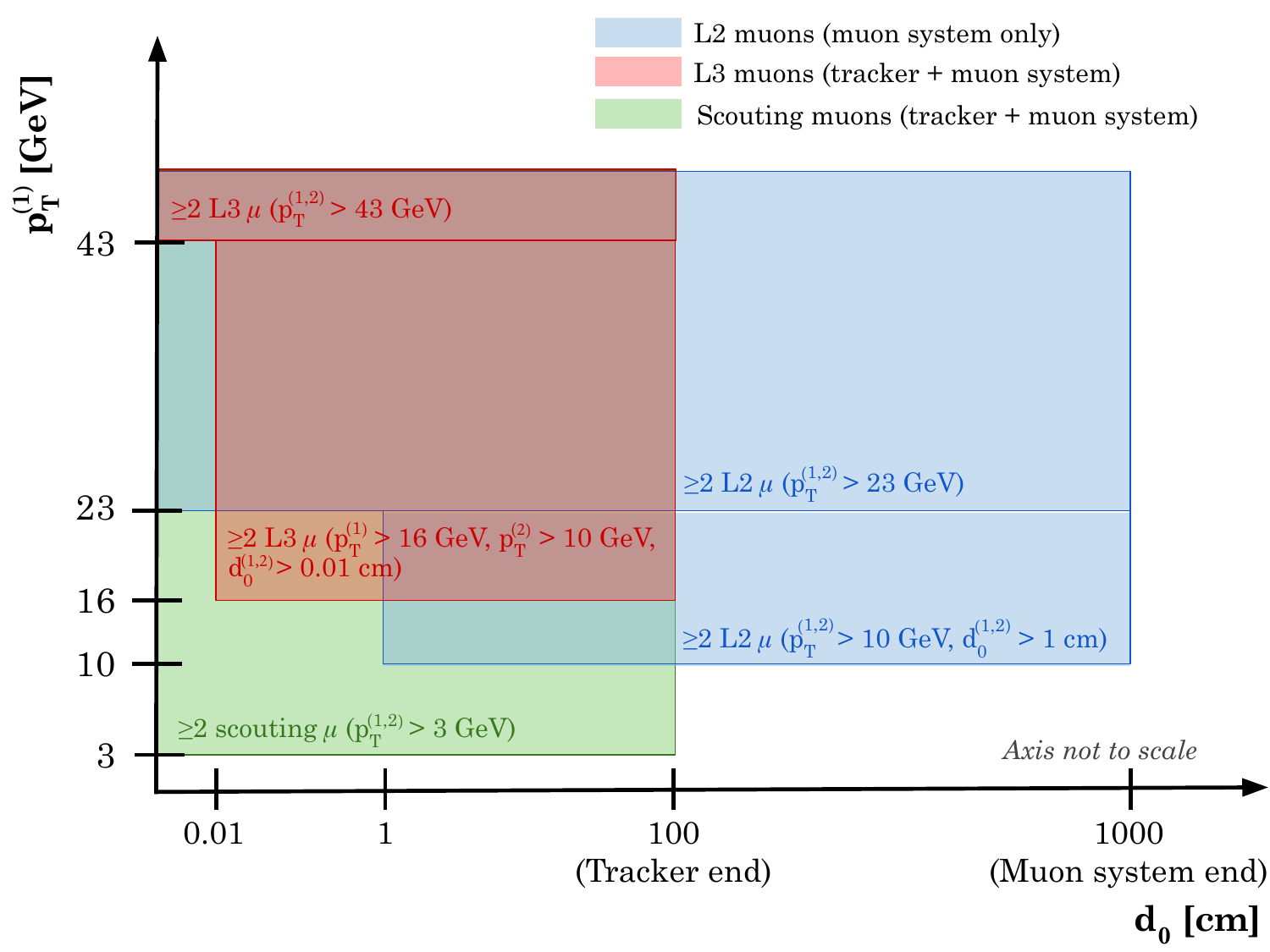}
\caption{Diagram of the overlapping coverage, in the HLT muon \pt-\dzero plane, of the dimuon triggers that target displaced signatures via L2 muons (shown in blue), L3 muons (shown in red), and scouting muons (shown in green). The displaced-dimuon triggers, the double displaced L3 muon triggers, and the dimuon scouting triggers are shown. The \pt of the highest \pt muon ($\pt^{1}$) is shown on the $y$-axis, and the minimum \pt thresholds on each triggered muon ($\pt^{1,2}$) are indicated within the colored parts of the diagram. The value of $100\,(1000)\cm$ given on the $x$ axis for the effective end of the tracker (muon system) is approximate.}
\label{fig:dimuon_coverage}
\end{figure}

Figure~\ref{fig:dimuon_coverage} shows the complementarity in coverage, in the HLT muon \pt-\dzero plane, of the various dimuon triggers that target displaced signatures. The displaced dimuon triggers (both L2 and L3 muon paths), the double displaced L3 muon triggers, and the dimuon scouting triggers are shown. The coverage of these triggers overlaps in this plane, but each trigger provides access to additional phase space.

\begin{figure}[htbp!]
\centering\includegraphics[width=0.49\textwidth]{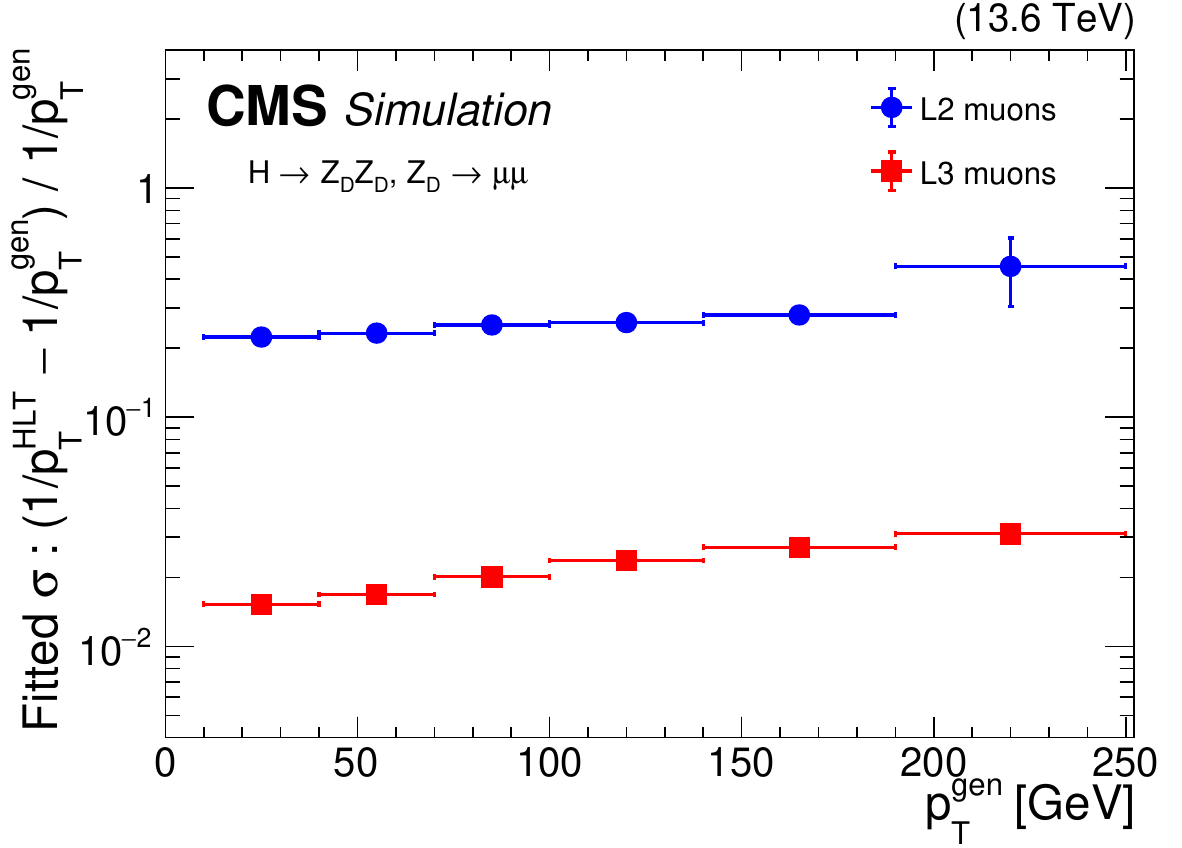}
\includegraphics[width=0.49\textwidth]{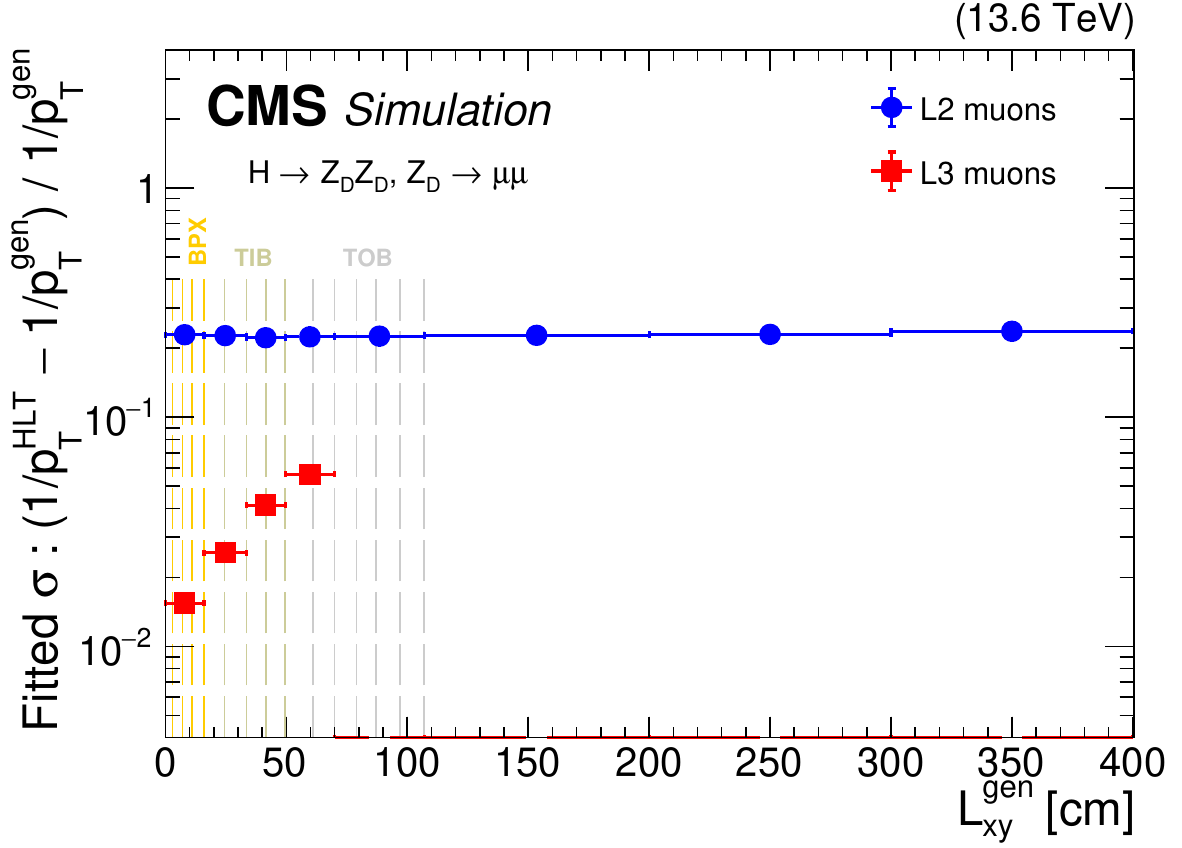}
\caption{Resolution of the inverse L2 (blue circles) and L3 (red squares) muon \pt ($(1/\pt^{\text{HLT}}-1/\pt^{\text{gen}})/(1/\pt^{\text{gen}})$) as a function of the generator-level muon \pt (\cmsLeft) and the generator-level \Lxy (\cmsRight), for simulated HAHM signal events, where the dark Higgs boson (\Hdark) mixes with the SM Higgs boson (\PH) and decays to a pair of long-lived dark photons (\PZD), for various values of \mZD and $\epsilon$. Conditions for 2022 data-taking are used. The muons must have $\pt>10\GeV$, and the L2 and L3 muons are geometrically matched to the generator-level muons. In the plot on the right, the dashed vertical lines indicate the radial positions of the layers of the tracking detectors, with BPX, TIB, and TOB denoting the barrel pixel, tracker inner barrel, and tracker outer barrel, respectively.}
\label{fig:HLTmuon_ptresolution}
\end{figure}

As explained in Section~\ref{sec:eventReconstruction}, while L2 muons provide higher efficiency than L3 muons for significantly displaced muons, L3 muons have a better \pt resolution. To study this matter, the resolution of the inverse HLT muon \pt, that is, $(1/\pt^{\text{HLT}}-1/\pt^{\text{gen}})/(1/\pt^{\text{gen}})$, is computed, where $\pt^{\text{HLT}}$ is the L2 or L3 muon \pt and $\pt^{\text{gen}}$ is the generator-level muon \pt. The generator-level muon is geometrically matched to an L2 or L3 muon. A Gaussian fit is performed for each distribution. The standard deviation of each Gaussian distribution, along with its uncertainty, is plotted in Fig.~\ref{fig:HLTmuon_ptresolution} as a function of the generator-level muon \pt (\cmsLeft) and of the generator-level \Lxy (\cmsRight), for L2 and L3 muons, in HAHM signal events. As shown in this figure, the \pt resolution of L3 muons (1.5--5\%, depending on the generator-level muon \pt or \Lxy) is about an order of magnitude better than that of L2 muons. Both the L2 and the L3 muon \pt resolutions degrade as the muon \pt increases, as expected. Also, as expected, the \pt resolution is relatively constant as a function of generator \Lxy for L2 muons. In contrast, for L3 muons, the \pt resolution degrades as \Lxy increases, within the tracker volume, \ie, up to around 70\cm.

The signal acceptances reported in this section are highly dependent on the trigger rate allocation, which must be determined in the context of the need for a balanced physics program and of the technical limitations of the readout bandwidth at CMS. For example, the delayed-jet triggers using ECAL timing (Section~\ref{sec:delayedjetsecal}) employ lower \pt and timing thresholds in the parked data-taking stream with respect to the standard data-taking stream, leading to rate increases by a factor of 2--3 and acceptance increases between 35 and 800\%, depending on the benchmark signal parameters. In many cases, relaxing trigger thresholds and thus increasing the trigger rate not only improves the acceptance for signals already passing that trigger but also expands the phase space coverage to new regimes. For example, lowering the minimum \pt thresholds for the displaced dimuon triggers from 43\GeV (Section~\ref{sec:delayedjetsecal}) to 3\GeV (Section~\ref{sec:delayedjetsecal}), which is done when going from the standard to the scouting data-taking stream, enables access to much lower mass LLPs.

\section{Long-lived particle triggers at the High-Luminosity LHC}
\label{sec:HL_LHC}

In the next few years, many more LLP analyses using these data and the triggers presented here will be completed. Future improvements, some potentially during Run~3, will further increase the sensitivity of searches for LLPs. The LLP triggers presented in this paper have laid the groundwork for these future expansions.

The HL-LHC, which is expected to begin operations in 2030, will include multiple CMS detector upgrades that will substantially enhance its performance~\cite{Contardo:2015bmq,CMS:2017lum,CERN-LHCC-2017-011,CMS:2017jpq,Hebbeker:2017bix,Zabi:2020gjd,CMS:2667167,CMS-DP-2022-025}. For example, timing resolutions of order $10\unit{ps}$ will be available across multiple upgraded and new subsystems, including the minimum ionizing particle timing detector and the high-granularity calorimeter, and this new forward calorimeter will also provide high-granularity energy and position information. In the L1T, tracking will be implemented, and sophisticated machine-learning algorithms will be deployed. In addition, the substantial increase in the integrated luminosity will enable us to probe rarer processes. Together, these improvements will significantly increase the sensitivity of the CMS experiment to LLPs. This improved sensitivity is demonstrated for the LLPs predicted by a wide range of BSM models, as shown in several studies of the physics performance at the HL-LHC~\cite{CMS:2022cju,Dainese:2019rgk,CMS:2022sfl}.

\section{Summary}
\label{sec:summary}

The CMS Run~3 long-lived particle (LLP) trigger program at the LHC, which includes many dedicated triggers designed to select different LLP signatures, has been presented. These dedicated LLP triggers extend the standard CMS trigger program and provide crucial access to the unconventional event signatures of LLPs. For 2022 data taking, some LLP triggers have been improved, such as the tracker-based displaced-jet, displaced-muon, and dimuon-scouting triggers. Other LLP triggers, such as the displaced-tau trigger, the delayed-jet trigger using the electromagnetic calorimeter timing, the displaced-jet trigger using the hadronic calorimeter, and the muon detector shower trigger, were newly added in 2022. The performance of these triggers has been shown for several new physics models and with 2022--2024 proton-proton collision data collected at $\sqrt{s}=13.6\TeV$.

These triggers greatly improve the CMS sensitivity to LLPs, including those predicted by theories of physics beyond the standard model, enhancing the ability of the experiment to search for new phenomena. At the time of writing, two CMS LLP analyses based on early Run~3 data collected with dedicated LLP triggers have been published: a search for displaced dimuons~\cite{EXO-23-014} and a search for displaced jets~\cite{displacedJets2022Data}. The Run~3 displaced dimuon search has comparable or better sensitivity than the previous Run~2 search at $\sqrt{s}=13\TeV$, with only 38\% of the integrated luminosity, and the Run~3 displaced jets search achieves an order-of-magnitude improvement over the previous Run~2 results. With the improvements in trigger performance and acceptance demonstrated in this paper, future Run~3 analyses are expected to outperform previous searches. The complementary nature of the trigger algorithms and their high signal acceptance will play a central role in probing a substantially larger LLP parameter space.

\begin{acknowledgments}
We congratulate our colleagues in the CERN accelerator departments for the excellent performance of the LHC and thank the technical and administrative staffs at CERN and at other CMS institutes for their contributions to the success of the CMS effort. In addition, we gratefully acknowledge the computing centers and personnel of the Worldwide LHC Computing Grid and other centers for delivering so effectively the computing infrastructure essential to our analyses. Finally, we acknowledge the enduring support for the construction and operation of the LHC, the CMS detector, and the supporting computing infrastructure provided by the following funding agencies: SC (Armenia), BMBWF and FWF (Austria); FNRS and FWO (Belgium); CNPq, CAPES, FAPERJ, FAPERGS, and FAPESP (Brazil); MES and BNSF (Bulgaria); CERN; CAS, MoST, and NSFC (China); MINCIENCIAS (Colombia); MSES and CSF (Croatia); RIF (Cyprus); SENESCYT (Ecuador); ERC PRG, TARISTU24-TK10 and MoER TK202 (Estonia); Academy of Finland, MEC, and HIP (Finland); CEA and CNRS/IN2P3 (France); SRNSF (Georgia); BMFTR, DFG, and HGF (Germany); GSRI (Greece); NKFIH (Hungary); DAE and DST (India); IPM (Iran); SFI (Ireland); INFN (Italy); MSIT and NRF (Republic of Korea); MES (Latvia); LMTLT (Lithuania); MOE and UM (Malaysia); BUAP, CINVESTAV, CONACYT, LNS, SEP, and UASLP-FAI (Mexico); MOS (Montenegro); MBIE (New Zealand); PAEC (Pakistan); MES, NSC, and NAWA (Poland); FCT (Portugal); MESTD (Serbia); MICIU/AEI and PCTI (Spain); MOSTR (Sri Lanka); Swiss Funding Agencies (Switzerland); MST (Taipei); MHESI (Thailand); TUBITAK and TENMAK (T\"{u}rkiye); NASU (Ukraine); STFC (United Kingdom); DOE and NSF (USA).

\hyphenation{Rachada-pisek} Individuals have received support from the Marie-Curie program and the European Research Council and Horizon 2020 Grant, contract Nos.\ 675440, 724704, 752730, 758316, 765710, 824093, 101115353, 101002207, 101001205, and COST Action CA16108 (European Union); the Leventis Foundation; the Alfred P.\ Sloan Foundation; the Alexander von Humboldt Foundation; the Science Committee, project no. 22rl-037 (Armenia); the Fonds pour la Formation \`a la Recherche dans l'Industrie et dans l'Agriculture (FRIA) and Fonds voor Wetenschappelijk Onderzoek contract No. 1228724N (Belgium); the Beijing Municipal Science \& Technology Commission, No. Z191100007219010, the Fundamental Research Funds for the Central Universities, the Ministry of Science and Technology of China under Grant No. 2023YFA1605804, the Natural Science Foundation of China under Grant No. 12535004, and USTC Research Funds of the Double First-Class Initiative No.\ YD2030002017 (China); the Ministry of Education, Youth and Sports (MEYS) of the Czech Republic; the Shota Rustaveli National Science Foundation, grant FR-22-985 (Georgia); the Deutsche Forschungsgemeinschaft (DFG), among others, under Germany's Excellence Strategy -- EXC 2121 ``Quantum Universe" -- 390833306, and under project number 400140256 - GRK2497; the Hellenic Foundation for Research and Innovation (HFRI), Project Number 2288 (Greece); the Hungarian Academy of Sciences, the New National Excellence Program - \'UNKP, the NKFIH research grants K 131991, K 133046, K 138136, K 143460, K 143477, K 146913, K 146914, K 147048, 2020-2.2.1-ED-2021-00181, TKP2021-NKTA-64, and 2025-1.1.5-NEMZ\_KI-2025-00004 (Hungary); the Council of Science and Industrial Research, India; ICSC -- National Research Center for High Performance Computing, Big Data and Quantum Computing, FAIR -- Future Artificial Intelligence Research, and CUP I53D23001070006 (Mission 4 Component 1), funded by the NextGenerationEU program (Italy); the Latvian Council of Science; the Ministry of Education and Science, project no. 2022/WK/14, and the National Science Center, contracts Opus 2021/41/B/ST2/01369, 2021/43/B/ST2/01552, 2023/49/B/ST2/03273, and the NAWA contract BPN/PPO/2021/1/00011 (Poland); the Funda\c{c}\~ao para a Ci\^encia e a Tecnologia, grant CEECIND/01334/2018 (Portugal); the National Priorities Research Program by Qatar National Research Fund; MICIU/AEI/10.13039/501100011033, ERDF/EU, ``European Union NextGenerationEU/PRTR", and Programa Severo Ochoa del Principado de Asturias (Spain); the Chulalongkorn Academic into Its 2nd Century Project Advancement Project, the National Science, Research and Innovation Fund program IND\_FF\_68\_369\_2300\_097, and the Program Management Unit for Human Resources \& Institutional Development, Research and Innovation, grant B39G680009 (Thailand); the Kavli Foundation; the Nvidia Corporation; the SuperMicro Corporation; the Welch Foundation, contract C-1845; and the Weston Havens Foundation (USA).  
\end{acknowledgments}

\bibliography{auto_generated}

\clearpage
\appendix

\section{Glossary of acronyms}
\label{sec:glossary}

\begin{longtable}[l]{ll}
3D & Three-dimensional \\
AMSB & Anomaly-mediated supersymmetry breaking\\
BMTF & Barrel muon track finder\\
BPTX & Beam pickup timing device\\
BSM & Beyond the standard model\\
BX & Bunch slot\\
CMS & Compact Muon Solenoid\\
CSC & Cathode strip chamber\\
CPU & Central processing unit\\
DT & Drift tube\\
ECAL & Electromagnetic calorimeter\\
EMTF & Endcap muon track finder\\
FPGA & Field-programmable gate array\\
GEM & Gas electron multiplier\\
GMSB & Gauge-mediated supersymmetry breaking\\
GPU & Graphical processing unit\\
HAHM & Hidden Abelian Higgs model\\
HCAL & Hadronic calorimeter\\
HB & Hadronic calorimeter barrel\\
HL-LHC & High-Luminosity LHC\\
HLT & High level trigger\\
HPS & Hadron plus strip\\
IP & Interaction point\\
ISR & Initial state radiation\\
L1 & First level\\
L1T & First level trigger\\
LHC & Large Hadron Collider\\
LLP & Long-lived particle\\
LO & Leading order\\
MDS & Muon detector shower\\
NHEF & Neutral hadron energy fraction\\
NLO & Next-to-leading order\\
NN & Neural network\\
OMTF & Overlap muon track finder\\
PDF & Parton distribution function\\
PF & Particle flow\\
PU & Pileup\\
PUPPI & Pileup per particle identification\\
PV & Primary vertex\\
QCD & Quantum chromodynamics\\
RPC & Resistive-plate chamber\\
SM & Standard model\\
STA & Muons reconstructed using only the muon system (standalone)\\
SUSY & Supersymmetry\\
TDC & Time-to-digital converter\\
TMS & Muons reconstructed using both the tracker and the muon system\\
\end{longtable}
\cleardoublepage \section{The CMS Collaboration \label{app:collab}}\begin{sloppypar}\hyphenpenalty=5000\widowpenalty=500\clubpenalty=5000\input{EXO-23-016-public-authorlist.tex}\end{sloppypar}
\end{document}

%% file: EXO-23-016-public-authorlist.tex
\cmsinstitute{Yerevan Physics Institute, Yerevan, Armenia}
{\tolerance=6000
A.~Hayrapetyan, V.~Makarenko\cmsorcid{0000-0002-8406-8605}, A.~Tumasyan\cmsAuthorMark{1}\cmsorcid{0009-0000-0684-6742}
\par}
\cmsinstitute{Institut f\"{u}r Hochenergiephysik, Vienna, Austria}
{\tolerance=6000
W.~Adam\cmsorcid{0000-0001-9099-4341}, J.W.~Andrejkovic, L.~Benato\cmsorcid{0000-0001-5135-7489}, T.~Bergauer\cmsorcid{0000-0002-5786-0293}, M.~Dragicevic\cmsorcid{0000-0003-1967-6783}, C.~Giordano, P.S.~Hussain\cmsorcid{0000-0002-4825-5278}, M.~Jeitler\cmsAuthorMark{2}\cmsorcid{0000-0002-5141-9560}, N.~Krammer\cmsorcid{0000-0002-0548-0985}, A.~Li\cmsorcid{0000-0002-4547-116X}, D.~Liko\cmsorcid{0000-0002-3380-473X}, M.~Matthewman, I.~Mikulec\cmsorcid{0000-0003-0385-2746}, J.~Schieck\cmsAuthorMark{2}\cmsorcid{0000-0002-1058-8093}, R.~Sch\"{o}fbeck\cmsAuthorMark{2}\cmsorcid{0000-0002-2332-8784}, D.~Schwarz\cmsorcid{0000-0002-3821-7331}, M.~Shooshtari\cmsorcid{0009-0004-8882-4887}, M.~Sonawane\cmsorcid{0000-0003-0510-7010}, W.~Waltenberger\cmsorcid{0000-0002-6215-7228}, C.-E.~Wulz\cmsAuthorMark{2}\cmsorcid{0000-0001-9226-5812}
% ...
\par}
\vskip\cmsinstskip
\dag:~Deceased\\
$^{1}$Also at Yerevan State University, Yerevan, Armenia\\
$^{2}$Also at TU Wien, Vienna, Austria\\
% ...
